\documentclass[aps,prx,superscriptaddress,amsfonts,amsmath,amssymb,showpacs,floatfix,reprint,longbibliography]{revtex4-1}

\usepackage{url}
\usepackage{bm}
\usepackage{graphicx}
\usepackage{amsmath}
\usepackage{amstext}
\usepackage{amssymb}
\usepackage{amsfonts}
\usepackage{amsbsy}
\usepackage{verbatim}
\usepackage{color}
\usepackage[colorlinks=true, urlcolor=blue, linkcolor=blue, citecolor=blue, pdftex]{hyperref}
\usepackage{multirow}
\usepackage{floatrow}
\usepackage{float}
\usepackage{gensymb}
\usepackage{textcomp}
\usepackage{enumitem}
\usepackage[version=3]{mhchem}
\usepackage{afterpage}
\usepackage{pbox}
\usepackage{makecell}
\usepackage{array,booktabs}
\usepackage{dsfont}
\usepackage{siunitx}

\newcommand{\na}{NaCaNi$_{2}$F$_{7}$}
\newcommand{\naco}{NaCaCo$_2$F$_7$}
\newcommand{\nafe}{NaCaFe$_2$F$_7$}
\newcommand{\srfe}{NaSrFe$_2$F$_7$}
\newcommand{\srmn}{NaSrMn$_2$F$_7$}

\begin{document}

\title{Quantum and Classical Phases of the Pyrochlore Heisenberg Model with Competing Interactions}

\author{Yasir Iqbal}
\email{yiqbal@physics.iitm.ac.in}
\affiliation{Department of Physics, Indian Institute of Technology Madras, Chennai 600036, India}
\author{Tobias M\"uller}
\affiliation{Institute for Theoretical Physics and Astrophysics, Julius-Maximilian's
University of W\"urzburg, Am Hubland, D-97074 W\"urzburg, Germany}
\author{Pratyay Ghosh}
\affiliation{Department of Physics, Indian Institute of Technology Madras, Chennai 600036, India}
\author{Michel J. P. Gingras}
\affiliation{Perimeter Institute for Theoretical Physics, Waterloo, Ontario, Canada N2L 5G7}
\affiliation{Department of Physics and Astronomy, University of Waterloo, Waterloo, Ontario, Canada N2L 3G1}
\affiliation{Quantum Materials Program, Canadian Institute for Advanced Research, MaRS Centre, West Tower 661 University Avenue, Suite 505, Toronto, Ontario, M5G 1M1, Canada}
\author{Harald O. Jeschke}
\affiliation{Research Institute for Interdisciplinary Science, Okayama University, 3-1-1 Tsushima-naka, Kita-ku, Okayama 700-8530, Japan}
\author{Stephan Rachel}
\affiliation{School of Physics, The University of Melbourne, Parkville, Victoria 3010, Australia}
\affiliation{Institut f\"ur Theoretische Physik, Technische Universit\"at Dresden, D-01069 Dresden, Germany}
\author{Johannes Reuther}
\affiliation{Dahlem Center for Complex Quantum Systems and Fachbereich Physik, Freie Universit{\"a}t Berlin, D-14195 Berlin, Germany}
\affiliation{Helmholtz-Zentrum Berlin f{\"u}r Materialien und Energie, D-14109 Berlin, Germany}
\author{Ronny Thomale}
\affiliation{Institute for Theoretical Physics and Astrophysics, Julius-Maximilian's
University of W\"urzburg, Am Hubland, D-97074 W\"urzburg, Germany}

\date{\today}

\begin{abstract}
We investigate the quantum Heisenberg model on the pyrochlore lattice for a generic spin $S$ in the presence of nearest-neighbor $J_{1}$ and second-nearest-neighbor $J_{2}$ exchange interactions. By employing the pseudofermion functional renormalization group method, we find, for $S=1/2$ and $S=1$, an extended quantum-spin-liquid phase centered around $J_{2}=0$, which is shown to be robust against the introduction of breathing anisotropy. The effects of temperature, quantum fluctuations, breathing anisotropies, and a $J_{2}$ coupling on the nature of the scattering profile, and the pinch points, in particular, are studied. For the magnetic phases of the $J_{1}$-$J_{2}$ model, quantum fluctuations are shown to renormalize phase boundaries compared to the classical model and to modify the ordering wave vectors of spiral magnetic states, while no new magnetic orders are stabilized. 
\end{abstract}

\maketitle

\section{Introduction} 
The classical nearest-neighbor Heisenberg antiferromagnet on the pyrochlore lattice stands as an epitome of geometric frustration in three dimensions as shown by its failure to develop magnetic long-range order down to absolute zero temperature, realizing what has been dubbed a ``cooperative paramagnet''~\cite{Villain-1979}. This failure is a consequence of the {\it extensive} classical ground-state degeneracy~\cite{Villain-1979,Reimers-1991a,Moessner-1998a,Moessner-1998b} which proves severe enough to prevent a thermal ``order-by-disorder'' mechanism~\cite{Villain-1980,Shender-1982,Henley-1989} from selecting a unique ground-state ordering pattern~\cite{Reimers-1992,Zinkin-1996,Moessner-1998a,Moessner-1998b}. In contrast to thermal fluctuations, the impact of quantum fluctuations remains much less understood and constitutes a critically outstanding problem. In the regime of large spin $S$, using an effective Hamiltonian approach~\cite{Henley-2001}, it is known that at harmonic order in $1/S$, the extensive classical ground-state degeneracy exp[$\mathcal{O}(L^{3})$] ($L$ is the linear dimension of the system) is {\it partly} lifted, yielding a subset of collinear states with a massive, albeit subextensive, degeneracy exp[$\mathcal{O}(L)$]~\cite{Sobral-1997,Tsunetsugu-2002,Henley-2006,Hizi-2006}. It turns out that the consideration of higher-order  terms in a  $1/S$ expansion also fails to select a unique ground state~\cite{Hizi-2007}. Indeed, while quartic corrections in boson operators do break the degeneracy of the harmonic ground states, there still remains a family of (almost) degenerate (exp[$\mathcal{O}(L)$]) states~\cite{Hizi-2009}. Thus, the fate of the semiclassical ($1/S$) approach remains unsettled due to weak selection effects at the anharmonic level. In the opposite extreme quantum limit of small $S$, there is reasonably strong evidence for a quantum paramagnetic ground state. Investigations of the $S=1/2$ antiferromagnet claim for either a valence-bond crystal~\cite{Harris-1991,Isoda-1998,Koga-2001,Tsunetsugu-2001a,Tsunetsugu-2001b,Berg-2003,Tchernyshyov-2006,Moessner-2006} or a quantum-spin-liquid~\cite{Canals-1998,Canals-2000,Canals-2001,Fouet-2003,Kim-2008,Burnell-2009,Huang-2016} ground state. We note that a $J_{1}$-$J_{2}$-$J_{3}$ $S=1/2$ model derived from a strong-coupling expansion of a one-band half-filled Hubbard model on the pyrochlore lattice has been proposed to host a quantum spin liquid~\cite{Normand-2014,Normand-2016}. In the much-less-investigated case of $S=1$~\cite{Garcia-2000,Yamashita-2000,Koga-2001,Tsunetsugu-2017}, there have been suggestions of a ground state with tetrahedral symmetry breaking~\cite{Yasufumi-2001}.  

The ``cooperative paramagnet'' ground state of the classical nearest-neighbor Heisenberg antiferromagnet is known to be extremely fragile, in that magnetic long-range order is induced upon the inclusion of various perturbations, such as further neighbor Heisenberg interactions~\cite{Reimers-1991a,Ioki-2007,Nakamura-2007,Chern-2008,Okubo-2011}, dipole interactions~\cite{Palmer-2000}, Dzyaloshinsky-Moriya anisotropy~\cite{Elhajal-2005,Chern-2010}, single-ion anisotropy~\cite{Bramwell-1994,Moessner-1998c}, lattice distortions~\cite{Terao-1996,Tchernyshyov-2002,Pinettes-2002,Tchernyshyov-2004,Chern-2006,Bergman-2006}, and bond disorder~\cite{Castella-2001,Saunders-Chalker,Andreanov-Chalker}. In particular, further neighbor Heisenberg interactions are found to stabilize a plethora of intricate classical magnetic orders~\cite{Lapa-2012,Tymoshenko-2017}. However, in the low-spin-$S$ regime, where the strong possibility of a quantum paramagnetic ground state for the nearest-neighbor \emph{quantum} Heisenberg antiferromagnet exists, the impact of the above-mentioned perturbations on the paramagnet remains largely unexplored. This topic is of high significance and importance when considering the behavior of real materials. In this paper, we carry out a broad investigation of the $J_{1}$-$J_{2}$ Heisenberg model for a generic spin $S$ on the pyrochlore lattice:

\begin{figure}
\includegraphics[width=\columnwidth]{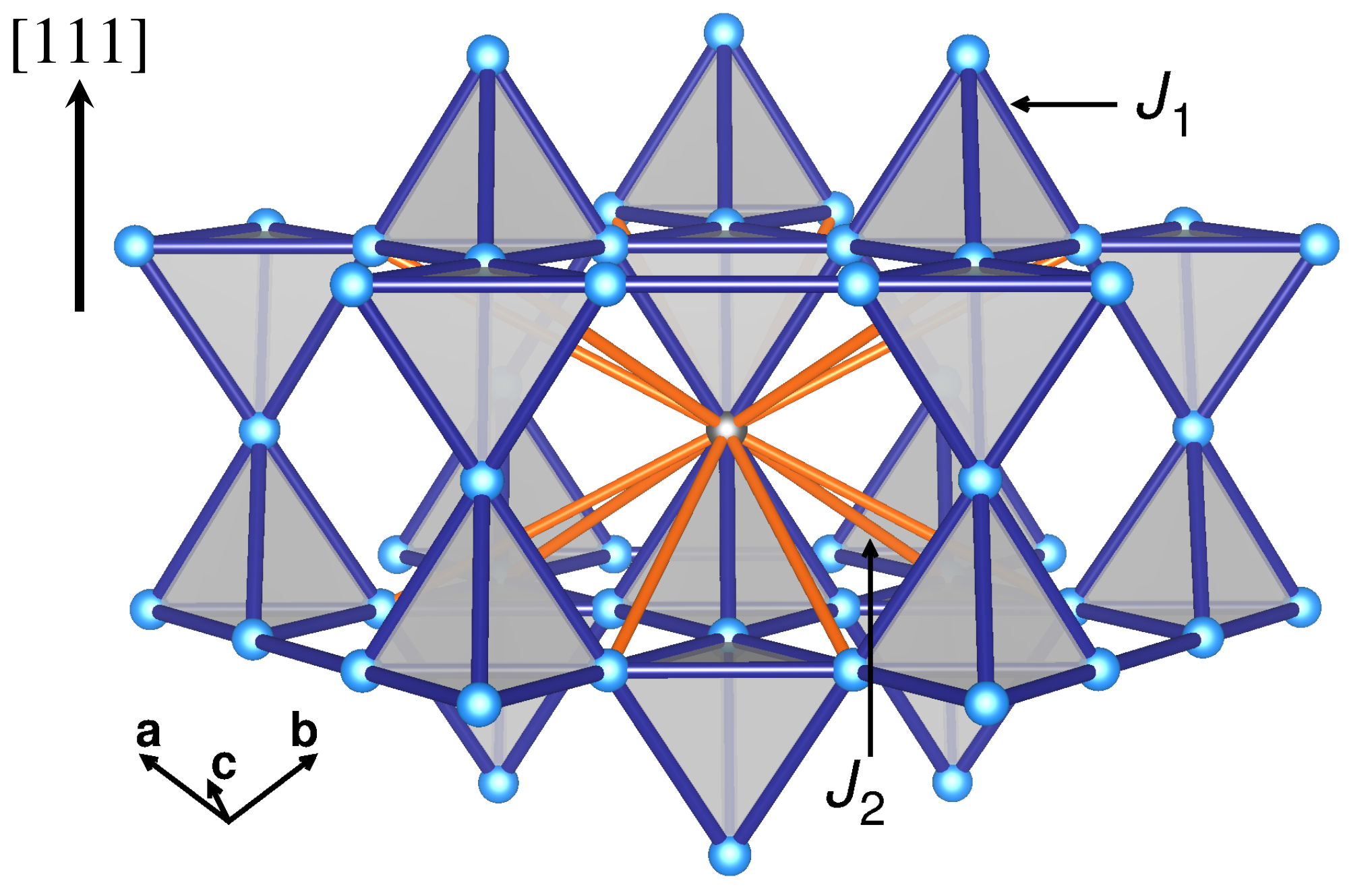}
\caption{The nearest-neighbor ($J_{1}$) and next-nearest-neighbor ($J_{2}$) bonds in the pyrochlore lattice.}\label{fig:structure}
\end{figure}

\begin{equation}\label{eqn:Ham1}
{\cal \hat{H}}=J_1 \sum_{{\langle i,j \rangle}} \mathbf{\hat{S}}_{i} \cdot \mathbf{\hat{S}}_{j}
+J_2 \sum_{{\langle\langle i,j \rangle\rangle}} \mathbf{\hat{S}}_{i} \cdot \mathbf{\hat{S}}_{j},
\end{equation}
where ${\mathbf{\hat{S}}}_{i}$ is a quantum spin-$S$ operator at a pyrochlore lattice site $i$. The indices $\langle i,j \rangle$ and $\langle\langle i,j \rangle\rangle$ denote sums over nearest-neighbor and second-nearest-neighbor pairs of sites, respectively [see Fig.~\ref{fig:structure}]. The investigation of the low-temperature properties of this Hamiltonian in the small-$S$ regime is notoriously difficult. This is a methodological challenge for which numerically exact and unbiased methods are not yet available. Indeed, traditional quantum many-body numerical methods such as density-matrix renormalization group and tensor network approaches~\cite{Schollwoeck-2005,Stoudenmire-2012}, while successful in one and two dimensions, become unfeasible in three dimensions due to entanglement scaling and system-size limitations. Quantum Monte Carlo methods~\cite{Reger-1988,Sandvik-1991}, while able to reach sufficiently large system sizes, are, in principle, restricted to unfrustrated systems, while variational Monte Carlo approaches~\cite{McMillan-1965,Ceperley-1977}, which are shown to be extremely successful in two dimensions~\cite{Iqbal-2013,Iqbal-2011b,Iqbal-2014}, require very large correlation volumes to extract reliable estimates in the thermodynamic limit. Finally, the bold diagrammatic Monte Carlo method can reach down only to moderately low temperatures~\cite{Huang-2016}. Thus, one is essentially left with only mean-field approaches based on Schwinger bosons~\cite{Arovas-1988}, semiclassical analysis based on spin waves, or linked-cluster expansion methods~\cite{Gelfand-2000}, which capture magnetic order accurately but are unsuitable for studying paramagnetic behavior deep in the collective paramagnetic (spin-liquid) regime. In this respect, the pseudofermion functional renormalization group (PFFRG) framework has an important feature in the form of a built-in balance towards the treatment of ordering and disordering tendencies for three-dimensional frustrated magnets~\cite{Iqbal-2016b}. 

By employing PFFRG for the spin-$S$ $J_{1}$-$J_{2}$ Heisenberg model, we find for $S=1/2$ an extended quantum-spin-liquid regime centered around $J_{2}=0$, with an extent of $-0.25(3)\leqslant J_{2}/J_{1}\leqslant 0.22(3)$ while, for $S=1$, its span is reduced by approximately a factor of 2, $-0.11(2)\leqslant J_{2}/J_{1}\leqslant0.09(2)$. For $S=1/2$ and $S=1$, the spin susceptibility profile of the nearest-neighbor antiferromagnet in the $[hhl]$ plane features a bow-tie pattern, characteristic of the well-known Coulomb spin-liquid phase~\cite{Zinkin-1997}. The bow ties are found to be robust up to temperatures $T/J_{1}\sim 1$. However, the inclusion of even a small $J_{2}$ coupling is shown to shift the spectral weight away from the pinch points, causing the bow ties to rapidly disappear upon cooling, similar to the findings for the corresponding classical model~\cite{Conlon-2010}. In the opposite limit of large $S$, quantum fluctuations lift the extensive degeneracy of the classical ground-state manifold either only partially to a subextensive one or completely (which would then potentially induce long-range magnetic ordering). The $J_{1}$-$J_{2}$ parameter space is known to host seven different classical magnetic orders~\cite{Lapa-2012}, which we also find in the $S=1/2$ model. Moreover, we show that quantum fluctuations do not stabilize any new phases, such as long-range dipolar or quadrupolar magnetic orders, and valence-bond-crystal states.
  
The paper is organized as follows: In Sec.~\ref{sec:methods}, we describe the PFFRG method (Sec.~\ref{sec:FRGA}) employed for the quantum treatment of the model, starting with a description of its formalism (Sec.~\ref{sec:formalism}) followed by details of its numerical implementation in Sec.~\ref{sec:numerical}. In Secs.~\ref{sec:LT} and \ref{sec:IMCH}, we discuss schemes used to obtain the ground state of classical spin models, namely, the Luttinger-Tisza method [Sec.~\ref{sec:LT}] and the iterative minimization of the energy [Sec.~\ref{sec:IMCH}] (the reader interested mainly in the results can directly jump to Secs.~\ref{sec:NNAF} and \ref{sec:J1J2}). Employing these methods, we begin with a treatment of the ground-state and low-energy physics of the nearest-neighbor Heisenberg antiferromagnet in Sec.~\ref{sec:NNAF}, starting first with a classical analysis [Sec.~\ref{sec:classical}] of the isotropic and breathing lattices and then moving on to the quantum treatment of the $S=1/2$ [Sec.~\ref{sec:S=1/2}] and $S=1$ [Sec.~\ref{sec:S=1}] models for both isotropic and breathing lattices. Finally, the section ends by addressing the problem of the ground state of the large-$S$ quantum Heisenberg antiferromagnet [Sec.~\ref{sec:large-S}]. Next, in Sec.~\ref{sec:J1J2}, we deal with the $J_{1}$-$J_{2}$ Heisenberg model, by first revisiting the classical phase diagram [Sec.~\ref{sec:J1J2-classical}], and subsequently present the results for the quantum model in Sec.~\ref{sec:J1J2-quantum}. We also discuss the impacts of quantum fluctuations on the nature of phases and phase boundaries. We end the paper with a summary of the results in Sec.~\ref{sec:summary}, followed by an outlook and discussion of future directions in Sec.~\ref{sec:outlook}.
  
\section{Methods}
\label{sec:methods}

\subsection{Pseudofermion functional renormalization group method}\label{sec:FRGA}

\subsubsection{Formalism}\label{sec:formalism}

\begin{figure*}[t]
\includegraphics[width=0.9\linewidth]{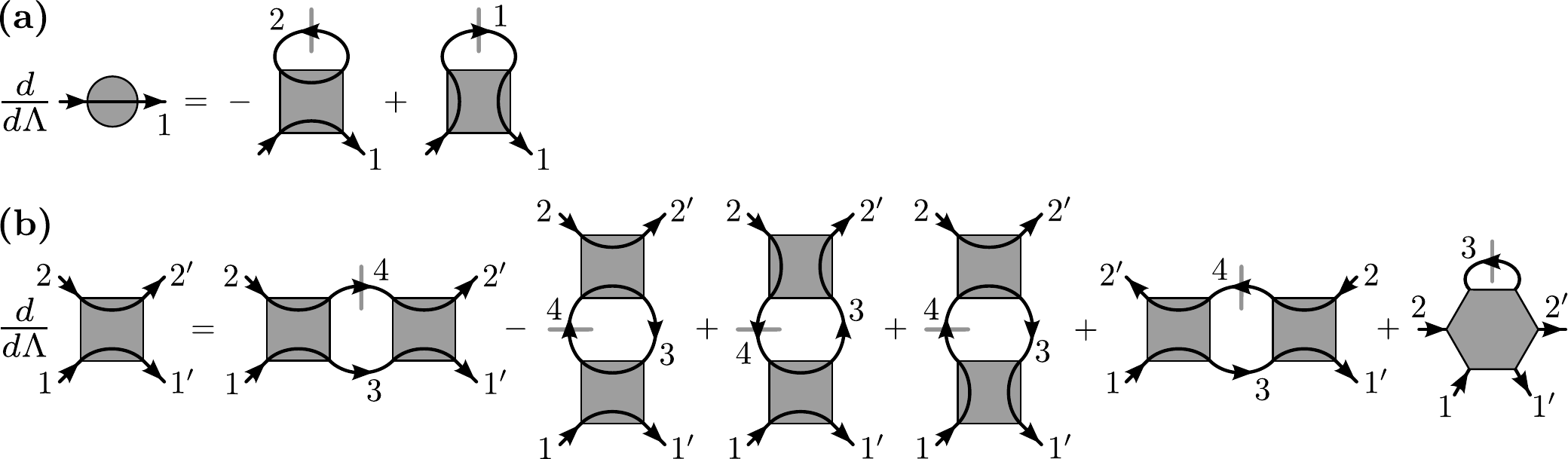}
\caption{Diagrammatic representation of the PFFRG equations for (a) the self-energy $\Sigma^\Lambda(i\omega)$ (gray disk) and (b) the two-particle vertex $\Gamma^\Lambda(1',2';1,2)$ (gray squares). Arrows denote the fully dressed propagator $G^\Lambda(i\omega)$, and slashed arrows denote the single-scale propagator $S^\Lambda(i\omega)$. The gray hexagon in (b) is the three-particle vertex. Note that the right-hand side of (b) contains additional terms where the slashes in the first to fifth terms appear in the respective other propagator. For a spin-$S$ generalization, the first term on the right-hand side of (a) and the second term in (b) are multiplied with a factor of $2S$. \label{fig:frg}}
\end{figure*}

The key idea of the PFFRG method~\cite{Reuther-2010} is to express the spin-1/2 operators in terms of pseudofermions~\cite{Abrikosov-1965},
\begin{equation}
 \hat{S}^{\mu}_i=\frac{1}{2}\sum\limits_{\alpha,\beta} \hat{f}^{\dagger}_{i\alpha}\sigma^{\mu}_{\alpha\beta}\hat{f}_{i\beta},\label{pseudo_fermions}
\end{equation}
where $\sigma^{\mu}_{\alpha\beta}$ are Pauli matrices ($\mu\in\{x,y,z\}$) and $\hat{f}_{i\alpha}$ ($\hat{f}^{\dagger}_{i\alpha}$) denote spin-$\alpha$ fermionic annihilation (creation) operators. For the implementation for spin systems with local $S>1/2$ spins, we adopt the approach of Ref.~\cite{Baez-2017}, where multiple copies of spin-1/2 degrees of freedom are introduced at each lattice site; i.e., the local spin operators are replaced by
\begin{equation}
\mathbf{\hat{S}}_i\to \sum_{\kappa=1}^M \mathbf{\hat{S}}_{i\kappa}\;,\label{sum}
\end{equation}
while the couplings $J_{ij}$ remain independent of the fermion ``flavor'' $\kappa$. If all individual $\mathbf{\hat{S}}_{i\kappa}$ ``spins'' ( $\kappa\in\{1,\ldots,M\}$) align ferromagnetically (see below for details), they realize the largest possible magnitude $S=M/2$ on each site, thus implementing the desired effective magnetic moment. In terms of pseudofermions, the substitution in Eq.~(\ref{sum}) amounts to equipping the fermion operators with an additional index $\kappa$:
\begin{equation}
\hat{S}_{i\kappa}^{\mu}=\frac{1}{2}\sum_{\alpha \beta}\hat{f}_{i\alpha\kappa}^{\dagger}\sigma_{\alpha\beta}^{\mu}\hat{f}_{i\beta\kappa}.\;\label{M_pseudo_fermions}
\end{equation}
Pseudofermionic representations for spin operators generally require some caution, since they introduce additional spurious states with zero ($Q_i\equiv f^{\dagger}_{i\uparrow}f_{i\uparrow}+f^{\dagger}_{i\downarrow}f_{i\downarrow}=0$) or two ($Q_i=2$) fermions at a site $i$. Such states carry no spin ($S=0$), and the physical spin-$1/2$ degrees of freedom are realized in the singly occupied subspace with $Q_i=1$. The pseudofermionic approach is guaranteed to be faithful only if the contribution from the $S=0$ states is negated. For a proper implementation of spins $S>1/2$, one additionally needs to ensure that the spin flavors $\kappa$ combine to the largest local moment $S=M/2$ while smaller spins with $S=M/2-1,\ldots$ are eliminated from the Hilbert space. A convenient approach that simultaneously fulfills both constraints is to add an on-site local level repulsion term $A (\sum_{\kappa=1}^M\mathbf{\hat{S}}_{i\kappa})^2$ to the Hamiltonian. For negative $A$, this term reduces the energies of all levels with finite magnetic moments, where the largest reduction occurs in the sector with the highest spin. An $|A|$ chosen sufficiently large guarantees that the low-energy subspace of the Hamiltonian is the one without any nonoccupied or doubly occupied states for each $\kappa$. Furthermore, the $M$ spin-$1/2$ copies combine into an effective spin $S=M/2$. We emphasize, however, that, for the ground states of generic Heisenberg spin models (such as the pyrochlore systems studied here), a vanishing level repulsion term $A=0$ turns out to be sufficient to fulfill both pseudoparticle constraints. This simplification is because, for two-body spin interactions, the energy naturally scales with the spin length squared such that the largest local moment is energetically favored even for $A=0$ (note, however, that counterexamples can be constructed~\footnote{ Heisenberg systems (on any lattice) with $S=1$ and single-ion anisotropies $\Delta \sum_i (S_i^z)^2$ provide a simple exception wherein if $\Delta$ is positive (and
sufficiently large) this term would always energetically prefer the
unphysical spin sector $S=0$ over all other sectors.}). 

Rewriting the spin Hamiltonian in terms of Eq.~(\ref{M_pseudo_fermions}), the resulting fermionic model is treated within the standard FRG framework for interacting fermion systems~\cite{Wetterich-1993,Metzner-2012,Platt-2013}. A somewhat unusual situation arises here: the system is purely quartic in the fermions without any quadratic kinetic terms that could be used as a noninteracting starting point in a perturbative expansion. Within FRG, this situation is addressed by summing up infinite-order diagrammatic contributions in different interaction channels as well as accounting for vertex corrections between them. Particularly, as explained in more detail below, the summation is such that, in the large-$S$ and the large-$N$ limits, where $N$ generalizes the spin symmetry group from SU$(2)$ to SU$(N)$, the leading diagrammatic contributions in $1/S$ and $1/N$ are both treated \emph{exactly}~\footnote{The class of diagrams representing the \emph{leading order} in $1/S$ contributions, which are thus of random phase approximation (RPA)-type and responsible for the formation of classical magnetic order, are summed up \emph{exactly}. Similarly, the class of diagrams capturing contributions to \emph{leading order} in $1/N$, and thus responsible for the formation of nonmagnetic states, are also summed up \emph{exactly}. However, an accurate treatment of the $S\to\infty$ limit may require a consideration of subleading terms in $1/S$, thus going beyond a bare RPA treatment [see Appendix~\ref{appendix1}]}. As a consequence, classical magnetically ordered states (typically favored at large $S$) and nonmagnetic spin liquids or dimerized states (as obtained at large $N$)~\cite{Sachdev-1991} may both be described within the same methodological framework.

Because of the absence of fermion kinetic hopping terms, the bare fermionic propagator is strictly local and takes the simple spin-independent form
\begin{equation}
G_0(i\omega)=\frac{1}{i\omega}\;,
\end{equation}
where $i\omega$ denotes a frequency on the imaginary Matsubara axis. Within the standard PFFRG scheme~\cite{Reuther-2010}, this propagator is dressed with an infrared steplike regulator function:
\begin{equation}
 \label{eq:LambdaGF}
G_0(i\omega)\longrightarrow G_0^{\Lambda}(i\omega)=\frac{\Theta\left ( \left | \omega  \right | - \Lambda \right) }{i\omega},
\end{equation}
which interpolates between the limits $\Lambda\to \infty$ (where the fermionic propagation is completely suppressed) and the original cutoff-free theory at $\Lambda=0$. This modification generates a $\Lambda$ dependence of all one-particle irreducible $m$-particle vertex functions as described by the FRG flow equations. For the self-energy $\Sigma^\Lambda(i\omega)$ and the two-particle vertex $\Gamma^\Lambda(1',2';1,2)$ (the label ``X'' stands for site, frequency, and spin variables, respectively, i.e., ${\rm X}\equiv\{i,i\omega,\alpha\})$. A diagrammatic version of these equations is illustrated in Fig.~\ref{fig:frg}, where the arrows denote dressed and $\Lambda$-dependent propagators
\begin{equation}
G^\Lambda(i\omega)=\frac{\Theta\left ( \left | \omega  \right | - \Lambda \right)}{i\omega-\Sigma^\Lambda(i\omega)}
\end{equation}
and slashed lines denote the single-scale propagator
\begin{equation}
S^\Lambda(i\omega)=\frac{\delta\left ( \left | \omega  \right | - \Lambda \right)}{i\omega-\Sigma^\Lambda(i\omega)}\;.
\end{equation}
Because of the locality of fermion propagators, the two-particle vertex $\Gamma^\Lambda(1',2';1,2)$ effectively depends on {\it two} site indices only, i.e., $\Gamma^\Lambda(1',2';1,2)\sim\delta_{i_1 i_{1'}}\delta_{i_2 i_{2'}}$. As illustrated in Fig.~\ref{fig:frg}, this restriction allows one to connect incoming and outgoing arrows of $\Gamma^\Lambda(1',2';1,2)$ in a way that on-site variables remain constant along fermion lines.

The FRG equations in Fig.~\ref{fig:frg} show a systematic interplay between the RG flows of different vertex functions where the $\Lambda$ derivative of each $m$-particle vertex couples to all $m'$-particle vertices with $m'\leqslant m+1$. To reduce this infinite hierarchy of intertwined equations to a finite and numerically solvable set, we neglect the three-particle vertex in Fig.~\ref{fig:frg}(b) albeit not in entirety, as certain three-loop terms obtained from the Katanin truncation scheme are included and which amount to self-energy corrections~\cite{Katanin-2004}, as described below; however, all higher vertices are completely discarded. However, this approximation effectively amounts to discarding three-body spin correlations such that the description of spin phases with chiral order parameters $\langle \mathbf{\hat{S}}_i \cdot (\mathbf{\hat{S}}_j\times\mathbf{\hat{S}}_k)\rangle$ is not possible~\cite{Reuther-2014b}. Still, parts of the three-particle vertex can be included by applying the so-called Katanin truncation~\cite{Katanin-2004}, which replaces the single scale propagator by the full $\Lambda$ derivative of the dressed propagator
\begin{equation}
 S^{\Lambda}\longrightarrow -\frac{d}{d\Lambda}G^{\Lambda}=S^{\Lambda}-\left(G^{\Lambda}\right)^2\frac{d}{d\Lambda}\Sigma^{\Lambda}\;. \label{Katanin}
\end{equation}
While the additional Katanin terms formally have the structure of the three-particle term [the last term in Fig.~\ref{fig:frg}(b)], they should rather be understood as self-energy corrections~\cite{Katanin-2004}. Indeed, the Katanin truncation ensures full self-consistency at the two-particle level in the sense that the self-energy is completely fed back into the flow of $\Gamma^\Lambda$. This feedback is particularly important for the description of strongly fluctuating spins which requires two-particle vertex renormalizations beyond the bare ladder summations. Together with the initial conditions defined in the limit $\Lambda\to \infty$ (where the self-energy vanishes and the two-particle vertex reduces to the bare couplings $J_{ij}$), the closed set of differential equations is now amenable to numerical treatment.

According to standard diagrammatic Feynman rules, the implementation of spins $S>1/2$ via the local replication of $S=1/2$ degrees of freedom [see Eq.~(\ref{sum})] introduces additional sums over flavor indices $\kappa$ for all closed fermion loops in the PFFRG equations. Since the bare couplings $J_{ij}$ are independent of $\kappa$, this summation simply leads to an extra factor $M=2S$ in the Hartree contribution for the self-energy [the first term on the right-hand side of Fig.~\ref{fig:frg}(a)] and in the RPA contribution for the two-particle vertex [the second term on the right-hand side of Fig.~\ref{fig:frg}(b)]. Increasing $S$ consequently strengthens the RPA term with respect to the other terms, indicating that these diagrams are responsible for the formation of classical magnetic long-range order. Indeed, one can show that, in the absence of finite-temperature divergencies of subleading $1/S$ diagrams, the bare RPA channel (which accounts for only leading $1/S$ diagrams) correctly reproduces the classical limit $S\to\infty$ where the PFFRG becomes identical to the Luttinger-Tisza method~\cite{Baez-2017}. We mention that a correct treatment of the classical nearest-neighbor Heisenberg antiferromagnet indeed requires accounting for the effects of subleading $1/S$ diagrams as discussed in Appendix~\ref{appendix1}. In a similar way, the PFFRG method can be generalized to treat SU$(N)$ spins with $N>2$. In such a scheme, the ladder channels [first and fifth terms on the right-hand side of Fig.~\ref{fig:frg}(b)] contribute with an additional factor of approximately $N$, indicating that these terms describe nonmagnetic spin liquids or dimerized states. In analogy to a large $S$ generalization, they become exact in the limit $N\to \infty$. This built-in balance between large-$S$ and large-$N$ terms represents the key property of the PFFRG that allows one to study magnetic order and disorder tendencies on fair footing. The PFFRG was initially developed in two dimensions~\cite{Reuther-2010}; however, subsequent refinements have made it capable of handling a wide spectrum of frustrated magnetic Hamiltonians for multilayer systems and in three dimensions~\cite{Reuther-2011a,Reuther-2011b,Reuther-2011c,Reuther-2011d,Singh-2012,Reuther-2014,Suttner-2014,Iqbal-2015,Balz-16,Iqbal-2016a,Iqbal-2016b,Iqbal-2016c,Buessen-16a,Buessen-2017,Roscher-2017,Buessen-2018,Hering-2017,Iqbal-2017,Chillal-2017,Keles-2018,Iqbal-2018}.

\subsubsection{Numerical solution of PFFRG flow equations and probing the nature of the ground state}\label{sec:numerical}

To solve the PFFRG equations numerically, we approximate the spatial dependence of $\Gamma^\Lambda(1',2';1,2)$ by discarding all vertices with a distance between sites $i_1$ and $i_2$ greater than some maximal value. In our calculations, we use a distance of approximately $11.5$ nearest-neighbor lattice spacings, which corresponds to a total volume of 2315 correlated spins. Likewise, the continuous frequency arguments of the vertices are approximated by discrete meshes, for which we typically use a combination of linear and logarithmic grids consisting of 64 discrete frequency points.  

By fusing the external legs $(1,1')$ and $(2,2')$ of the two-particle vertex $\Gamma^{\Lambda}\left( 1',2';1,2\right)$, one can calculate the static spin-spin correlator
\begin{equation}
\chi_{ij}^{zz}=\int_0^{\infty} d\tau \left < T_{\tau}S^z_i(\tau)S^{z}_{j}(0) \right >\;,\label{correlator}
\end{equation}
where $T_{\tau}$ (with $\tau$ being the imaginary time) is the imaginary time-ordering operator. 

Transforming $\chi_{ij}^{zz}$ into $\mathbf{k}$ space yields the wave-vector-resolved susceptibility $\chi(\mathbf{k})$:

\begin{equation}\label{eqn:suscep}
    \chi(\mathbf{k})=\frac{1}{4}\sum_{i=1}^{4}\sum_{j}\chi_{ij}^{zz}e^{i\mathbf{k}\cdot({\mathbf{r}_i}-{\mathbf{r}_j})},
\end{equation}
which is the central outcome of the PFFRG to probe the system's magnetic properties. Note that, since in the Heisenberg case the susceptibility is always isotropic, we omit the component indices $xx/yy/zz$ in the susceptibility $\chi(\mathbf{k})$. Here, the first summation is carried out over the four sites of a given primitive unit cell, and the prefactor of $1/4$ is the inverse of the total number of sites in the unit cell. This quantity has the periodicity of the extended Brillouin zone but not of the first Brillouin zone, and thus the susceptibilities are always presented in the former. Henceforth, all wave vectors $\mathbf{k}$ are expressed in units where the edge length of the pyrochlore cubic unit cell is one. The onset of long-range dipolar magnetic order is signaled by a divergence in the $\Lambda$ flow of the susceptibility as observed in the thermodynamic limit. This divergence is a manifestation of the fact that the spin-spin correlations do not decay in the limit of long distances, which would ultimately cause the Fourier transform $\chi(\mathbf{k})$ to diverge. However, in the numerical calculations, we employ a frequency discretization and keep only a limited spatial range of the two-particle vertices; hence, the Fourier transform amounts to a finite site summation that no longer diverges. Thus, these divergences end up being regularized, manifesting themselves as kinks or cusps at some critical $\Lambda_{c}$ in the $\Lambda$ evolution of the susceptibility (henceforth referred to as ``breakdown of the RG flow") [see Appendix~\ref{appendix2} for a discussion on the detection of magnetic instabilities in the RG flow]. 

The type of magnetic order is characterized by the wave vector at which the breakdown of the RG flow occurs. In $3$D, the PFFRG ordering scales, i.e., $\Lambda_{c}$, are directly related to the ordering temperatures $T_{c}$ via $\frac{T_{c}}{J}=\Big{(}\frac{2\pi S(S+1)}{3}\Big{)}\frac{\Lambda_{c}}{J}$~\cite{Iqbal-2016b}. The conversion factor $2\pi S(S+1)/3$ between the RG scale $\Lambda$ and the temperature $T$ can be obtained by comparing the limit of PFFRG where only the RPA diagrams contribute~\cite{Baez-2017}, i.e., a mean-field description, and the conventional spin mean-field theory which is formulated in terms of the temperature instead of $\Lambda$~\cite{Khomskii-2010}. On the other hand, nonmagnetic (absence of dipolar magnetic order) ground states are signaled by a susceptibility flow that \emph{continues to evolve smoothly} down to the (numerical) limit $\Lambda\to 0$. Even in the absence of long-range dipolar magnetic order, the momentum profile of $\chi(\mathbf{k})$ at $\Lambda\ll 1$ allows one to determine the dominant types of short-range spin correlations or to identify competing ordering tendencies.

In the absence of long-range dipolar magnetic order in the ground state, we can further probe for possible spin-nematic~\cite{Moessner-1998a,Moessner-1998b,Shannon-2010} and valence-bond-crystal orders~\cite{Harris-1991,Isoda-1998,Koga-2001,Tsunetsugu-2001a,Tsunetsugu-2001b,Berg-2003,Tchernyshyov-2006,Moessner-2006} by computing the corresponding nematic and dimer response functions. Here, we are particularly interested in studying the tendency of the quantum paramagnet towards spontaneous breaking of either spin rotation symmetry, i.e., nematic order, \emph{or} translational symmetry, i.e., dimer order. The onset of these orders is marked by the divergence of the corresponding order-parameter susceptibility, which is given by a four-spin correlator. For spin-nematic order, this correlator is the standard nematic correlation function $\sum_{\mu,\nu}\langle \mathcal{O}_{ij}^{\mu\nu} \mathcal{O}_{kl}^{\mu\nu} \rangle$, where $\mathcal{O}_{ij}^{\mu\nu}=\hat{S}_i^\mu \hat{S}_j^\nu-(\delta_{\mu\nu}/3)\hat{\bf S}_i\cdot\hat{\bf S}_j$~\cite{Andreev-1984,Taillefumier-2017} (with $\mu$, $\nu=x,y,z$ denoting the three directions in spin space and $i$, $j$ representing the lattice sites) is a symmetric traceless tensor. For dimer order, it is the singlet-singlet correlation function $D_{ijkl}=\langle(\mathbf{\hat{S}}_{i} \cdot \mathbf{\hat{S}}_{j})(\mathbf{\hat{S}}_{k} \cdot \mathbf{\hat{S}}_{l})\rangle-\langle\mathbf{\hat{S}}_{i} \cdot \mathbf{\hat{S}}_{j}\rangle^{2}$. In PFFRG, such correlators are represented by the fermionic four-particle vertex, and, while the PFFRG formalism could, in principle, be straightforwardly extended to obtain the RG flow equation for the \emph{four}-particle vertex, their numerical solution is, at present, not feasible due to limitations posed by computational complexity limitations and memory requirements. The fact that the four-particle vertex is {\emph a priori} excluded from the RG equations implies that the RG flow of the spin susceptibility [Eq.~\eqref{eqn:suscep}] is unaffected by the possible presence of competing nematic and dimer orders. Hence, we adopt a simple recipe within the PFFRG framework to calculate the nematic (dimer) response function $\eta_{\rm SN}$ ($\eta_{\rm VBC}$) which measures the propensity of the system to support nematic (valence-bond-crystal) order. It amounts to adding a small perturbation to the bare Hamiltonian which enters the flow equations as the initial condition for the two-particle vertex. The perturbing term for probing spin-nematic order is
\begin{equation}\label{eqn:nem}
\hat{\mathcal{H}}_{\text{SN}}=\delta\sum_{\langle ij\rangle}(\hat{S}_i^x\hat{S}_{j}^x+\hat{S}_i^y\hat{S}_{j}^y)-\delta\sum_{\langle ij\rangle}\hat{S}_i^z\hat{S}_{j}^z
\end{equation}
which strengthens (weakens) the $xx$ and $yy$ ($zz$) component of the couplings $J_{ij}$ on all nearest-neighbor bonds and where $0<|\delta|\ll J$. This term induces a small bias towards the lowering of spin-rotational symmetry in such a way that spin isotropy is always retained for spin rotations in the $xy$ plane; i.e., the spin-rotational symmetry is broken down from SU(2) to U(1). Similarly, the perturbing term for probing dimer order is 
\begin{equation}\label{eqn:vbc}
\hat{{\cal H}}_{\rm VBC}=\delta\sum_{\langle i,j\rangle\in {S}}\mathbf{\hat{S}}_{i}\cdot \mathbf{\hat{S}}_{j}- 
                   \delta\sum_{\langle i,j\rangle\in {W}}\mathbf{\hat{S}}_{i}\cdot \mathbf{\hat{S}}_{j},
\end{equation}
which strengthens the couplings $J_{ij}$ on all bonds in $S$ [$J_{ij}\to  J_{ij}+\delta$ for $\langle i,j\rangle\in S$] and weakens 
the couplings in $W$ [$J_{ij}\to  J_{ij}-\delta$ for $\langle i,j\rangle\in W$]. The bond pattern $P\equiv\{S_{p},W_{p}\}$ (the subscript ``$p$'' labels the strong and weak bonds corresponding to a pattern ``$P$'') employed here specifies the spatial pattern of symmetry breaking one wishes to probe. 

These modifications amount to changing the initial conditions of the RG flow at large cutoff scales $\Lambda$. As $\Lambda$ is lowered, we keep track of the evolution of all nearest-neighbor spin susceptibilities $\chi_{ij}$. We then define the nematic response function for a given pair of nearest-neighbor sites by 
\begin{equation} 
\eta_\text{SN}=\frac{J}{\delta}\frac{(\chi_{ij}^{xx})_{\Lambda} - (\chi_{ij}^{zz})_{\Lambda}}{(\chi_{ij}^{xx})_{\Lambda} + (\chi_{ij}^{zz})_{\Lambda}}\;,\label{nematic_susceptibility}
\end{equation}
where $\chi_{ij}^{xx}$ ($\chi_{ij}^{zz}$) are the correlators on the strengthened (weakened) bonds. Similarly, the dimer response function is given by
\begin{equation}\label{eqn:dimer-response}
\eta^{P}_{\rm VBC}=\frac{J}{\delta}\frac{(\chi_{S_{P}})_{\Lambda}-(\chi_{W_{P}})_{\Lambda}}{(\chi_{S_{P}})_{\Lambda}+(\chi_{W_{P}})_{\Lambda}},
\end{equation}
where, $\chi_{S_{p}}$ ($\chi_{W_{p}}$) denotes $\chi_{ij}\in S_{p}$ ($\chi_{ij}\in W_{p}$). The normalization factor $J/\delta$ ensures that the RG flow starts with an initial value of $\eta_{\rm SN/VBC}=1$. If the absolute value $\eta_{\rm SN/VBC}$ decreases or remains small under the RG flow, the system tends to equalize, i.e., to reject the perturbation on that link, while, if $\eta_{\rm SN/VBC}$ develops a large value under the RG flow, it indicates that the system is tending to develop an instability towards the probed nematic or valence-bond-crystal order. 

\subsection{Luttinger-Tisza method}
\label{sec:LT}
The classical limit of a system of $n$ quantum spins described by a Heisenberg model is achieved by first normalizing the spin operators by dividing them by their angular momentum $S$ and then taking the limit $S \to \infty$~\cite{Millard-1971,Lieb-1973}. This procedure yields the corresponding classical spin system wherein the spin operators in Eq.~(\ref{eqn:Ham1}) are replaced by ordinary vectors of unit length at each lattice site $i$. For general interactions, the classical Hamiltonian to be minimized reads as
\begin{equation}\label{eqn:Hamclass}
 {\cal H} = \sum_{i,j, \alpha, \beta} J_{\alpha \beta}(\mathbf{R}_{ij}) \mathbf{S}_{i, \alpha} \cdot \mathbf{S}_{j, \beta},
\end{equation}
where by $i/j$ we denote the primitive lattice site separated by the lattice translation vectors $\mathbf{R}_{ij}$ and $\alpha/\beta$ denotes the sublattice site index. The underlying primitive lattice of the pyrochlore lattice is the face-centered cubic lattice,  and the pyrochlore structure is composed of four interpenetrating face-centered cubic lattices. The Luttinger-Tisza method~\cite{Luttinger-1946, Luttinger-1951, Kaplan-2007} attempts to find a ground state of Eq.~(\ref{eqn:Hamclass}) by enforcing the spin-length constraint only globally, $\sum_{i}|\mathbf{S}_{i}^{2}|=S^{2}n$, where $n$ is the total number of lattice sites, which is termed the \emph{weak constraint}. This relaxed constraint implies that site-dependent average local moments are now permissible, which, strictly speaking, take us beyond the classical limit by approximately incorporating some aspects of quantum fluctuations~\cite{Kimchi-2014}.

To solve this relaxed problem, we decompose the spin configuration into its Fourier modes $\mathbf{\tilde{S}}_{\alpha}(\mathbf{k})$ on the four sublattices of the pyrochlore lattice
\begin{equation}
 \mathbf{S}_{i, \alpha} = \frac{1}{\sqrt{N/4}} \sum_{\mathbf{k}} \mathbf{\tilde{S}}_{\alpha}(\mathbf{k}) e^{\imath \mathbf{k} \cdot \mathbf{r}_{i,\alpha}}.
\end{equation}
Inserting this equation into Eq.~(\ref{eqn:Hamclass}) results in
\begin{equation}
 {\cal H} = \sum_{\mathbf{k}}\sum_{\alpha, \beta} \tilde{J}_{\alpha \beta}(\mathbf{k}) \mathbf{\tilde{S}}_{\alpha}(\mathbf{k})\cdot \mathbf{\tilde{S}}_{\beta}(-\mathbf{k}),
\end{equation}
with the interaction matrix given by
\begin{equation}\label{eqn:ltmatrix}
  \tilde{J}_{\alpha \beta}(\mathbf{k}) = \sum_{i,j} J_{\alpha \beta}(\mathbf{R}_{ij}) e^{\imath \mathbf{k} \cdot \mathbf{R}_{ij}}.
\end{equation}

The optimal modes satisfying the weak constraint are then given by the wave vector $\mathbf{k}$, for which the \emph{lowest} eigenvalue of Eq.~(\ref{eqn:ltmatrix}) has its \emph{minimum}. The eigenvector corresponding to this eigenvalue gives the relative weight of the modes on the sublattices~\cite{Bertaut-1961}, which means that the optimal modes do not fulfill the strong constraint ($|\mathbf{S}_{i}^{2}|=S^{2}$, i.e., fixed spin-length constraint on every site) if the components of the eigenvector do not have the same magnitude. If, however, this condition is met, the true ground state of the classical model is a coplanar spiral determined by the optimal Luttinger-Tisza wave vector~\cite{Nussinov-2004}. There are also cases where one can construct an explicit parametrization of the ground state purely from the optimal modes in the pyrochlore lattice, as is the case with the cuboctahedral stack state described in Sec.~\ref{sec:classical}.

\subsection{Iterative minimization of the classical Hamiltonian}\label{sec:IMCH}

To find the ground state of the classical Heisenberg Hamiltonian in parameter regions where the Luttinger-Tisza method is not exact\textemdash i.e., a state constructed solely from the optimal modes does not fulfill the strong constraint\textemdash we employ an iterative minimization scheme which preserves the fixed spin-length (strong) constraint at every site~\cite{Lapa-2012}. Starting from a random spin configuration on a lattice with periodic boundary conditions, we choose a random lattice point and rotate its spin to point antiparallel to its local field defined by
\begin{equation}
 \mathbf{h}_i = \frac{\partial {\cal H}}{\partial \mathbf{S}_i} = \sum\limits_{j} J_{ij} \mathbf{S}_j.
\end{equation}
This rotation results in the energy being minimized for every spin update and thereby converging to a local minimum. We choose a lattice with $L=32$ cubic unit cells in each direction, and thus a single iteration consists of $16 L^3$ sequential single-spin updates. One can therefore view this scheme as a variant of classical Monte Carlo with Metropolis updates at zero temperature, where we accept only optimal updates. This iterative scheme is carried out starting from ten up to 50 different random initial configurations per parameter set to maximize the likelihood of having found a global energy minimum. The exact number depends on convergence of the resulting energies. From the minimal energy spin configuration, the spin structure factor 
\begin{equation}
\mathcal{F}(\mathbf{k}) = \frac{1}{16 L^3} \Big{|}\sum\limits_i \mathbf{S}_i e^{\imath\mathbf{k} \cdot \mathbf{r}_{i}}\Big{|}^{2}
\end{equation}
is computed, which is, up to a normalization constant, the same as the susceptibility defined in Eq.~\eqref{eqn:suscep}, but now for a finite system. Although it is not guaranteed that this scheme ends up in the global energy minimum, we find that, in all cases where an exact ground state is known, the iterative minimization scheme recovers the ground state, even when there exist nonoptimal states corresponding to local energy minima and having the same wave-vector content as the true ground state. This scheme also provides us with the opportunity to use spin configurations built from various (which can be arbitrarily chosen) parametrizations as a starting point of the minimization to check the quality of these parametrizations and also compare the competition between two states directly at a phase boundary.

As the iterative minimization works in direct space, we naturally see lattice symmetry breaking inherent to the ordered ground state, which cannot be captured by symmetry-preserving Fourier-space-based methods such as Luttinger-Tisza.
 
In the following section, we investigate the ground state of the general $J_{1}$-$J_{2}$ Heisenberg model, both in the small spin-$S$ regime (employing PFFRG) as well as the corresponding classical model using a combination of the Luttinger-Tisza method and iterative energy minimization schemes. We first begin with a discussion of the nearest-neighbor Heisenberg antiferromagnet.
 
 \begin{figure*}
\includegraphics[width=0.75\columnwidth]{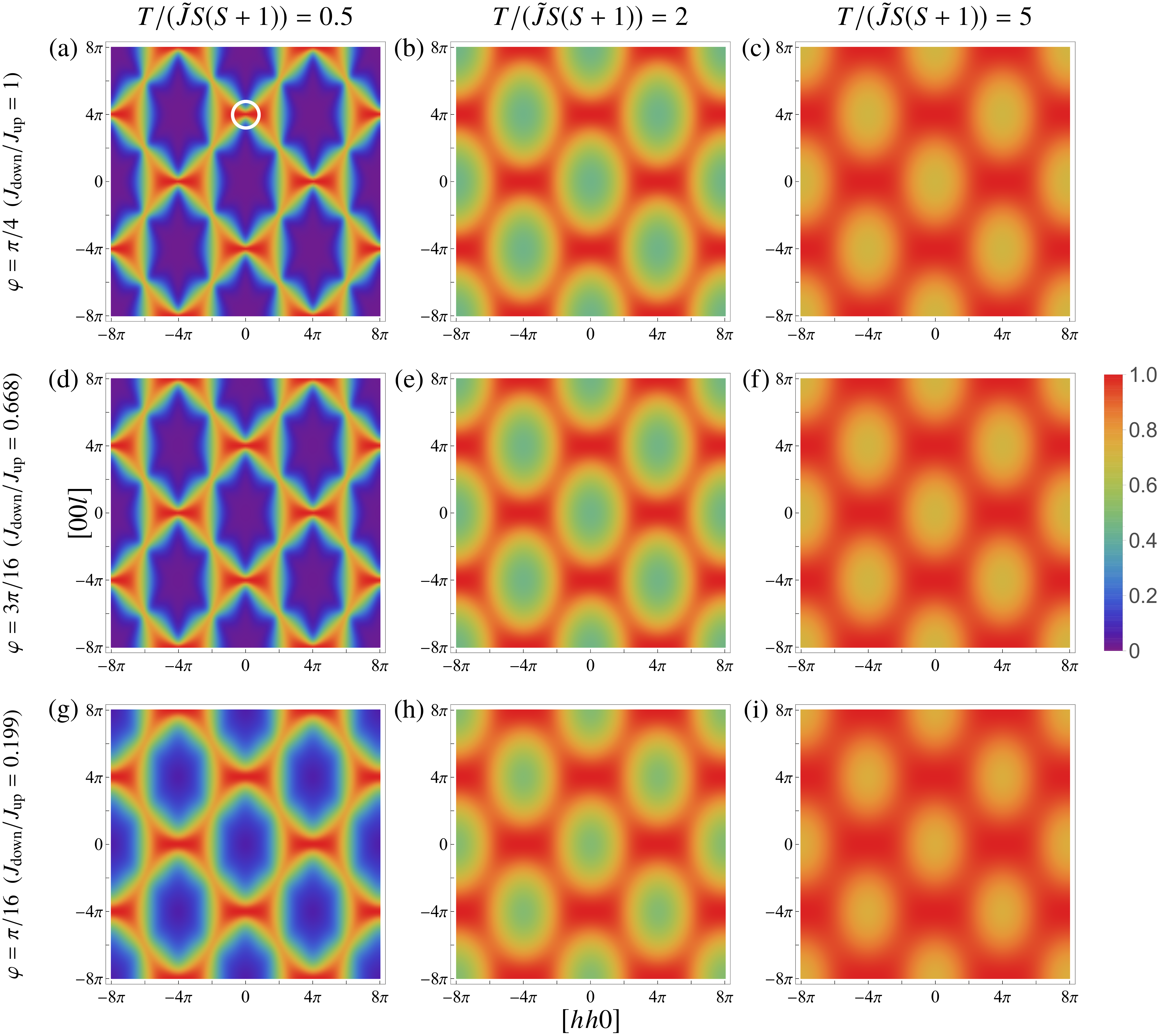}
\caption{For the classical ($S\to\infty$) nearest-neighbor Heisenberg antiferromagnet, the spin susceptibility profile (in units of $1/[\tilde{J}S(S+1)]$) in the $[hhl]$ plane obtained using PFFRG and evaluated for (a)\textendash(c) the isotropic model, (d)\textendash(f) the breathing model for $\varphi=3\pi/16$, and (g)\textendash(i) the breathing model for $\varphi=\pi/16$ at three different temperatures: $T/[\tilde{J}S(S+1)]=0.5$ [(a), (d), (g)], $T/[\tilde{J}S(S+1)]=2$ [(b), (e), (h)], $T/[\tilde{J}S(S+1)]=5$ [(c), (f), (i)]. In (a), we encircle the pinch point at $\mathbf{k}=(0,0,4\pi)$. Each plot has its own color scale, where the red corresponds to the maximum of each plot and blue is fixed to zero.}\label{fig:breathing-1}
\end{figure*}
 
\section{The Nearest-neighbor Heisenberg antiferromagnet}
\label{sec:NNAF}

We begin by investigating the ground state and behavior of the spin-spin correlation functions of the Heisenberg model with only a nearest-neighbor antiferromagnetic interaction and for a general pyrochlore lattice with nonzero breathing anisotropy
\begin{equation}\label{eqn:Ham2}
{\cal \hat{H}}=J_{\rm up} \sum_{{\langle i,j \rangle_{\rm up}}} \mathbf{\hat{S}}_{i} \cdot \mathbf{\hat{S}}_{j} + J_{\rm down} \sum_{{\langle i,j \rangle_{\rm down}}} \mathbf{\hat{S}}_{i} \cdot \mathbf{\hat{S}}_{j},
\end{equation}
where $J_{\rm up}>0$ and $J_{\rm down}>0$ are two different antiferromagnetic couplings on the nearest-neighbor bonds within the up and down tetrahedra, i.e., $\langle i,j\rangle_{\rm up}$ and $\langle i,j\rangle_{\rm down}$, respectively. Hereafter, we parametrize these couplings in terms of a single angle $\varphi$ and an overall energy scale $\tilde{J}$:  

\begin{equation}\label{eqn:couplings}
J_{\rm up}=\tilde{J}\cos(\varphi),~~~~~J_{\rm down}=\tilde{J}\sin(\varphi).
\end{equation}

From a material perspective, the isotropic version of the model, i.e., $\varphi=\pi/4$, proves to be of relevance in understanding the low-temperature dynamics in chromium spinels~\cite{Yaresko-2008,Tymoshenko-2017}. On the other hand, the spatially anisotropic version of the model, wherein the up and down tetrahedra feature different exchange couplings, i.e., $J_{\rm down}/J_{\rm up}\neq 1$, the so-called breathing pyrochlore is realized in the recently synthesized spinels LiGaCr$_{4}$O$_{8}$, LiInCr$_{4}$O$_{8}$, LiInCr$_{4}$S$_{8}$, LiGaCr$_{4}$S$_{8}$, CuInCr$_{4}$S$_{8}$, and CuInCr$_{4}$Se$_{8}$~\cite{Unger-1975,Okamoto-2013,Tanaka-2014,Okamoto-2015,Nilsen-2015,Li-2016,Lee-2016,Saha-2016,Aoyama-2016,Wawrzy-2017,Takeyama-2017,Okamoto-2018,Pokharel-2018,Ezawa-2018,Benton-2015} and in a pseudospin $S=1/2$ Yb-based compound Ba$_{3}$Yb$_{2}$Zn$_{5}$O$_{11}$~\cite{Kimura-2014,Rau-2016,Savary-2016}. In these compounds, the magnetic Cr$^{3+}$ (Yb$^{3+}$) ions, which carry $S=3/2$ ($S=1/2$), form an alternating array of small and large tetrahedra, resulting in different exchange couplings for the two sets of tetrahedra. We begin by reviewing the established results for the classical Heisenberg antiferromagnet on the isotropic and breathing~\cite{Benton-2015} pyrochlore lattices. While a number of the results given below have previously been published in the literature, reestablishing them here sets the stage for our own original results.

\subsection{Classical model}
\subsubsection{Isotropic case}
\label{sec:classical}
At the isotropic point of Eq.~\eqref{eqn:Ham2}, we have $J_{\rm up}=J_{\rm down}=\tilde{J}/\sqrt 2\equiv J_{1}$. Henceforth, all temperatures for the isotropic classical and quantum models are expressed in units of $J_{1}S(S+1)$ and  $J_{1}$, respectively (and we omit the factor of $\sqrt{2}$), while for the breathing model they are expressed in units of $\tilde{J}S(S+1)$ and $\tilde{J}$ for the classical and quantum models, respectively. In the classical limit of Eq.~\eqref{eqn:Ham2}, the Heisenberg spin operators $\mathbf{\hat{S}}_{i}$ reduce to standard three-component vectors $\mathbf{S}_{i}$. In the ensuing analysis, it proves convenient to introduce the magnetization $\mathbf{M}_{\mathcal{T}}$ of the $\mathcal{T}$th tetrahedron,
\begin{equation}
\label{eqn:mag}
\mathbf{M}_{\mathcal{T}}=\sum_{\alpha=1}^{4}\mathbf{S}_{\mathcal{T},\alpha},
\end{equation} 
where the index $\alpha=1,2,3,$ and 4 labels the four spins within the $\mathcal{T}$th tetrahedron. In terms of $\mathbf{M}_\mathcal{T}$, the Heisenberg Hamiltonian can be recast as a {\it disjoint} sum of the square of the magnetizations $\mathbf{M}_\mathcal{T}$ over the ``up'' and ``down'' tetrahedra,
\begin{equation}
\label{eqn:Ham3}
\mathcal{H}_{\rm isotropic}=\frac{J_{1}}{2}\sum_{\mathcal{T}}\mathbf{M}_{\mathcal{T}}^{2} - {\rm const.}
\end{equation} 

\begin{figure}
\includegraphics[width=1.0\columnwidth]{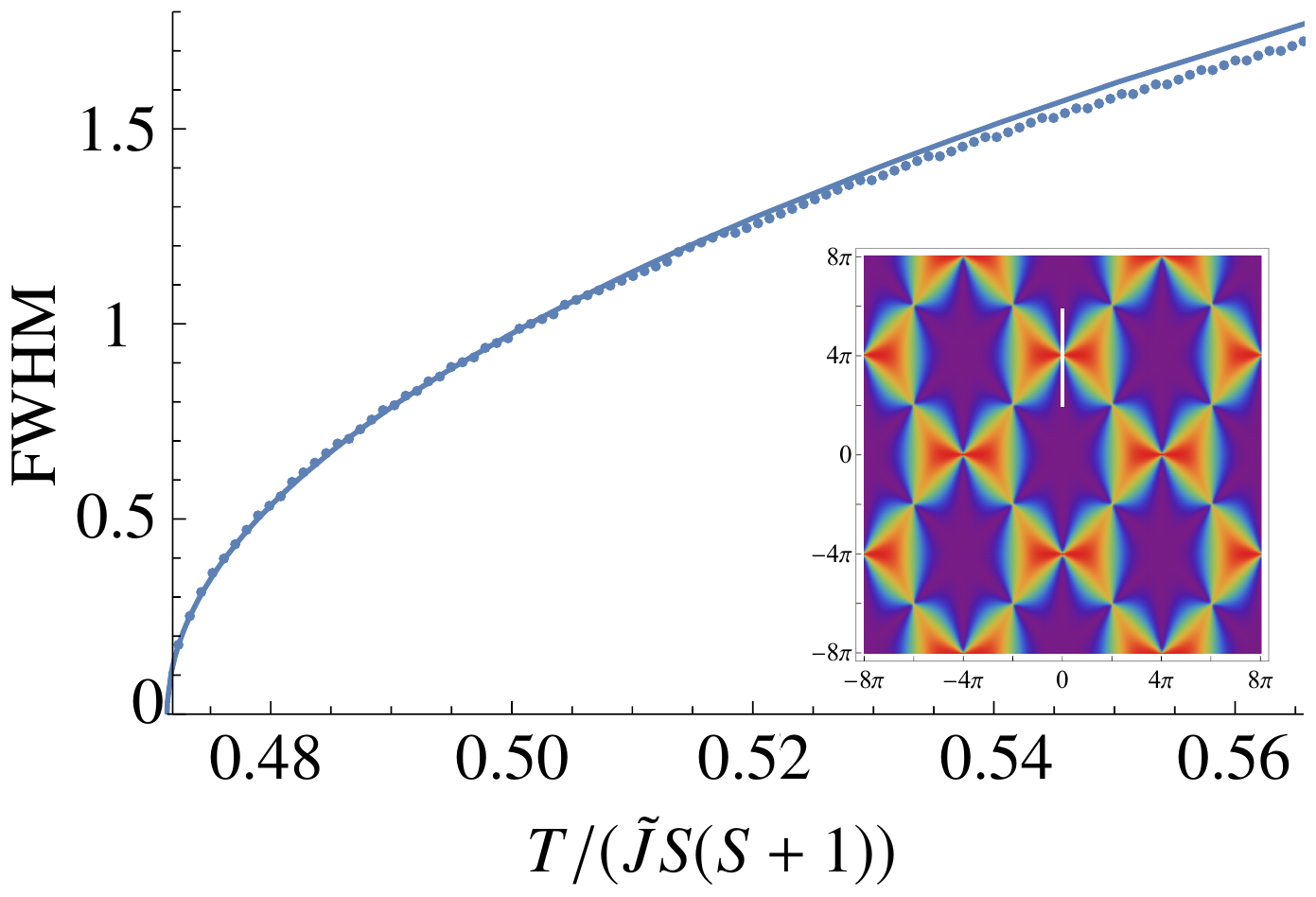}
\caption{The PFFRG data (dotted curve) showing the full width at half maximum (FWHM) (along the $[00l]$ cut, white line in the inset) of the pinch point as a function of the temperature in the classical isotropic nearest-neighbor Heisenberg antiferromagnet. The calculation is done in the bare RPA limit [see Appendix~\ref{appendix1}], wherein the exact pinch-point pattern shown in the inset [plotted using the numerator in Eq.~\eqref{chi2}] naturally occurs due to the flat modes in the interaction matrix [see Eq.~\eqref{eqn:ltmatrix}]. However, this approximation (which accounts for only leading $1/S$ diagrams) contains a methodological artifact which manifests in the form of a divergence of the susceptibility at a finite temperature $T/[\tilde{J}S(S+1)]=\sqrt{2}/3$ [produced by the denominator of Eq.~\eqref{chi2}], at which the $[hhl]$ plane susceptibility shown in the inset is evaluated. Above this temperature, the width of the pinch points is seen to reproduce the $T^{1/2}$ behavior~\cite{Moessner-1998b}. In Appendix~\ref{appendix1}, we show how the inclusion of higher-order diagrammatic contributions in $1/S$ cure this spurious divergence.}\label{fig:ppw}
\end{figure}

From Eq.~\eqref{eqn:Ham3}, it follows that any state which satisfies the condition $\mathbf{M}_\mathcal{T}=\mathbf{0}$ on each tetrahedron $\mathcal{T}$ is a classical ground state. The dimension of the ground-state manifold turns out to be {\it countably} infinite, which is best illustrated via a ``Maxwellian counting argument''~\cite{Moessner-1998a,Moessner-1998b}, which proceeds as follows: For a system of $N_{s}$ classical Heisenberg spins, we have the number of degrees of freedom $F=2N_{s}$ (three degrees of freedom with one spin-length normalization constraint). In the ground state, all three components of $\mathbf{M}_{\mathcal{T}}$ should be zero on every tetrahedron, which gives the number of constraints $K=3N_{c}$, where $N_{c}$ is the number of tetrahedral clusters, and $N_{s}=2N_{c}$ (each tetrahedron has four spins, but each spin is shared between two tetrahedra). Hence, under the assumption that all constraints can be satisfied simultaneously and are all linearly independent, we arrive at the number of ground-state degrees of freedom $D=F-K=4N_{c}-3N_{c}=N_{c}$ which is an extensive quantity. If the constraints are not all linearly independent, then one underestimates $D$; however, for the pyrochlore Heisenberg antiferromagnet, it is known~\cite{Moessner-1998a,Moessner-1998b} that the corrections to the estimate for $D$ are at most subextensive. The extensive (exp[$\mathcal{O}(L^{3})$]) degeneracy of the ground-state manifold proves severe enough to preclude a finite-temperature phase transition, thus realizing a zero-temperature ``cooperative paramagnet''~\cite{Villain-1979} with nonzero entropy~\cite{Anderson-1956}, referred to as a ``classical spin liquid''~\cite{Reimers-1992,Zinkin-1996,Moessner-1998a,Moessner-1998b,Henley-2010}. Indeed, at low temperatures, the Heisenberg model not only fails to develop long-range dipolar magnetic order of the N\'eel type but also does not have conventional nematic order~\cite{Moessner-1998a,Moessner-1998b} of the type characterized by an order parameter which takes on its maximal value in a perfectly collinear state~\cite{Shannon-2010}. At $T=0$, the classical spin liquid features \emph{critical}, i.e., algebraic, spin-spin correlations of dipolar character~\cite{Stillinger-1973}, which is a consequence of the local constraint that the magnetization $\mathbf{M}_{\mathcal{T}}$ on each tetrahedron is identically zero for any ground state~\cite{Youngblood-1981,Henley-1992,Huse-2003,Isakov-2004,Hopkinson-2007}. These dipolar correlations most visibly show up in the Fourier transform of the two-spin correlator, where they form a pattern of bow ties [see Fig.~\ref{fig:breathing-1}(a)] with sharp singularities termed pinch points [see the  encircled point in Fig.~\ref{fig:breathing-1}(a)]~\cite{Zinkin-1996,Zinkin-1997,Moessner-1998a,Canals-2001,Fennell-2007}. The dipolar nature of the correlations in the $T\to 0$ regime is, in fact, a common feature of all classical O$(N)$ nearest-neighbor antiferromagnets for which the system remains paramagnetic down to $T=0$~\cite{Isakov-2004}. This feature excludes the $N=2$ ($XY$-spins) case, as this case is known to show a thermal order-by-disorder transition to collinear ordering for spins which have a global easy plane~\cite{Moessner-1998a,Moessner-1998b} as well as those with local sublattice-dependent easy planes which are perpendicular to the local $\langle111\rangle$ axes~\cite{Bramwell-1994,Champion-2001,Champion-2003,Champion-2004,Zhitomirsky-2012,McClarty-2014}. The limit $N=1$ (Ising spins) is realized in various spin-ice materials $A_{2}B_{2}$O$_{7}$ ($A\equiv {\rm Dy},~{\rm Ho}$ and $B\equiv {\rm Ti},~{\rm Sn}$) which, at low but nonzero temperatures, host a classical spin liquid featuring dipolar correlations and the associated pinch points~\cite{Castelnovo-2012}. Coming back to the case of $N=3$ (Heisenberg spins) at finite temperatures, we note that thermal fluctuations lead to violations of the $\mathbf{M}_\mathcal{T}=\mathbf{0}$ constraint and generate a {\em finite} correlation length $\xi$ which, at low temperatures, diverges as $T^{-1/2}$~\cite{Moessner-1998b}. At distances $r\gg\xi$, the algebraic nature of the real-space spin-spin correlations changes into an exponential. Consequently, at finite temperatures the pinch points acquire a finite width $\sim 1/\xi$~\cite{Conlon-2010} [see Figs.~\ref{fig:breathing-1}(a)\textendash\ref{fig:breathing-1}(c)] which, at low temperatures, goes to zero as $T^{1/2}$~\cite{Moessner-1998b} [see Fig.~\ref{fig:ppw}].

\subsubsection{Breathing case}
\label{sec:classical-breathing}

As in the case of the isotropic pyrochlore lattice, the Heisenberg Hamiltonian in the presence of breathing anisotropy [Eq.~\eqref{eqn:Ham2}] can be straightforwardly recast as a {\it disjoint} sum of terms, each involving the magnetization $\mathbf{M}_{\mathcal{T}}$ [Eq.~\eqref{eqn:mag}] of a tetrahedron $\mathcal{T}$:

\begin{equation}
\label{eqn:Ham4}
\mathcal{H}_{\rm breathing}=J_{\rm up}\sum_{\mathcal{T}\in{\rm up}}\mathbf{M}_{\mathcal{T}}^{2}+
J_{\rm down}\sum_{\mathcal{T}\in{\rm down}}\mathbf{M}_{\mathcal{T}}^{2}- {\rm const.}
\end{equation} 

It is clear that when $J_{\rm up}$ and $J_{\rm down}$ are both antiferromagnetic, any state in which $\mathbf{M}_\mathcal{T}=\mathbf{0}$ on every up and down tetrahedron $\mathcal{T}$ is a classical ground state. Thus, in the presence of a breathing anisotropy, the extensive degeneracy of the isotropic model remains intact, and, consequently, the ground state at low temperatures remains a classical spin liquid~\cite{Benton-2015}. However, as one moves away from the isotropic point $\varphi=\pi/4$, the appearance of the bow-tie pattern with a decreasing temperature, and the development of the pinch-point singularities in the limit $T\to 0$, becomes progressively slower on approaching the decoupled tetrahedron limit, which is because the correlation length is proportional to the product $J_{\rm up}J_{\rm down}/\tilde{J}^2 =\cos\phi \sin\phi$~\cite{Benton-2015}, and, hence, the development of the correlations is slower when closer to the decoupled tetrahedron limit. In Fig.~\ref{fig:breathing-1}, we show the spin susceptibility profile for two values of the breathing anisotropy, $\varphi=3\pi/16$ and $\varphi=\pi/16$, to enable a comparison with Fig.~8 in Ref.~\cite{Benton-2015}. As expected, the development of the bow-tie pattern of scattering with sharp singularities as $T\to 0$ becomes progressively slower as one moves towards the decoupled tetrahedron limit.      

In the following section, we consider the regime of small spin $S$ where strong quantum fluctuations are expected to significantly alter the ground state and nature of the spin-spin correlations.

\begin{figure}
\includegraphics[width=\columnwidth]{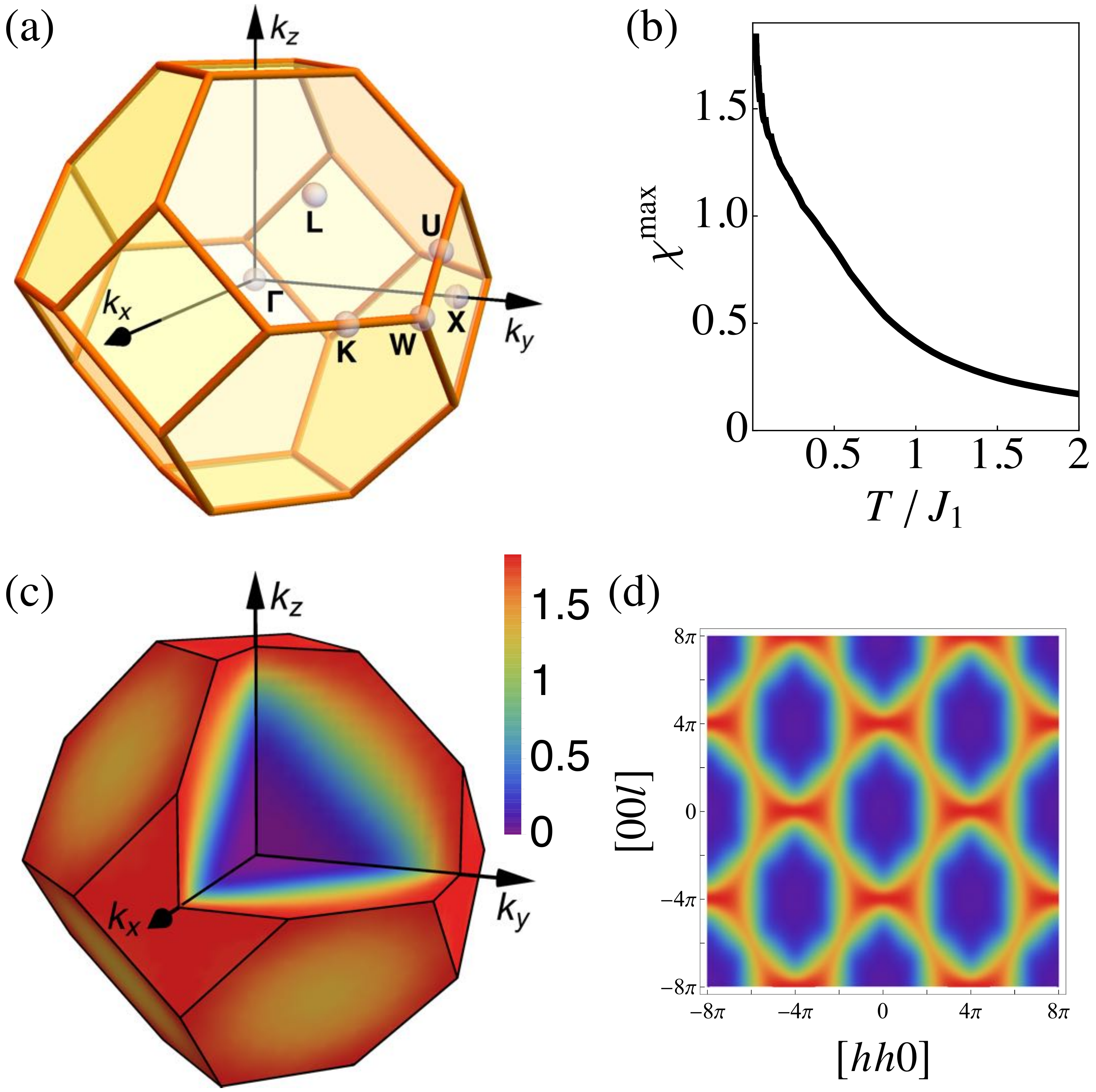}
\caption{(a) The EBZ (a truncated octahedron) of the pyrochlore lattice labeled with the high-symmetry points. (b)\textendash(d) For the $S=1/2$ isotropic nearest-neighbor Heisenberg antiferromagnet, the RG flow of the susceptibility evaluated at the $W$ point (b), the $\mathbf{k}$-space-resolved magnetic susceptibility profiles (in units of $1/J_{1}$) evaluated at $T/J_{1}=1/100$ and shown in the EBZ (c) and projected onto the $[hhl]$ plane (d).}\label{fig:J2=0}
\end{figure}

\begin{figure*}[h!tb]
\includegraphics[width=\columnwidth]{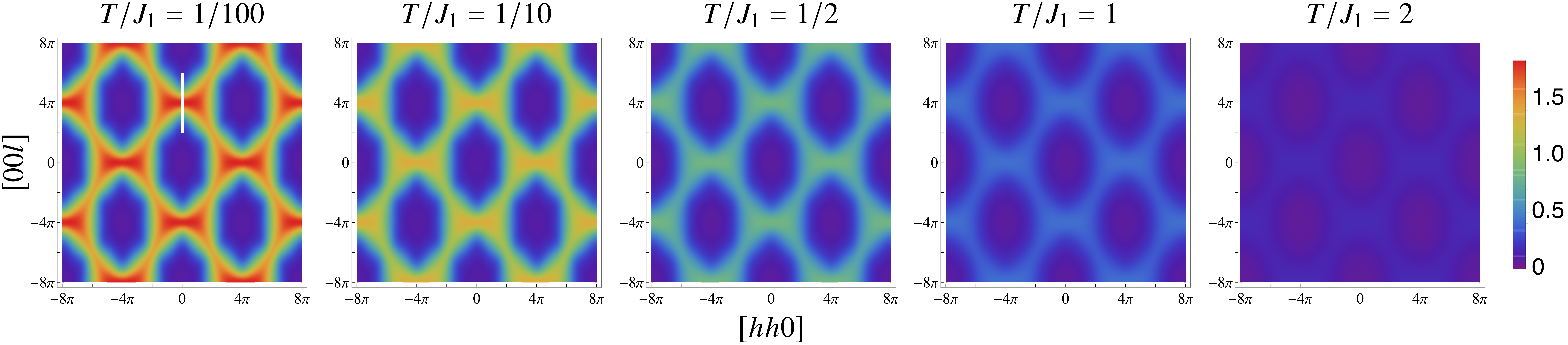}
\caption{The spin susceptibility profile (in units of $1/J_{1}$) in the $[hhl]$ plane at different temperatures for the $S=1/2$ isotropic nearest-neighbor Heisenberg antiferromagnet.}\label{fig:bowtie-1}
\end{figure*}

\subsection{Spin-$1/2$ model}\label{sec:S=1/2}

\subsubsection{Isotropic case}\label{sec:isotropic_s12}

The investigation of the low-temperature ($T\ll J_{1}$) physics of Eq.~(\ref{eqn:Ham2}) in the small spin-$S$ regime proves to be of utmost physical interest by virtue of the fact that in this limit the model harbors strong correlations which conspire with amplified quantum fluctuations to set the stage for a potential realization of a quantum spin liquid. However, it is precisely in this regime that the model acquires a notorious reputation for difficulties due to its nonperturbative character which makes the conclusions obtained from perturbative approaches unreliable~\cite{Harris-1991,Isoda-1998,Koga-2001,Tsunetsugu-2001a,Tsunetsugu-2001b,Berg-2003,Tchernyshyov-2006,Moessner-2006,Canals-1998,Canals-2000,Canals-2001,Fouet-2003,Kim-2008,Burnell-2009,Huang-2016,Yan-2017}. Herein, we address this problem within the PFFRG framework, which is particularly suited for addressing this regime due to its nonperturbative character. 

To probe the propensity of the system towards developing long-range magnetic order at any wave vector $\mathbf{k}$, we track the evolution of the susceptibility $\chi(\mathbf{k})$ with $\Lambda$ for all wave vectors $\mathbf{k}$ in the extended Brillouin zone (EBZ) of the pyrochlore lattice. As discussed in Sec.~\ref{sec:FRGA}, the onset of magnetic long-range order at a particular $\mathbf{k}$ is signaled by the presence of kinks or cusps in the $\Lambda$ flow of $\chi(\mathbf{k})$, whereas a smooth monotonically increasing behavior of $\chi(\mathbf{k})$ down to $\Lambda\to 0$ points to a quantum-disordered ground state. For $S=1/2$, we observe that the $\Lambda$ evolution of the susceptibility $\chi(\mathbf{k})~{\forall}~\mathbf{k}~{\in}~{\rm EBZ}$ [see Fig.~\ref{fig:J2=0}(a) for the EBZ] is smooth and displays a monotonically increasing behavior down to $\Lambda\to 0$ with no detectable signatures of an instability or a kink [see also Appendix~\ref{appendix2}]. A numerical maximization of the susceptibility function in the EBZ finds feeble maxima at the high-symmetry $W$ points, i.e., at $\mathbf{k}=2\pi(2,1,0)$ [see Fig.~\ref{fig:J2=0}(a)]. The RG flow of the susceptibility evaluated at the $W$ point is shown in Fig.~\ref{fig:J2=0}(b), wherein the smooth nature of the flow gives strong evidence in favor of a quantum paramagnetic ground state of the $S=1/2$ quantum Heisenberg antiferromagnet on the pyrochlore lattice, in agreement with previous works~\cite{Harris-1991,Isoda-1998,Canals-1998,Canals-2000,Canals-2001,Koga-2001,Tsunetsugu-2001a,Tsunetsugu-2001b,Berg-2003,Tchernyshyov-2006,Moessner-2006,Kim-2008,Burnell-2009}. 

\begin{figure}[b]
\includegraphics[width=\columnwidth]{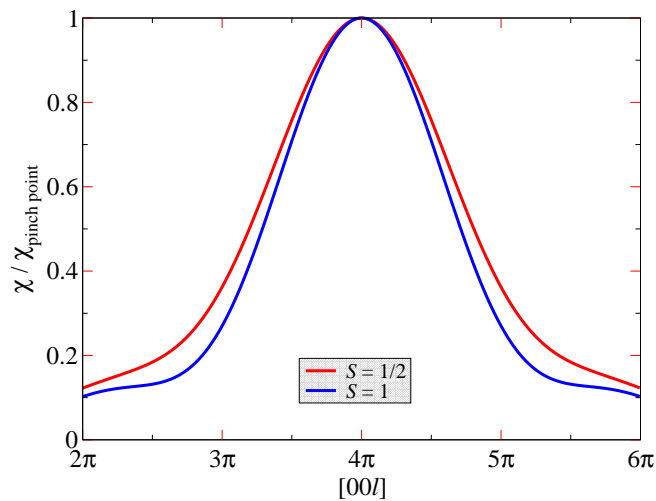}
\caption{The susceptibilities of the $S=1/2$ and $S=1$ isotropic nearest-neighbor Heisenberg antiferromagnet plotted along the $1$D cut (see the white line in the $T/J_{1}=1/100$ plot in Fig.~\ref{fig:bowtie-1}) across the bow-tie width at the lowest simulated temperature $T/J_{1}=1/100$.}\label{fig:bwVsS}
\end{figure}

The corresponding reciprocal space spin susceptibility profile in the EBZ evaluated at the lowest simulated temperature $T/J_{1}=1/100$ is shown in Fig.~\ref{fig:J2=0}(c). The profile appears to be of a highly diffusive character along the edges and surfaces of the EBZ. So as to reveal the nature of the correlations, we plot $\chi(\mathbf{k})$ in the $[hhl]$ plane (i.e., $k_{x}=k_{y}$ plane) [see Fig.~\ref{fig:J2=0}(d)], wherein one clearly sees the characteristic bow-tie pattern, albeit with a softening and broadening of the pinch points due to quantum fluctuations~\cite{Benton-2012,Benton-2016,Huang-2016,Onoda-2011,Lee-2012,Yan-2017}. Indeed, in the small spin-$S$ regime, the spin-flip exchange processes in the Heisenberg Hamiltonian become important and generate quantum fluctuations which dynamically violate the zero magnetization per tetrahedron constraint. Since it is this constraint which is ultimately responsible for the singular and perfectly sharp pinch points observed in the classical model, its violation in the quantum spin-$1/2$ model leads to a regularization \emph{or} a softening of the pinch-point amplitude as their singular character disappears. In addition, quantum fluctuations also generate a finite correlation length $\xi$ for the direct-space spin-spin correlations, such that at distances $r\gg\xi$ the dipolar nature of the correlations changes into an exponential. Consequently, the pinch points undergo ``broadening,'' which can be quantified by their FWHM. Indeed, the FWHM is determined by the inverse of this correlation length, i.e., FWHM $\sim 1/\xi$. In Fig.~\ref{fig:bwVsS}, we show the variation of $\chi(\mathbf{k})$ along the width of the pinch point, i.e., along the white vertical line in Fig.~\ref{fig:bowtie-1}(a), and for $S=1/2$ the FWHM of the pinch point is determined to be $1.6\pi$ at the lowest simulated temperature $T/J_{1}=1/100$. 

\begin{figure}[t]
\includegraphics[width=\columnwidth]{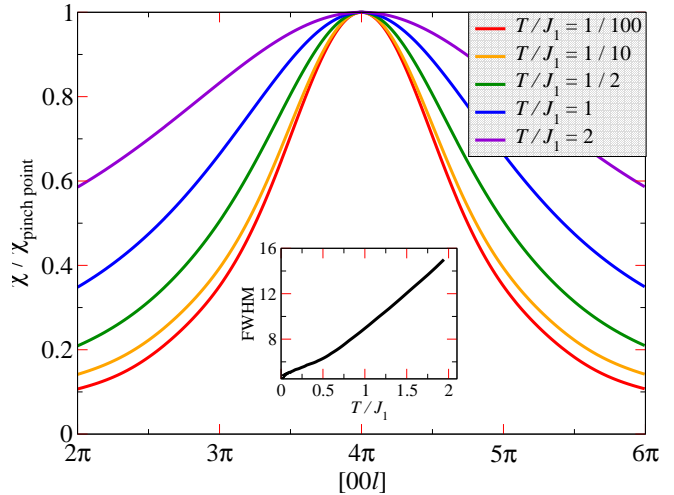}
\caption{The susceptibility of the $S=1/2$ isotropic nearest-neighbor Heisenberg antiferromagnet plotted along the $1$D cut (see the white line in the $T/J_{1}=1/100$ plot in Fig.~\ref{fig:bowtie-1}) across the bow-tie width at different temperatures. The inset shows the FWHM of the curves as a function of the temperature.}\label{fig:fwhm}
\end{figure}

Our finding of relatively rounded pinch points is in agreement with the results of  Refs.~\cite{Canals-1998,Canals-2000,Huang-2016}, which also observe pinch points of a similar nature. The fact that the overall bow-tie pattern of susceptibility appears rather intact (despite relatively rounded pinch points) lends support to the view that the low-temperature ($T/J_{1}=1/100$) paramagnetic phase of the $S=1/2$ nearest-neighbor Heisenberg antiferromagnet respects to a good degree the zero net magnetic moment per tetrahedron constraint, i.e., the ``ice rules''\textemdash as also found in Ref.~\cite{Huang-2016}. The temperature evolution of the susceptibility in the $[hhl]$ plane is shown in Fig.~\ref{fig:bowtie-1}. To obtain a quantitative picture, we plot in Fig.~\ref{fig:fwhm} the susceptibility along a $1$D cut (white line in the $T/J_{1}=1/100$ plot of Fig.~\ref{fig:bowtie-1}) across the width of the pinch point in the bow-tie structure. On increasing the temperature by even an order of magnitude, i.e., up to $T/J_{1}=1/10$, it is found that the susceptibility profile and the width of the pinch points remain essentially unchanged. In the temperature range $T/J_{1}=1/10$ till $T/J_{1}\sim1$, the pinch-point width is seen to increase (approximately) linearly (see the inset of Fig.~\ref{fig:fwhm}) in contrast to the $T^{1/2}$ behavior expected classically (see Fig.~\ref{fig:ppw}). However, the fact that the overall bow-tie structure remains relatively intact up till $T\sim J_{1}$ seems to suggest that the ice rules govern the physics (to a good degree of accuracy) over a surprisingly large temperature range as also found in Ref.~\cite{Huang-2016}. We also study the behavior of the direct-space spin-spin correlations with the temperature and find that, for any given distance, it is only their amplitude that varies with the temperature, while their signs remain constant over the entire temperature range, in agreement with the findings of Ref.~\cite{Canals-2000}. Also, the signs of {\it all} correlators up to the 16th neighbor as obtained from PFFRG agree with those obtained in Table~I of Ref.~\cite{Canals-2000}. This agreement is interesting in light of the fact that Ref.~\cite{Canals-2000} evaluates the equal-time spin-spin correlators, i.e., $S(q,\omega)$ integrated over the frequency, whereas we compute only the $\omega=0$ correlator, which implies that an integration over frequencies does not change the sign. 

\begin{figure}[t]
\includegraphics[width=\columnwidth]{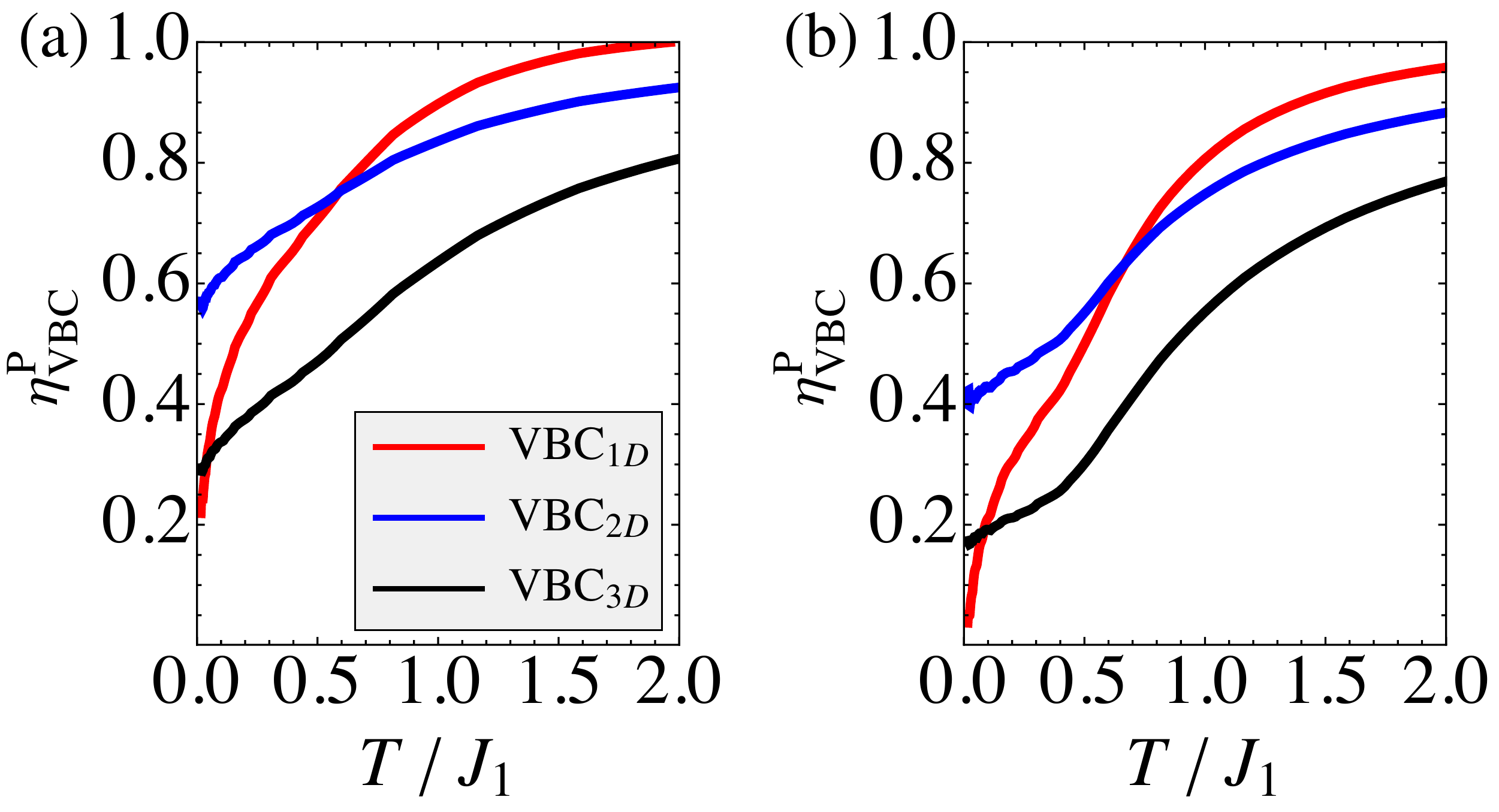}
\caption{The RG flows of the dimer response functions $\eta_{\rm VBC}^{P}$ [Eq.~\eqref{eqn:dimer-response}] of different valence-bond crystals for (a) $S=1/2$ and (b) $S=1$ isotropic nearest-neighbor Heisenberg antiferromagnets.}\label{fig:RG-Flows-VBC}\end{figure}

Early investigations into the nature of the ground state of the $S=1/2$ nearest-neighbor Heisenberg antiferromagnet, predominantly based on perturbative approaches in the intertetrahedra coupling, found the ground-state to be a valence-bond crystal~\cite{Harris-1991,Isoda-1998,Koga-2001,Tsunetsugu-2001a,Tsunetsugu-2001b,Berg-2003,Tchernyshyov-2006,Moessner-2006}. Using PFFRG, we probe for possible instabilities of the quantum paramagnet towards valence-bond-crystal formation. We consider three simple dimerization patterns which, respectively, break the translational symmetry along (i) all three tetrahedral axis directions (VBC$_{3\rm D}$), (ii) two tetrahedral axis directions (VBC$_{2\rm D}$), and (iii) one tetrahedral axis direction (VBC$_{1\rm D}$). The dimer response functions $\eta_{\rm VBC}^{P}$ [Eq.~\eqref{eqn:dimer-response}] of all three VBCs are found to decrease under the RG flow [see Fig.~\ref{fig:RG-Flows-VBC}(a) for the RG flow of $\eta_{\rm VBC}^{P}$] which lends support towards the scenario of a symmetric quantum-spin-liquid ground state as opposed to the previously proposed scenario of a VBC ground state. The disagreement between our findings and those of previous studies~\cite{Harris-1991,Isoda-1998,Koga-2001,Tsunetsugu-2001a,Tsunetsugu-2001b,Berg-2003,Tchernyshyov-2006,Moessner-2006}, which argue for a VBC ground state, is likely explained by the fact that a common thread of these approaches is the inherent symmetry breaking already built in to the scheme considered therein, which then biases the conclusion towards a VBC ground state. That being said, here we investigate VBCs only up to an eight-site unit cell, and the possibility of VBCs with larger unit cells cannot, in principle, be ruled out. 

\begin{figure}
\includegraphics[width=\columnwidth]{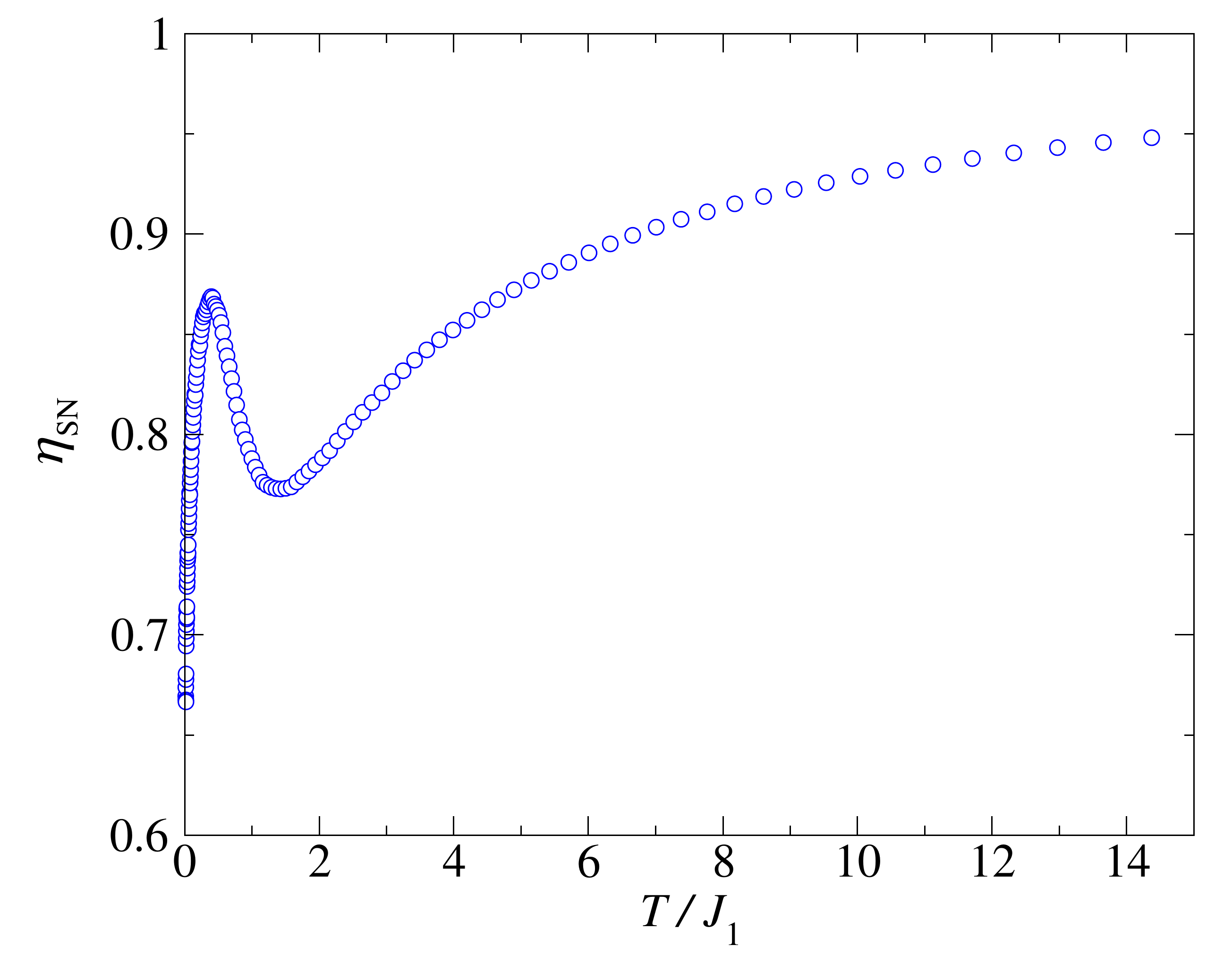}
\caption{The RG flows of the spin-nematic response function $\eta_{\rm SN}$ [Eq.~\eqref{nematic_susceptibility}] for the $S=1/2$ isotropic nearest-neighbor Heisenberg antiferromagnet.}\label{fig:SN}
\end{figure}

The possibility of the occurrence of spin-nematic order in the classical nearest-neighbor Heisenberg antiferromagnet is discussed in Refs.~\cite{Moessner-1998a,Moessner-1998b}, wherein it is found that the system evades such nematic order~\cite{Shannon-2010}. Here, we investigate for the possibility of nematic order [see Sec.~\ref{sec:numerical}] in the $S=1/2$ nearest-neighbor Heisenberg antiferromagnetic model. We plot the RG flow of the nematic response function $\eta_{\rm SN}$ [Eq.~\eqref{nematic_susceptibility}] in Fig.~\ref{fig:SN}, wherein one observes that $\eta_{\rm SN}$ remains less than one throughout the RG flow (albeit displaying nonmonotonic behavior) and sharply decreases at low temperatures ($T\ll J_{1}$). This result indicates that the system tends to reject spontaneous breaking of SU(2) spin rotational symmetry via a quadrupolar order parameter in the ground state of the $S=1/2$ nearest-neighbor isotropic Heisenberg antiferromagnet. Though our results are at variance with Ref.~\cite{Benton-2018}, which argues for a nematic quantum spin liquid featuring spin-nematic order in the $S=1/2$ nearest-neighbor Heisenberg antiferromagnetic model, we mention that, since we \emph{a priori} exclude the fermionic four-particle vertex from the RG equations and hence we cannot calculate the nematic susceptibility, our calculation of the nematic response function by applying symmetry breaking is approximative in character. Thus, we do not definitively exclude the possibility of the realization of a nematic quantum-spin-liquid ground state.

\subsubsection{Breathing case}\label{sec:bs12}

\begin{figure}[b]
\includegraphics[width=\columnwidth]{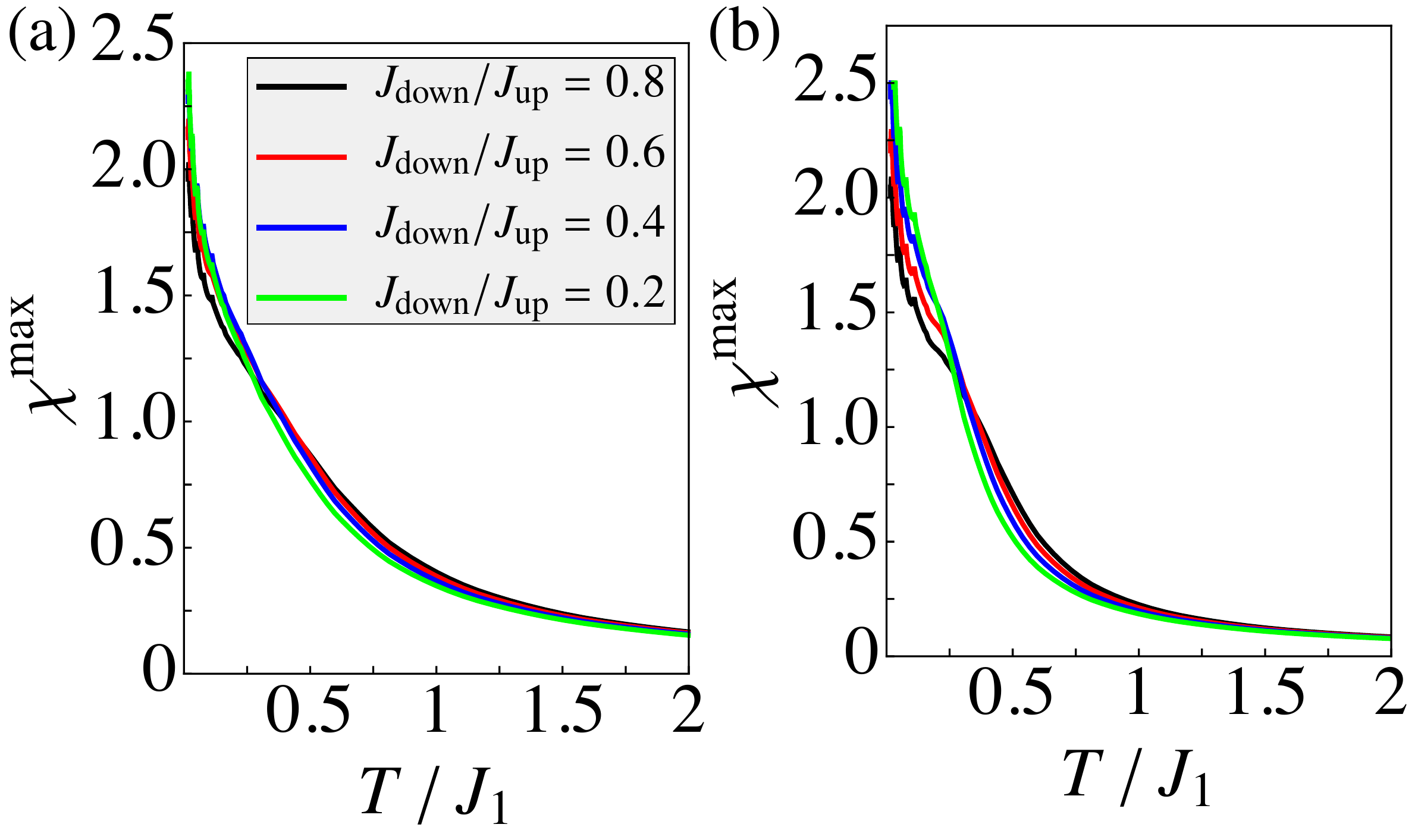}
\caption{The RG flow of the susceptibility tracked at the dominant wave vector for different values of the breathing anisotropy for (a) $S=1/2$ and (b) $S=1$ nearest-neighbor Heisenberg antiferromagnet.}\label{fig:RG-Flows-Breathing}\end{figure}

\begin{figure}[t]
\includegraphics[width=\columnwidth]{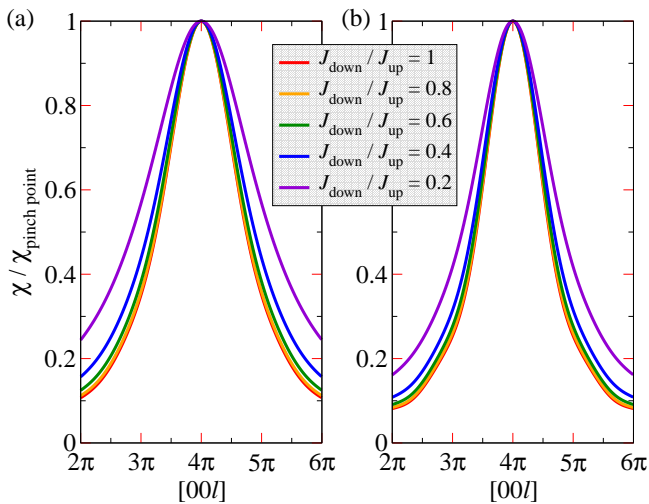}
\caption{For the breathing nearest-neighbor Heisenberg antiferromagnet, we plot the susceptibility along the $1$D cut (see the white line in the $T/J_{1}=1/100$ panel in Fig.~\ref{fig:bowtie-1}) at the lowest simulated temperature $T/J_{1}=1/100$ for (a) $S=1/2$ and (b) $S=1$.}\label{fig:bwba}
\end{figure}

In a breathing pyrochlore system, the ratio of the inter- to intratetrahedra coupling $J_{\rm down}/J_{\rm up}$ provides a convenient interpolation parameter which connects the decoupled tetrahedron and the isotropic limits. It is of interest to investigate the stability of the isotropic model ground state and the evolution of the spin-spin correlations as a function of $J_{\rm down}/J_{\rm up}$. The RG flow of the dominant susceptibility for different values of the breathing anisotropy is shown in Fig.~\ref{fig:RG-Flows-Breathing}(a), wherein we observe a smooth flow down to $\Lambda\to0$, in similarity with the finding for the isotropic model [see Fig.~\ref{fig:J2=0}(b)]. Our results thus point to an extended region of parameter space (accessible by tuning $J_{\rm down}/J_{\rm up}$) over which a quantum paramagnetic phase is stabilized. We also assess the stability of the paramagnetic phase against dimerization into the type of VBC orders considered for the isotropic model and find that the system rejects the applied symmetry breaking under the RG flow, hinting at a possible quantum-spin-liquid state. In the strongly anisotropic limit, we cannot totally exclude the possible scenario of a ground state with more involved patterns of symmetry breaking, e.g., lattice nematic order or VBC with a larger unit cell. Indeed, in the $S=1/2$ breathing kagome Heisenberg antiferromagnet, the situation is contentious: with one work finding VBC~\cite{Iqbal-2018breathing} while the other finds lattice nematic order~\cite{Repellin-2017}. So, further work on the (strongly) anisotropic breathing pyrochlore is probably warranted to ascertain whether it remains without VBC or lattice nematic order down to the limit of the decoupled tetrahedron. Furthermore, we find that the bow-tie pattern of scattering seen in the $[hhl]$ plane is remarkably robust with regard to the introduction of breathing anisotropy, and the width of the bow tie increases only marginally even for strong values of anisotropy [see Fig.~\ref{fig:bwba}(a)]. This result  shows that in the quantum paramagnetic ground state the low-energy physics is approximately governed by the ice rules.  

\subsection{Spin-$1$ model}\label{sec:S=1}

\subsubsection{Isotropic case}\label{sec:isotropic_s1}

Increasing the spin $S$ from $S=1/2$ to $S=1$ renders the effects of quantum fluctuations less pronounced, thus favoring conditions amenable for stabilizing long-range magnetic order. Previous investigations of the $S=1$ Heisenberg antiferromagnet have not been able to reach an unambiguous conclusion regarding the presence or absence of magnetic order~\cite{Koga-2001,Chen-2017}. The $\Lambda$ evolution of the susceptibility at the $\mathbf{k}$ vector where it has its maximum value, i.e., the high-symmetry $W$ point, is shown in Fig.~\ref{fig:higher-spin}(a). The RG flow is not seen to exhibit any instabilities as would be signaled by the presence of kinks and, on the contrary, appears to be of a smooth character [see Appendix~\ref{appendix2} for an analysis on the detection of possible magnetic instabilities in the $S=1$ RG flow]. Similar flow behaviors of the susceptibility are exhibited \emph{for all} wave vectors $\mathbf{k}\in{\rm EBZ}$. These observations lead us to the interesting conclusion that in increased spatial dimensionality (here, $3$D) if geometric frustration is severe enough, such as on the pyrochlore lattice, then even for $S=1$ quantum fluctuations are able to prevent the onset of long-range magnetic order in the Heisenberg antiferromagnet, thereby stabilizing a quantum paramagnetic ground state. The susceptibility profile in the $[hhl]$ plane is qualitatively similar to the one obtained for $S=1/2$; however, the pinch points become slightly sharper as reflected by the decrease in FWHM to $1.42\pi$ compared to $1.6\pi$ for $S=1/2$, evaluated at the lowest simulated temperature $T/J_{1}=1/100$ [see Fig.~\ref{fig:bwVsS}].

To assess the stability of this paramagnetic phase against spontaneous dimerization, we study the dimer response functions of three candidate VBC states described in Sec.~\ref{sec:S=1/2}. The $\Lambda$ evolution of the dimer response functions for the three VBCs [see Fig.~\ref{fig:RG-Flows-VBC}(b)] shows that, similar to the $S=1/2$ case, the system strongly rejects the corresponding applied symmetry breaking. With the present data, we cannot, as in the $S=1/2$ case, rule out the possibility of VBCs with larger unit cells and more complicated patterns of symmetry breaking being stabilized. Nonetheless, from the current PFFRG results, the predicted ground state would be a quantum spin liquid. 

\subsubsection{Breathing case}\label{sec:bs1}
Upon tuning a breathing anisotropy, i.e., $J_{\rm down}/J_{\rm up}\neq 1$, we observe that the RG flows [see Fig.~\ref{fig:RG-Flows-Breathing}(b)] do not develop any signatures of a kink or an instability [as inferred from an analysis based on the method of detection of instabilities as explained in Appendix~\ref{appendix2}] down to the strongly anisotropic limit and remain smooth as $\Lambda\to0$, pointing to the absence of magnetic long-range order. Thus, our results show that even for $S=1$, where quantum fluctuations are expected to be less pronounced, there exists an extended region in parameter space hosting a quantum paramagnet which can be accessed from the isotropic point ($J_{\rm down}/J_{\rm up}=1$) by tuning the breathing anisotropy. We probe this paramagnetic phase for possible VBC instabilities, and find that the system rejects the applied symmetry breaking; however, as in the case of $S=1/2$, we do not exclude the possibility of a ground state featuring a more elaborate pattern of symmetry breaking~\cite{Yasufumi-2001}. We also observe that the bow-tie pattern and the pinch-point width remain essentially unchanged compared to the isotropic model [see Fig.~\ref{fig:bwba}(b)], indicating that the ice rules continue to dictate the low-energy physics of the quantum paramagnetic ground state even for strong breathing anisotropy. 

\subsection{Large spin-$S$ regime}\label{sec:large-S}

\begin{figure}[t]
\includegraphics[width=\columnwidth]{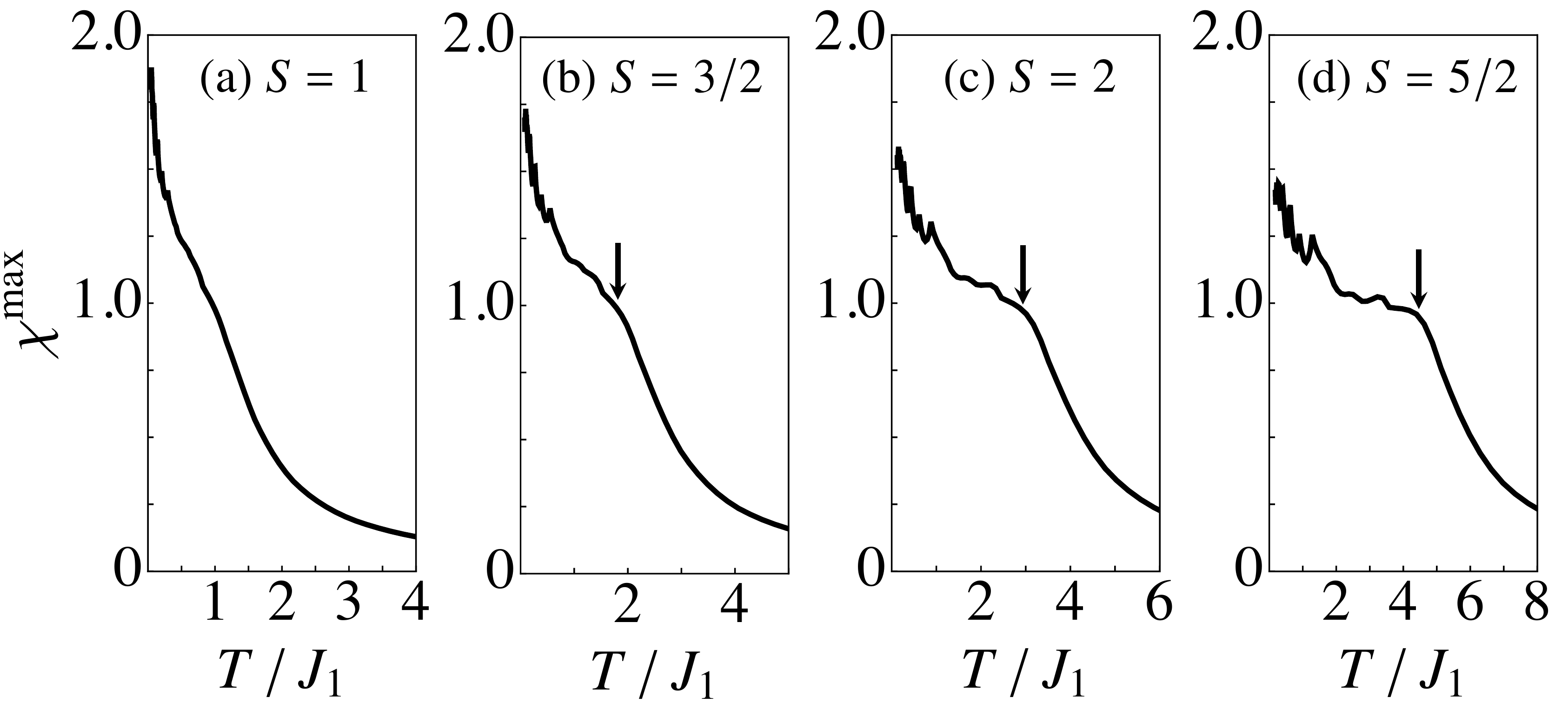}
\caption{For the isotropic nearest-neighbor Heisenberg antiferromagnet, we show, for different values of the spin $S$, the RG flow of the susceptibility tracked at the dominant wave vector.}\label{fig:higher-spin}
\end{figure}

As quantum fluctuations decrease in strength with increasing spin $S$, magnetic long-range order might be expected to ultimately prevail. Indeed, we find that, for $S=3/2$, the RG flow of the dominant susceptibility [see Fig.~\ref{fig:higher-spin}(b)] shows feeble signatures of the development of an instability or kink at the point marked by an arrow. This faint feature, appearing in the $S=3/2$ RG flow, develops into a pronounced kink (marking the breakdown of the RG flow) for increasing values of $S$ [see Figs.~\ref{fig:higher-spin}(c) and \ref{fig:higher-spin}(d)]. The details of the scheme employed to detect the instability or kink are given in Appendix~\ref{appendix2}. Based on this analysis [see Fig.~\ref{fig:maxmethod}], we conclude that for $S=3/2$ and beyond there is an onset of magnetic long-range order in the nearest-neighbor isotropic Heisenberg antiferromagnet. It is worth emphasizing that, for the finite $S$ values studied in our manuscript, the correct balance between leading $1/S$ terms and subleading contributions is
already incorporated in the PFFRG [see Sec.~\ref{sec:FRGA}]. For this reason, the PFFRG at any finite $S$ is still well justified even if plain RPA in the large-$S$ limit, i.e., treating only leading $1/S$ diagrams, produces the aforementioned artifact of finite-temperature divergence of the susceptibility [see Fig.~\ref{fig:ppw}]. However, with increasing $S$, the PFFRG becomes numerically more challenging (and also more sensitive to errors), because it becomes progressively difficult to account for the proper interplay between (large) leading $1/S$ and (much smaller but still important) subleading terms in our numerical algorithm. For this reason, we applied the PFFRG only to ``moderate'' spin magnitudes smaller than eight and use plain RPA in the infinite-$S$ limit [see Appendix~\ref{appendix1}]. Therefore, we are unable to comment on the long-standing issue of the presence or absence of long-range magnetic order in the large-$S$ quantum Heisenberg antiferromagnet.

\begin{figure}[t]
\includegraphics[width=\columnwidth]{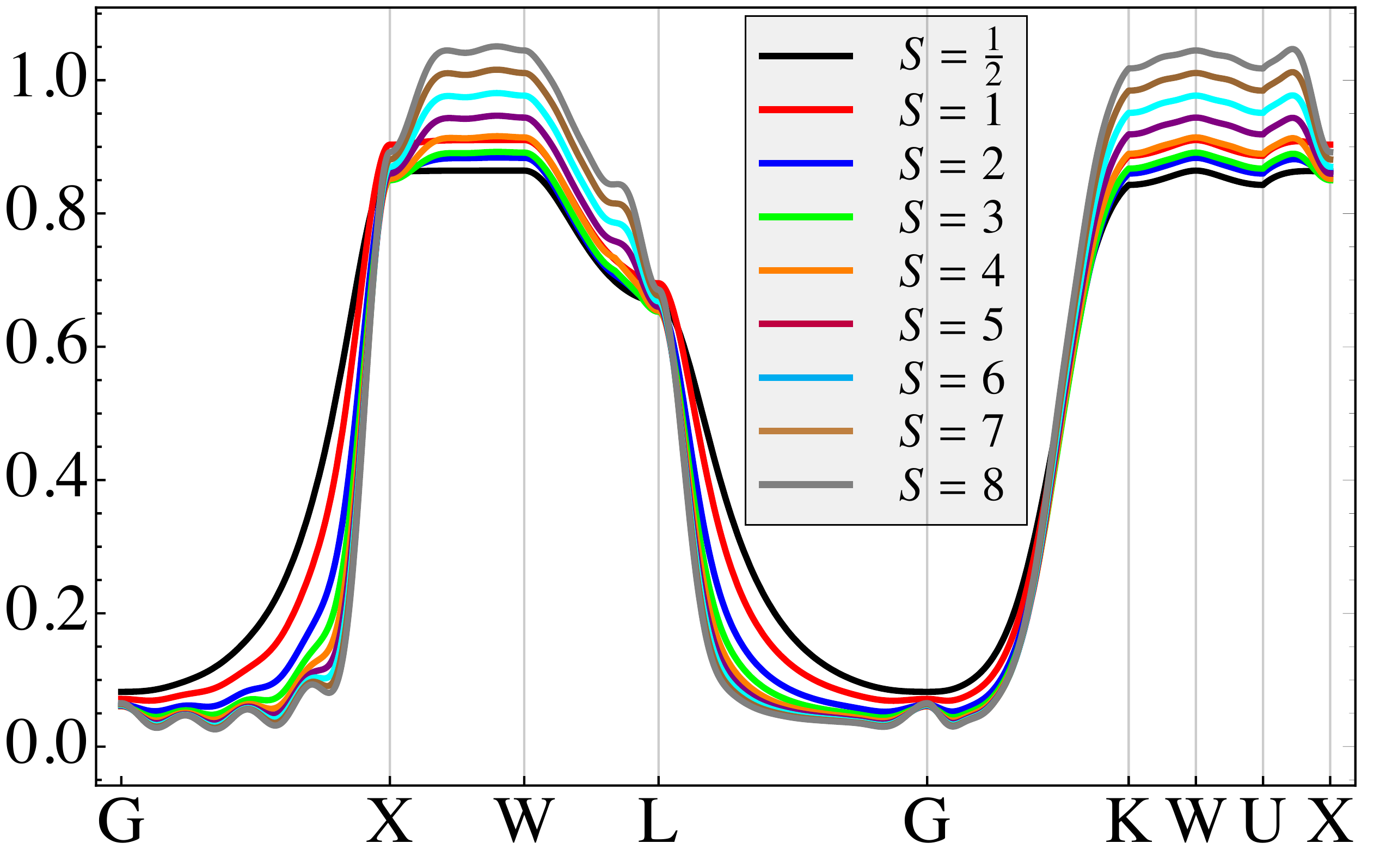}
\caption{The susceptibility of the isotropic nearest-neighbor Heisenberg antiferromagnet plotted along the high-symmetry path evaluated at the lowest simulated temperature ($T/J_{1}=1/100$) for $S=1/2$ and $S=1$ and at the critical breakdown temperature ($T_{c}$) (marked by arrows) in the RG flows in Fig.~\ref{fig:higher-spin} for $S>1$.}\label{fig:HSP}
\end{figure}

Determining the precise nature of the magnetic order (if any) for intermediate values of $S$ constitutes an intriguing and challenging question which has remained unanswered to date. The problem of the ground state of the large-$S$ quantum antiferromagnet on the pyrochlore lattice is addressed extensively using effective Hamiltonian approaches~\cite{Henley-2001,Sobral-1997,Tsunetsugu-2002,Henley-2006,Hizi-2006,Hizi-2007,Hizi-2009}. However, due to the weak selection effects operating at both the harmonic and anharmonic level, no definitive conclusion on the nature of the ground state has yet been reached. Addressing this problem within the PFFRG scheme, we study the evolution of the spin susceptibility profile with increasing values of $S$ in order to figure out whether quantum fluctuations are successful in distilling a unique (magnetically ordered) ground state with a given wave vector $\mathbf{k}\in {\rm EBZ}$ out of the extensively degenerate classical ground-state manifold. In Fig.~\ref{fig:HSP}, we show the variation in the susceptibility along a path passing through the high-symmetry points [see Fig.~\ref{fig:J2=0}(a)] for increasing $S$ values. One observes that, while the susceptibility increases with increasing $S$, there is no clear enhancement at any given wave vector, and the susceptibility profile evaluated at and above the critical breakdown temperature in Figs.~\ref{fig:HSP}(b)-(d) remains essentially unchanged compared to that of the $S=1/2$ and $S=1$ paramagnetic phase, with just an overall enhancement. The absence of pronounced Lorentzian peaks points to the fact that the quantum order-by-disorder selection effects as captured by one-loop PFFRG~\cite{Reuther-2010} may be extremely feeble down to the lowest cutoff or temperature considered, even upon the inclusion of higher orders in $1/S$ embedded within the PFFRG calculation framework~\cite{Baez-2017}. It will be interesting to investigate the large-$S$ limit beyond one loop formulations of PFFRG, e.g., by employing the recently formulated multiloop PFFRG which sums up all parquet diagrams to arbitrary order in the interaction~\cite{Kugler-2018a,Kugler-1998b,Kugler-2018c}.


\section{$J_{1}$-$J_{2}$ Heisenberg model}\label{sec:J1J2}
\subsection{Classical phase diagram}\label{sec:J1J2-classical}

\floatsetup[table]{capposition=bottom}
\begin{table*}
\centering
\renewcommand{\arraystretch}{1.4}
\setlength{\tabcolsep}{8pt}
\begin{tabular}{lllll}
 \hline \hline
       \multicolumn{1}{l}{State}
    & \multicolumn{1}{l}{Wave vector}
    & \multicolumn{1}{l}{Ordering} 
    & \multicolumn{1}{l}{Classical domain}
    & \multicolumn{1}{l}{Quantum $S=1/2$ domain} \\ \hline
       
\multirow{1}{*}{Paramagnet} &  & & & $[345.6\degree\pm1.8\degree,12.6\degree\pm1.8\degree]$   \\ 
                                                                                                                     
\multirow{1}{*}{$\mathbf{k}=\mathbf{0}$} & $2\pi(2,0,0)$ & Coplanar & $(0\degree,26.56\degree]$ & $[12.6\degree\pm1.8\degree,26.56\degree]$   \\ 
                                                                              
\multirow{1}{*}{Planar Spiral} & $2\pi(k,0,0)$ & Coplanar & $[26.56\degree,145.78\degree]$ & $[26.56\degree,151.74\degree\pm0.36\degree]$  \\

\multirow{1}{*}{Double-Twist} & $2\pi(\frac{3}{4},\frac{3}{4},0)$ & Noncoplanar & $[145.78\degree,154.59\degree]$ & $[151.74\degree\pm0.36\degree,160.83\degree\pm0.09\degree]$  \\ 

\multirow{1}{*}{\makecell{Multiply Modulated Spiral}} & $2\pi(\frac{3}{4}^{*},\frac{1}{2},\frac{1}{4}^{*})$ & Noncoplanar & $[154.59\degree,158.37\degree]$ & $[160.83\degree\pm0.09\degree,161.91\degree\pm0.09\degree]$  \\

\multirow{1}{*}{Cuboctahedral stack} & $2\pi(\frac{1}{2},\frac{1}{2},\frac{1}{2})$ & Noncoplanar & $[158.37\degree,170.30\degree]$ & $[161.91\degree\pm0.09\degree,171.27\degree\pm0.27\degree]$   \\

\multirow{1}{*}{Ferromagnet} & $2\pi(0,0,0)$ & Coplanar & $[170.30\degree,312.53\degree]$ & $[171.27\degree\pm0.27\degree,308.61\degree\pm0.27\degree]$  \\

\multirow{1}{*}{Kawamura} & $2\pi(\frac{5}{4}^{*},\frac{5}{4}^{*},0)$ & Noncoplanar & $[312.53\degree,0\degree)$ & $[308.61\degree\pm0.27\degree,345.6\degree\pm1.8\degree]$  \\ \hline \hline

\end{tabular}

\caption{Classical magnetic long-range ordered phases stabilized in the $J_{1}$-$J_{2}$ Heisenberg model. The ordering is labeled as coplanar if there exists a subset of states which are coplanar. The wave-vector components marked by an asterisk have slight incommensurate deviations within the phase away from the given rational values [see the text for details].}
\label{tab:states}
\end{table*}

\begin{figure}[b]
\includegraphics[width=\columnwidth]{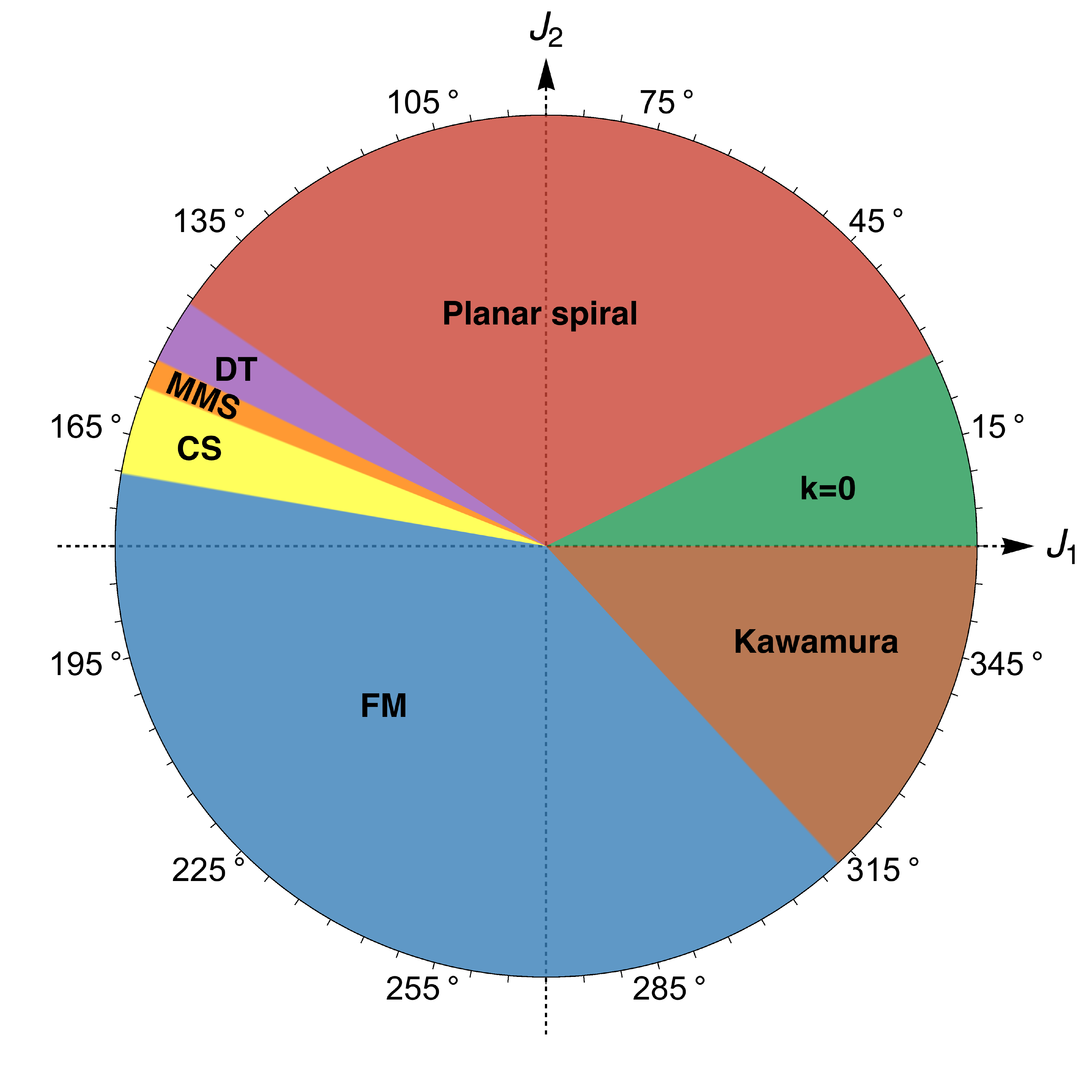}
\caption{The classical phase diagram of the $J_{1}$-$J_{2}$ Heisenberg model on the pyrochlore lattice. The couplings are parametrized as $J_{1}=J{\cos}(\theta$) and $J_{2}=J{\sin}(\theta$) with $J$ an overall energy scale. See Table~\ref{tab:states} for a description of the phases and the location of the phase boundaries.}\label{fig:PhaseDiag_J1J2-Classical}
\end{figure}

Given the absence of long-range order at a nonzero temperature in the classical nearest-neighbor Heisenberg pyrochlore antiferromagnet, any weak perturbations to that model have strong effects on the thermodynamic and magnetic properties of the system that may result in, e.g., magnetic long-range ordering. Indeed, the inclusion of a second-nearest-neighbor Heisenberg coupling $J_2$ to the classical nearest-neighbor Heisenberg model on the pyrochlore lattice is known to stabilize a plethora of intricate magnetic orders [see Table~\ref{tab:states} and Fig.~\ref{fig:PhaseDiag_J1J2-Classical}], part of which is investigated in Refs.~\cite{Ioki-2007,Nakamura-2007,Chern-2008,Okubo-2011}, with a full exploration of the $J_{1}$-$J_{2}$ parameter space reported in Ref.~\cite{Lapa-2012}. Despite the fair number of results available in the literature for this classical $J_1$-$J_2$ model, we find and report below some corrections and/or amendments to the current knowledge about the classical phases of this system.

We find the $J_{1}$-$J_{2}$ model to host seven different classical magnetic orders, in addition to a classical spin-liquid (cooperative paramagnetic) phase found for the nearest-neighbor antiferromagnetic model. Employing an approach which combines a Luttinger-Tisza analysis with an iterative energy minimization on large system sizes of $32{\times}32{\times}32$ cubic unit cells (i.e., \num{524288} spins), we present a refined analysis of the classical phase diagram and the nature of its magnetic orders. The principal differences in our findings compared to those presented in Ref.~\cite{Lapa-2012} can be attributed to the substantially reduced finite-size effects in our calculations compared to those of Ref.~\cite{Lapa-2012}, which are based on a $4\times 4\times 4$ cubic unit cell ($1024$ sites) system. In addition, we identify within the EBZ of the pyrochlore lattice the ordering wave vectors of the classical magnetic orders [see Table~\ref{tab:states}] as would be determined in neutron-scattering experiments. It is important to discuss these states in detail here, since,  as we will see in the next section, the quantum ($S=1/2$ and $S=1$) models harbor the same long-range ordered states.

The pure nearest-neighbor Heisenberg antiferromagnet ($J_2=0$) features an extensively degenerate manifold of classical ground states whose sole shared  feature is that the sum of the spins on every tetrahedron is identically zero [see Sec.~\ref{sec:classical}]. It is shown in Ref.~\cite{Reimers-1991a} that an infinitesimal amount of antiferromagnetic second-nearest-neighbor coupling $J_2>0$ proves sufficient to partially lift this degeneracy by selecting a nonextensive subset of the ground states of the pure nearest-neighbor antiferromagnet. These states are such  that the spins within each of the four face-centered cubic (fcc) sublattices of the pyrochlore lattice order ferromagnetically, and therefore this state is dubbed $\mathbf{k}=\mathbf{0}$. However, the sublattices are not aligned parallel to each other, but the state preserves the constraint of zero spin sum per tetrahedron, resulting in an ordering wave vector at $\mathbf{k}=2\pi(2,0,0)$ and symmetry-related points in the EBZ. This result can perhaps be most easily understood by noting that a second-nearest-neighbor interaction $J_2$ is equivalent to a third-nearest-neighbor interaction $J_{3}$ of the opposite sign, i.e., $J_{3}=-J_{2}$, as long as every tetrahedron satisfies the zero spin sum (``ice rule'') constraint~\cite{Chern-2008}. Since $J_3$ couples only spins on the same sublattice, it is straightforwardly optimized by selecting states with ferromagnetic ordering within each sublattice. This state turns out to be an \emph{exact} Luttinger-Tisza eigenstate of the $\tilde{J}_{\alpha\beta}^{\mathbf{k}}$ matrix in Eq.~(\ref{eqn:ltmatrix}) with an energy per spin $E=-2J_{1}-4J_{2}$. Given that the ordering is fixed {\it only} within each sublattice separately, there remains the freedom of choosing the relative orientation of the individual ferromagnetically aligned sublattices while respecting the zero spin sum per tetrahedron constraint. Hence, at $T=0$ there exists a ground-state degeneracy characterized by three angular degrees of freedom. Therefore, the distribution of spectral weight between the dominant $\mathbf{k}=2\pi(2,0,0)$-type vectors is \emph{not} fixed. At $T=0$, the breaking of the cubic pyrochlore symmetry is not energetically determined by the interactions; however, for finite temperatures entropic effects could select a unique ground state. The relative weights of the dominant peaks in the structure factor then serve as a measure of the collinearity of the sublattices, with the case of only one of them being present corresponding to a fully collinear state. Irrespective of the relative orientation of the sublattices, the ferromagnetic correlations within each of these manifest themselves in the spin structure factor by subdominant peaks of equal intensity at {\it all} of the $2\pi(1,1,1)$ points at the edge of the EBZ. The spectral weight of any one of the given subdominant peaks is exactly one-eighth of the total weight of the dominant peaks.

The aforementioned $\mathbf{k}=\mathbf{0}$ state minimizes the energy only in the regime where antiferromagnetic $J_1>0$ is dominant over sufficiently weak antiferromagnetic $J_2$. Since the $J_{2}$ bonds are twice as many as the $J_{1}$ bonds, the $J_{2}$ interaction becomes dominant when $J_{2}/J_{1}>1/2$
($\theta\gtrsim 26.56\degree$), resulting in a phase transition to a {\it planar} spiral ground state with one of the symmetry-related $\mathbf{k}=2\pi(k,0,0)$-type wave vector as the ordering wave vector. This state is also an eigenstate of the Luttinger-Tisza matrix Eq.~\eqref{eqn:ltmatrix}, thus giving the exact expression $k=(2/\pi) \arccos[-J_{1}/(4 J_{2}) - 1/2]$ for the wave vector and an energy per spin of $E=-J_{1}^{2}/(2 J_{2}) - 6 J_{2}$. This wave vector differs from the one given in Ref.~\cite{Lapa-2012} by a factor of 2, which is due to the fact that the transformation done on this state to map it into an equivalent spin-chain model~\cite{Sklan-2013} was apparently not performed correctly. The pure second-nearest-neighbor antiferromagnet ($J_1=0$, $J_{2}=1$) also falls into this region and has a $120\degree$ spiral structure on each fcc sublattice. Taking into account the relative phases of the spirals between the sublattices, we find a resulting ordering wave vector $\mathbf{k}=2\pi(4/3,0,0)$ in the EBZ of the pyrochlore lattice. In the planar spiral, and corresponding to the aforementioned dominant peaks at $\mathbf{k}=2\pi(k,0,0)$-type ordering wave vectors, there also exist subdominant peaks at ordering wave vectors of the $\mathbf{k}=2\pi(3-k,1,1)$ type in the EBZ. The $k$ and $3-k$ entries of the dominant and subdominant ordering wave vectors, respectively, always appear in the same component for each of these wave-vector pairs. The subdominant peaks are a signature of the correlations within the fcc sublattices of the pyrochlore lattice and have a fixed relative amplitude of one-quarter of the dominant peak. 

The \emph{planar} spiral order is stable against $J_1 <0$ now becoming ferromagnetic (keeping $J_2>0$ antiferromagnetic), up to $J_2/J_1=-0.68$ ($\theta\approx 145.78\degree$). Beyond that point, the ground state changes to a noncoplanar structure, the so-called double-twist (DT) state, first uncovered in a frustrated antiferromagnet on an octahedral lattice~\cite{Sklan-2013}. Its name derives from the fact that the spins form two different kinds of spirals in two perpendicular directions but both governed by the same type of wave vector. 
In reciprocal space, this state features two pairs of $\mathbf{k}=2\pi(3/4,3/4,0)$-type wave vectors on different reciprocal space planes; the first pair, e.g., could be located in the $k_{x}$-$k_{y}$ plane with $\mathbf{k}=2\pi(3/4,3/4,0)$ and $\mathbf{k}=2\pi(3/4,-3/4,0)$, while the second pair, e.g., could be located in the $k_{y}$-$k_{z}$ plane with $\mathbf{k}=2\pi(0,3/4,3/4)$ and $\mathbf{k}=2\pi(0,3/4,-3/4)$. In the first plane, e.g., the $k_{x}$-$k_{y}$ plane, two dominant peaks in the structure factor are located at the aforementioned wave vectors and have identical spectral weight. In the second plane, e.g., the $k_{y}$-$k_{z}$ plane, subdominant peaks with approximately $59\%$ of the spectral weight of the dominant ones are located at the aforementioned wave vectors. An approximate parametrization of such a state is given in Ref.~\cite{Lapa-2012}. Both pairs of wave vectors control the ordering on the individual fcc sublattices. The relative orientations of the spins on the sublattices lead to the appearance of additional subdominant peaks at  $\mathbf{k}=2\pi(5/4,5/4,0)$-type wave vectors. For example, corresponding to the pair of dominant peaks in the $k_{x}$-$k_{y}$ plane, there appear a pair of subdominant peaks at wave vectors $\mathbf{k}=2\pi(5/4,5/4,0)$ and $\mathbf{k}=2\pi(5/4,-5/4,0)$ carrying approximately $29\%$ of the amplitude of the dominant peaks. Similarly, corresponding to the pair of subdominant peaks in the $k_{y}$-$k_{z}$ plane, there appear a pair of weaker peaks at wave vectors $\mathbf{k}=2\pi(0,5/4,5/4)$ and $\mathbf{k}=2\pi(0,5/4,-5/4)$ carrying approximately $13\%$ of the amplitude of the dominant peaks (in the $k_{x}$-$k_{y}$ plane). The particular choice of planes chosen for the dominant and subdominant planes is not fixed by the Heisenberg model, but is determined by the spatial symmetry breaking when entering this phase.

Decreasing antiferromagnetic $J_2>0$ further, we encounter a phase transition at $J_{2}/J_{1}\approx-0.475(5)$ ($\theta\approx 154.59\degree$) to a state which is similar to the \emph{multiply modulated commensurate spiral} of Ref.~\cite{Lapa-2012}, for which the transition point is estimated to be $J_{2}/J_{1} \approx -0.43$. In reciprocal space, this state is characterized by the presence of four dominant commensurate ordering wave vectors of the $\mathbf{k}=2\pi(3/4,1/2,1/4)$ type in the EBZ, for all of which the $1/2$ component is in a common direction. We also find subdominant ordering vectors of the $\mathbf{k}=2\pi(3/4,0,-3/4)$ type; the zero component is the one which is $1/2$ in the dominant $\mathbf{k}=2\pi(3/4,1/2,1/4)$ wave vectors. This result is a consequence of a magnetic structure wherein the spins trace out multiple spirals in different directions in direct space which are controlled by the above wave vectors. Our refined analysis reveals that the observed commensurability of the wave vectors found in Ref.~\cite{Lapa-2012} is an artifact of large finite-size effects at play in that work. The imposition of periodic boundary conditions in the simulation of a $L\times L\times L$ cubic unit cell system allows only those $\mathbf{k}$ vectors whose components are integer multiples of $2\pi/L$. This restriction implies that an incommensurate ordering wave vector which is proximate to a commensurate one leads to an observed peak at the commensurate position. Indeed, we find that, for $J_{2}/J_{1}\approx -0.47$, the four incommensurate ordering wave vectors of $\mathbf{k}=2\pi[0.81(2), 0.50(2), 0.19(2)]$ type evolve continuously (at least within the used $k$-space numerical resolution of $2\pi/32$) towards the commensurate values which are taken on at the transition point to the cuboctohedral stack (CS) state in Fig.~\ref{fig:PhaseDiag_J1J2-Classical}. At the same time, the subdominant ordering vector stays unchanged, but its weight relative to the weight of the dominant peak varies from approximately $26\%$ at its border with the DT state to approximately $32\%$ at its border to the CS state. Our calculations show that, while the manner in which dominant and subdominant wave vectors control this state does not change, the dominant wave vector it is composed of does evolve as a function of $J_{2}/J_{1}$. Our findings are also supported by a Luttinger-Tisza analysis, which shows that there are incommensurate wave vectors with slightly lower energy close to the commensurate point. In this parameter regime, the Luttinger-Tisza  state does not fulfill the strong spin-length constraint [see Sec.~\ref{sec:LT}] but needs to be supported by the subdominant wave vectors we find, in order to be able to construct a normalized state. Because of the incommensurability of the dominant wave vector, we simply refer to this state as a multiply modulated spiral (MMS).

At $J_{2}/J_{1}=-0.3965(5)$ ($\theta\approx 158.37\degree$), the MMS state evolves into the CS state~\cite{Lapa-2012,Sklan-2013}. Its name derives from the fact that, in a construction of the pyrochlore lattice as a stacking of alternating kagome lattice and triangular lattice layers in a $[111]$ direction, the spins in each kagome layer are arranged such that they point towards the 12 vertices of a cuboctahedron, forming a 12-sublattice magnetic structure first found on the kagome lattice~\cite{Domenge-2005,Messio-2011}. At the same time, the spins on the triangular layers point to the eight midpoints of the triangular faces of the same cuboctahedron. This noncoplanar state is built up from any three wave vectors of the $\mathbf{k}=2\pi(1/2,1/2,1/2)$ type, e.g., $\mathbf{k}=2\pi(1/2,1/2,1/2)$, $\mathbf{k}=2\pi(-1/2,1/2,1/2)$, and $\mathbf{k}=2\pi(1/2,-1/2,1/2)$ with identical spectral weight, and is stacked along the $[111]$ direction parallel to the fourth wave vector of this type, e.g., $\mathbf{k}=2\pi(1/2,1/2,-1/2)$. The spin configuration in this state can be expressed analytically (see Ref.~\cite{Lapa-2012}). Each of the dominant ordering vectors is accompanied by a subdominant wave vector of $\mathbf{k}=2\pi(1/2,1/2,3/2)$ type with approximately $18\%$ of the spectral weight of the dominant vectors. From the parametrization, it follows that the average energy per spin, $E=J_{1}(3/4+\sqrt{6}/2)$, is independent of $J_2$ (an extensive discussion how this originates from the state can be found in Ref.~\cite{Lapa-2012}). Thus, decreasing $J_2$ further does not change the energy of this state but, rather, lowers the energy of competing states.

At $J_{2}/J_{1}=(-3/8+\sqrt{6}/12)$ ($\theta\approx 170.30\degree$), the energy of the ferromagnet becomes lower than that of the CS state and occupies the largest extent of the $J_1$-$J_2$ parameter space. Just as for the $\mathbf{k}=\mathbf{0}$ state, the ferromagnetic ordering within the sublattices features subdominant ordering wave vectors at all the $\mathbf{k}=2\pi(1,1,1)$-type points in the EBZ, which have a spectral weight of one-quarter of the dominant $\mathbf{k}=2\pi(0,0,0)$ vector. The pure $J_{2}$ ferromagnet proves to be fairly robust against moderately strong antiferromagnetic $J_{1}$ coupling.

For $J_{2}/J_{1} \gtrsim -1.09$ ($\theta\approx 312.53\degree$), the antiferromagnetic $J_1$ exchange destroys the ferromagnetic order, and a phase transition occurs to a family of states dubbed the Kawamura states after the group which investigated them in great detail~\cite{Ioki-2007}. This phase is made up of a family of degenerate ground states with dominant incommensurate wave vectors around the $\mathbf{k}=(k,k,0)$ points with $k\approx 2\pi(5/4)$ and subdominant ones at $k \approx 2\pi(3/4)$ having approximately $22\%$ of the spectral weight of the dominant vectors. In addition, we find stronger subdominant ordering at $\mathbf{k}\approx2\pi(1,1/4,7/4)$-type vectors with approximately $55\%$ spectral weight. There are two classes of ground states, composed of either four or all six of the ordering wave vectors, the latter therefore respecting the cubic symmetry of the pyrochlore lattice. In the case of a ground state composed of four of the six wave vectors, the Heisenberg model \emph{a priori} does not determine which four are selected. A common feature of both these states is that they are superpositions of spirals with the pertinent wave vectors which, when combined, realize a noncoplanar state. The parameter $k$ for the dominant ordering starts with a value $k\approx2\pi (1.31)$ at the phase boundary to the ferromagnetic state $J_{2}/J_{1}=-1.09$ and approaches $k=2\pi(5/4)$ as $J_2 \to 0$. The Kawamura states also approximately fulfill the zero spin sum per tetrahedron constraint, so they can likewise be considered as perturbed eigenstates of the pure $J_{1}$-only antiferromagnetic model.

\subsection{Quantum Phase Diagram}\label{sec:J1J2-quantum}

\begin{figure*}
\includegraphics[width=\columnwidth]{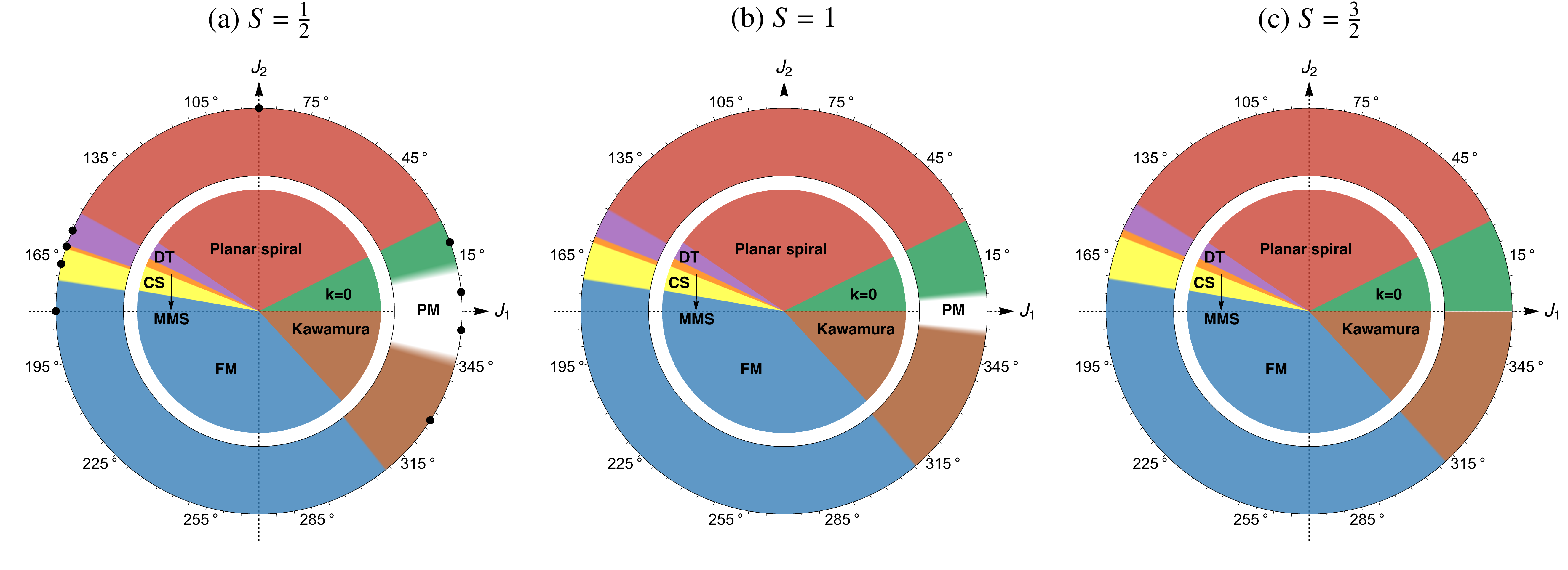}
\caption{The outer rings show the quantum phase diagrams of the $J_{1}$-$J_{2}$ Heisenberg model on the pyrochlore lattice for different values of the spin $S$. An extended quantum paramagnetic regime is stabilized for $S=1/2$ and $S=1$. The inner rings correspond to the classical phase diagram. The Heisenberg couplings are parametrized as $(J,\theta)$ defined by $J_{1}=J{\cos}(\theta$) and $J_{2}=J{\sin}(\theta$).}\label{fig:PhaseDiag_J1J2}
\end{figure*}

\begin{figure}[b]
\includegraphics[width=\columnwidth]{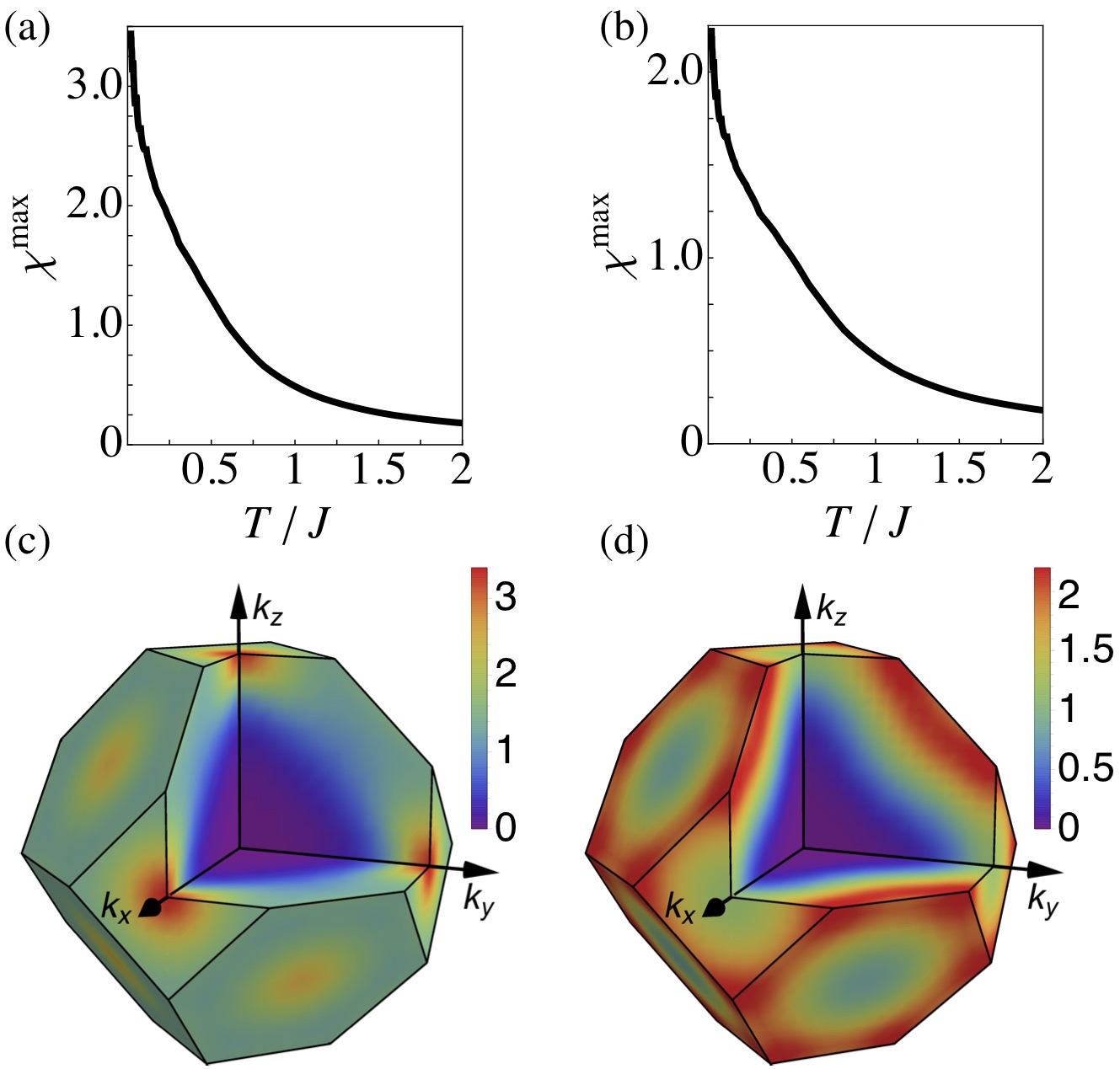}
\caption{The RG flow of the dominant susceptibility (in units of $1/J$) inside the paramagnetic regime of the $S=1/2$ $J_{1}$-$J_{2}$ model shown for (a) $J_{2}/J_{1}=0.1$, (b) $J_{2}/J_{1}=-0.1$ [marked by black circles in Fig.~\ref{fig:PhaseDiag_J1J2}(a)], and (c),(d) their respective spin susceptibility profiles evaluated at the lowest simulated temperature $T/J=1/100$.}\label{fig:J2-AF-FM}
\end{figure}

\begin{figure*}
\includegraphics[width=\columnwidth]{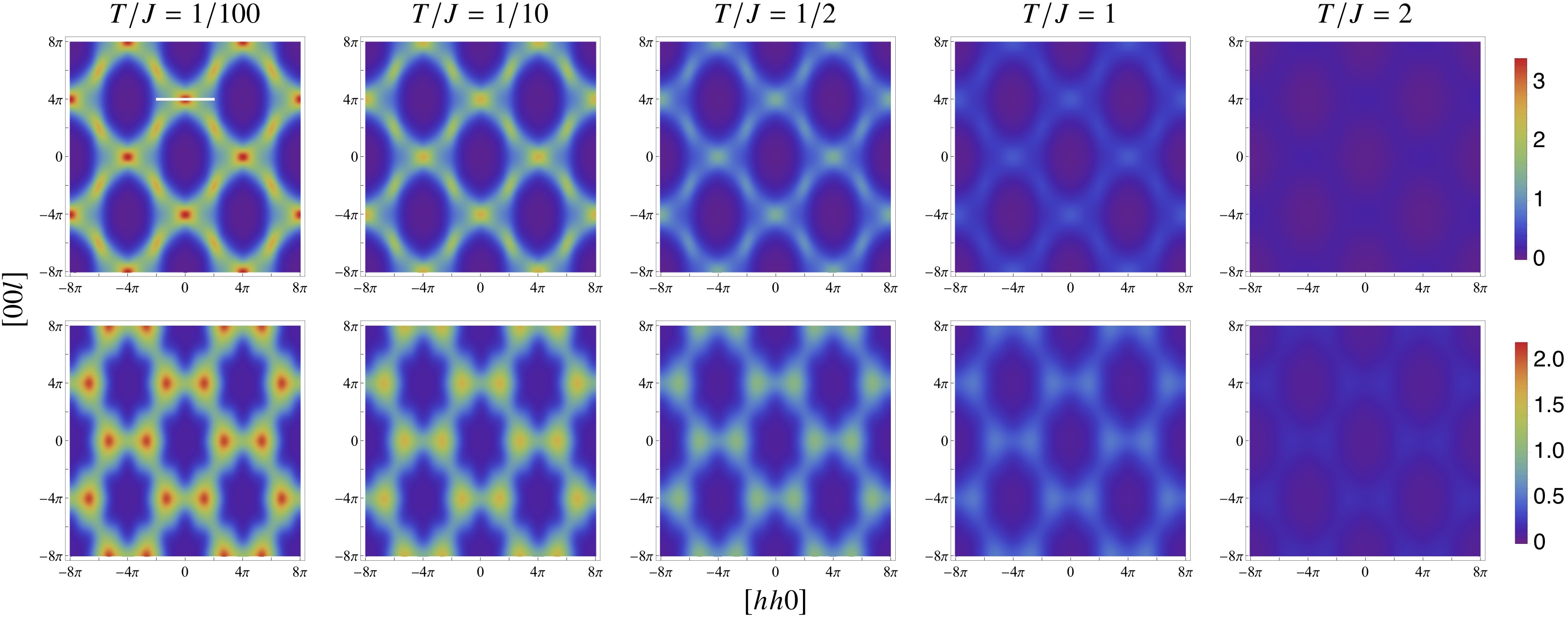}
\caption{The spin susceptibility profile (in units of $1/J$) in the $[hhl]$ plane shown at different temperatures for the $S=1/2$ $J_{1}$-$J_{2}$ Heisenberg model. The first row is for antiferromagnetic $J_{2}$ (evaluated at $J_{2}/J_{1}=0.1$) and the second row for ferromagnetic $J_{2}$ (evaluated at $J_{2}/J_{1}=-0.1$).\label{fig:bowtie-2}}
\end{figure*}

\begin{figure}
\includegraphics[width=\columnwidth]{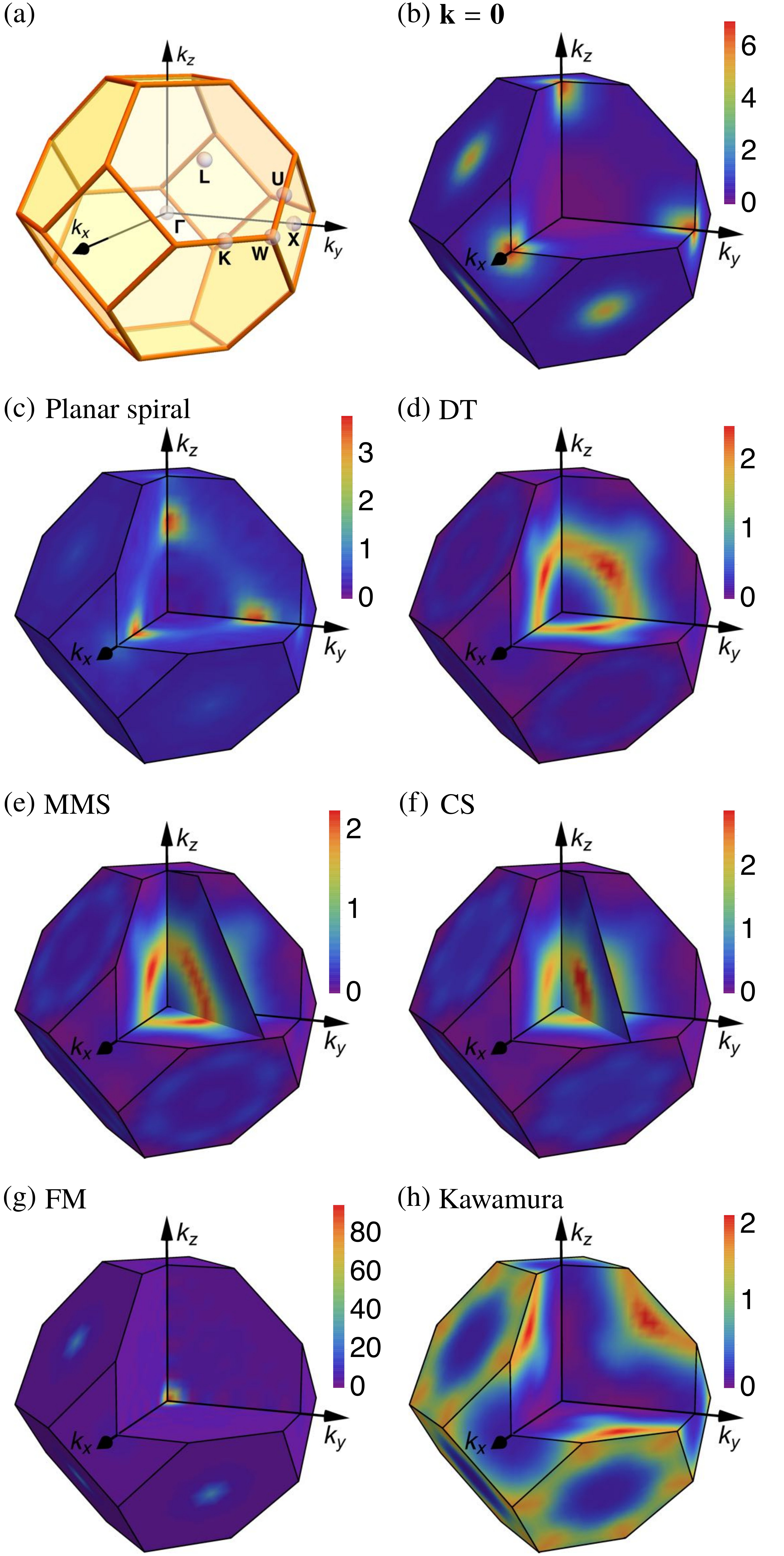}
\caption{Representative reciprocal-space-resolved magnetic susceptibility profiles (in units of $1/J$) for different magnetic orders evaluated at the data points marked by black dots in the $S=1/2$ quantum phase diagram in Fig.~\ref{fig:PhaseDiag_J1J2}(a). Also shown, the Brillouin zone, a ``truncated octahedron,'' with the high-symmetry points labeled.}\label{fig:susc_MO}
\end{figure}

The regime of small spin $S$ in highly frustrated magnets harbors strong quantum fluctuations which display intriguing effects such as (i) melting magnetic orders to potentially realize a quantum spin liquid, (ii) fostering the birth of new kinds of magnetic orders, (iii) shifting the pitch vector of spiral magnetic states, and (iv) shifting the phase boundaries relative to that found for the same Hamiltonian in its classical $S\to\infty$ limit. With the aim of investigating these possibilities, we carry out a study of the quantum phase diagram of the $J_{1}$-$J_{2}$ Heisenberg pyrochlore model for low values of spin $S$, which, to the best of our knowledge, had not been performed before the present work. We first address the important question concerning the possibility of stabilizing a quantum paramagnetic phase in the presence of a $J_{2}$ coupling. At the classical level, and as discussed in the previous section, it is shown~\cite{Reimers-1991a} that the presence of an infinitesimal further neighbor $J_{2}$ coupling induces long-range magnetic order at low temperatures. However, strong quantum fluctuations in the small-$S$ regime may destabilize those classical magnetic orders. Therefore, the question arises, in what range of $J_2/|J_1|$, with either antiferromagnetic $J_1>0$ or possibly even ferromagnetic $J_1<0$, may a quantum-spin-liquid phase be potentially realized.

By employing PFFRG, we map out the full $J_{1}$-$J_{2}$ quantum phase diagram for $S=1/2$, $S=1$, and $S=3/2$, which is shown in Fig.~\ref{fig:PhaseDiag_J1J2}. Our most important finding, which is the main result of our work, is the presence of an extended quantum paramagnetic phase for the $S=1/2$ model [see Fig.~\ref{fig:PhaseDiag_J1J2}(a)] and, perhaps surprisingly, also for the $S=1$ model [see Fig.~\ref{fig:PhaseDiag_J1J2}(b)]. In Figs.~\ref{fig:PhaseDiag_J1J2}(a) and \ref{fig:PhaseDiag_J1J2}(b), quantum fluctuations are seen to melt away a significant portion (around $J_{2}=0$) of the classical domain of existence of the $\mathbf{k}=\mathbf{0}$ and Kawamura magnetic orders. For $S=1/2$, the paramagnet ranges from $-0.25(3)\leqslant J_{2}/J_{1}\leqslant 0.22(3)$, while, for $S=1$, its span is reduced by half to $-0.11(2)\leqslant J_{2}/J_{1}\leqslant 0.09(2)$ but remains nonetheless appreciable. For $S=1/2$, we show the representative RG flows within the paramagnetic regime for a point in the antiferromagnetic $J_{2}$ regime [Fig.~\ref{fig:J2-AF-FM}(a)] and one in the ferromagnetic $J_{2}$ regime [Fig.~\ref{fig:J2-AF-FM}(b)]. These display a smooth and monotonically increasing behavior with no signatures of a kink, pointing to the absence of magnetic long-range order. The paramagnetic character of the ground state also shows up in the spin susceptibility profile in the form of an absence of sharp maxima in the EBZ which would be a signature of incipient Bragg peaks (IBPs) marking the onset of magnetic long-range order, along with a diffuse spectral weight caused by quantum fluctuations. Indeed, the antiferromagnetic $J_{2}$ spin susceptibility profile [see Fig.~\ref{fig:J2-AF-FM}(c) for the $S=1/2$ result] displays weak maxima at $\mathbf{k}=2\pi(2,0,0)$ (and symmetry-related points), which correspond to the dominant Bragg peak wave vectors of the underlying $\mathbf{k}=\mathbf{0}$ parent classical magnetic order (see Sec.~\ref{sec:J1J2-classical}). Similarly, the spin susceptibility profile for ferromagnetic $J_{2}$ [see Fig.~\ref{fig:J2-AF-FM}(d) for the $S=1/2$ result] features a smeared distribution of spectral weight forming homogeneous ringlike features on the surface of the Brillouin zone (see Fig.~\ref{fig:bowtie-2} for the $[hhl]$ plane scattering profiles). Classically, this parameter regime hosts the Kawamura magnetic order with dominant and subdominant Bragg peaks at $\mathbf{k}\approx2\pi(5/4,5/4,0)$ and $\mathbf{k}\approx2\pi(3/4,3/4,0)$ (and symmetry-related points). A comparison of the $S=1/2$ paramagnetic spin susceptibility profiles, i.e., Fig.~\ref{fig:J2-AF-FM}(c) for $J_{2}$ antiferromagnetic and Fig.~\ref{fig:J2-AF-FM}(d) for $J_{2}$ ferromagnetic, with those of the respective parent classical magnetic orders, i.e., $\mathbf{k}=\mathbf{0}$ [Fig.~\ref{fig:susc_MO}(b)] and Kawamura [Fig.~\ref{fig:susc_MO}(h)] states, lends support to the view that the quantum paramagnetic ground state may be viewed as a molten version of the parent magnetic orders under the action of quantum fluctuations.

\begin{figure}[t]
\includegraphics[width=\columnwidth]{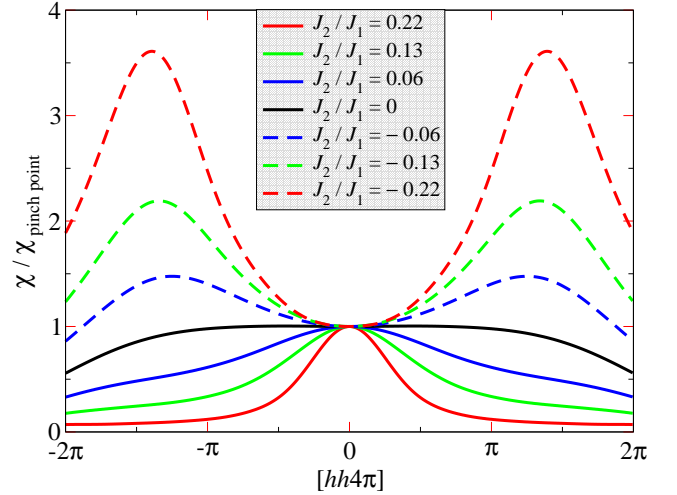}
\caption{For the $S=1/2$ $J_{1}$-$J_{2}$ model, the susceptibility plotted along the $[hh4\pi]$ cut (white line in Fig.~\ref{fig:bowtie-2}) evaluated at $T/J=1/100$ for different $J_{2}$.}\label{fig:hh4pi-cut}
\end{figure}

\begin{figure}[b]
\includegraphics[width=\columnwidth]{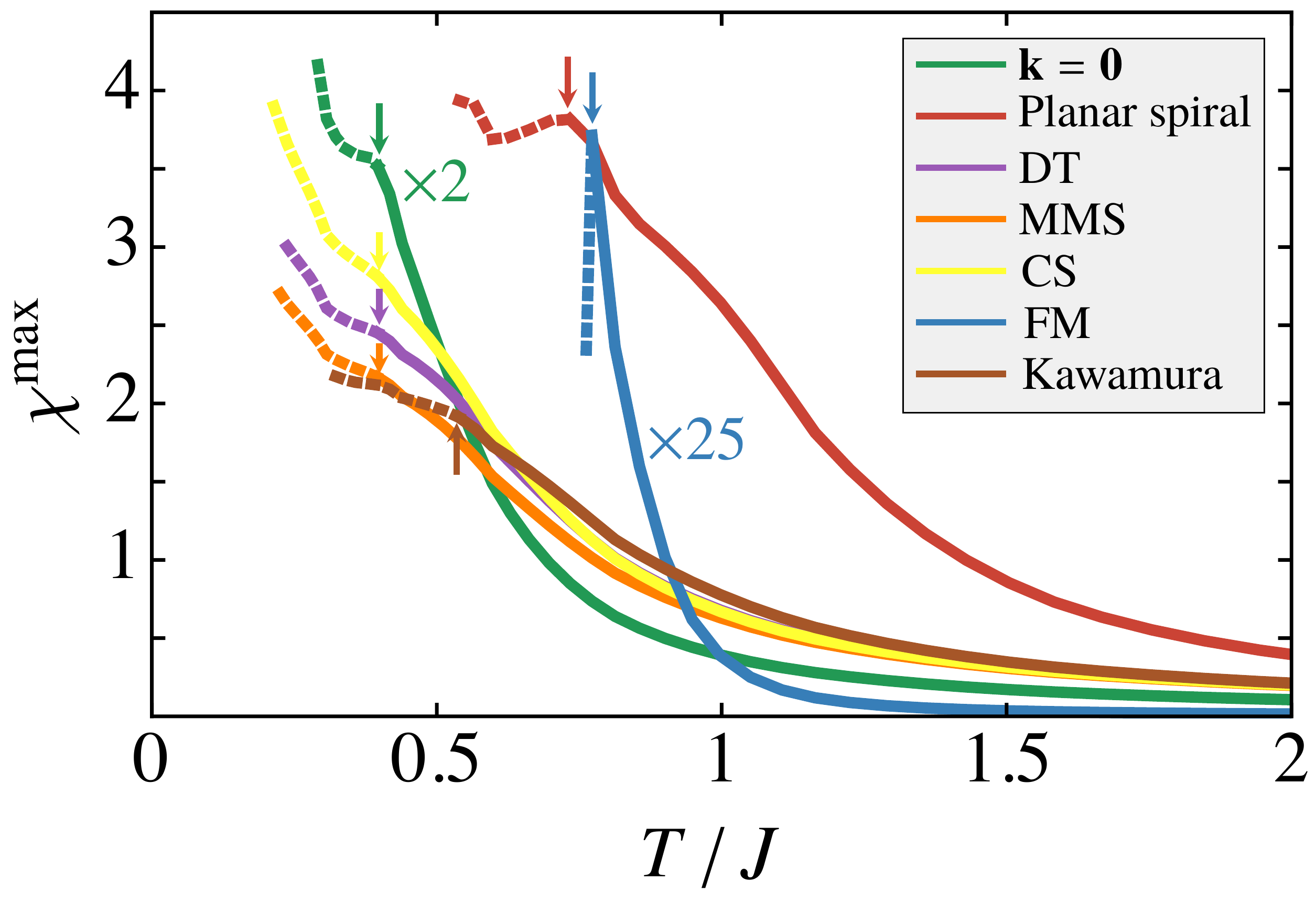}
\caption{RG flows of the spin susceptibility for $S=1/2$ at the ordering wave vectors of the seven magnetically ordered phases evaluated at the data points marked by black disks in Fig.~\ref{fig:PhaseDiag_J1J2}(a). The points at which the solid lines become dashed (marked by arrows) indicate an instability in the flow, indicating an onset of magnetic order.}\label{fig:RGFlow_MO}
\end{figure}

The inclusion of a $J_{2}$ coupling also substantially modifies the nature of the paramagnetic scattering profile at low temperatures (see Fig.~\ref{fig:bowtie-2} for the $[hhl]$ plane scattering). We find that for antiferromagnetic $J_{2}>0$ there is an enhancement of the pinch-point scattering amplitude as found in the corresponding classical model~\cite{Conlon-2010}, while for ferromagnetic $J_{2}$ the scattering intensity at the pinch points is strongly suppressed and instead redistributes to form a hexagonal cluster pattern of scattering~\cite{Conlon-2010}. In Fig.~\ref{fig:hh4pi-cut}, we plot the relative weight of the susceptibility (at $T/J=1/100$) with respect to its value at the pinch point, i.e., ($\chi/\chi_{\rm pinch~point}$) along a $1$D cut (marked by a white line in Fig.~\ref{fig:bowtie-2}). This plot clearly reveals the degree of enhancement at the pinch point as an antiferromagnetic $J_{2}$ coupling is cranked up, while, for ferromagnetic $J_{2}$, we see clearly the drifting of the maxima of susceptibility away from the pinch point and its enhancement at the wave vectors of the Kawamura state. The overall structure of the paramagnetic scattering profile is seen to be robust up to high temperatures $T/J\sim1$ [see Fig.~\ref{fig:bowtie-2}]. Although the above results and discussions are for the quantum paramagnet in the $S=1/2$ model, the findings for the $S=1$ model differ only quantitatively, and the entire discussion for $S=1/2$ holds true for $S=1$, albeit for the smaller collective paramagnetic regime of the $S=1$ model. 

\begin{figure}
\includegraphics[width=\columnwidth]{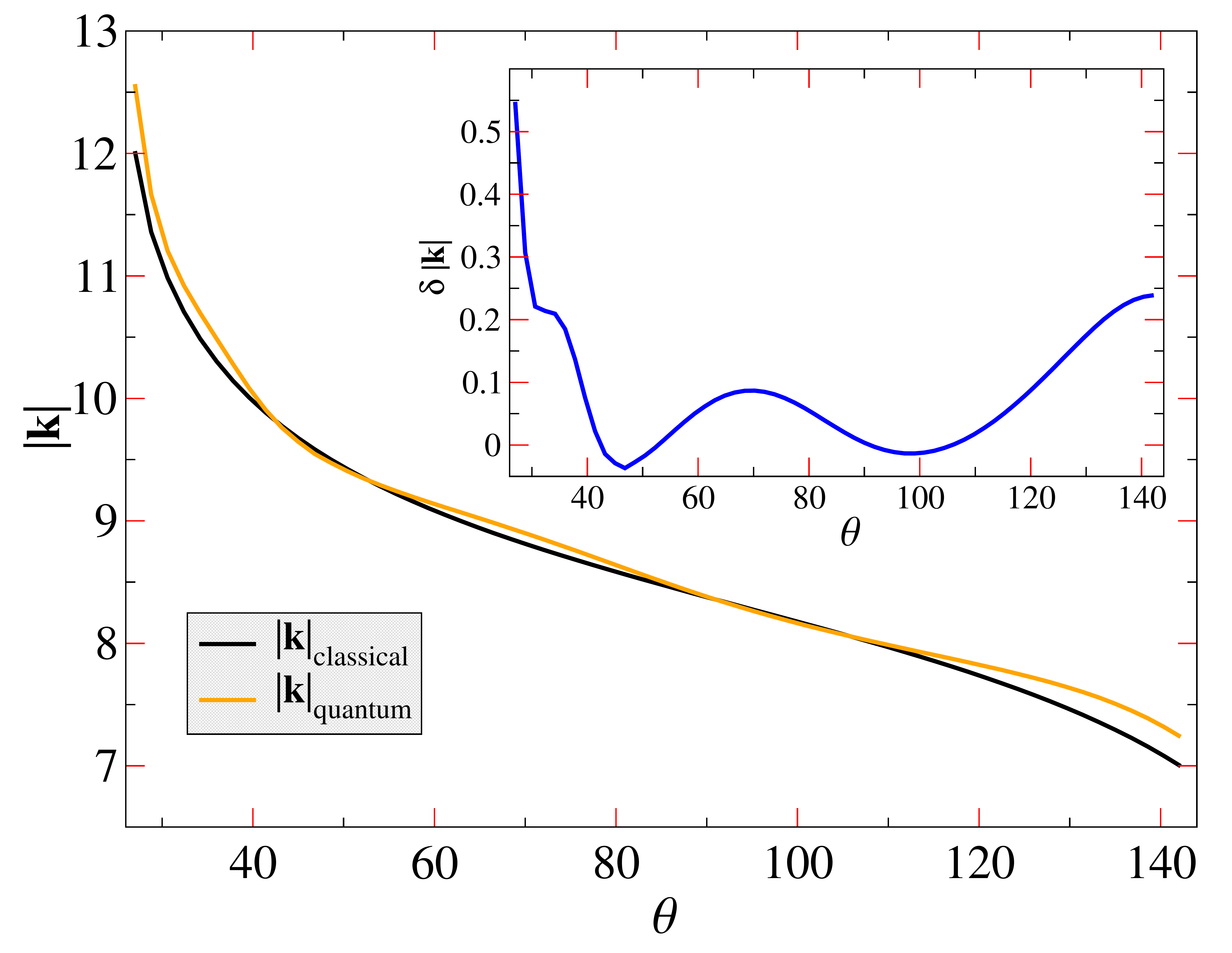}
\caption{The behavior of the classical $|\mathbf{k}_{\rm cl}|$ and quantum $|\mathbf{k}_{\rm qu}|$ ordering wave vectors as a function of $\theta=\arctan(J_{2}/J_{1}$). Inset: Deviation $|\delta\mathbf{k}|=|\mathbf{k}_{\rm qu}|-|\mathbf{k}_{\rm cl}|$ of the ordering wave vector $\mathbf{k}$ from its classical value as a function of $\theta$.}\label{fig:shift}
\end{figure}

We now move on to the discussion of the magnetically ordered phases in the low-spin regime of the $J_{1}$-$J_{2}$ model. A comparison of the classical and quantum phase diagrams in Fig.~\ref{fig:PhaseDiag_J1J2} shows that {\it all} the classical magnetic orders are present in the low-spin regime of the model and that \emph{no new} magnetic orders are found to be stabilized by quantum fluctuations, as is found for the Heisenberg model on the square lattice~\cite{Sindzingre-2009}. Starting our discussion with the $\mathbf{k}=\mathbf{0}$ order, we find that its span is considerably diminished for the $S=1/2$ model [see Table~\ref{tab:states} for phase boundaries], due to the fact that it gives way to an extended spin liquid phase around the $J_{2}=0$ point. The RG flow of the dominant susceptibility evaluated in the middle of the $\mathbf{k}=\mathbf{0}$ phase [$J_{2}/J_{1} \approx 0.36$, marked by a black disk in Fig.~\ref{fig:PhaseDiag_J1J2}(a)] clearly shows signature of an instability [see Fig.~\ref{fig:RGFlow_MO}], indicating the onset of $\mathbf{k}=\mathbf{0}$ magnetic order with a N\'eel temperature of $T_{c}/J \approx 0.39(2)$ which is given by the position of the instability, marked by an arrow in Fig.~\ref{fig:RGFlow_MO}. The spin susceptibility profile evaluated for $J_{2}/J_{1}=0.36$ at the instability point is shown in Fig.~\ref{fig:susc_MO}(b), wherein one observes the dominant IBP at the high-symmetry $X$ points [Fig.~\ref{fig:susc_MO}(a)], i.e., $\mathbf{k}=2\pi(2,0,0)$ (and symmetry-related points), and the subdominant peaks at the $L$ points [Fig.~\ref{fig:susc_MO}(a)], i.e., $\mathbf{k}=2\pi(1,1,1)$, and symmetry-related points, are also seen to be clearly resolved. Although both thermal and quantum order from fluctuation effects (order by disorder) are in principle captured in our simulations~\cite{Iqbal-2018}, we cannot make a statement about the collinearity of the ground state, as the PFFRG in its current formulation does not allow for lattice symmetry breaking; i.e., all symmetry-related IBPs have the same height. As discussed in Sec.~\ref{sec:J1J2-classical}, classically, the collinear $\mathbf{k}=\mathbf{0}$ state is selected by thermal fluctuations~\cite{Chern-2008}, and quantum fluctuations are likely to select the same state~\cite{Henley-1989}.

The $\mathbf{k}=\mathbf{0}$ state undergoes a phase transition at $J_{2}/J_{1}=1/2$ to an incommensurate planar spiral magnetic order. The RG flow of the dominant susceptibility evaluated deep inside the spiral ordered phase [$J_{2}=1$, marked by a black disk in Fig.~\ref{fig:PhaseDiag_J1J2}(a)] features an instability at $T_{c}/J_{2}\approx 0.73(3)$ [marked by an arrow in Fig.~\ref{fig:RGFlow_MO}] pointing to the onset of magnetic order at this temperature. The corresponding spin susceptibility profile evaluated at the instability point is shown in Fig.~\ref{fig:susc_MO}(c). We find that the effect of quantum fluctuations on the planar spiral order is twofold: (i) it leads to a shift of the spiral wave vector compared to its classical value~\cite{Chubukov-1984} and (ii) is found to increase the region of stability of the planar spiral beyond its classical domain. First, concerning the shift in the spiral wave vector, we show in Fig.~\ref{fig:shift} its evolution across its domain of existence for the classical and the quantum models. The wave vector is found to decrease monotonically as one traverses the spiral domain starting from its boundary with the $\mathbf{k}=\mathbf{0}$ to the DT magnetic order. Meanwhile, the shift 
$|\delta\mathbf{k}|\equiv|\mathbf{k}_{\rm qu}|-|\mathbf{k}_{\rm cl}|$ from the classical
$\mathbf{k}_{\rm cl}$ wave vector to the quantum $\mathbf{k}_{\rm qu}$ wave vector changes nonmonotonically across the domain of the planar spiral ordered phase [see the inset in Fig.~\ref{fig:shift}]. For the most part of the spiral ordered regime, we find that quantum fluctuations always \emph{increase} the wave-vector value, leading to more antiferromagnetic types of order. The shift $\delta\mathbf{k}$ achieves a maximal value of approximately $4\%$ of the classical value near the boundary to the $\mathbf{k}$=$\mathbf{0}$ order. Second, concerning the increase in the region of stability of the planar spiral order, we find that there is a strong renormalization of the phase boundary of the planar spiral with the DT order, which gets shifted from its classical value of $J_{2}/J_{1} \approx -0.68$ to $J_{2}/J_{1} \approx -0.537(6)$ for the $S=1/2$ model [see Fig.~\ref{fig:PhaseDiag_J1J2}(a) and Table~\ref{tab:states}], implying a significant enhancement of the domain of existence of the planar spiral order. 

At $J_{2}/J_{1} = -0.537(6)$, the planar spiral gives way to the DT magnetic order, whose RG flow evaluated at $J_{2}/J_{1} \approx -0.43$ [marked by a black disk in Fig.~\ref{fig:PhaseDiag_J1J2}(a)] and tracked at the dominant wave vector becomes unstable at $T/J \approx 0.39(2)$ [marked by an arrow in Fig.~\ref{fig:RGFlow_MO}]. The corresponding spin susceptibility profile is shown in Fig.~\ref{fig:susc_MO}(d), wherein, besides the dominant one, the subdominant peaks are also clearly resolved. We find that the DT phase in the $S=1/2$ model occupies a similar extent in parameter space as in the classical model, albeit with displaced phase boundaries. As the ratio $J_{2}/J_{1}$ is lowered, we find that at $J_{2}/J_{1}=-0.347(2)$ the susceptibility at the ordering wave vectors of the MMS phase becomes stronger compared to that at the DT ordering wave vectors, and the MMS order is stabilized. However, the extent of the MMS phase in the $S=1/2$ model is reduced to approximately one-third of its classical extent and thus now occupies only a tiny sliver in parameter space. Just as in the classical model, the IBPs of the quantum model are still located at incommensurate wave vectors, which are, however, shifted compared to those  of the classical model. In Fig.~\ref{fig:RGFlow_MO}, we show the RG flow evaluated at the optimal quantum wave vectors for $J_{2}/J_{1} \approx -0.335$ [marked by a black disk in Fig.~\ref{fig:PhaseDiag_J1J2}(a)], which reveals the onset of magnetic order at a N\'eel temperature of $T_{c}/J=0.39(2)$. The associated spin susceptibility profile is shown in Fig.~\ref{fig:susc_MO}(e). At $J_{2}/J_{1}=-0.326(2)$, the MMS phase ends and the susceptibility at the CS order wave vectors becomes dominant. The CS phase for $S=1/2$ has an appreciable extent in parameter space comparable to the classical model but with shifted phase boundaries. The spin susceptibility profile evaluated for $J_{2}/J_{1}=-0.24$ [see Fig.~\ref{fig:susc_MO}(f)] shows that the dominant IBP is located along the line joining the origin and the high-symmetry $L$ point, and that the peak undergoes substantial smearing due to quantum fluctuations. The instability feature at $T/J=0.39(2)$ in the RG flow [Fig.~\ref{fig:RGFlow_MO}] appears feeble, possibly hinting at the ``weakness'' of the CS magnetic order. It is of interest to note that the analogous cuboctohedral kagome orders~\cite{Domenge-2005,Messio-2011} found in Heisenberg models with long-range interactions also display an extremely feeble signal of an instability in their RG flow~\cite{Suttner-2014,Iqbal-2015}. 

\floatsetup[table]{capposition=top}
\begin{table*}
\centering
\begin{tabular}{lllllll}
 \hline \hline
      \multicolumn{1}{l}{Method}
    & \multicolumn{1}{l}{Pyrochlore}
    & \multicolumn{1}{l}{Simple cubic}
    & \multicolumn{1}{l}{$T_{\rm c}^{\rm Pyro}/T_{\rm c}^{\rm SC}$}
    & \multicolumn{1}{l}{Pyrochlore}
    & \multicolumn{1}{l}{Simple cubic} 
    & \multicolumn{1}{l}{$T_{\rm c}^{\rm Pyro}/T_{\rm c}^{\rm SC}$} \\ \hline
    
\multirow{1}{*}{PFFRG} & $0.77(4)$ & $0.90(4)$ & $0.86$ & &  &      \\
\multirow{1}{*}{QMC/CMC} & $0.7182(3)$~\cite{Mueller-2017} & $0.839(1)$~\cite{Troyer-2004,Wessel-2010} & $0.86$ & $\num{1.31695(2)}$~\cite{Soldatov-2017} & $1.443$~\cite{Peczak-1991,Chen-1993} & $0.91$ \\ 
\multirow{1}{*}{HTE (Pad\'e)} & $0.724{-}0.754$~\cite{Mueller-2017} & $0.827$~\cite{Lohmann-2014} & $0.88$ & $1.316{-}1.396$~\cite{Mueller-2017} & $1.438$~\cite{Lohmann-2014} & $0.92$  \\ 
\multirow{1}{*}{RGM} & $0.778$~\cite{Mueller-2017} & $0.926$~\cite{Mueller-2017} & $0.84$ &$1.172$~\cite{Mueller-2017} & $1.330$~\cite{Mueller-2017}  & $0.88$ \\
\multirow{1}{*}{RPA} & $0.872$~\cite{Hutak-2018} & $0.989$~\cite{TyablikovBook} & $0.88$ &$$ & $$ & $$\\ 
\multirow{1}{*}{MFA~\footnote{We adopt the convention of single-counting of bonds in Eq.~\eqref{eqn:Ham1}, and thus employ the formula $T_{c}/|J_{1}|=\frac{1}{3}zS(S+1)$, where $z$ is the coordination number.}} & $3/2$ & $3/2$ & $1$ &$2$ & $2$ & $1$
\\ \hline \hline

\end{tabular}
\caption{The ordering (Curie) temperatures for the $S=1/2$ nearest-neighbor quantum Heisenberg ferromagnet (in units $T_{c}/|J_{1}|$) (columns $2$ and $3$) and its corresponding classical ($S\to\infty$) model (in units of $T_{c}/[|J_{1}|S(S+1)]$) (columns $5$ and $6$) on the pyrochlore and simple cubic lattices as obtained by PFFRG and compared with estimates obtained from quantum Monte Carlo (QMC), classical Monte Carlo (CMC), high-temperature expansion (HTE), rotation-invariant Green's function method (RGM), and random-phase-approximation (RPA). The fact that $T_{c}^{\rm pyro}/T_{c}^{\rm SC}<1$ can be attributed to finite-temperature frustration effects~\cite{Mueller-2017}. We also quote the result in the mean-field approximation (MFA), which is insensitive to the difference between the pyrochlore and simple-cubic lattice, since it depends only on the coordination number.}
\label{tab:t-curie}
\end{table*}

Finally, as we lower $J_{2}/J_{1}$ further, the ferromagnetic $J_{1}$ coupling becomes dominant enough to drive the system into a ferromagnetic ordered state which onsets at $J_{2}/J_{1}=-0.153(5)$. On comparison with the classical transition boundary at $J_{2}/J_{1} \approx -0.171$, we see that the antiferromagnetic CS order intrudes into a portion of the phase diagram
occupied by the ferromagnetic order at the classical level, as expected from general considerations~\cite{Nagaev-1984,Iqbal-2016c}. For the $J_{1}=-1$-only model [marked by a black disk in Fig.~\ref{fig:PhaseDiag_J1J2}(a)], we show the RG flow of the $\mathbf{k}=(0,0,0)$ susceptibility in Fig.~\ref{fig:RGFlow_MO}, wherein we observe a strong signal of an  instability. We obtain an estimate of the critical (Curie) temperature $T_{c}/|J_{1}|=0.77(4)$, which is equal within two error bars to the Quantum Monte Carlo value of $T/|J_{1}|=0.718$~\cite{Mueller-2017} [see Table~\ref{tab:t-curie} for a comparison with other methods]. In Table~\ref{tab:t-curie}, we also provide for a comparison the Curie temperatures of the simple cubic lattice which has the same coordination number $z=6$ as the pyrochlore lattice but is bipartite. It is of interest to observe that, for both the $S=1/2$ and classical ($S\to \infty$) models, the Curie temperature of the pyrochlore lattice is lower compared to the simple cubic lattice, a fact which can be attributed to \emph{finite temperature} frustration effects~\cite{Lohmann-2014,Schmalfuss-2005,Mueller-2015,Mueller-2017a,Mueller-2017}. 

The spin susceptibility profile [see Fig.~\ref{fig:susc_MO}(g)] also reveals the presence of subdominant IBPs at the $L$ point besides the dominant peak at the $\Gamma$ point. As expected, the ferromagnetic phase occupies an entire quadrant of the phase diagram spanning from the limit $J_{1}=-1$ till $J_{2}=-1$ and gets destabilized only when a significant antiferromagnetic $J_{1}$ coupling is added to the $J_{2}=-1$ ferromagnetic model. Our PFFRG calculations identify the value of $J_{2}/J_{1}=-1.252(5)$ when the ferromagnetic order gives way to the antiferromagnetic Kawamura state, whereas classically the transition occurs at $J_{2}/J_{1}\approxeq -1.09$. Herein, similar to the CS state, we observe that quantum fluctuations extend the region of stability of the antiferromagnetic Kawamura order at the cost of the ferromagnetic state~\cite{Nagaev-1984,Iqbal-2016c}. The optimal wave vectors of the Kawamura state evolve within the region it occupies in the phase diagram; however, their value remains close to $2\pi(5/4,5/4,0)$. In Fig.~\ref{fig:RGFlow_MO}, we show the RG flow of the susceptibility evaluated at the optimal wave vectors for $J_{2}/J_{1}=-0.634(4)$ [marked by a black disk in Fig.~\ref{fig:PhaseDiag_J1J2}(a)]. The signature of an instability is not very pronounced and appears to be located around $T_{c}/J=0.54(2)$. The corresponding spin susceptibility profile is shown in Fig.~\ref{fig:susc_MO}(h), wherein one observes that quantum fluctuations cause a significant diffusing of the spectral weight for both the dominant and subdominant IBPs~\cite{Iqbal-2017}. 

The quantum phase diagram for the $S=1$ model [see Fig.~\ref{fig:PhaseDiag_J1J2}(b)] appears qualitatively similar to the one for $S=1/2$, with the only differences being quantitative ones, such as the location of the phase boundaries, value of optimal wave vectors, etc. As we gradually increase the value of the spin $S$, we see that the quantum phase diagram starts going over into the classical one, as is already manifestly apparent for $S=3/2$ [see Fig.~\ref{fig:PhaseDiag_J1J2}(c)].

\section{Summary}
\label{sec:summary}

In this paper, we employed the PFFRG method to investigate the long-standing problem of the effects of quantum fluctuations on the pyrochlore lattice for generic spin $S$ in a Heisenberg model with nearest-neighbor $J_{1}$ and second-nearest-neighbor $J_{2}$ couplings. For the spin $S=1/2$ nearest-neighbor Heisenberg antiferromagnetic model with spatially isotropic couplings, we find a quantum paramagnetic ground state [Sec.~\ref{sec:isotropic_s12}]. The paramagnet appears robust against potential instabilities towards the formation of either a valence-bond crystal [Fig.~\ref{fig:RG-Flows-VBC}(a)] or spin-nematic order [Fig.~\ref{fig:SN}], thus providing evidence in support of a quantum-spin-liquid ground state. The reciprocal space susceptibility plotted in the $[hhl]$ plane displays the characteristic bow-tie pattern [Fig.~\ref{fig:J2=0}(d)]. However, the dynamic violation of the zero magnetization per tetrahedron constraint due to quantum fluctuations manifests itself as (i) a regularization or softening of the pinch-point amplitude which loses its singular character and (ii) the generation of a finite-correlation length $\xi$ which endows the pinch points with a finite width $\sim 1/\xi$ [Fig.~\ref{fig:bwVsS}]. The fact that the bow-tie structure of susceptibility appears intact indicates that the low-temperature phase of the $S=1/2$ nearest-neighbor Heisenberg antiferromagnet respects the ice rules to a good degree of accuracy. An increase in temperature is seen to be associated with an overall decrease in the scattering intensity, while the bow-tie pattern appears to be remarkably robust up till $T\sim J_{1}$ [Figs.~\ref{fig:bowtie-1} and \ref{fig:fwhm}], suggesting that the ice rules govern the physics over a surprisingly large temperature range. We find that, within a significant segment of this temperature range up till $T\sim J_{1}$, the width of the bow tie as measured by its full width at half maximum increases (approximately) linearly [Fig.~\ref{fig:fwhm}]. 

For the spin $S=1$ nearest-neighbor Heisenberg antiferromagnet with spatially isotropic couplings [Sec.~\ref{sec:isotropic_s1}], we find that, strikingly, the ground state remains magnetically disordered [Fig.~\ref{fig:higher-spin}(a)] with no instability towards dimerizing into a valence-bond-crystal structure [Fig.~\ref{fig:RG-Flows-VBC}(b)], pointing to the realization of a rare scenario of a $S=1$ quantum spin liquid in three dimensions. The formation of the bow-tie pattern of scattering now features relatively sharper pinch points, as seen by a decrease in their full width at half maximum compared to $S=1/2$ [Fig.~\ref{fig:bwVsS}]. This decrease is as expected, since with increasing spin, quantum fluctuations decrease in strength, and the ice rules are better fulfilled. We find that the bow-tie structure remains robust up till $T\sim J_{1}$, similar to what is observed for $S=1/2$.

In the presence of breathing anisotropy (of arbitrary strength) in the nearest-neighbor Heisenberg antiferromagnet, we find that, for both $S=1/2$ [Sec.~\ref{sec:bs12}] and $S=1$ [Sec.~\ref{sec:bs1}], the quantum paramagnetic nature of the ground state remains intact [Fig.~\ref{fig:RG-Flows-Breathing}]. The reciprocal space spin susceptibility profile is still characterized by bow ties and the associated ``rounded'' pinch points, whose width is found to remain essentially unchanged from the isotropic point down to the strongly anisotropic limit [Fig.~\ref{fig:bwba}]. Our results thus point to the presence of an enlarged region in parameter space over which the low-temperature physics is approximately governed by the ice rules.

For the nearest-neighbor isotropic Heisenberg antiferromagnetic model with spin $S>1$ [Sec.~\ref{sec:large-S}], we find that for $S=3/2$ and beyond long-range dipolar magnetic order finally sets in [see Figs.~\ref{fig:higher-spin} and \ref{fig:maxmethod}]. We mention that, for the finite $S$ values studied in our manuscript, the correct balance between leading $1/S$ terms and subleading contributions is already incorporated in the PFFRG [see Sec.~\ref{sec:FRGA}]. However, with increasing $S$, the PFFRG becomes numerically more challenging (and also more sensitive to errors), because it becomes progressively difficult to account for the proper interplay between (large) leading $1/S$ and (much smaller but still important) subleading terms in our numerical algorithm. For this reason, we applied the PFFRG only to moderate spin magnitudes smaller than eight and use plain RPA in the infinite $S$ limit [see Appendix~\ref{appendix1}]. Therefore, we are unable to unambiguously address the question of the nature of the ground state (presence or absence of long-range magnetic order) in the large-$S$ nearest-neighbor quantum Heisenberg antiferromagnet.

Upon inclusion of a $J_{2}$ coupling [Sec.~\ref{sec:J1J2}], the complete parameter space of the $J_{1}$-$J_{2}$ Heisenberg model is shown to host seven different kinds of magnetic orders in the classical model [Fig.~\ref{fig:PhaseDiag_J1J2-Classical}]. We have reported some corrections and/or amendments to previously known results~\cite{Lapa-2012} concerning the nature of the magnetic orders and the classical phase diagram [Table~\ref{tab:states}]. For low values of spin, i.e., $S=1/2$ and $S=1$, quantum fluctuations are shown to stabilize an extended domain of quantum-spin-liquid behavior centered around the point $J_{1}>0$ and $J_{2}=0$, i.e., the nearest-neighbor Heisenberg antiferromagnet [Fig.~\ref{fig:PhaseDiag_J1J2}]. For $S=1/2$, the quantum spin liquid ranges from $-0.25(3)\leqslant J_{2}/J_{1}\leqslant 0.22(3)$, while for $S=1$, its span is reduced by half to $-0.11(2)\leqslant J_{2}/J_{1}\leqslant 0.09(2)$ but remains nonetheless appreciable. The introduction of even a small $J_{2}$ coupling is seen to substantially modify the reciprocal space scattering profile at low temperatures such that the bow-tie structure becomes quickly obliviated accompanied by an enhancement (decrement) for antiferromagnetic (ferromagnetic) $J_{2}$ in the spectral weight at the wave vector ($\mathbf{k}=(0,0,4\pi)$) where the pinch-point did exist [see Fig.~\ref{fig:bowtie-2}]. Indeed, we find that for antiferromagnetic $J_{2}>0$ there is an enhancement of the pinch-point scattering amplitude as found in the corresponding classical model~\cite{Conlon-2010} [Fig.~\ref{fig:bowtie-2} (first row) and Fig.~\ref{fig:hh4pi-cut}], while for ferromagnetic $J_{2}$ the scattering intensity at the pinch points is strongly suppressed and instead redistributes to form a hexagonal cluster pattern of scattering [Fig.~\ref{fig:bowtie-2} (second row) and Fig.~\ref{fig:hh4pi-cut}]~\cite{Conlon-2010}. Interestingly, we do not observe the stabilization of a paramagnetic phase by frustrating the nearest-neighbor Heisenberg ferromagnet, i.e., in the regime $J_{1}<0$ (FM) and $J_{2}>0$ (AFM). The phase boundaries between magnetically ordered phases get significantly modified compared to the classical model [Fig.~\ref{fig:PhaseDiag_J1J2}], and the wave vectors of spiral orders get shifted by quantum fluctuations [Fig.~\ref{fig:shift}]. Finally, we provide the N\'eel and Curie temperatures for different magnetically ordered phases, and for the $S=1/2$ nearest-neighbor Heisenberg ferromagnet we benchmark our PFFRG results with available numerically exact quantum Monte Carlo and other methods [Table~\ref{tab:t-curie}].

\section{Outlook and future directions}\label{sec:outlook}

Our analysis of quantum effects on the pyrochlore lattice lays new avenues towards further exploration in search of novel quantum phases in a more generic symmetry-allowed Hamiltonian~\cite{Curnoe-2007,Ross-2011,Onoda-2011,Petit-2016,Lee-2012,Onoda-2010} relevant for a large class of materials. Indeed, it has been shown at the classical level that anisotropic nearest-neighbor spin interactions can stabilize novel phases such as spin liquids and spin nematics and a plethora of intricate magnetic orders~\cite{Onoda-2011,Benton-2014,Petit-2016,Yan-2017,Taillefumier-2017,Essafi-2017}. The simplest extension to an $XXZ$ model has been argued to serve as a minimal model of quantum spin ice~\cite{Ross-2011} and has recently been shown to host spin nematic order and a variety of spin-liquid phases, albeit considered \emph{only} at the classical level~\cite{Taillefumier-2017,Essafi-2017}. Surprisingly, little is known about the role of quantum fluctuations beyond a perturbative treatment~\cite{Hermele-2004,Onoda-2011,Lee-2012,Savary-2012a,Savary-2012b,Savary-2013,Hao-2014,Fu-2017,Chen-2017}. In particular, the nature of the competing ordered or disordered quantum phases in the low spin-$S$ regime of the $XXZ$ model remain open questions, and it will be interesting to investigate if, and to what extent, the quantum-spin-liquid phase of the isotropic model~\cite{Huang-2016} found in this work remains stable in the presence of $XXZ$ anisotropy.

Our identification of extended regimes of quantum spin liquid and, in general, quantum paramagnetic behavior in the $S=1/2$ and $S=1$ models in the presence of breathing anisotropy or $J_{2}$ coupling sets the stage for future theoretical and numerical studies aimed at identifying the precise nature of the quantum-spin-liquid phase, e.g., gapped or gapless spin liquid, and its associated gauge structure, SU($2$), U$(1)$, $\mathbb{Z}_{2}$, etc. One promising approach would be to carry out a fermionic projective symmetry group (PSG) classification~\cite{Wen-2002,Wen-1991,Wen-1990} of the mean-field spin-liquid states on the pyrochlore lattice for both symmetric~\cite{Burnell-2009} and chiral spin liquids~\cite{Bieri-2016} similar to what has been accomplished on other lattices~\cite{Huang-2017,Huang-2018,Lu-2011,Lu-2016}. The ground-state energies of the corresponding projected variational wave functions could then be calculated from variational Monte Carlo methods~\cite{Yunoki-2006,BeccaBook}, enabling one to identify the most competitive variational ground state, which could then be improved by a subsequent application of Lanczos steps to obtain an estimate of the true ground-state energy~\cite{Iqbal-2013,Hu-2013,Iqbal-2014,Iqbal-2015b,Iqbal-2016a}. Recently, the PFFRG method has been successfully combined with a self-consistent Fock-like mean-field scheme to calculate low-energy effective theories for emergent spinon excitations in spin-1/2 quantum spin liquids~\cite{Hering-2018}. In this approach, the two particle vertices, i.e., the effective spin interactions from PFFRG, are taken as an input for the Fock equation yielding a self-consistent scheme to determine spinon band structures beyond mean field. The precise forms of
such free spinon Ans\"atze are dictated by a PSG classification of quantum spin liquids~\cite{Wen-2002}, allowing for a systematic investigation of kinetic spinon properties. It would be of interest and importance to apply this scheme to the pyrochlore Heisenberg antiferromagnet and compare the findings with those of variational Monte Carlo calculations. 
To address the issue of the nature of the elementary excitations and, in particular, to reveal the possible presence of a spinon continuum which is a manifestation of fractionalization and a hallmark of a quantum-spin-liquid phase, one needs a knowledge of the dynamical structure factor $S(q,\omega)$. The PFFRG framework can also be formulated directly in the real frequency domain employing the Keldysh formalism, which would allow one to obtain the complete $S(q,\omega)$. We leave the treatment of the Keldysh formalism and its application to the pyrochlore Heisenberg antiferromagnet as an important and exciting future endeavor.  

From a materials perspective, a fascinating class of transition-metal-based fluorides with the pyrochlore structure have recently come into the limelight. This family of materials is at the boundary between quantum spin liquid, magnetic order, and magnetic freezing (or glassy regime). Their importance stems from the availability of large high-quality single crystals. Prominent candidate spin-liquid examples include the $S=1$ {\na}~\cite{Krizan-2015}, which may be a first realization of a $S=1$ quantum spin liquid in three dimensions~\cite{Plumb-2017}, and the related higher-spin fluoride compounds featuring a high frustration index ($f=\Theta_{\rm CW}/T_{c}$), such as {\naco}~\cite{Krizan-2014,Ross-2017,Sarkar-2017}, {\nafe}, {\srfe}, and {\srmn}~\cite{Sanders-2016}, which, nonetheless, either show signs of long-range magnetic order at low temperatures or undergo spin freezing~\cite{Andrade-2017}. With the PFFRG formalism in place, it would be useful in such a material context to extend the mapping of the quantum phase diagram in the presence of longer-ranged Heisenberg couplings which will most likely give rise to additional novel phases compared to the seven phases of the classical $J_{1}$-$J_{2}$ Heisenberg model, as, for instance, shown in Ref.~\cite{Tymoshenko-2017} for classical spins. It would seem likely that most of the above-mentioned materials could be placed to a good degree of approximation in the extended phase diagram so determined. 

Given that frustrated quantum spin systems are challenging to deal with theoretically and, in three dimensions, pose a formidable barrier to most quantum many-body numerical methods, PFFRG is one of the very few methods that can be used to shed light on the physics at play in these systems, with the field now poised to benefit from the arrival of more materials. It is in this broader context that we investigated and presented in this paper the rich example of the $J_{1}$-$J_{2}$ Heisenberg model on the pyrochlore lattice.

\section{Acknowledgments}
Y. I. and R. T. thank F. Becca and S. Bieri for useful discussions. Y.I. acknowledges helpful discussions with J. Richter and thanks O. Derzhko for providing details of the pyrochlore ferromagnet QMC calculations. S.R. acknowledges discussions with D. Inosov, E. Andrade, J. Hoyos, and M. Vojta. The work was supported by the European Research Council through ERC-StG-TOPOLECTRICS-Thomale-336012. T.M. and R.T. thank the DFG (Deutsche Forschungsgemeinschaft) for financial support through SFB 1170 (project B04). J.R. is supported by the Freie Universit\"at Berlin within the Excellence Initiative of the German Research Foundation. S.R. acknowledges support from the DFG  through SFB 1143 and from an Australian Research Council Future Fellowship (FT180100211). The work at the University of Waterloo was supported by the Canada Research Chair program (M.G., tier 1) and by the Perimeter Institute (PI) for Theoretical Physics. Research at the Perimeter Institute is supported by the Government of Canada through Innovation, Science and Economic Development Canada and by the Province of Ontario through the Ministry of Research, Innovation and Science. Y.I. acknowledges the kind hospitality of the Helmholtz-Zentrum f\"ur Materialien und Energie, Berlin, where part of the work was carried out. We gratefully acknowledge the Gauss Centre for Supercomputing e.V. for funding this project by providing computing time on the GCS Supercomputer SuperMUC at Leibniz Supercomputing Centre (LRZ). 

\appendix

\section{Diagrammatic investigation of the nearest neighbor pyrochlore Heisenberg model in the large $S$ limit}\label{appendix1}

In this Appendix, we present further details about how the susceptibility of the nearest-neighbor pyrochlore Heisenberg model in the infinite-$S$ limit as depicted in Fig.~\ref{fig:ppw} is computed. Particularly, we explain why the simple RPA-type summation which we use to obtain these results reproduces the correct pinch-point singularity but also results in a spurious divergence of the susceptibility at a finite temperature $T$ which is not expected from the exact (numerical)  solution~\cite{Moessner-1998a,Reimers-1992}. We further present analytical arguments why the summation of further diagram classes can cure this artifact by regularizing the divergence.

\begin{figure}
\includegraphics[width=\columnwidth]{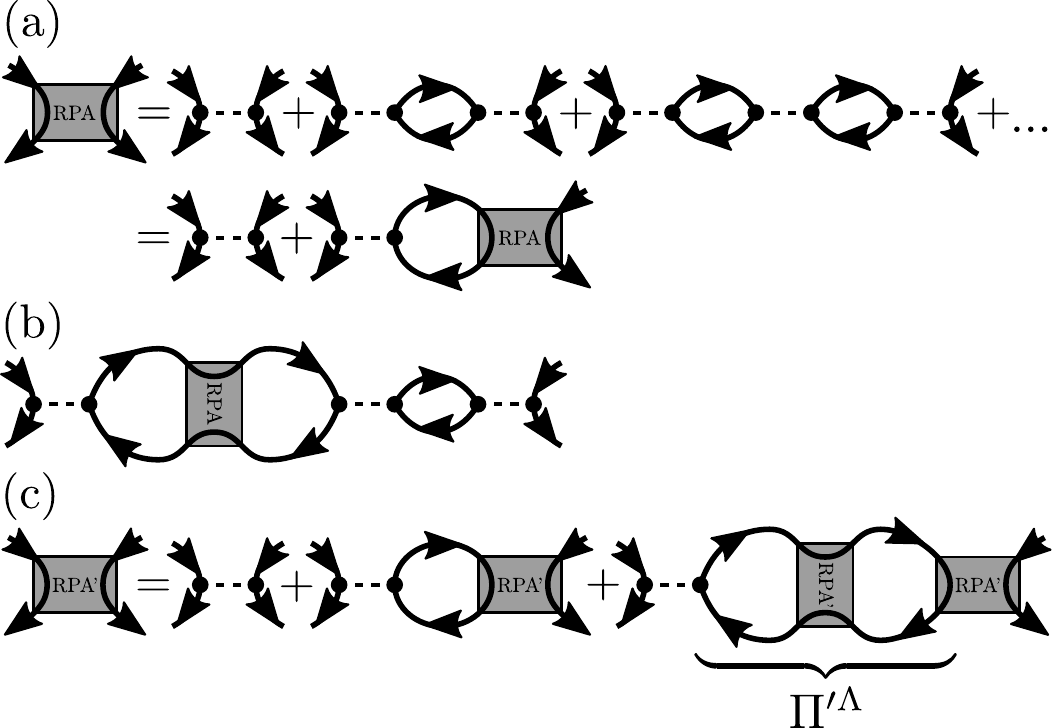}
\caption{RPA-type approximations for the two-particle vertex in the large-$S$ limit. Dashed lines are the bare interactions $J_{ij}$, and lines with an arrow are the bare and $\Lambda$-regularized pseudofermion propagators from Eq.~(\ref{eq:LambdaGF}). Gray boxes denote the two-particle vertex in different approximations. (a) Plain RPA scheme summing up diagrammatic terms of the order of $1/S$. (b) An example of a contribution to the two-particle vertex of the order of $(1/S)^2$. (c) Improved RPA scheme, RPA', regularizing the divergence of the two-particle vertex occurring in plain RPA. See the text for details.}
\label{fig:diagrams_rpa}
\end{figure}

We begin by reviewing the PFFRG scheme in the large-$S$ limit and explain that, to leading order when $S\rightarrow\infty$, the PFFRG becomes identical to a simple RPA-type approximation (for further details, see Ref.~\cite{Baez-2017}). As briefly mentioned in Sec.~\ref{sec:formalism}, the generalization of the PFFRG for arbitrary spin $S$ amounts to introducing fermion flavors $f_{i\uparrow\kappa}$, $f_{i\downarrow\kappa}$ with $\kappa=\{1,\ldots,2S\}$ on each lattice site $i$ which add up to a total spin $S$. Furthermore, to avoid diverging energy scales in the large-$S$ limit, it is convenient to renormalize all interactions via $J_{ij}\rightarrow J_{ij}/(2S)$. As a consequence of the additional flavor index $\kappa$, the Feynman diagrams acquire an extra factor of $2S$ for each closed fermion loop. Hence, when formulating the PFFRG equations for arbitrary $S$, the second term on the right-hand side in Fig.~\ref{fig:frg}(b) (the so-called RPA channel) acquires a prefactor of $2S$, indicating that, among all interaction channels in Fig.~\ref{fig:frg}(b), this term is singled out at large $S$. The flow equation for the two-particle vertex at $S\rightarrow\infty$, where only the RPA term contributes on the right-hand side, can be readily solved~\cite{Baez-2017} and leads to the RPA-type diagram series shown in Fig.~\ref{fig:diagrams_rpa}(a). These two-particle vertex diagrams are precisely the ones, and no others, of leading order in $1/S$. This result is evident from the fact that, for a given number of interaction lines, they each maximize the number of loops. Specifically, each term of the series has $n$ bare interaction lines and $n-1$ fermion loops, resulting in an overall order of $1/S$.

Having established that, to leading order in $1/S$, the PFFRG generically reduces to an RPA-type approximation, we now study the structure of this approximation in the context of the nearest-neighbor pyrochlore Heisenberg model. In the following, we are interested only in the static frequency components ($\omega=0$) of the two-particle vertex $\Gamma^\Lambda(1',2';1,2)$. We thus omit the arguments $1'$, $2'$, $\ldots$ and write $\Gamma^\Lambda(1',2';1,2)\rightarrow\Gamma_{ij}^\Lambda$ with the site indices $i,j$ as subscripts. Furthermore, the propagators considered are the bare (i.e., without self-energy corrections) and $\Lambda$-regularized ones from Eq.~(\ref{eq:LambdaGF}). The RPA diagram series may be expressed in a self-consistent form [second line of Fig.~\ref{fig:diagrams_rpa}(a)], leading to
\begin{equation}
\Gamma_{ij}^\Lambda=-\frac{J_{ij}}{2S}-\sum_{l}\frac{J_{il}}{2S}\Pi^\Lambda\Gamma_{lj}^\Lambda.\label{rpa}
\end{equation}
Here, $\Pi^\Lambda$ is the $\omega=0$ component of the bare fermion loop and is given by $\Pi^\Lambda=S/(\pi\Lambda)$. The solution of Eq.~\eqref{rpa} can be obtained via a Fourier transform, giving 
\begin{equation}
\tilde{\Gamma}^\Lambda({\mathbf k})=-[\Pi^\Lambda\mathds{1}+2S\tilde{J}^{-1}({\mathbf k})]^{-1}\label{rpa2},
\end{equation}
where $\tilde{J}({\mathbf k})$ is the interaction matrix in sublattice space as given by Eq.~\eqref{eqn:ltmatrix}. $\tilde{\Gamma}^\Lambda({\mathbf k})$ is also analogously defined in sublattice space, and $\mathds{1}$ denotes the identity matrix in the same space. To better understand the physical implications of Eq.~\eqref{rpa2}, we diagonalize $\tilde{J}({\mathbf k})$ via $M^\dagger({\mathbf k})\tilde{J}({\mathbf k})M({\mathbf k})=\tilde{J}_{d}({\mathbf k})$, where $M({\mathbf k})$ is a unitary matrix and $\tilde{J}_{d}({\mathbf k})$ is a diagonal matrix whose elements are the eigenvalues of $\tilde{J}({\mathbf k})$. It follows that
\begin{equation}
\tilde{\Gamma}^\Lambda({\mathbf k})=-M({\mathbf k})[\Pi^\Lambda\mathds{1}+2S\tilde{J}_{d}^{-1}({\mathbf k})]^{-1}M^\dagger({\mathbf k}).\label{gamma}
\end{equation}
For the nearest-neighbor pyrochlore Heisenberg antiferromagnetic model, the lowest bands of $\tilde{J}_{d}({\mathbf k})$ take the form of two degenerate flat modes with an energy $-2J_1$~\cite{Bertaut-1961,Reimers-1991a}. As a result of these flat modes, the matrix $\Pi^\Lambda\mathds{1}+2S\tilde{J}_{d}^{-1}({\mathbf k})$ in Eq.~\eqref{gamma} becomes singular at $\Lambda=J_1/\pi$ for {\it all} wave vectors ${\mathbf k}$, which leads to a diverging susceptibility at the corresponding (finite) temperature $T=\frac{2\pi}{3}S(S+1)\Lambda$. However, as explained further below, this divergence is a methodological artifact of the plain RPA treatment within which \emph{only} the leading $1/S$ diagrammatic contributions are considered.  

The flat modes in $\tilde{J}_{d}({\mathbf k})$ are also responsible for the pinch-point singularities in the susceptibility~\cite{Isakov-2004}. To see this, we first note that (up to irrelevant overall factors from fusing external fermion lines) the susceptibility $\chi^\Lambda({\mathbf{k}})$ of Eqs.~(\ref{correlator}) and (\ref{eqn:suscep}), rewritten in sublattice coordinates, is related to the two-particle vertex $\tilde{\Gamma}^\Lambda({\mathbf k})$ via
\begin{equation}
\chi^\Lambda({\mathbf k})\sim\sum_{\alpha\beta}e^{i{\mathbf k}(\bm{\xi}_\alpha-\bm{\xi}_\beta)}\tilde{\Gamma}^\Lambda_{\alpha\beta}({\mathbf k}).\label{chi}
\end{equation}
Here, $\alpha$, $\beta$ are sublattice indices and $\bm{\xi}_\alpha$ denote the sublattice displacements, i.e., site coordinates ${\mathbf r}_i$, unit cell coordinates ${\mathbf R}_i$, and displacements $\bm{\xi}_\alpha$ fulfilling ${\mathbf r}_i={\mathbf R}_i+\bm{\xi}_\alpha$. Since the lowest (flat) modes give the dominant contribution to the susceptibility and also describe the physics of pinch points we are interested in, we may approximate Eq.~(\ref{gamma}) by neglecting higher-energy bands in $\tilde{J}_{d}({\mathbf k})$. Using Eqs.~\eqref{gamma} and \eqref{chi}, one then obtains  
\begin{equation}
\chi^\Lambda({\mathbf k})\sim\frac{\sum_{\alpha\beta}\sum_{\gamma=\text{fm}}e^{i{\mathbf k}(\bm{\xi}_\alpha-\bm{\xi}_\beta)}M_{\alpha\gamma}({\mathbf k})M^\dagger_{\gamma\beta}({\mathbf k})}{\frac{S}{\pi\Lambda}-\frac{S}{J_1}},\label{chi2}
\end{equation}
where $\gamma=\text{fm}$ only sums over the flat modes (fm). The numerator in this expression (which is  used to plot the inset in Fig.~\ref{fig:ppw}) contains the pinch-point pattern, while the denominator produces the aforementioned singularity at finite $\Lambda$. This analysis shows that in plain RPA, as obtained from PFFRG in leading order in $1/S$,  the pinch points are correctly reproduced. However, their manifestation within this plain RPA scheme is implicitly tied with a divergence of the $\mathbf{k}$-dependent susceptibility for all $\mathbf{k}$ that define the flat modes. Thus, the physically correct paramagnetic (broadened) pinch points observed in plain RPA exist only above the instability, and so their discussion in plain RPA is bounded from below by the instability at $\Lambda=J_{1}/\pi$.

We now investigate how Eq.~\eqref{chi2} is modified when adding diagrams of order higher than $1/S$. Within PFFRG, such higher orders are generally described by the other interaction channels on the right-hand side in Fig.~\ref{fig:frg}(b), i.e., those corrections to RPA which do not contain a fermion loop. In contrast to the leading order in $1/S$ discussed above, where all diagrammatic contributions to the two-particle vertex are exactly included in the PFFRG, higher orders are treated only approximately. A thorough analytical discussion of all subleading diagrams \emph{implicitly included} within the PFFRG computational scheme is, admittedly, very challenging, because, already to the order of $(1/S)^2$, they may not be represented by a simple series of diagrams such as the one shown in Fig.~\ref{fig:diagrams_rpa}(a). Furthermore, from a more technical perspective, it is a rather involved task to apply the PFFRG at large but finite $S$ and systematically explore the effects of different diagrammatic orders in $1/S$. This hurdle arises because of numerical difficulties in capturing the subtle competition between large leading $1/S$ and much smaller, but still important and possibly singular subleading terms, when the frequency dependence of the vertex functions is approximated by a finite grid (which is a computational necessity within PFFRG).

To still be able to investigate general properties of higher diagrammatic orders in $1/S$, we, therefore, use a different strategy. We take as a starting point the $S\to\infty$ limit (as described above) and then incorporate ``by hand'' subleading diagrams to study their effects on the spurious divergence encountered in a plain RPA treatment. Subleading diagrams of the order of $(1/S)^2$ are obtained by feeding back the RPA two-particle vertex into a fermion loop of the RPA series as shown in Fig.~\ref{fig:diagrams_rpa}(b). In the following, we discuss a generalization of such terms (dubbed RPA$^{\prime}$) where (i) the feedback of the RPA takes place in {\it every} fermion loop and (ii) the insertion is performed self-consistently as shown in Fig.~\ref{fig:diagrams_rpa}(c). The resummation of such diagram classes also involves contributions from orders higher than $(1/S)^2$. This type of approximation first amounts to replacing the bare fermion loop $\Pi^\Lambda$ by $\Pi^\Lambda+\Pi'^\Lambda$, where $\Pi'^\Lambda$ is the loop diagram with the RPA series reinserted as depicted in Fig.~\ref{fig:diagrams_rpa}(c). Using the fact that only the {\it local} two-particle vertex $\Gamma^\Lambda_{ii}$ contributes to this diagram, one finds
\begin{equation}
\Pi'^\Lambda=\frac{S}{4\pi\Lambda^2}\Gamma^\Lambda_{ii}=\frac{S}{4\pi\Lambda^2}\frac{1}{(2\pi)^3}\int_{\text{BZ}}d^3k\tilde{\Gamma}^\Lambda_{11}({\mathbf k}).
\end{equation}  

Without the loss of generality, we choose the ``11''-sublattice component of the two-particle vertex, since all sublattices are equivalent in the paramagnetic regime. Also note that, in order for the calculation to be analytically tractable, we perform a static approximation where the two-particle vertex is assumed to be $\omega$ independent. The self-consistency for $\Pi'^\Lambda$ is closed using Eq.~(\ref{gamma}) and replacing $\Pi^\Lambda\rightarrow\Pi^\Lambda+\Pi'^\Lambda$, giving
\begin{equation}
\tilde{\Gamma}^\Lambda_{11}({\mathbf k})=\left\{-M({\mathbf k})\left[\Pi^\Lambda\mathds{1}+\Pi'^\Lambda\mathds{1}+2SJ_{d}^{-1}({\mathbf k})\right]^{-1}M^\dagger({\mathbf k})\right\}_{11}.
\end{equation}
Here again, we consider only the contribution from the flat modes in $J_{d}({\mathbf k})$ and neglect higher-energy bands. Furthermore, we write the momentum integral (which is a {\it positive} dimensionless number) as
\begin{equation}
x\equiv\frac{1}{(2\pi)^3}\int_{\text{BZ}}d^3k \sum_{\gamma=\text{fm}}M_{1\gamma}({\mathbf k})M_{\gamma 1}^\dagger({\mathbf k})
\end{equation}
and again use $\Pi^\Lambda=S/(\pi\Lambda)$, leading to
\begin{equation}
\Pi'^\Lambda=-\frac{Sx}{4\pi\Lambda^2}\frac{1}{\frac{S}{\pi\Lambda}+\Pi'^\Lambda-\frac{S}{J_1}}.\label{quadratic}
\end{equation}
This is a quadratic equation for $\Pi'^\Lambda$ which can be solved to yield the susceptibility
\begin{equation}
\chi^\Lambda({\mathbf k})\sim\frac{\sum_{\alpha\beta}\sum_{\gamma=\text{fm}}e^{i{\mathbf k}(\bm{\xi}_\alpha-\bm{\xi}_\beta)}M_{\alpha\gamma}({\mathbf k})M^\dagger_{\gamma\beta}({\mathbf k})}{\frac{S}{\pi\Lambda}+\Pi'^\Lambda-\frac{S}{J_1}},
\end{equation}
where, when compared to Eq.~(\ref{chi2}), an additional contribution from $\Pi'^\Lambda$ appears in the denominator. From the two solutions for $\Pi'^\Lambda$ following from Eq. (\ref{quadratic}), the correct one is identified by the condition that the leading order at large $S$ must be a contribution $\sim 1/S$ as is the case for the bare RPA. One then obtains
\begin{align}
&\chi^\Lambda({\mathbf k})\sim\sum_{\alpha\beta}\sum_{\gamma=\text{fm}}e^{-i{\mathbf k}(\bm{\xi}_\alpha-\bm{\xi}_\beta)}M_{\alpha\gamma}({\mathbf k})M^\dagger_{\gamma\beta}({\mathbf k})\notag\\
&\times\frac{2\Lambda}{x}\left[\left(\frac{\pi\Lambda}{J_1}-1\right)-\text{sgn}\left(\frac{\pi\Lambda}{J_1}-1\right)\right.\notag\\
&\times\left.\sqrt{\left(\frac{\pi\Lambda}{J_1}-1\right)^2-\frac{\pi x}{S}}\right].\label{chi_result}
\end{align}

\begin{figure}
\includegraphics[width=\columnwidth]{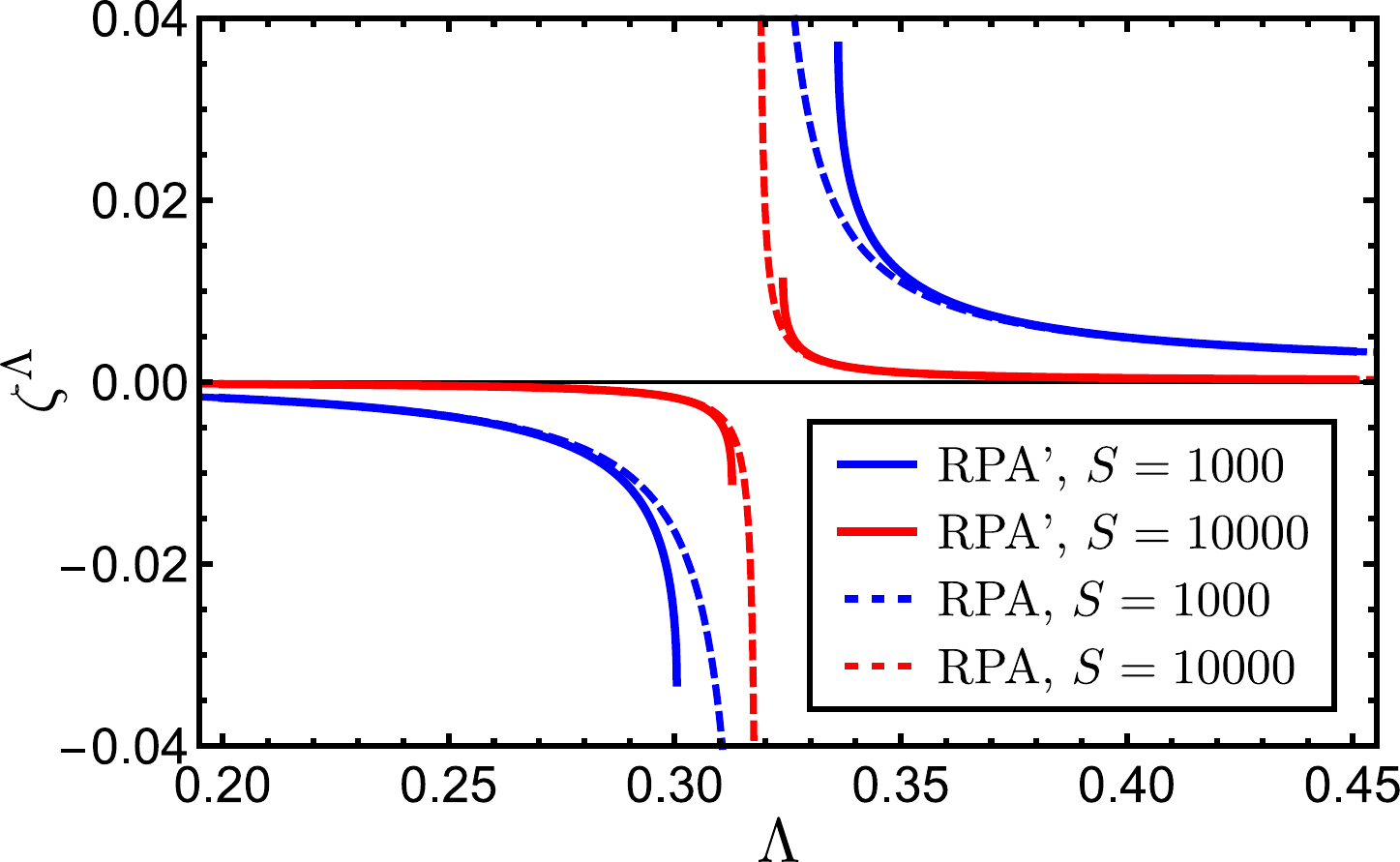}
\caption{Plot of the $\Lambda$ flow of $\zeta^{\Lambda}$ which, in the case of (i) RPA, refers to the denominator of Eq.~(\ref{chi2}) and, in the case of (ii) RPA$^{\prime}$, refers to the $\mathbf{k}$-independent expression in the second and third line of Eq.~\eqref{chi_result} with $J_1=x=1$. Blue and red curves denote spin $S=1000$ and $S=10000$, respectively. The divergence in RPA at $\Lambda=J_1/\pi$ is regularized in the RPA$^{\prime}$ scheme. No data are plotted in the interval where the susceptibility becomes imaginary.}
\label{fig:plot_rpa}
\end{figure}

Most importantly, this expression \emph{no longer} has a divergence in $\Lambda$ while the pinch-point pattern given by the ${\mathbf k}$-dependent term [first line of Eq.~(\ref{chi_result}) and numerator of Eq.~(\ref{chi2}) which generate the pinch points] persists. The $\Lambda$-dependent second and third line of Eq.~(\ref{chi_result}) is plotted in Fig.~\ref{fig:plot_rpa} for $S=1000$ and $S=\num{10000}$. It can be seen that the diverging susceptibility of the RPA scheme is regularized by the higher-order terms such that $\chi^\Lambda({\mathbf k})$ becomes bounded in the vicinity of the singularity. Yet, certain artifacts still remain in the RPA$^{\prime}$ scheme such as a steplike behavior of the susceptibility and a finite interval where $\chi^\Lambda({\mathbf k})$ becomes imaginary (the size of this interval shrinks with increasing $S$). We expect that such spurious behavior would become further regularized upon including more diagrammatic contributions.

In summary, even though this analysis is based on an approximate resummation of a certain class of diagrams, it demonstrates that higher-order terms have a significant effect even in the large-$S$ limit and may counteract the diverging susceptibility observed in the bare RPA calculation leading to Eq.~(\ref{chi2}). This calculation also shows that\textemdash even though counterintuitive at first sight\textemdash leading $1/S$ diagrams are not sufficient to treat the classical limit $S\to\infty$ exactly. One may, therefore, conclude that, while the spatial structure of the spin correlations at large $S$ is already correctly described by plain RPA, thermal fluctuations are much more intricate in pseudofermionic formulation. This conclusion may possibly indicate that pseudofermions are not ideally suited to describe the thermodynamics of spin systems in the classical large-$S$ limit. We also emphasize, however, that such methodological subtleties do not affect the PFFRG at finite (but not too large) $S$, where the correct balance between classical magnetic phenomena and quantum fluctuations is captured by the interplay between leading $1/S$ and leading $1/N$ diagrammatic contributions [where $N$ generalizes the spin symmetry group from SU$(2)$ to SU$(N)$; see Sec.~\ref{sec:formalism} for details].

\section{Detecting a magnetic instability in the RG flow}\label{appendix2}

\begin{figure*}
\includegraphics[width=\columnwidth]{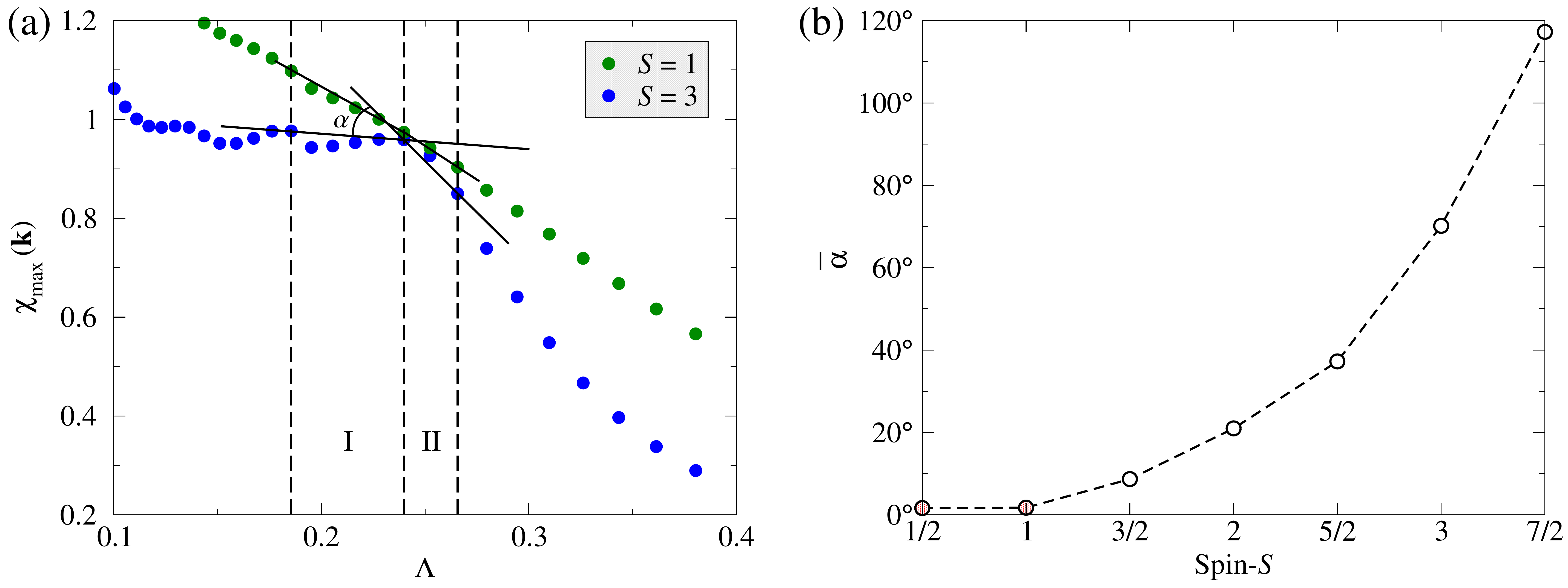}
\caption{(a) Illustration of the scheme for determining kinks in the $\Lambda$-dependent susceptibility flow: One divides a fixed $\Lambda$ interval into two regions, I and II, which by construction lie between two adjacent pairs of susceptibility kinks or peaks. Within each region, $\chi(\mathbf{k})$ is approximated by a tangent connecting the neighboring kinks. We find that, while the angle between the two tangents is negligible for $S=1$, it acquires a sizable finite value for $S=3$, implying that two curves represent different phases. To obtain a more quantitatively robust measure for the size of the kink, we consider additional pairs of adjacent peaks, compute the angles between their tangents, and calculate the average angle $\bar{\alpha}$. (b) Averaged angle $\bar{\alpha}$ as a function of spin $S$. While $\bar{\alpha}$ is negligible and almost constant for $S=1/2$ and $S=1$ (labeled by filled circles), there is a pronounced increase for higher values of spin $S\geqslant 3/2$ (empty circles). Based on this behavior, we estimate the phase transition of the spin-$S$ nearest-neighbor Heisenberg antiferromagnet to occur for $S=3/2$.}
\label{fig:maxmethod}
\end{figure*}

Here, we present the details of the numerical procedure~\cite{Hering-2017} used to detect the onset of long-range magnetic order in the RG flow. The expected divergence of the spin susceptibility [Eq.~(\ref{eqn:suscep})] at a critical $\Lambda$ which would signal the spontaneous breaking of SU(2) spin-rotation symmetry towards long-range dipolar magnetic order is, in practice, regularized due to two numerical approximations in the PFFRG method: (i) the discretization of the frequencies in the arguments of the vertex functions and (ii) the finite spatial extent of the two-particle vertex function. Both these approximations regularize the divergence to a finite maximum, or a feeble kinklike feature when the ordered magnetic moment is small. In addition, the discretization of the frequencies induces the artifact of oscillations in the susceptibility flow, especially at small $\Lambda$. The distinct advantage of the method presented here lies in its ability to detect such kinks even in the presence of pronounced frequency oscillations and a small ordered magnetic moment. To illustrate the method, we focus on the transition with increasing spin $S$, from the paramagnetic into the magnetically ordered phase, for the nearest-neighbor pyrochlore Heisenberg antiferromagnet. 

The appearance of a finite maxima or a kinklike feature in the susceptibility evolution with decreasing $\Lambda$ is marked by a change in the slope of the RG flow. However, as the susceptibility flow is plagued by oscillations due to frequency discretization, one encounters a difficulty in defining the slope. As each discrete frequency grid point produces a small peak or an upturn in the susceptibility flow, we compute the slope in a manner that averages out these oscillations. To this effect, one connects two adjacent peaks via a straight line which represents a tangent of the susceptibility and approximates $\chi(\mathbf{k})$ between the two peaks. A kink in the RG flow now manifests as a change in the slope, i.e., a finite-angle $\alpha$, between the two neighboring tangents, as shown in Fig.~\ref{fig:maxmethod}(a), which then serves as a measure of the size of the kink. We first choose a fixed $\Lambda$ interval wherein multiple kinks, potentially representing magnetic instabilities, appear to be located. We then consider tangents between different pairs of adjacent peaks and calculate the average $\bar{\alpha}$ of the absolute value of these angles within a given $\Lambda$ interval. The angle $\bar{\alpha}$ then serves as a relatively robust quantitative measure of the change in slope (i.e., the size of the kink) within this $\Lambda$ interval; a larger $\bar{\alpha}$ implying a more pronounced kink. To locate the phase transition, we plot $\bar{\alpha}$ as a function of the spin $S$ [see Fig.~\ref{fig:maxmethod}(b)]. For $S=1/2$ and $S=1$, we observe a small and constant value of $\bar{\alpha}\approx 1.6\degree$, followed by a sudden increase at $S=3/2$ indicating a transition point to magnetic long-range ordered state.


\begin{thebibliography}{210}%
\makeatletter
\providecommand \@ifxundefined [1]{%
 \@ifx{#1\undefined}
}%
\providecommand \@ifnum [1]{%
 \ifnum #1\expandafter \@firstoftwo
 \else \expandafter \@secondoftwo
 \fi
}%
\providecommand \@ifx [1]{%
 \ifx #1\expandafter \@firstoftwo
 \else \expandafter \@secondoftwo
 \fi
}%
\providecommand \natexlab [1]{#1}%
\providecommand \enquote  [1]{``#1''}%
\providecommand \bibnamefont  [1]{#1}%
\providecommand \bibfnamefont [1]{#1}%
\providecommand \citenamefont [1]{#1}%
\providecommand \href@noop [0]{\@secondoftwo}%
\providecommand \href [0]{\begingroup \@sanitize@url \@href}%
\providecommand \@href[1]{\@@startlink{#1}\@@href}%
\providecommand \@@href[1]{\endgroup#1\@@endlink}%
\providecommand \@sanitize@url [0]{\catcode `\\12\catcode `\$12\catcode
  `\&12\catcode `\#12\catcode `\^12\catcode `\_12\catcode `\%12\relax}%
\providecommand \@@startlink[1]{}%
\providecommand \@@endlink[0]{}%
\providecommand \url  [0]{\begingroup\@sanitize@url \@url }%
\providecommand \@url [1]{\endgroup\@href {#1}{\urlprefix }}%
\providecommand \urlprefix  [0]{URL }%
\providecommand \Eprint [0]{\href }%
\providecommand \doibase [0]{http://dx.doi.org/}%
\providecommand \selectlanguage [0]{\@gobble}%
\providecommand \bibinfo  [0]{\@secondoftwo}%
\providecommand \bibfield  [0]{\@secondoftwo}%
\providecommand \translation [1]{[#1]}%
\providecommand \BibitemOpen [0]{}%
\providecommand \bibitemStop [0]{}%
\providecommand \bibitemNoStop [0]{.\EOS\space}%
\providecommand \EOS [0]{\spacefactor3000\relax}%
\providecommand \BibitemShut  [1]{\csname bibitem#1\endcsname}%
\let\auto@bib@innerbib\@empty
\bibitem [{\citenamefont {Villain}(1979)}]{Villain-1979}%
  \BibitemOpen
  \bibfield  {author} {\bibinfo {author} {\bibfnamefont {Jacques}\ \bibnamefont
  {Villain}},\ }\bibfield  {title} {\enquote {\bibinfo {title} {{\it Insulating
  Spin Glasses}},}\ }\href {\doibase 10.1007/BF01325811} {\bibfield  {journal}
  {\bibinfo  {journal} {Z. Phys. B}\ }\textbf {\bibinfo {volume} {33}},\
  \bibinfo {pages} {31} (\bibinfo {year} {1979})}\BibitemShut {NoStop}%
\bibitem [{\citenamefont {Reimers}\ \emph {et~al.}(1991)\citenamefont
  {Reimers}, \citenamefont {Berlinsky},\ and\ \citenamefont
  {Shi}}]{Reimers-1991a}%
  \BibitemOpen
  \bibfield  {author} {\bibinfo {author} {\bibfnamefont {J.~N.}\ \bibnamefont
  {Reimers}}, \bibinfo {author} {\bibfnamefont {A.~J.}\ \bibnamefont
  {Berlinsky}}, \ and\ \bibinfo {author} {\bibfnamefont {A.-C.}\ \bibnamefont
  {Shi}},\ }\bibfield  {title} {\enquote {\bibinfo {title} {{\it Mean-Field
  Approach to Magnetic Ordering in Highly Frustrated Pyrochlores}},}\ }\href
  {\doibase 10.1103/PhysRevB.43.865} {\bibfield  {journal} {\bibinfo  {journal}
  {Phys. Rev. B}\ }\textbf {\bibinfo {volume} {43}},\ \bibinfo {pages}
  {865--878} (\bibinfo {year} {1991})}\BibitemShut {NoStop}%
\bibitem [{\citenamefont {Moessner}\ and\ \citenamefont
  {Chalker}(1998{\natexlab{a}})}]{Moessner-1998a}%
  \BibitemOpen
  \bibfield  {author} {\bibinfo {author} {\bibfnamefont {R.}~\bibnamefont
  {Moessner}}\ and\ \bibinfo {author} {\bibfnamefont {J.~T.}\ \bibnamefont
  {Chalker}},\ }\bibfield  {title} {\enquote {\bibinfo {title} {{\it Properties
  of a Classical Spin Liquid: The \mbox{Heisenberg} Pyrochlore
  Antiferromagnet}},}\ }\href {\doibase 10.1103/PhysRevLett.80.2929} {\bibfield
   {journal} {\bibinfo  {journal} {Phys. Rev. Lett.}\ }\textbf {\bibinfo
  {volume} {80}},\ \bibinfo {pages} {2929--2932} (\bibinfo {year}
  {1998}{\natexlab{a}})}\BibitemShut {NoStop}%
\bibitem [{\citenamefont {Moessner}\ and\ \citenamefont
  {Chalker}(1998{\natexlab{b}})}]{Moessner-1998b}%
  \BibitemOpen
  \bibfield  {author} {\bibinfo {author} {\bibfnamefont {R.}~\bibnamefont
  {Moessner}}\ and\ \bibinfo {author} {\bibfnamefont {J.~T.}\ \bibnamefont
  {Chalker}},\ }\bibfield  {title} {\enquote {\bibinfo {title} {{\it
  Low-Temperature Properties of Classical Geometrically Frustrated
  Antiferromagnets}},}\ }\href {\doibase 10.1103/PhysRevB.58.12049} {\bibfield
  {journal} {\bibinfo  {journal} {Phys. Rev. B}\ }\textbf {\bibinfo {volume}
  {58}},\ \bibinfo {pages} {12049--12062} (\bibinfo {year}
  {1998}{\natexlab{b}})}\BibitemShut {NoStop}%
\bibitem [{\citenamefont {{J. Villain}}\ \emph {et~al.}(1980)\citenamefont {{J.
  Villain}}, \citenamefont {{R. Bidaux}}, \citenamefont {{J.-P. Carton}},\ and\
  \citenamefont {{R. Conte}}}]{Villain-1980}%
  \BibitemOpen
  \bibfield  {author} {\bibinfo {author} {\bibnamefont {{J. Villain}}},
  \bibinfo {author} {\bibnamefont {{R. Bidaux}}}, \bibinfo {author}
  {\bibnamefont {{J.-P. Carton}}}, \ and\ \bibinfo {author} {\bibnamefont {{R.
  Conte}}},\ }\bibfield  {title} {\enquote {\bibinfo {title} {{\it Order as an
  Effect of Disorder}},}\ }\href {\doibase 10.1051/jphys:0198000410110126300}
  {\bibfield  {journal} {\bibinfo  {journal} {J. Phys. II (France)}\ }\textbf
  {\bibinfo {volume} {41}},\ \bibinfo {pages} {1263} (\bibinfo {year}
  {1980})}\BibitemShut {NoStop}%
\bibitem [{\citenamefont {Shender}(1982)}]{Shender-1982}%
  \BibitemOpen
  \bibfield  {author} {\bibinfo {author} {\bibfnamefont {E.~F.}\ \bibnamefont
  {Shender}},\ }\bibfield  {title} {\enquote {\bibinfo {title} {{\it
  Antiferromagnetic Garnets with Fluctuationally Interacting Sublattices}},}\
  }\href {http://www.jetp.ac.ru/cgi-bin/e/index/e/56/1/p178?a=list} {\bibfield
  {journal} {\bibinfo  {journal} {Zh. Eksp. Teor. Fiz.}\ }\textbf {\bibinfo
  {volume} {83}},\ \bibinfo {pages} {326} (\bibinfo {year} {1982})}\BibitemShut
  {NoStop}%
\bibitem [{\citenamefont {Henley}(1989)}]{Henley-1989}%
  \BibitemOpen
  \bibfield  {author} {\bibinfo {author} {\bibfnamefont {Christopher~L.}\
  \bibnamefont {Henley}},\ }\bibfield  {title} {\enquote {\bibinfo {title}
  {{\it Ordering due to Disorder in a Frustrated Vector Antiferromagnet}},}\
  }\href {\doibase 10.1103/PhysRevLett.62.2056} {\bibfield  {journal} {\bibinfo
   {journal} {Phys. Rev. Lett.}\ }\textbf {\bibinfo {volume} {62}},\ \bibinfo
  {pages} {2056--2059} (\bibinfo {year} {1989})}\BibitemShut {NoStop}%
\bibitem [{\citenamefont {Reimers}(1992)}]{Reimers-1992}%
  \BibitemOpen
  \bibfield  {author} {\bibinfo {author} {\bibfnamefont {J.~N.}\ \bibnamefont
  {Reimers}},\ }\bibfield  {title} {\enquote {\bibinfo {title} {{\it Absence of
  Long-Range Order in a Three-Dimensional Geometrically Frustrated
  Antiferromagnet}},}\ }\href {\doibase 10.1103/PhysRevB.45.7287} {\bibfield
  {journal} {\bibinfo  {journal} {Phys. Rev. B}\ }\textbf {\bibinfo {volume}
  {45}},\ \bibinfo {pages} {7287--7294} (\bibinfo {year} {1992})}\BibitemShut
  {NoStop}%
\bibitem [{\citenamefont {Zinkin}(1996)}]{Zinkin-1996}%
  \BibitemOpen
  \bibfield  {author} {\bibinfo {author} {\bibfnamefont {M.~P.}\ \bibnamefont
  {Zinkin}},\ }\href@noop {} {\emph {\bibinfo {title} {\rm D. Phil. thesis}}}\
  (\bibinfo  {publisher} {University of Oxford},\ \bibinfo {year}
  {1996})\BibitemShut {NoStop}%
\bibitem [{\citenamefont {Henley}(2001)}]{Henley-2001}%
  \BibitemOpen
  \bibfield  {author} {\bibinfo {author} {\bibfnamefont {C~L}\ \bibnamefont
  {Henley}},\ }\bibfield  {title} {\enquote {\bibinfo {title} {{\it Effective
  Hamiltonians and Dilution Effects in Kagome and Related
  Anti-Ferromagnets}},}\ }\href {\doibase 10.1139/p01-097} {\bibfield
  {journal} {\bibinfo  {journal} {Can. J. Phys.}\ }\textbf {\bibinfo {volume}
  {79}},\ \bibinfo {pages} {1307} (\bibinfo {year} {2001})}\BibitemShut
  {NoStop}%
\bibitem [{\citenamefont {Sobral}\ and\ \citenamefont
  {Lacroix}(1997)}]{Sobral-1997}%
  \BibitemOpen
  \bibfield  {author} {\bibinfo {author} {\bibfnamefont {R.R.}\ \bibnamefont
  {Sobral}}\ and\ \bibinfo {author} {\bibfnamefont {C.}~\bibnamefont
  {Lacroix}},\ }\bibfield  {title} {\enquote {\bibinfo {title} {{\it Order by
  Disorder in the Pyrochlore Antiferromagnets}},}\ }\href {\doibase
  10.1016/S0038-1098(97)00212-3} {\bibfield  {journal} {\bibinfo  {journal}
  {Solid State Commun.}\ }\textbf {\bibinfo {volume} {103}},\ \bibinfo {pages}
  {407} (\bibinfo {year} {1997})}\BibitemShut {NoStop}%
\bibitem [{\citenamefont {Tsunetsugu}(2002)}]{Tsunetsugu-2002}%
  \BibitemOpen
  \bibfield  {author} {\bibinfo {author} {\bibfnamefont {Hirokazu}\
  \bibnamefont {Tsunetsugu}},\ }\bibfield  {title} {\enquote {\bibinfo {title}
  {{\it Quantum Fluctuations in Geometrically Frustrated Antiferromagnet}},}\
  }\href {\doibase http://dx.doi.org/10.1016/S0022-3697(02)00038-0} {\bibfield
  {journal} {\bibinfo  {journal} {J. Phys. Chem. Solids}\ }\textbf {\bibinfo
  {volume} {63}},\ \bibinfo {pages} {1325} (\bibinfo {year}
  {2002})}\BibitemShut {NoStop}%
\bibitem [{\citenamefont {Henley}(2006)}]{Henley-2006}%
  \BibitemOpen
  \bibfield  {author} {\bibinfo {author} {\bibfnamefont {Christopher~L.}\
  \bibnamefont {Henley}},\ }\bibfield  {title} {\enquote {\bibinfo {title}
  {{\it Order by Disorder and Gaugelike Degeneracy in a Quantum Pyrochlore
  Antiferromagnet}},}\ }\href {\doibase 10.1103/PhysRevLett.96.047201}
  {\bibfield  {journal} {\bibinfo  {journal} {Phys. Rev. Lett.}\ }\textbf
  {\bibinfo {volume} {96}},\ \bibinfo {pages} {047201} (\bibinfo {year}
  {2006})}\BibitemShut {NoStop}%
\bibitem [{\citenamefont {Hizi}\ and\ \citenamefont
  {Henley}(2006)}]{Hizi-2006}%
  \BibitemOpen
  \bibfield  {author} {\bibinfo {author} {\bibfnamefont {U.}~\bibnamefont
  {Hizi}}\ and\ \bibinfo {author} {\bibfnamefont {C.~L.}\ \bibnamefont
  {Henley}},\ }\bibfield  {title} {\enquote {\bibinfo {title} {{\it Effective
  Hamiltonian for the Pyrochlore Antiferromagnet: Semiclassical Derivation and
  Degeneracy}},}\ }\href {\doibase 10.1103/PhysRevB.73.054403} {\bibfield
  {journal} {\bibinfo  {journal} {Phys. Rev. B}\ }\textbf {\bibinfo {volume}
  {73}},\ \bibinfo {pages} {054403} (\bibinfo {year} {2006})}\BibitemShut
  {NoStop}%
\bibitem [{\citenamefont {Hizi}\ and\ \citenamefont
  {Henley}(2007)}]{Hizi-2007}%
  \BibitemOpen
  \bibfield  {author} {\bibinfo {author} {\bibfnamefont {Uzi}\ \bibnamefont
  {Hizi}}\ and\ \bibinfo {author} {\bibfnamefont {Christopher~L}\ \bibnamefont
  {Henley}},\ }\bibfield  {title} {\enquote {\bibinfo {title} {{\it Effective
  Hamiltonians for Large-S Pyrochlore Antiferromagnets}},}\ }\href
  {http://stacks.iop.org/0953-8984/19/i=14/a=145268} {\bibfield  {journal}
  {\bibinfo  {journal} {J. Phys. Condens. Matter}\ }\textbf {\bibinfo {volume}
  {19}},\ \bibinfo {pages} {145268} (\bibinfo {year} {2007})}\BibitemShut
  {NoStop}%
\bibitem [{\citenamefont {Hizi}\ and\ \citenamefont
  {Henley}(2009)}]{Hizi-2009}%
  \BibitemOpen
  \bibfield  {author} {\bibinfo {author} {\bibfnamefont {U.}~\bibnamefont
  {Hizi}}\ and\ \bibinfo {author} {\bibfnamefont {C.~L.}\ \bibnamefont
  {Henley}},\ }\bibfield  {title} {\enquote {\bibinfo {title} {{\it Anharmonic
  Ground State Selection in the Pyrochlore Antiferromagnet}},}\ }\href
  {\doibase 10.1103/PhysRevB.80.014407} {\bibfield  {journal} {\bibinfo
  {journal} {Phys. Rev. B}\ }\textbf {\bibinfo {volume} {80}},\ \bibinfo
  {pages} {014407} (\bibinfo {year} {2009})}\BibitemShut {NoStop}%
\bibitem [{\citenamefont {Harris}\ \emph {et~al.}(1991)\citenamefont {Harris},
  \citenamefont {Berlinsky},\ and\ \citenamefont {Bruder}}]{Harris-1991}%
  \BibitemOpen
  \bibfield  {author} {\bibinfo {author} {\bibfnamefont {A.~B.}\ \bibnamefont
  {Harris}}, \bibinfo {author} {\bibfnamefont {A.~J.}\ \bibnamefont
  {Berlinsky}}, \ and\ \bibinfo {author} {\bibfnamefont {C.}~\bibnamefont
  {Bruder}},\ }\bibfield  {title} {\enquote {\bibinfo {title} {{\it Ordering by
  Quantum Fluctuations in a Strongly Frustrated Heisenberg Antiferromagnet}},}\
  }\href {\doibase 10.1063/1.348098} {\bibfield  {journal} {\bibinfo  {journal}
  {J. Appl. Phys.}\ }\textbf {\bibinfo {volume} {69}},\ \bibinfo {pages} {5200}
  (\bibinfo {year} {1991})}\BibitemShut {NoStop}%
\bibitem [{\citenamefont {Isoda}\ and\ \citenamefont
  {Mori}(1998)}]{Isoda-1998}%
  \BibitemOpen
  \bibfield  {author} {\bibinfo {author} {\bibfnamefont {Makoto}\ \bibnamefont
  {Isoda}}\ and\ \bibinfo {author} {\bibfnamefont {Shigeyoshi}\ \bibnamefont
  {Mori}},\ }\bibfield  {title} {\enquote {\bibinfo {title} {{\it Valence-Bond
  Crystal and Anisotropic Excitation Spectrum on 3-Dimensionally Frustrated
  Pyrochlore}},}\ }\href {\doibase 10.1143/JPSJ.67.4022} {\bibfield  {journal}
  {\bibinfo  {journal} {J. Phys. Soc. Jpn.}\ }\textbf {\bibinfo {volume}
  {67}},\ \bibinfo {pages} {4022} (\bibinfo {year} {1998})}\BibitemShut
  {NoStop}%
\bibitem [{\citenamefont {Koga}\ and\ \citenamefont
  {Kawakami}(2001)}]{Koga-2001}%
  \BibitemOpen
  \bibfield  {author} {\bibinfo {author} {\bibfnamefont {Akihisa}\ \bibnamefont
  {Koga}}\ and\ \bibinfo {author} {\bibfnamefont {Norio}\ \bibnamefont
  {Kawakami}},\ }\bibfield  {title} {\enquote {\bibinfo {title} {{\it
  Frustrated Heisenberg Antiferromagnet on the Pyrochlore Lattice}},}\ }\href
  {\doibase 10.1103/PhysRevB.63.144432} {\bibfield  {journal} {\bibinfo
  {journal} {Phys. Rev. B}\ }\textbf {\bibinfo {volume} {63}},\ \bibinfo
  {pages} {144432} (\bibinfo {year} {2001})}\BibitemShut {NoStop}%
\bibitem [{\citenamefont {Tsunetsugu}(2001{\natexlab{a}})}]{Tsunetsugu-2001a}%
  \BibitemOpen
  \bibfield  {author} {\bibinfo {author} {\bibfnamefont {Hirokazu}\
  \bibnamefont {Tsunetsugu}},\ }\bibfield  {title} {\enquote {\bibinfo {title}
  {{\it Antiferromagnetic Quantum Spins on the Pyrochlore Lattice}},}\ }\href
  {\doibase 10.1143/JPSJ.70.640} {\bibfield  {journal} {\bibinfo  {journal} {J.
  Phys. Soc. Jpn.}\ }\textbf {\bibinfo {volume} {70}},\ \bibinfo {pages} {640}
  (\bibinfo {year} {2001}{\natexlab{a}})}\BibitemShut {NoStop}%
\bibitem [{\citenamefont {Tsunetsugu}(2001{\natexlab{b}})}]{Tsunetsugu-2001b}%
  \BibitemOpen
  \bibfield  {author} {\bibinfo {author} {\bibfnamefont {Hirokazu}\
  \bibnamefont {Tsunetsugu}},\ }\bibfield  {title} {\enquote {\bibinfo {title}
  {{\it Spin-Singlet Order in a Pyrochlore Antiferromagnet}},}\ }\href
  {\doibase 10.1103/PhysRevB.65.024415} {\bibfield  {journal} {\bibinfo
  {journal} {Phys. Rev. B}\ }\textbf {\bibinfo {volume} {65}},\ \bibinfo
  {pages} {024415} (\bibinfo {year} {2001}{\natexlab{b}})}\BibitemShut
  {NoStop}%
\bibitem [{\citenamefont {Berg}\ \emph {et~al.}(2003)\citenamefont {Berg},
  \citenamefont {Altman},\ and\ \citenamefont {Auerbach}}]{Berg-2003}%
  \BibitemOpen
  \bibfield  {author} {\bibinfo {author} {\bibfnamefont {Erez}\ \bibnamefont
  {Berg}}, \bibinfo {author} {\bibfnamefont {Ehud}\ \bibnamefont {Altman}}, \
  and\ \bibinfo {author} {\bibfnamefont {Assa}\ \bibnamefont {Auerbach}},\
  }\bibfield  {title} {\enquote {\bibinfo {title} {{\it Singlet Excitations in
  Pyrochlore: A Study of Quantum Frustration}},}\ }\href {\doibase
  10.1103/PhysRevLett.90.147204} {\bibfield  {journal} {\bibinfo  {journal}
  {Phys. Rev. Lett.}\ }\textbf {\bibinfo {volume} {90}},\ \bibinfo {pages}
  {147204} (\bibinfo {year} {2003})}\BibitemShut {NoStop}%
\bibitem [{\citenamefont {Tchernyshyov}\ \emph {et~al.}(2006)\citenamefont
  {Tchernyshyov}, \citenamefont {Moessner},\ and\ \citenamefont
  {Sondhi}}]{Tchernyshyov-2006}%
  \BibitemOpen
  \bibfield  {author} {\bibinfo {author} {\bibfnamefont {O.}~\bibnamefont
  {Tchernyshyov}}, \bibinfo {author} {\bibfnamefont {R.}~\bibnamefont
  {Moessner}}, \ and\ \bibinfo {author} {\bibfnamefont {S.~L.}\ \bibnamefont
  {Sondhi}},\ }\bibfield  {title} {\enquote {\bibinfo {title} {{\it Flux
  Expulsion and Greedy Bosons: Frustrated Magnets at Large N}},}\ }\href
  {http://stacks.iop.org/0295-5075/73/i=2/a=278} {\bibfield  {journal}
  {\bibinfo  {journal} {Europhys. Lett.}\ }\textbf {\bibinfo {volume} {73}},\
  \bibinfo {pages} {278} (\bibinfo {year} {2006})}\BibitemShut {NoStop}%
\bibitem [{\citenamefont {Moessner}\ \emph {et~al.}(2006)\citenamefont
  {Moessner}, \citenamefont {Sondhi},\ and\ \citenamefont
  {Goerbig}}]{Moessner-2006}%
  \BibitemOpen
  \bibfield  {author} {\bibinfo {author} {\bibfnamefont {R.}~\bibnamefont
  {Moessner}}, \bibinfo {author} {\bibfnamefont {S.~L.}\ \bibnamefont
  {Sondhi}}, \ and\ \bibinfo {author} {\bibfnamefont {M.~O.}\ \bibnamefont
  {Goerbig}},\ }\bibfield  {title} {\enquote {\bibinfo {title} {{\it Quantum
  Dimer Models and Effective Hamiltonians on the Pyrochlore Lattice}},}\ }\href
  {\doibase 10.1103/PhysRevB.73.094430} {\bibfield  {journal} {\bibinfo
  {journal} {Phys. Rev. B}\ }\textbf {\bibinfo {volume} {73}},\ \bibinfo
  {pages} {094430} (\bibinfo {year} {2006})}\BibitemShut {NoStop}%
\bibitem [{\citenamefont {Canals}\ and\ \citenamefont
  {Lacroix}(1998)}]{Canals-1998}%
  \BibitemOpen
  \bibfield  {author} {\bibinfo {author} {\bibfnamefont {B.}~\bibnamefont
  {Canals}}\ and\ \bibinfo {author} {\bibfnamefont {C.}~\bibnamefont
  {Lacroix}},\ }\bibfield  {title} {\enquote {\bibinfo {title} {{\it Pyrochlore
  Antiferromagnet: A Three-Dimensional Quantum Spin Liquid}},}\ }\href
  {\doibase 10.1103/PhysRevLett.80.2933} {\bibfield  {journal} {\bibinfo
  {journal} {Phys. Rev. Lett.}\ }\textbf {\bibinfo {volume} {80}},\ \bibinfo
  {pages} {2933} (\bibinfo {year} {1998})}\BibitemShut {NoStop}%
\bibitem [{\citenamefont {Canals}\ and\ \citenamefont
  {Lacroix}(2000)}]{Canals-2000}%
  \BibitemOpen
  \bibfield  {author} {\bibinfo {author} {\bibfnamefont {B.}~\bibnamefont
  {Canals}}\ and\ \bibinfo {author} {\bibfnamefont {C.}~\bibnamefont
  {Lacroix}},\ }\bibfield  {title} {\enquote {\bibinfo {title} {{\it Quantum
  Spin Liquid: The Heisenberg Antiferromagnet on the Three-Dimensional
  Pyrochlore Lattice}},}\ }\href {\doibase 10.1103/PhysRevB.61.1149} {\bibfield
   {journal} {\bibinfo  {journal} {Phys. Rev. B}\ }\textbf {\bibinfo {volume}
  {61}},\ \bibinfo {pages} {1149} (\bibinfo {year} {2000})}\BibitemShut
  {NoStop}%
\bibitem [{\citenamefont {Canals}\ and\ \citenamefont
  {Garanin}(2001)}]{Canals-2001}%
  \BibitemOpen
  \bibfield  {author} {\bibinfo {author} {\bibfnamefont {B.}~\bibnamefont
  {Canals}}\ and\ \bibinfo {author} {\bibfnamefont {D.~A.}\ \bibnamefont
  {Garanin}},\ }\bibfield  {title} {\enquote {\bibinfo {title} {{\it
  Spin-Liquid Phase in the Pyrochlore Anti-Ferromagnet}},}\ }\href {\doibase
  10.1139/p01-101} {\bibfield  {journal} {\bibinfo  {journal} {Can. J. Phys.}\
  }\textbf {\bibinfo {volume} {79}},\ \bibinfo {pages} {1323} (\bibinfo {year}
  {2001})}\BibitemShut {NoStop}%
\bibitem [{\citenamefont {Fouet}\ \emph {et~al.}(2003)\citenamefont {Fouet},
  \citenamefont {Mambrini}, \citenamefont {Sindzingre},\ and\ \citenamefont
  {Lhuillier}}]{Fouet-2003}%
  \BibitemOpen
  \bibfield  {author} {\bibinfo {author} {\bibfnamefont {J.-B.}\ \bibnamefont
  {Fouet}}, \bibinfo {author} {\bibfnamefont {M.}~\bibnamefont {Mambrini}},
  \bibinfo {author} {\bibfnamefont {P.}~\bibnamefont {Sindzingre}}, \ and\
  \bibinfo {author} {\bibfnamefont {C.}~\bibnamefont {Lhuillier}},\ }\bibfield
  {title} {\enquote {\bibinfo {title} {{\it Planar Pyrochlore: A Valence-Bond
  Crystal}},}\ }\href {\doibase 10.1103/PhysRevB.67.054411} {\bibfield
  {journal} {\bibinfo  {journal} {Phys. Rev. B}\ }\textbf {\bibinfo {volume}
  {67}},\ \bibinfo {pages} {054411} (\bibinfo {year} {2003})}\BibitemShut
  {NoStop}%
\bibitem [{\citenamefont {Kim}\ and\ \citenamefont {Han}(2008)}]{Kim-2008}%
  \BibitemOpen
  \bibfield  {author} {\bibinfo {author} {\bibfnamefont {J.~H.}\ \bibnamefont
  {Kim}}\ and\ \bibinfo {author} {\bibfnamefont {J.~H.}\ \bibnamefont {Han}},\
  }\bibfield  {title} {\enquote {\bibinfo {title} {{\it Chiral Spin States in
  the Pyrochlore Heisenberg Magnet: Fermionic Mean-Field Theory and Variational
  Monte Carlo Calculations}},}\ }\href {\doibase 10.1103/PhysRevB.78.180410}
  {\bibfield  {journal} {\bibinfo  {journal} {Phys. Rev. B}\ }\textbf {\bibinfo
  {volume} {78}},\ \bibinfo {pages} {180410} (\bibinfo {year}
  {2008})}\BibitemShut {NoStop}%
\bibitem [{\citenamefont {Burnell}\ \emph {et~al.}(2009)\citenamefont
  {Burnell}, \citenamefont {Chakravarty},\ and\ \citenamefont
  {Sondhi}}]{Burnell-2009}%
  \BibitemOpen
  \bibfield  {author} {\bibinfo {author} {\bibfnamefont {F.~J.}\ \bibnamefont
  {Burnell}}, \bibinfo {author} {\bibfnamefont {Shoibal}\ \bibnamefont
  {Chakravarty}}, \ and\ \bibinfo {author} {\bibfnamefont {S.~L.}\ \bibnamefont
  {Sondhi}},\ }\bibfield  {title} {\enquote {\bibinfo {title} {{\it Monopole
  Flux State on the Pyrochlore Lattice}},}\ }\href {\doibase
  10.1103/PhysRevB.79.144432} {\bibfield  {journal} {\bibinfo  {journal} {Phys.
  Rev. B}\ }\textbf {\bibinfo {volume} {79}},\ \bibinfo {pages} {144432}
  (\bibinfo {year} {2009})}\BibitemShut {NoStop}%
\bibitem [{\citenamefont {Huang}\ \emph {et~al.}(2016)\citenamefont {Huang},
  \citenamefont {Chen}, \citenamefont {Deng}, \citenamefont {Prokof'ev},\ and\
  \citenamefont {Svistunov}}]{Huang-2016}%
  \BibitemOpen
  \bibfield  {author} {\bibinfo {author} {\bibfnamefont {Y.}~\bibnamefont
  {Huang}}, \bibinfo {author} {\bibfnamefont {K.}~\bibnamefont {Chen}},
  \bibinfo {author} {\bibfnamefont {Y.}~\bibnamefont {Deng}}, \bibinfo {author}
  {\bibfnamefont {N.}~\bibnamefont {Prokof'ev}}, \ and\ \bibinfo {author}
  {\bibfnamefont {B.}~\bibnamefont {Svistunov}},\ }\bibfield  {title} {\enquote
  {\bibinfo {title} {{\it Spin-Ice State of the Quantum Heisenberg
  Antiferromagnet on the Pyrochlore Lattice}},}\ }\href {\doibase
  10.1103/PhysRevLett.116.177203} {\bibfield  {journal} {\bibinfo  {journal}
  {Phys. Rev. Lett.}\ }\textbf {\bibinfo {volume} {116}},\ \bibinfo {pages}
  {177203} (\bibinfo {year} {2016})}\BibitemShut {NoStop}%
\bibitem [{\citenamefont {Normand}\ and\ \citenamefont
  {Nussinov}(2014)}]{Normand-2014}%
  \BibitemOpen
  \bibfield  {author} {\bibinfo {author} {\bibfnamefont {B.}~\bibnamefont
  {Normand}}\ and\ \bibinfo {author} {\bibfnamefont {Z.}~\bibnamefont
  {Nussinov}},\ }\bibfield  {title} {\enquote {\bibinfo {title} {{\it Hubbard
  Model on the Pyrochlore Lattice: A 3D Quantum Spin Liquid}},}\ }\href
  {\doibase 10.1103/PhysRevLett.112.207202} {\bibfield  {journal} {\bibinfo
  {journal} {Phys. Rev. Lett.}\ }\textbf {\bibinfo {volume} {112}},\ \bibinfo
  {pages} {207202} (\bibinfo {year} {2014})}\BibitemShut {NoStop}%
\bibitem [{\citenamefont {Normand}\ and\ \citenamefont
  {Nussinov}(2016)}]{Normand-2016}%
  \BibitemOpen
  \bibfield  {author} {\bibinfo {author} {\bibfnamefont {B.}~\bibnamefont
  {Normand}}\ and\ \bibinfo {author} {\bibfnamefont {Z.}~\bibnamefont
  {Nussinov}},\ }\bibfield  {title} {\enquote {\bibinfo {title} {{\it Fermionic
  Spinon and Holon Statistics in the Pyrochlore Quantum Spin Liquid}},}\ }\href
  {\doibase 10.1103/PhysRevB.93.115122} {\bibfield  {journal} {\bibinfo
  {journal} {Phys. Rev. B}\ }\textbf {\bibinfo {volume} {93}},\ \bibinfo
  {pages} {115122} (\bibinfo {year} {2016})}\BibitemShut {NoStop}%
\bibitem [{\citenamefont {Garc\'{\i}a-Adeva}\ and\ \citenamefont
  {Huber}(2000)}]{Garcia-2000}%
  \BibitemOpen
  \bibfield  {author} {\bibinfo {author} {\bibfnamefont {A.~J.}\ \bibnamefont
  {Garc\'{\i}a-Adeva}}\ and\ \bibinfo {author} {\bibfnamefont {D.~L.}\
  \bibnamefont {Huber}},\ }\bibfield  {title} {\enquote {\bibinfo {title} {{\it
  Quantum Tetrahedral Mean Field Theory of the Magnetic Susceptibility for the
  Pyrochlore Lattice}},}\ }\href {\doibase 10.1103/PhysRevLett.85.4598}
  {\bibfield  {journal} {\bibinfo  {journal} {Phys. Rev. Lett.}\ }\textbf
  {\bibinfo {volume} {85}},\ \bibinfo {pages} {4598} (\bibinfo {year}
  {2000})}\BibitemShut {NoStop}%
\bibitem [{\citenamefont {Yamashita}\ and\ \citenamefont
  {Ueda}(2000)}]{Yamashita-2000}%
  \BibitemOpen
  \bibfield  {author} {\bibinfo {author} {\bibfnamefont {Y.}~\bibnamefont
  {Yamashita}}\ and\ \bibinfo {author} {\bibfnamefont {K.}~\bibnamefont
  {Ueda}},\ }\bibfield  {title} {\enquote {\bibinfo {title} {{\it Spin-Driven
  Jahn-Teller Distortion in a Pyrochlore System}},}\ }\href {\doibase
  10.1103/PhysRevLett.85.4960} {\bibfield  {journal} {\bibinfo  {journal}
  {Phys. Rev. Lett.}\ }\textbf {\bibinfo {volume} {85}},\ \bibinfo {pages}
  {4960} (\bibinfo {year} {2000})}\BibitemShut {NoStop}%
\bibitem [{\citenamefont {Tsunetsugu}(2017)}]{Tsunetsugu-2017}%
  \BibitemOpen
  \bibfield  {author} {\bibinfo {author} {\bibfnamefont {H.}~\bibnamefont
  {Tsunetsugu}},\ }\bibfield  {title} {\enquote {\bibinfo {title} {{\it Theory
  of Antiferromagnetic Heisenberg Spins on a Breathing Pyrochlore Lattice}},}\
  }\href {\doibase 10.1093/ptep/ptx023} {\bibfield  {journal} {\bibinfo
  {journal} {Prog. Theor. Exp. Phys.}\ }\textbf {\bibinfo {volume} {2017}},\
  \bibinfo {pages} {033I01} (\bibinfo {year} {2017})}\BibitemShut {NoStop}%
\bibitem [{\citenamefont {Yamashita}\ \emph {et~al.}(2001)\citenamefont
  {Yamashita}, \citenamefont {Ueda},\ and\ \citenamefont
  {Sigrist}}]{Yasufumi-2001}%
  \BibitemOpen
  \bibfield  {author} {\bibinfo {author} {\bibfnamefont {Y.}~\bibnamefont
  {Yamashita}}, \bibinfo {author} {\bibfnamefont {K.}~\bibnamefont {Ueda}}, \
  and\ \bibinfo {author} {\bibfnamefont {M.}~\bibnamefont {Sigrist}},\
  }\bibfield  {title} {\enquote {\bibinfo {title} {{\it Parity-Broken Ground
  State for the Spin-1 Pyrochlore Antiferromagnet}},}\ }\href
  {http://stacks.iop.org/0953-8984/13/i=50/a=102} {\bibfield  {journal}
  {\bibinfo  {journal} {J. Phys. Condens. Matter}\ }\textbf {\bibinfo {volume}
  {13}},\ \bibinfo {pages} {L961} (\bibinfo {year} {2001})}\BibitemShut
  {NoStop}%
\bibitem [{\citenamefont {Tsuneishi}\ \emph {et~al.}(2007)\citenamefont
  {Tsuneishi}, \citenamefont {Ioki},\ and\ \citenamefont
  {Kawamura}}]{Ioki-2007}%
  \BibitemOpen
  \bibfield  {author} {\bibinfo {author} {\bibfnamefont {D.}~\bibnamefont
  {Tsuneishi}}, \bibinfo {author} {\bibfnamefont {M.}~\bibnamefont {Ioki}}, \
  and\ \bibinfo {author} {\bibfnamefont {H.}~\bibnamefont {Kawamura}},\
  }\bibfield  {title} {\enquote {\bibinfo {title} {{\it Novel Ordering of the
  Pyrochlore Heisenberg Antiferromagnet with the Ferromagnetic
  Next-Nearest-Neighbour Interaction}},}\ }\href
  {http://stacks.iop.org/0953-8984/19/i=14/a=145273} {\bibfield  {journal}
  {\bibinfo  {journal} {J. Phys. Condens. Matter}\ }\textbf {\bibinfo {volume}
  {19}},\ \bibinfo {pages} {145273} (\bibinfo {year} {2007})}\BibitemShut
  {NoStop}%
\bibitem [{\citenamefont {Nakamura}\ and\ \citenamefont
  {Hirashima}(2007)}]{Nakamura-2007}%
  \BibitemOpen
  \bibfield  {author} {\bibinfo {author} {\bibfnamefont {T.}~\bibnamefont
  {Nakamura}}\ and\ \bibinfo {author} {\bibfnamefont {D.}~\bibnamefont
  {Hirashima}},\ }\bibfield  {title} {\enquote {\bibinfo {title} {{\it
  Classical Antiferromagnet on the Pyrochlore Lattice}},}\ }\href {\doibase
  10.1016/j.jmmm.2006.10.473} {\bibfield  {journal} {\bibinfo  {journal} {J.
  Magn. Magn. Mater.}\ }\textbf {\bibinfo {volume} {310}},\ \bibinfo {pages}
  {1297} (\bibinfo {year} {2007})}\BibitemShut {NoStop}%
\bibitem [{\citenamefont {Chern}\ \emph {et~al.}(2008)\citenamefont {Chern},
  \citenamefont {Moessner},\ and\ \citenamefont {Tchernyshyov}}]{Chern-2008}%
  \BibitemOpen
  \bibfield  {author} {\bibinfo {author} {\bibfnamefont {G.-W.}\ \bibnamefont
  {Chern}}, \bibinfo {author} {\bibfnamefont {R.}~\bibnamefont {Moessner}}, \
  and\ \bibinfo {author} {\bibfnamefont {O.}~\bibnamefont {Tchernyshyov}},\
  }\bibfield  {title} {\enquote {\bibinfo {title} {{\it Partial Order from
  Disorder in a Classical Pyrochlore Antiferromagnet}},}\ }\href {\doibase
  10.1103/PhysRevB.78.144418} {\bibfield  {journal} {\bibinfo  {journal} {Phys.
  Rev. B}\ }\textbf {\bibinfo {volume} {78}},\ \bibinfo {pages} {144418}
  (\bibinfo {year} {2008})}\BibitemShut {NoStop}%
\bibitem [{\citenamefont {Okubo}\ \emph {et~al.}(2011)\citenamefont {Okubo},
  \citenamefont {Nguyen},\ and\ \citenamefont {Kawamura}}]{Okubo-2011}%
  \BibitemOpen
  \bibfield  {author} {\bibinfo {author} {\bibfnamefont {T.}~\bibnamefont
  {Okubo}}, \bibinfo {author} {\bibfnamefont {T.~H.}\ \bibnamefont {Nguyen}}, \
  and\ \bibinfo {author} {\bibfnamefont {H.}~\bibnamefont {Kawamura}},\
  }\bibfield  {title} {\enquote {\bibinfo {title} {{\it Cubic and Noncubic
  Multiple-$q$ States in the Heisenberg Antiferromagnet on the Pyrochlore
  Lattice}},}\ }\href {\doibase 10.1103/PhysRevB.84.144432} {\bibfield
  {journal} {\bibinfo  {journal} {Phys. Rev. B}\ }\textbf {\bibinfo {volume}
  {84}},\ \bibinfo {pages} {144432} (\bibinfo {year} {2011})}\BibitemShut
  {NoStop}%
\bibitem [{\citenamefont {Palmer}\ and\ \citenamefont
  {Chalker}(2000)}]{Palmer-2000}%
  \BibitemOpen
  \bibfield  {author} {\bibinfo {author} {\bibfnamefont {S.~E.}\ \bibnamefont
  {Palmer}}\ and\ \bibinfo {author} {\bibfnamefont {J.~T.}\ \bibnamefont
  {Chalker}},\ }\bibfield  {title} {\enquote {\bibinfo {title} {{\it Order
  Induced by Dipolar Interactions in a Geometrically Frustrated
  Antiferromagnet}},}\ }\href {\doibase 10.1103/PhysRevB.62.488} {\bibfield
  {journal} {\bibinfo  {journal} {Phys. Rev. B}\ }\textbf {\bibinfo {volume}
  {62}},\ \bibinfo {pages} {488--492} (\bibinfo {year} {2000})}\BibitemShut
  {NoStop}%
\bibitem [{\citenamefont {Elhajal}\ \emph {et~al.}(2005)\citenamefont
  {Elhajal}, \citenamefont {Canals}, \citenamefont {Sunyer},\ and\
  \citenamefont {Lacroix}}]{Elhajal-2005}%
  \BibitemOpen
  \bibfield  {author} {\bibinfo {author} {\bibfnamefont {M.}~\bibnamefont
  {Elhajal}}, \bibinfo {author} {\bibfnamefont {B.}~\bibnamefont {Canals}},
  \bibinfo {author} {\bibfnamefont {R.}~\bibnamefont {Sunyer}}, \ and\ \bibinfo
  {author} {\bibfnamefont {C.}~\bibnamefont {Lacroix}},\ }\bibfield  {title}
  {\enquote {\bibinfo {title} {{\it Ordering in the Pyrochlore Antiferromagnet
  due to Dzyaloshinsky-Moriya Interactions}},}\ }\href {\doibase
  10.1103/PhysRevB.71.094420} {\bibfield  {journal} {\bibinfo  {journal} {Phys.
  Rev. B}\ }\textbf {\bibinfo {volume} {71}},\ \bibinfo {pages} {094420}
  (\bibinfo {year} {2005})}\BibitemShut {NoStop}%
\bibitem [{\citenamefont {Chern}(2010)}]{Chern-2010}%
  \BibitemOpen
  \bibfield  {author} {\bibinfo {author} {\bibfnamefont {G.-W.}\ \bibnamefont
  {Chern}},\ }\bibfield  {title} {\enquote {\bibinfo {title} {{\it Noncoplanar
  Magnetic Ordering Driven by Itinerant Electrons on the Pyrochlore
  Lattice}},}\ }\href {\doibase 10.1103/PhysRevLett.105.226403} {\bibfield
  {journal} {\bibinfo  {journal} {Phys. Rev. Lett.}\ }\textbf {\bibinfo
  {volume} {105}},\ \bibinfo {pages} {226403} (\bibinfo {year}
  {2010})}\BibitemShut {NoStop}%
\bibitem [{\citenamefont {Bramwell}\ \emph {et~al.}(1994)\citenamefont
  {Bramwell}, \citenamefont {Gingras},\ and\ \citenamefont
  {Reimers}}]{Bramwell-1994}%
  \BibitemOpen
  \bibfield  {author} {\bibinfo {author} {\bibfnamefont {S.~T.}\ \bibnamefont
  {Bramwell}}, \bibinfo {author} {\bibfnamefont {M.~J.~P.}\ \bibnamefont
  {Gingras}}, \ and\ \bibinfo {author} {\bibfnamefont {J.~N.}\ \bibnamefont
  {Reimers}},\ }\bibfield  {title} {\enquote {\bibinfo {title} {{\it Order by
  Disorder in an Anisotropic Pyrochlore Lattice Antiferromagnet}},}\ }\href
  {\doibase 10.1063/1.355676} {\bibfield  {journal} {\bibinfo  {journal} {J.
  Appl. Phys.}\ }\textbf {\bibinfo {volume} {75}},\ \bibinfo {pages} {5523}
  (\bibinfo {year} {1994})}\BibitemShut {NoStop}%
\bibitem [{\citenamefont {Moessner}(1998)}]{Moessner-1998c}%
  \BibitemOpen
  \bibfield  {author} {\bibinfo {author} {\bibfnamefont {R.}~\bibnamefont
  {Moessner}},\ }\bibfield  {title} {\enquote {\bibinfo {title} {{\it Relief
  and Generation of Frustration in Pyrochlore Magnets by Single-Ion
  Anisotropy}},}\ }\href {\doibase 10.1103/PhysRevB.57.R5587} {\bibfield
  {journal} {\bibinfo  {journal} {Phys. Rev. B}\ }\textbf {\bibinfo {volume}
  {57}},\ \bibinfo {pages} {R5587} (\bibinfo {year} {1998})}\BibitemShut
  {NoStop}%
\bibitem [{\citenamefont {Terao}(1996)}]{Terao-1996}%
  \BibitemOpen
  \bibfield  {author} {\bibinfo {author} {\bibfnamefont {K.}~\bibnamefont
  {Terao}},\ }\bibfield  {title} {\enquote {\bibinfo {title} {{\it Effect of
  Lattice Distortions upon the Spin Configuration of Antiferromagnetic
  $\mathrm{Y}\mathrm{Mn}_{2}$ with $\mathrm{C15}$ Structure}},}\ }\href
  {\doibase 10.1143/JPSJ.65.1413} {\bibfield  {journal} {\bibinfo  {journal}
  {J. Phys. Soc. Jpn.}\ }\textbf {\bibinfo {volume} {65}},\ \bibinfo {pages}
  {1413} (\bibinfo {year} {1996})}\BibitemShut {NoStop}%
\bibitem [{\citenamefont {Tchernyshyov}\ \emph {et~al.}(2002)\citenamefont
  {Tchernyshyov}, \citenamefont {Moessner},\ and\ \citenamefont
  {Sondhi}}]{Tchernyshyov-2002}%
  \BibitemOpen
  \bibfield  {author} {\bibinfo {author} {\bibfnamefont {O.}~\bibnamefont
  {Tchernyshyov}}, \bibinfo {author} {\bibfnamefont {R.}~\bibnamefont
  {Moessner}}, \ and\ \bibinfo {author} {\bibfnamefont {S.~L.}\ \bibnamefont
  {Sondhi}},\ }\bibfield  {title} {\enquote {\bibinfo {title} {{\it Order by
  Distortion and String Modes in Pyrochlore Antiferromagnets}},}\ }\href
  {\doibase 10.1103/PhysRevLett.88.067203} {\bibfield  {journal} {\bibinfo
  {journal} {Phys. Rev. Lett.}\ }\textbf {\bibinfo {volume} {88}},\ \bibinfo
  {pages} {067203} (\bibinfo {year} {2002})}\BibitemShut {NoStop}%
\bibitem [{\citenamefont {Pinettes}\ \emph {et~al.}(2002)\citenamefont
  {Pinettes}, \citenamefont {Canals},\ and\ \citenamefont
  {Lacroix}}]{Pinettes-2002}%
  \BibitemOpen
  \bibfield  {author} {\bibinfo {author} {\bibfnamefont {C.}~\bibnamefont
  {Pinettes}}, \bibinfo {author} {\bibfnamefont {B.}~\bibnamefont {Canals}}, \
  and\ \bibinfo {author} {\bibfnamefont {C.}~\bibnamefont {Lacroix}},\
  }\bibfield  {title} {\enquote {\bibinfo {title} {{\it Classical Heisenberg
  Antiferromagnet away from the Pyrochlore Lattice limit: Entropic versus
  Energetic Selection}},}\ }\href {\doibase 10.1103/PhysRevB.66.024422}
  {\bibfield  {journal} {\bibinfo  {journal} {Phys. Rev. B}\ }\textbf {\bibinfo
  {volume} {66}},\ \bibinfo {pages} {024422} (\bibinfo {year}
  {2002})}\BibitemShut {NoStop}%
\bibitem [{\citenamefont {Tchernyshyov}(2004)}]{Tchernyshyov-2004}%
  \BibitemOpen
  \bibfield  {author} {\bibinfo {author} {\bibfnamefont {O.}~\bibnamefont
  {Tchernyshyov}},\ }\bibfield  {title} {\enquote {\bibinfo {title} {{\it
  Structural, Orbital, and Magnetic Order in Vanadium Spinels}},}\ }\href
  {\doibase 10.1103/PhysRevLett.93.157206} {\bibfield  {journal} {\bibinfo
  {journal} {Phys. Rev. Lett.}\ }\textbf {\bibinfo {volume} {93}},\ \bibinfo
  {pages} {157206} (\bibinfo {year} {2004})}\BibitemShut {NoStop}%
\bibitem [{\citenamefont {Chern}\ \emph {et~al.}(2006)\citenamefont {Chern},
  \citenamefont {Fennie},\ and\ \citenamefont {Tchernyshyov}}]{Chern-2006}%
  \BibitemOpen
  \bibfield  {author} {\bibinfo {author} {\bibfnamefont {G.-W.}\ \bibnamefont
  {Chern}}, \bibinfo {author} {\bibfnamefont {C.~J.}\ \bibnamefont {Fennie}}, \
  and\ \bibinfo {author} {\bibfnamefont {O.}~\bibnamefont {Tchernyshyov}},\
  }\bibfield  {title} {\enquote {\bibinfo {title} {{\it Broken Parity and a
  Chiral Ground State in the Frustrated Magnet
  $\mathrm{Cd}{\mathrm{Cr}}_{2}{\mathrm{O}}_{4}$}},}\ }\href {\doibase
  10.1103/PhysRevB.74.060405} {\bibfield  {journal} {\bibinfo  {journal} {Phys.
  Rev. B}\ }\textbf {\bibinfo {volume} {74}},\ \bibinfo {pages} {060405}
  (\bibinfo {year} {2006})}\BibitemShut {NoStop}%
\bibitem [{\citenamefont {Bergman}\ \emph {et~al.}(2006)\citenamefont
  {Bergman}, \citenamefont {Shindou}, \citenamefont {Fiete},\ and\
  \citenamefont {Balents}}]{Bergman-2006}%
  \BibitemOpen
  \bibfield  {author} {\bibinfo {author} {\bibfnamefont {D.~L.}\ \bibnamefont
  {Bergman}}, \bibinfo {author} {\bibfnamefont {R.}~\bibnamefont {Shindou}},
  \bibinfo {author} {\bibfnamefont {G.~A.}\ \bibnamefont {Fiete}}, \ and\
  \bibinfo {author} {\bibfnamefont {L.}~\bibnamefont {Balents}},\ }\bibfield
  {title} {\enquote {\bibinfo {title} {{\it Models of Degeneracy Breaking in
  Pyrochlore Antiferromagnets}},}\ }\href {\doibase 10.1103/PhysRevB.74.134409}
  {\bibfield  {journal} {\bibinfo  {journal} {Phys. Rev. B}\ }\textbf {\bibinfo
  {volume} {74}},\ \bibinfo {pages} {134409} (\bibinfo {year}
  {2006})}\BibitemShut {NoStop}%
\bibitem [{\citenamefont {Bellier-Castella}\ \emph {et~al.}(2001)\citenamefont
  {Bellier-Castella}, \citenamefont {Gingras}, \citenamefont {Holdsworth},\
  and\ \citenamefont {Moessner}}]{Castella-2001}%
  \BibitemOpen
  \bibfield  {author} {\bibinfo {author} {\bibfnamefont {L.}~\bibnamefont
  {Bellier-Castella}}, \bibinfo {author} {\bibfnamefont {M.~J.~P.}\
  \bibnamefont {Gingras}}, \bibinfo {author} {\bibfnamefont {P.~C.~W.}\
  \bibnamefont {Holdsworth}}, \ and\ \bibinfo {author} {\bibfnamefont
  {R.}~\bibnamefont {Moessner}},\ }\bibfield  {title} {\enquote {\bibinfo
  {title} {{\it Frustrated Order by Disorder: The Pyrochlore Anti-Ferromagnet
  with Bond Disorder}},}\ }\href {\doibase 10.1139/p01-098} {\bibfield
  {journal} {\bibinfo  {journal} {Can. J. Phys.}\ }\textbf {\bibinfo {volume}
  {79}},\ \bibinfo {pages} {1365} (\bibinfo {year} {2001})}\BibitemShut
  {NoStop}%
\bibitem [{\citenamefont {Saunders}\ and\ \citenamefont
  {Chalker}(2007)}]{Saunders-Chalker}%
  \BibitemOpen
  \bibfield  {author} {\bibinfo {author} {\bibfnamefont {T.~E.}\ \bibnamefont
  {Saunders}}\ and\ \bibinfo {author} {\bibfnamefont {J.~T.}\ \bibnamefont
  {Chalker}},\ }\bibfield  {title} {\enquote {\bibinfo {title} {{\it Spin
  Freezing in Geometrically Frustrated Antiferromagnets with Weak Disorder}},}\
  }\href {\doibase 10.1103/PhysRevLett.98.157201} {\bibfield  {journal}
  {\bibinfo  {journal} {Phys. Rev. Lett.}\ }\textbf {\bibinfo {volume} {98}},\
  \bibinfo {pages} {157201} (\bibinfo {year} {2007})}\BibitemShut {NoStop}%
\bibitem [{\citenamefont {Andreanov}\ \emph {et~al.}(2010)\citenamefont
  {Andreanov}, \citenamefont {Chalker}, \citenamefont {Saunders},\ and\
  \citenamefont {Sherrington}}]{Andreanov-Chalker}%
  \BibitemOpen
  \bibfield  {author} {\bibinfo {author} {\bibfnamefont {A.}~\bibnamefont
  {Andreanov}}, \bibinfo {author} {\bibfnamefont {J.~T.}\ \bibnamefont
  {Chalker}}, \bibinfo {author} {\bibfnamefont {T.~E.}\ \bibnamefont
  {Saunders}}, \ and\ \bibinfo {author} {\bibfnamefont {D.}~\bibnamefont
  {Sherrington}},\ }\bibfield  {title} {\enquote {\bibinfo {title} {{\it
  Spin-Glass Transition in Geometrically Frustrated Antiferromagnets with Weak
  Disorder}},}\ }\href {\doibase 10.1103/PhysRevB.81.014406} {\bibfield
  {journal} {\bibinfo  {journal} {Phys. Rev. B}\ }\textbf {\bibinfo {volume}
  {81}},\ \bibinfo {pages} {014406} (\bibinfo {year} {2010})}\BibitemShut
  {NoStop}%
\bibitem [{\citenamefont {{Lapa}}\ and\ \citenamefont
  {{Henley}}(2012)}]{Lapa-2012}%
  \BibitemOpen
  \bibfield  {author} {\bibinfo {author} {\bibfnamefont {M.~F.}\ \bibnamefont
  {{Lapa}}}\ and\ \bibinfo {author} {\bibfnamefont {C.~L.}\ \bibnamefont
  {{Henley}}},\ }\bibfield  {title} {\enquote {\bibinfo {title} {{Ground States
  of the Classical Antiferromagnet on the Pyrochlore Lattice}},}\ }\href@noop
  {} {\bibfield  {journal} {\bibinfo  {journal} {ArXiv e-prints}\ } (\bibinfo
  {year} {2012})},\ \Eprint {http://arxiv.org/abs/1210.6810} {arXiv:1210.6810
  [cond-mat.str-el]} \BibitemShut {NoStop}%
\bibitem [{\citenamefont {Tymoshenko}\ \emph {et~al.}(2017)\citenamefont
  {Tymoshenko}, \citenamefont {Onykiienko}, \citenamefont {M\"uller},
  \citenamefont {Thomale}, \citenamefont {Rachel}, \citenamefont {Cameron},
  \citenamefont {Portnichenko}, \citenamefont {Efremov}, \citenamefont
  {Tsurkan}, \citenamefont {Abernathy}, \citenamefont {Ollivier}, \citenamefont
  {Schneidewind}, \citenamefont {Piovano}, \citenamefont {Felea}, \citenamefont
  {Loidl},\ and\ \citenamefont {Inosov}}]{Tymoshenko-2017}%
  \BibitemOpen
  \bibfield  {author} {\bibinfo {author} {\bibfnamefont {Y.~V.}\ \bibnamefont
  {Tymoshenko}}, \bibinfo {author} {\bibfnamefont {Y.~A.}\ \bibnamefont
  {Onykiienko}}, \bibinfo {author} {\bibfnamefont {T.}~\bibnamefont
  {M\"uller}}, \bibinfo {author} {\bibfnamefont {R.}~\bibnamefont {Thomale}},
  \bibinfo {author} {\bibfnamefont {S.}~\bibnamefont {Rachel}}, \bibinfo
  {author} {\bibfnamefont {A.~S.}\ \bibnamefont {Cameron}}, \bibinfo {author}
  {\bibfnamefont {P.~Y.}\ \bibnamefont {Portnichenko}}, \bibinfo {author}
  {\bibfnamefont {D.~V.}\ \bibnamefont {Efremov}}, \bibinfo {author}
  {\bibfnamefont {V.}~\bibnamefont {Tsurkan}}, \bibinfo {author} {\bibfnamefont
  {D.~L.}\ \bibnamefont {Abernathy}}, \bibinfo {author} {\bibfnamefont
  {J.}~\bibnamefont {Ollivier}}, \bibinfo {author} {\bibfnamefont
  {A.}~\bibnamefont {Schneidewind}}, \bibinfo {author} {\bibfnamefont
  {A.}~\bibnamefont {Piovano}}, \bibinfo {author} {\bibfnamefont
  {V.}~\bibnamefont {Felea}}, \bibinfo {author} {\bibfnamefont
  {A.}~\bibnamefont {Loidl}}, \ and\ \bibinfo {author} {\bibfnamefont {D.~S.}\
  \bibnamefont {Inosov}},\ }\bibfield  {title} {\enquote {\bibinfo {title}
  {{\it Pseudo-Goldstone Magnons in the Frustrated $S=3/2$ Heisenberg
  Helimagnet ${\mathrm{ZnCr}}_{2}{\mathrm{Se}}_{4}$ with a Pyrochlore Magnetic
  Sublattice}},}\ }\href {\doibase 10.1103/PhysRevX.7.041049} {\bibfield
  {journal} {\bibinfo  {journal} {Phys. Rev. X}\ }\textbf {\bibinfo {volume}
  {7}},\ \bibinfo {pages} {041049} (\bibinfo {year} {2017})}\BibitemShut
  {NoStop}%
\bibitem [{\citenamefont {Schollw\"ock}(2005)}]{Schollwoeck-2005}%
  \BibitemOpen
  \bibfield  {author} {\bibinfo {author} {\bibfnamefont {U.}~\bibnamefont
  {Schollw\"ock}},\ }\bibfield  {title} {\enquote {\bibinfo {title} {{\it The
  Density-Matrix Renormalization Group}},}\ }\href {\doibase
  10.1103/RevModPhys.77.259} {\bibfield  {journal} {\bibinfo  {journal} {Rev.
  Mod. Phys.}\ }\textbf {\bibinfo {volume} {77}},\ \bibinfo {pages} {259}
  (\bibinfo {year} {2005})}\BibitemShut {NoStop}%
\bibitem [{\citenamefont {Stoudenmire}\ and\ \citenamefont
  {White}(2012)}]{Stoudenmire-2012}%
  \BibitemOpen
  \bibfield  {author} {\bibinfo {author} {\bibfnamefont {E.M.}\ \bibnamefont
  {Stoudenmire}}\ and\ \bibinfo {author} {\bibfnamefont {Steven~R.}\
  \bibnamefont {White}},\ }\bibfield  {title} {\enquote {\bibinfo {title} {{\it
  Studying Two-Dimensional Systems with the Density Matrix Renormalization
  Group}},}\ }\href {\doibase 10.1146/annurev-conmatphys-020911-125018}
  {\bibfield  {journal} {\bibinfo  {journal} {Annu. Rev. Condens. Matter
  Phys.}\ }\textbf {\bibinfo {volume} {3}},\ \bibinfo {pages} {111--128}
  (\bibinfo {year} {2012})}\BibitemShut {NoStop}%
\bibitem [{\citenamefont {Reger}\ and\ \citenamefont
  {Young}(1988)}]{Reger-1988}%
  \BibitemOpen
  \bibfield  {author} {\bibinfo {author} {\bibfnamefont {J.~D.}\ \bibnamefont
  {Reger}}\ and\ \bibinfo {author} {\bibfnamefont {A.~P.}\ \bibnamefont
  {Young}},\ }\bibfield  {title} {\enquote {\bibinfo {title} {{\it Monte Carlo
  Simulations of the Spin-(1/2) Heisenberg Antiferromagnet on a Square
  Lattice}},}\ }\href {\doibase 10.1103/PhysRevB.37.5978} {\bibfield  {journal}
  {\bibinfo  {journal} {Phys. Rev. B}\ }\textbf {\bibinfo {volume} {37}},\
  \bibinfo {pages} {5978} (\bibinfo {year} {1988})}\BibitemShut {NoStop}%
\bibitem [{\citenamefont {Sandvik}\ and\ \citenamefont
  {Kurkij\"arvi}(1991)}]{Sandvik-1991}%
  \BibitemOpen
  \bibfield  {author} {\bibinfo {author} {\bibfnamefont {A.~W.}\ \bibnamefont
  {Sandvik}}\ and\ \bibinfo {author} {\bibfnamefont {J.}~\bibnamefont
  {Kurkij\"arvi}},\ }\bibfield  {title} {\enquote {\bibinfo {title} {{\it
  Quantum Monte Carlo Simulation Method for Spin Systems}},}\ }\href {\doibase
  10.1103/PhysRevB.43.5950} {\bibfield  {journal} {\bibinfo  {journal} {Phys.
  Rev. B}\ }\textbf {\bibinfo {volume} {43}},\ \bibinfo {pages} {5950}
  (\bibinfo {year} {1991})}\BibitemShut {NoStop}%
\bibitem [{\citenamefont {McMillan}(1965)}]{McMillan-1965}%
  \BibitemOpen
  \bibfield  {author} {\bibinfo {author} {\bibfnamefont {W.~L.}\ \bibnamefont
  {McMillan}},\ }\bibfield  {title} {\enquote {\bibinfo {title} {{\it Ground
  State of Liquid ${\mathrm{He}}^{4}$}},}\ }\href {\doibase
  10.1103/PhysRev.138.A442} {\bibfield  {journal} {\bibinfo  {journal} {Phys.
  Rev.}\ }\textbf {\bibinfo {volume} {138}},\ \bibinfo {pages} {A442} (\bibinfo
  {year} {1965})}\BibitemShut {NoStop}%
\bibitem [{\citenamefont {Ceperley}\ \emph {et~al.}(1977)\citenamefont
  {Ceperley}, \citenamefont {Chester},\ and\ \citenamefont
  {Kalos}}]{Ceperley-1977}%
  \BibitemOpen
  \bibfield  {author} {\bibinfo {author} {\bibfnamefont {D.}~\bibnamefont
  {Ceperley}}, \bibinfo {author} {\bibfnamefont {G.~V.}\ \bibnamefont
  {Chester}}, \ and\ \bibinfo {author} {\bibfnamefont {M.~H.}\ \bibnamefont
  {Kalos}},\ }\bibfield  {title} {\enquote {\bibinfo {title} {{\it Monte Carlo
  Simulation of a Many-Fermion Study}},}\ }\href {\doibase
  10.1103/PhysRevB.16.3081} {\bibfield  {journal} {\bibinfo  {journal} {Phys.
  Rev. B}\ }\textbf {\bibinfo {volume} {16}},\ \bibinfo {pages} {3081}
  (\bibinfo {year} {1977})}\BibitemShut {NoStop}%
\bibitem [{\citenamefont {Iqbal}\ \emph {et~al.}(2013)\citenamefont {Iqbal},
  \citenamefont {Becca}, \citenamefont {Sorella},\ and\ \citenamefont
  {Poilblanc}}]{Iqbal-2013}%
  \BibitemOpen
  \bibfield  {author} {\bibinfo {author} {\bibfnamefont {Y.}~\bibnamefont
  {Iqbal}}, \bibinfo {author} {\bibfnamefont {F.}~\bibnamefont {Becca}},
  \bibinfo {author} {\bibfnamefont {S.}~\bibnamefont {Sorella}}, \ and\
  \bibinfo {author} {\bibfnamefont {D.}~\bibnamefont {Poilblanc}},\ }\bibfield
  {title} {\enquote {\bibinfo {title} {{\it Gapless Spin-Liquid Phase in the
  Kagome Spin-$\frac{1}{2}$ Heisenberg Antiferromagnet}},}\ }\href {\doibase
  10.1103/PhysRevB.87.060405} {\bibfield  {journal} {\bibinfo  {journal} {Phys.
  Rev. B}\ }\textbf {\bibinfo {volume} {87}},\ \bibinfo {pages} {060405}
  (\bibinfo {year} {2013})}\BibitemShut {NoStop}%
\bibitem [{\citenamefont {Iqbal}\ \emph {et~al.}(2011)\citenamefont {Iqbal},
  \citenamefont {Becca},\ and\ \citenamefont {Poilblanc}}]{Iqbal-2011b}%
  \BibitemOpen
  \bibfield  {author} {\bibinfo {author} {\bibfnamefont {Y.}~\bibnamefont
  {Iqbal}}, \bibinfo {author} {\bibfnamefont {F.}~\bibnamefont {Becca}}, \ and\
  \bibinfo {author} {\bibfnamefont {D.}~\bibnamefont {Poilblanc}},\ }\bibfield
  {title} {\enquote {\bibinfo {title} {{\it Projected Wave Function Study of
  ${\mathbb{Z}}_{2}$ Spin Liquids on the Kagome Lattice for the
  Spin-$\frac{1}{2}$ Quantum Heisenberg Antiferromagnet}},}\ }\href {\doibase
  10.1103/PhysRevB.84.020407} {\bibfield  {journal} {\bibinfo  {journal} {Phys.
  Rev. B}\ }\textbf {\bibinfo {volume} {84}},\ \bibinfo {pages} {020407}
  (\bibinfo {year} {2011})}\BibitemShut {NoStop}%
\bibitem [{\citenamefont {Iqbal}\ \emph {et~al.}(2014)\citenamefont {Iqbal},
  \citenamefont {Poilblanc},\ and\ \citenamefont {Becca}}]{Iqbal-2014}%
  \BibitemOpen
  \bibfield  {author} {\bibinfo {author} {\bibfnamefont {Y.}~\bibnamefont
  {Iqbal}}, \bibinfo {author} {\bibfnamefont {D.}~\bibnamefont {Poilblanc}}, \
  and\ \bibinfo {author} {\bibfnamefont {F.}~\bibnamefont {Becca}},\ }\bibfield
   {title} {\enquote {\bibinfo {title} {{\it Vanishing Spin Gap in a Competing
  Spin-Liquid Phase in the Kagome Heisenberg Antiferromagnet}},}\ }\href
  {\doibase 10.1103/PhysRevB.89.020407} {\bibfield  {journal} {\bibinfo
  {journal} {Phys. Rev. B}\ }\textbf {\bibinfo {volume} {89}},\ \bibinfo
  {pages} {020407} (\bibinfo {year} {2014})}\BibitemShut {NoStop}%
\bibitem [{\citenamefont {Arovas}\ and\ \citenamefont
  {Auerbach}(1988)}]{Arovas-1988}%
  \BibitemOpen
  \bibfield  {author} {\bibinfo {author} {\bibfnamefont {D.~P.}\ \bibnamefont
  {Arovas}}\ and\ \bibinfo {author} {\bibfnamefont {A.}~\bibnamefont
  {Auerbach}},\ }\bibfield  {title} {\enquote {\bibinfo {title} {{\it
  Functional Integral Theories of Low-Dimensional Quantum Heisenberg
  Models}},}\ }\href {\doibase 10.1103/PhysRevB.38.316} {\bibfield  {journal}
  {\bibinfo  {journal} {Phys. Rev. B}\ }\textbf {\bibinfo {volume} {38}},\
  \bibinfo {pages} {316} (\bibinfo {year} {1988})}\BibitemShut {NoStop}%
\bibitem [{\citenamefont {Gelfand}\ and\ \citenamefont
  {Singh}(2000)}]{Gelfand-2000}%
  \BibitemOpen
  \bibfield  {author} {\bibinfo {author} {\bibfnamefont {M.~P.}\ \bibnamefont
  {Gelfand}}\ and\ \bibinfo {author} {\bibfnamefont {R.~R.~P.}\ \bibnamefont
  {Singh}},\ }\bibfield  {title} {\enquote {\bibinfo {title} {{\it High-Order
  Convergent Expansions for Quantum Many Particle Systems}},}\ }\href {\doibase
  10.1080/000187300243390} {\bibfield  {journal} {\bibinfo  {journal} {Adv.
  Phys.}\ }\textbf {\bibinfo {volume} {49}},\ \bibinfo {pages} {93} (\bibinfo
  {year} {2000})}\BibitemShut {NoStop}%
\bibitem [{\citenamefont {Iqbal}\ \emph
  {et~al.}(2016{\natexlab{a}})\citenamefont {Iqbal}, \citenamefont {Thomale},
  \citenamefont {Toldin}, \citenamefont {Rachel},\ and\ \citenamefont
  {Reuther}}]{Iqbal-2016b}%
  \BibitemOpen
  \bibfield  {author} {\bibinfo {author} {\bibfnamefont {Y.}~\bibnamefont
  {Iqbal}}, \bibinfo {author} {\bibfnamefont {R.}~\bibnamefont {Thomale}},
  \bibinfo {author} {\bibfnamefont {F.~P.}\ \bibnamefont {Toldin}}, \bibinfo
  {author} {\bibfnamefont {S.}~\bibnamefont {Rachel}}, \ and\ \bibinfo {author}
  {\bibfnamefont {J.}~\bibnamefont {Reuther}},\ }\bibfield  {title} {\enquote
  {\bibinfo {title} {{\it Functional Renormalization Group for
  Three-Dimensional Quantum Magnetism}},}\ }\href {\doibase
  10.1103/PhysRevB.94.140408} {\bibfield  {journal} {\bibinfo  {journal} {Phys.
  Rev. B}\ }\textbf {\bibinfo {volume} {94}},\ \bibinfo {pages} {140408}
  (\bibinfo {year} {2016}{\natexlab{a}})}\BibitemShut {NoStop}%
\bibitem [{\citenamefont {Zinkin}\ \emph {et~al.}(1997)\citenamefont {Zinkin},
  \citenamefont {Harris},\ and\ \citenamefont {Zeiske}}]{Zinkin-1997}%
  \BibitemOpen
  \bibfield  {author} {\bibinfo {author} {\bibfnamefont {M.~P.}\ \bibnamefont
  {Zinkin}}, \bibinfo {author} {\bibfnamefont {M.~J.}\ \bibnamefont {Harris}},
  \ and\ \bibinfo {author} {\bibfnamefont {T.}~\bibnamefont {Zeiske}},\
  }\bibfield  {title} {\enquote {\bibinfo {title} {{\it Short-Range Magnetic
  Order in the Frustrated Pyrochlore Antiferromagnet
  ${\mathrm{CsNiCrF}}_{6}$}},}\ }\href {\doibase 10.1103/PhysRevB.56.11786}
  {\bibfield  {journal} {\bibinfo  {journal} {Phys. Rev. B}\ }\textbf {\bibinfo
  {volume} {56}},\ \bibinfo {pages} {11786} (\bibinfo {year}
  {1997})}\BibitemShut {NoStop}%
\bibitem [{\citenamefont {Conlon}\ and\ \citenamefont
  {Chalker}(2010)}]{Conlon-2010}%
  \BibitemOpen
  \bibfield  {author} {\bibinfo {author} {\bibfnamefont {P.~H.}\ \bibnamefont
  {Conlon}}\ and\ \bibinfo {author} {\bibfnamefont {J.~T.}\ \bibnamefont
  {Chalker}},\ }\bibfield  {title} {\enquote {\bibinfo {title} {{\it Absent
  Pinch Points and Emergent Clusters: Further Neighbor Interactions in the
  Pyrochlore Heisenberg Antiferromagnet}},}\ }\href {\doibase
  10.1103/PhysRevB.81.224413} {\bibfield  {journal} {\bibinfo  {journal} {Phys.
  Rev. B}\ }\textbf {\bibinfo {volume} {81}},\ \bibinfo {pages} {224413}
  (\bibinfo {year} {2010})}\BibitemShut {NoStop}%
\bibitem [{\citenamefont {Reuther}\ and\ \citenamefont
  {W\"olfle}(2010)}]{Reuther-2010}%
  \BibitemOpen
  \bibfield  {author} {\bibinfo {author} {\bibfnamefont {J.}~\bibnamefont
  {Reuther}}\ and\ \bibinfo {author} {\bibfnamefont {P.}~\bibnamefont
  {W\"olfle}},\ }\bibfield  {title} {\enquote {\bibinfo {title} {{\it
  ${J}_{1}\text{-}{J}_{2}$ Frustrated Two-Dimensional \mbox{Heisenberg} Model:
  Random Phase Approximation and Functional Renormalization Group}},}\ }\href
  {\doibase 10.1103/PhysRevB.81.144410} {\bibfield  {journal} {\bibinfo
  {journal} {Phys. Rev. B}\ }\textbf {\bibinfo {volume} {81}},\ \bibinfo
  {pages} {144410} (\bibinfo {year} {2010})}\BibitemShut {NoStop}%
\bibitem [{\citenamefont {Abrikosov}(1965)}]{Abrikosov-1965}%
  \BibitemOpen
  \bibfield  {author} {\bibinfo {author} {\bibfnamefont {A.~A.}\ \bibnamefont
  {Abrikosov}},\ }\bibfield  {title} {\enquote {\bibinfo {title} {{\it Electron
  Scattering on Magnetic Impurities in Metals and Anomalous Resistivity
  Effects}},}\ }\href {\doibase 10.1103/PhysicsPhysiqueFizika.2.5} {\bibfield
  {journal} {\bibinfo  {journal} {Physics}\ }\textbf {\bibinfo {volume} {2}},\
  \bibinfo {pages} {5} (\bibinfo {year} {1965})}\BibitemShut {NoStop}%
\bibitem [{\citenamefont {Baez}\ and\ \citenamefont
  {Reuther}(2017)}]{Baez-2017}%
  \BibitemOpen
  \bibfield  {author} {\bibinfo {author} {\bibfnamefont {M.~L.}\ \bibnamefont
  {Baez}}\ and\ \bibinfo {author} {\bibfnamefont {J.}~\bibnamefont {Reuther}},\
  }\bibfield  {title} {\enquote {\bibinfo {title} {{\it Numerical Treatment of
  Spin Systems with Unrestricted Spin Length $S$: A Functional Renormalization
  Group Study}},}\ }\href {\doibase 10.1103/PhysRevB.96.045144} {\bibfield
  {journal} {\bibinfo  {journal} {Phys. Rev. B}\ }\textbf {\bibinfo {volume}
  {96}},\ \bibinfo {pages} {045144} (\bibinfo {year} {2017})}\BibitemShut
  {NoStop}%
\bibitem [{Note1()}]{Note1}%
  \BibitemOpen
  \bibinfo {note} {Heisenberg systems (on any lattice) with $S=1$ and
  single-ion anisotropies $\Delta \DOTSB \sum@ \slimits@ _i (S_i^z)^2$ provide
  a simple exception wherein if $\Delta $ is positive (and sufficiently large)
  this term would always energetically prefer the unphysical spin sector $S=0$
  over all other sectors.}\BibitemShut {Stop}%
\bibitem [{\citenamefont {Wetterich}(1993)}]{Wetterich-1993}%
  \BibitemOpen
  \bibfield  {author} {\bibinfo {author} {\bibfnamefont {C.}~\bibnamefont
  {Wetterich}},\ }\bibfield  {title} {\enquote {\bibinfo {title} {{\it Exact
  Evolution Equation for the Effective Potential}},}\ }\href {\doibase
  http://dx.doi.org/10.1016/0370-2693(93)90726-X} {\bibfield  {journal}
  {\bibinfo  {journal} {Phys. Lett. B}\ }\textbf {\bibinfo {volume} {301}},\
  \bibinfo {pages} {90} (\bibinfo {year} {1993})}\BibitemShut {NoStop}%
\bibitem [{\citenamefont {Metzner}\ \emph {et~al.}(2012)\citenamefont
  {Metzner}, \citenamefont {Salmhofer}, \citenamefont {Honerkamp},
  \citenamefont {Meden},\ and\ \citenamefont {Sch\"onhammer}}]{Metzner-2012}%
  \BibitemOpen
  \bibfield  {author} {\bibinfo {author} {\bibfnamefont {W.}~\bibnamefont
  {Metzner}}, \bibinfo {author} {\bibfnamefont {M.}~\bibnamefont {Salmhofer}},
  \bibinfo {author} {\bibfnamefont {C.}~\bibnamefont {Honerkamp}}, \bibinfo
  {author} {\bibfnamefont {V.}~\bibnamefont {Meden}}, \ and\ \bibinfo {author}
  {\bibfnamefont {K.}~\bibnamefont {Sch\"onhammer}},\ }\bibfield  {title}
  {\enquote {\bibinfo {title} {{\it Functional Renormalization Group Approach
  to Correlated Fermion Systems}},}\ }\href {\doibase
  10.1103/RevModPhys.84.299} {\bibfield  {journal} {\bibinfo  {journal} {Rev.
  Mod. Phys.}\ }\textbf {\bibinfo {volume} {84}},\ \bibinfo {pages} {299}
  (\bibinfo {year} {2012})}\BibitemShut {NoStop}%
\bibitem [{\citenamefont {Platt}\ \emph {et~al.}(2013)\citenamefont {Platt},
  \citenamefont {Hanke},\ and\ \citenamefont {Thomale}}]{Platt-2013}%
  \BibitemOpen
  \bibfield  {author} {\bibinfo {author} {\bibfnamefont {C.}~\bibnamefont
  {Platt}}, \bibinfo {author} {\bibfnamefont {W.}~\bibnamefont {Hanke}}, \ and\
  \bibinfo {author} {\bibfnamefont {R.}~\bibnamefont {Thomale}},\ }\bibfield
  {title} {\enquote {\bibinfo {title} {{\it Functional Renormalization Group
  for Multi-Orbital Fermi Surface Instabilities}},}\ }\href {\doibase
  10.1080/00018732.2013.862020} {\bibfield  {journal} {\bibinfo  {journal}
  {Adv. Phys.}\ }\textbf {\bibinfo {volume} {62}},\ \bibinfo {pages} {453}
  (\bibinfo {year} {2013})}\BibitemShut {NoStop}%
\bibitem [{Note2()}]{Note2}%
  \BibitemOpen
  \bibinfo {note} {The class of diagrams representing the \protect \emph
  {leading order} in $1/S$ contributions, which are thus of random phase
  approximation (RPA)-type and responsible for the formation of classical
  magnetic order, are summed up \protect \emph {exactly}. Similarly, the class
  of diagrams capturing contributions to \protect \emph {leading order} in
  $1/N$, and thus responsible for the formation of nonmagnetic states, are also
  summed up \protect \emph {exactly}. However, an accurate treatment of the
  $S\to \infty $ limit may require a consideration of subleading terms in
  $1/S$, thus going beyond a bare RPA treatment [see Appendix~\ref
  {appendix1}]}\BibitemShut {NoStop}%
\bibitem [{\citenamefont {Sachdev}\ and\ \citenamefont
  {Read}(1991)}]{Sachdev-1991}%
  \BibitemOpen
  \bibfield  {author} {\bibinfo {author} {\bibfnamefont {S.}~\bibnamefont
  {Sachdev}}\ and\ \bibinfo {author} {\bibfnamefont {N.}~\bibnamefont {Read}},\
  }\bibfield  {title} {\enquote {\bibinfo {title} {{\it Large N Expansion for
  Frustrated and Doped Quantum Antiferromagnets}},}\ }\href {\doibase
  10.1142/S0217979291000158} {\bibfield  {journal} {\bibinfo  {journal} {Int.
  J. Mod. Phys. B}\ }\textbf {\bibinfo {volume} {05}},\ \bibinfo {pages} {219}
  (\bibinfo {year} {1991})}\BibitemShut {NoStop}%
\bibitem [{\citenamefont {Katanin}(2004)}]{Katanin-2004}%
  \BibitemOpen
  \bibfield  {author} {\bibinfo {author} {\bibfnamefont {A.~A.}\ \bibnamefont
  {Katanin}},\ }\bibfield  {title} {\enquote {\bibinfo {title} {{\it
  Fulfillment of Ward Identities in the Functional Renormalization Group
  Approach}},}\ }\href {\doibase 10.1103/PhysRevB.70.115109} {\bibfield
  {journal} {\bibinfo  {journal} {Phys. Rev. B}\ }\textbf {\bibinfo {volume}
  {70}},\ \bibinfo {pages} {115109} (\bibinfo {year} {2004})}\BibitemShut
  {NoStop}%
\bibitem [{\citenamefont {Reuther}\ and\ \citenamefont
  {Thomale}(2014)}]{Reuther-2014b}%
  \BibitemOpen
  \bibfield  {author} {\bibinfo {author} {\bibfnamefont {J.}~\bibnamefont
  {Reuther}}\ and\ \bibinfo {author} {\bibfnamefont {R.}~\bibnamefont
  {Thomale}},\ }\bibfield  {title} {\enquote {\bibinfo {title} {{\it Cluster
  Functional Renormalization Group}},}\ }\href {\doibase
  10.1103/PhysRevB.89.024412} {\bibfield  {journal} {\bibinfo  {journal} {Phys.
  Rev. B}\ }\textbf {\bibinfo {volume} {89}},\ \bibinfo {pages} {024412}
  (\bibinfo {year} {2014})}\BibitemShut {NoStop}%
\bibitem [{\citenamefont {Reuther}\ and\ \citenamefont
  {Thomale}(2011)}]{Reuther-2011a}%
  \BibitemOpen
  \bibfield  {author} {\bibinfo {author} {\bibfnamefont {J.}~\bibnamefont
  {Reuther}}\ and\ \bibinfo {author} {\bibfnamefont {R.}~\bibnamefont
  {Thomale}},\ }\bibfield  {title} {\enquote {\bibinfo {title} {{\it Functional
  Renormalization Group for the Anisotropic Triangular Antiferromagnet}},}\
  }\href {\doibase 10.1103/PhysRevB.83.024402} {\bibfield  {journal} {\bibinfo
  {journal} {Phys. Rev. B}\ }\textbf {\bibinfo {volume} {83}},\ \bibinfo
  {pages} {024402} (\bibinfo {year} {2011})}\BibitemShut {NoStop}%
\bibitem [{\citenamefont {Reuther}\ \emph
  {et~al.}(2011{\natexlab{a}})\citenamefont {Reuther}, \citenamefont {Abanin},\
  and\ \citenamefont {Thomale}}]{Reuther-2011b}%
  \BibitemOpen
  \bibfield  {author} {\bibinfo {author} {\bibfnamefont {J.}~\bibnamefont
  {Reuther}}, \bibinfo {author} {\bibfnamefont {D.~A.}\ \bibnamefont {Abanin}},
  \ and\ \bibinfo {author} {\bibfnamefont {R.}~\bibnamefont {Thomale}},\
  }\bibfield  {title} {\enquote {\bibinfo {title} {{\it Magnetic Order and
  Paramagnetic Phases in the Quantum ${J}_{1}$-${J}_{2}$-${J}_{3}$ Honeycomb
  Model}},}\ }\href {\doibase 10.1103/PhysRevB.84.014417} {\bibfield  {journal}
  {\bibinfo  {journal} {Phys. Rev. B}\ }\textbf {\bibinfo {volume} {84}},\
  \bibinfo {pages} {014417} (\bibinfo {year} {2011}{\natexlab{a}})}\BibitemShut
  {NoStop}%
\bibitem [{\citenamefont {Reuther}\ \emph
  {et~al.}(2011{\natexlab{b}})\citenamefont {Reuther}, \citenamefont
  {W\"olfle}, \citenamefont {Darradi}, \citenamefont {Brenig}, \citenamefont
  {Arlego},\ and\ \citenamefont {Richter}}]{Reuther-2011c}%
  \BibitemOpen
  \bibfield  {author} {\bibinfo {author} {\bibfnamefont {J.}~\bibnamefont
  {Reuther}}, \bibinfo {author} {\bibfnamefont {P.}~\bibnamefont {W\"olfle}},
  \bibinfo {author} {\bibfnamefont {R.}~\bibnamefont {Darradi}}, \bibinfo
  {author} {\bibfnamefont {W.}~\bibnamefont {Brenig}}, \bibinfo {author}
  {\bibfnamefont {M.}~\bibnamefont {Arlego}}, \ and\ \bibinfo {author}
  {\bibfnamefont {J.}~\bibnamefont {Richter}},\ }\bibfield  {title} {\enquote
  {\bibinfo {title} {{\it Quantum Phases of the Planar Antiferromagnetic
  ${J}_{1}$-${J}_{2}$-${J}_{3}$ \mbox{Heisenberg} Model}},}\ }\href {\doibase
  10.1103/PhysRevB.83.064416} {\bibfield  {journal} {\bibinfo  {journal} {Phys.
  Rev. B}\ }\textbf {\bibinfo {volume} {83}},\ \bibinfo {pages} {064416}
  (\bibinfo {year} {2011}{\natexlab{b}})}\BibitemShut {NoStop}%
\bibitem [{\citenamefont {Reuther}\ \emph
  {et~al.}(2011{\natexlab{c}})\citenamefont {Reuther}, \citenamefont
  {Thomale},\ and\ \citenamefont {Trebst}}]{Reuther-2011d}%
  \BibitemOpen
  \bibfield  {author} {\bibinfo {author} {\bibfnamefont {J.}~\bibnamefont
  {Reuther}}, \bibinfo {author} {\bibfnamefont {R.}~\bibnamefont {Thomale}}, \
  and\ \bibinfo {author} {\bibfnamefont {S.}~\bibnamefont {Trebst}},\
  }\bibfield  {title} {\enquote {\bibinfo {title} {{\it Finite-Temperature
  Phase Diagram of the \mbox{Heisenberg}-\mbox{Kitaev} Model}},}\ }\href
  {\doibase 10.1103/PhysRevB.84.100406} {\bibfield  {journal} {\bibinfo
  {journal} {Phys. Rev. B}\ }\textbf {\bibinfo {volume} {84}},\ \bibinfo
  {pages} {100406} (\bibinfo {year} {2011}{\natexlab{c}})}\BibitemShut
  {NoStop}%
\bibitem [{\citenamefont {Singh}\ \emph {et~al.}(2012)\citenamefont {Singh},
  \citenamefont {Manni}, \citenamefont {Reuther}, \citenamefont {Berlijn},
  \citenamefont {Thomale}, \citenamefont {Ku}, \citenamefont {Trebst},\ and\
  \citenamefont {Gegenwart}}]{Singh-2012}%
  \BibitemOpen
  \bibfield  {author} {\bibinfo {author} {\bibfnamefont {Y.}~\bibnamefont
  {Singh}}, \bibinfo {author} {\bibfnamefont {S.}~\bibnamefont {Manni}},
  \bibinfo {author} {\bibfnamefont {J.}~\bibnamefont {Reuther}}, \bibinfo
  {author} {\bibfnamefont {T.}~\bibnamefont {Berlijn}}, \bibinfo {author}
  {\bibfnamefont {R.}~\bibnamefont {Thomale}}, \bibinfo {author} {\bibfnamefont
  {W.}~\bibnamefont {Ku}}, \bibinfo {author} {\bibfnamefont {S.}~\bibnamefont
  {Trebst}}, \ and\ \bibinfo {author} {\bibfnamefont {P.}~\bibnamefont
  {Gegenwart}},\ }\bibfield  {title} {\enquote {\bibinfo {title} {{\it
  Relevance of the Heisenberg-Kitaev Model for the Honeycomb Lattice Iridates
  ${A}_{2}{\mathrm{IrO}}_{3}$}},}\ }\href {\doibase
  10.1103/PhysRevLett.108.127203} {\bibfield  {journal} {\bibinfo  {journal}
  {Phys. Rev. Lett.}\ }\textbf {\bibinfo {volume} {108}},\ \bibinfo {pages}
  {127203} (\bibinfo {year} {2012})}\BibitemShut {NoStop}%
\bibitem [{\citenamefont {Reuther}\ \emph {et~al.}(2014)\citenamefont
  {Reuther}, \citenamefont {Thomale},\ and\ \citenamefont
  {Rachel}}]{Reuther-2014}%
  \BibitemOpen
  \bibfield  {author} {\bibinfo {author} {\bibfnamefont {J.}~\bibnamefont
  {Reuther}}, \bibinfo {author} {\bibfnamefont {R.}~\bibnamefont {Thomale}}, \
  and\ \bibinfo {author} {\bibfnamefont {S.}~\bibnamefont {Rachel}},\
  }\bibfield  {title} {\enquote {\bibinfo {title} {{\it Spiral Order in the
  Honeycomb Iridate $\mathrm{Li_2IrO_3}$}},}\ }\href {\doibase
  10.1103/PhysRevB.90.100405} {\bibfield  {journal} {\bibinfo  {journal} {Phys.
  Rev. B}\ }\textbf {\bibinfo {volume} {90}},\ \bibinfo {pages} {100405}
  (\bibinfo {year} {2014})}\BibitemShut {NoStop}%
\bibitem [{\citenamefont {Suttner}\ \emph {et~al.}(2014)\citenamefont
  {Suttner}, \citenamefont {Platt}, \citenamefont {Reuther},\ and\
  \citenamefont {Thomale}}]{Suttner-2014}%
  \BibitemOpen
  \bibfield  {author} {\bibinfo {author} {\bibfnamefont {R.}~\bibnamefont
  {Suttner}}, \bibinfo {author} {\bibfnamefont {C.}~\bibnamefont {Platt}},
  \bibinfo {author} {\bibfnamefont {J.}~\bibnamefont {Reuther}}, \ and\
  \bibinfo {author} {\bibfnamefont {R.}~\bibnamefont {Thomale}},\ }\bibfield
  {title} {\enquote {\bibinfo {title} {{\it Renormalization Group Analysis of
  Competing Quantum Phases in the ${J}_{1}$-${J}_{2}$ Heisenberg Model on the
  Kagome Lattice}},}\ }\href {\doibase 10.1103/PhysRevB.89.020408} {\bibfield
  {journal} {\bibinfo  {journal} {Phys. Rev. B}\ }\textbf {\bibinfo {volume}
  {89}},\ \bibinfo {pages} {020408} (\bibinfo {year} {2014})}\BibitemShut
  {NoStop}%
\bibitem [{\citenamefont {Iqbal}\ \emph {et~al.}(2015)\citenamefont {Iqbal},
  \citenamefont {Jeschke}, \citenamefont {Reuther}, \citenamefont
  {Valent\'{\i}}, \citenamefont {Mazin}, \citenamefont {Greiter},\ and\
  \citenamefont {Thomale}}]{Iqbal-2015}%
  \BibitemOpen
  \bibfield  {author} {\bibinfo {author} {\bibfnamefont {Y.}~\bibnamefont
  {Iqbal}}, \bibinfo {author} {\bibfnamefont {H.~O.}\ \bibnamefont {Jeschke}},
  \bibinfo {author} {\bibfnamefont {J.}~\bibnamefont {Reuther}}, \bibinfo
  {author} {\bibfnamefont {R.}~\bibnamefont {Valent\'{\i}}}, \bibinfo {author}
  {\bibfnamefont {I.~I.}\ \bibnamefont {Mazin}}, \bibinfo {author}
  {\bibfnamefont {M.}~\bibnamefont {Greiter}}, \ and\ \bibinfo {author}
  {\bibfnamefont {R.}~\bibnamefont {Thomale}},\ }\bibfield  {title} {\enquote
  {\bibinfo {title} {{\it Paramagnetism in the Kagome Compounds
  $(\mathrm{Zn},\mathrm{Mg},\mathrm{Cd}){\mathrm{Cu}}_{3}{(\mathrm{OH})}_{6}{\mathrm{Cl}}_{2}$}},}\
  }\href {\doibase 10.1103/PhysRevB.92.220404} {\bibfield  {journal} {\bibinfo
  {journal} {Phys. Rev. B}\ }\textbf {\bibinfo {volume} {92}},\ \bibinfo
  {pages} {220404} (\bibinfo {year} {2015})}\BibitemShut {NoStop}%
\bibitem [{\citenamefont {Balz}\ \emph {et~al.}(2016)\citenamefont {Balz},
  \citenamefont {Lake}, \citenamefont {Reuther}, \citenamefont {Luetkens},
  \citenamefont {Sch{\"o}nemann}, \citenamefont {Herrmannsd{\"o}rfer},
  \citenamefont {Singh}, \citenamefont {Islam}, \citenamefont {Wheeler},
  \citenamefont {Rodriguez-Rivera}, \citenamefont {Guidi}, \citenamefont
  {Simeoni}, \citenamefont {Baines},\ and\ \citenamefont {Ryll}}]{Balz-16}%
  \BibitemOpen
  \bibfield  {author} {\bibinfo {author} {\bibfnamefont {C.}~\bibnamefont
  {Balz}}, \bibinfo {author} {\bibfnamefont {B.}~\bibnamefont {Lake}}, \bibinfo
  {author} {\bibfnamefont {J.}~\bibnamefont {Reuther}}, \bibinfo {author}
  {\bibfnamefont {H.}~\bibnamefont {Luetkens}}, \bibinfo {author}
  {\bibfnamefont {R.}~\bibnamefont {Sch{\"o}nemann}}, \bibinfo {author}
  {\bibfnamefont {T.}~\bibnamefont {Herrmannsd{\"o}rfer}}, \bibinfo {author}
  {\bibfnamefont {Y.}~\bibnamefont {Singh}}, \bibinfo {author} {\bibfnamefont
  {A.~T. M.~N.}\ \bibnamefont {Islam}}, \bibinfo {author} {\bibfnamefont
  {E.~M.}\ \bibnamefont {Wheeler}}, \bibinfo {author} {\bibfnamefont {J.~A.}\
  \bibnamefont {Rodriguez-Rivera}}, \bibinfo {author} {\bibfnamefont
  {T.}~\bibnamefont {Guidi}}, \bibinfo {author} {\bibfnamefont {G.~G.}\
  \bibnamefont {Simeoni}}, \bibinfo {author} {\bibfnamefont {C.}~\bibnamefont
  {Baines}}, \ and\ \bibinfo {author} {\bibfnamefont {H.}~\bibnamefont
  {Ryll}},\ }\bibfield  {title} {\enquote {\bibinfo {title} {{\it Physical
  Realization of a Quantum Spin Liquid Based on a Complex Frustration
  Mechanism}},}\ }\href {http://dx.doi.org/10.1038/nphys3826} {\bibfield
  {journal} {\bibinfo  {journal} {Nat. Phys.}\ }\textbf {\bibinfo {volume}
  {12}},\ \bibinfo {pages} {942} (\bibinfo {year} {2016})}\BibitemShut
  {NoStop}%
\bibitem [{\citenamefont {Iqbal}\ \emph
  {et~al.}(2016{\natexlab{b}})\citenamefont {Iqbal}, \citenamefont {Hu},
  \citenamefont {Thomale}, \citenamefont {Poilblanc},\ and\ \citenamefont
  {Becca}}]{Iqbal-2016a}%
  \BibitemOpen
  \bibfield  {author} {\bibinfo {author} {\bibfnamefont {Y.}~\bibnamefont
  {Iqbal}}, \bibinfo {author} {\bibfnamefont {W.-J.}\ \bibnamefont {Hu}},
  \bibinfo {author} {\bibfnamefont {R.}~\bibnamefont {Thomale}}, \bibinfo
  {author} {\bibfnamefont {D.}~\bibnamefont {Poilblanc}}, \ and\ \bibinfo
  {author} {\bibfnamefont {F.}~\bibnamefont {Becca}},\ }\bibfield  {title}
  {\enquote {\bibinfo {title} {{\it Spin Liquid Nature in the Heisenberg
  ${J}_{1}$-${J}_{2}$ Triangular Antiferromagnet}},}\ }\href {\doibase
  10.1103/PhysRevB.93.144411} {\bibfield  {journal} {\bibinfo  {journal} {Phys.
  Rev. B}\ }\textbf {\bibinfo {volume} {93}},\ \bibinfo {pages} {144411}
  (\bibinfo {year} {2016}{\natexlab{b}})}\BibitemShut {NoStop}%
\bibitem [{\citenamefont {Iqbal}\ \emph
  {et~al.}(2016{\natexlab{c}})\citenamefont {Iqbal}, \citenamefont {Ghosh},
  \citenamefont {Narayanan}, \citenamefont {Kumar}, \citenamefont {Reuther},\
  and\ \citenamefont {Thomale}}]{Iqbal-2016c}%
  \BibitemOpen
  \bibfield  {author} {\bibinfo {author} {\bibfnamefont {Y.}~\bibnamefont
  {Iqbal}}, \bibinfo {author} {\bibfnamefont {P.}~\bibnamefont {Ghosh}},
  \bibinfo {author} {\bibfnamefont {R.}~\bibnamefont {Narayanan}}, \bibinfo
  {author} {\bibfnamefont {B.}~\bibnamefont {Kumar}}, \bibinfo {author}
  {\bibfnamefont {J.}~\bibnamefont {Reuther}}, \ and\ \bibinfo {author}
  {\bibfnamefont {R.}~\bibnamefont {Thomale}},\ }\bibfield  {title} {\enquote
  {\bibinfo {title} {{\it Intertwined Nematic Orders in a Frustrated
  Ferromagnet}},}\ }\href {\doibase 10.1103/PhysRevB.94.224403} {\bibfield
  {journal} {\bibinfo  {journal} {Phys. Rev. B}\ }\textbf {\bibinfo {volume}
  {94}},\ \bibinfo {pages} {224403} (\bibinfo {year}
  {2016}{\natexlab{c}})}\BibitemShut {NoStop}%
\bibitem [{\citenamefont {Buessen}\ and\ \citenamefont
  {Trebst}(2016)}]{Buessen-16a}%
  \BibitemOpen
  \bibfield  {author} {\bibinfo {author} {\bibfnamefont {F.~L.}\ \bibnamefont
  {Buessen}}\ and\ \bibinfo {author} {\bibfnamefont {S.}~\bibnamefont
  {Trebst}},\ }\bibfield  {title} {\enquote {\bibinfo {title} {{\it Competing
  Magnetic Orders and Spin Liquids in Two- and Three-Dimensional Kagome
  Systems: Pseudofermion Functional Renormalization Group Perspective}},}\
  }\href {\doibase 10.1103/PhysRevB.94.235138} {\bibfield  {journal} {\bibinfo
  {journal} {Phys. Rev. B}\ }\textbf {\bibinfo {volume} {94}},\ \bibinfo
  {pages} {235138} (\bibinfo {year} {2016})}\BibitemShut {NoStop}%
\bibitem [{\citenamefont {Buessen}\ \emph
  {et~al.}(2018{\natexlab{a}})\citenamefont {Buessen}, \citenamefont {Roscher},
  \citenamefont {Diehl},\ and\ \citenamefont {Trebst}}]{Buessen-2017}%
  \BibitemOpen
  \bibfield  {author} {\bibinfo {author} {\bibfnamefont {F.~L.}\ \bibnamefont
  {Buessen}}, \bibinfo {author} {\bibfnamefont {D.}~\bibnamefont {Roscher}},
  \bibinfo {author} {\bibfnamefont {S.}~\bibnamefont {Diehl}}, \ and\ \bibinfo
  {author} {\bibfnamefont {S.}~\bibnamefont {Trebst}},\ }\bibfield  {title}
  {\enquote {\bibinfo {title} {{\it Functional Renormalization Group Approach
  to $\mathrm{SU}(N)$ Heisenberg Models: Real-Space Renormalization Group at
  Arbitrary $N$}},}\ }\href {\doibase 10.1103/PhysRevB.97.064415} {\bibfield
  {journal} {\bibinfo  {journal} {Phys. Rev. B}\ }\textbf {\bibinfo {volume}
  {97}},\ \bibinfo {pages} {064415} (\bibinfo {year}
  {2018}{\natexlab{a}})}\BibitemShut {NoStop}%
\bibitem [{\citenamefont {Roscher}\ \emph {et~al.}(2018)\citenamefont
  {Roscher}, \citenamefont {Buessen}, \citenamefont {Scherer}, \citenamefont
  {Trebst},\ and\ \citenamefont {Diehl}}]{Roscher-2017}%
  \BibitemOpen
  \bibfield  {author} {\bibinfo {author} {\bibfnamefont {D.}~\bibnamefont
  {Roscher}}, \bibinfo {author} {\bibfnamefont {F.~L.}\ \bibnamefont
  {Buessen}}, \bibinfo {author} {\bibfnamefont {M.~M.}\ \bibnamefont
  {Scherer}}, \bibinfo {author} {\bibfnamefont {S.}~\bibnamefont {Trebst}}, \
  and\ \bibinfo {author} {\bibfnamefont {S.}~\bibnamefont {Diehl}},\ }\bibfield
   {title} {\enquote {\bibinfo {title} {{\it Functional Renormalization Group
  Approach to $\mathrm{SU}(N)$ Heisenberg Models: Momentum-Space
  Renormalization Group for the Large-$N$ Limit}},}\ }\href {\doibase
  10.1103/PhysRevB.97.064416} {\bibfield  {journal} {\bibinfo  {journal} {Phys.
  Rev. B}\ }\textbf {\bibinfo {volume} {97}},\ \bibinfo {pages} {064416}
  (\bibinfo {year} {2018})}\BibitemShut {NoStop}%
\bibitem [{\citenamefont {Buessen}\ \emph
  {et~al.}(2018{\natexlab{b}})\citenamefont {Buessen}, \citenamefont {Hering},
  \citenamefont {Reuther},\ and\ \citenamefont {Trebst}}]{Buessen-2018}%
  \BibitemOpen
  \bibfield  {author} {\bibinfo {author} {\bibfnamefont {F.~L.}\ \bibnamefont
  {Buessen}}, \bibinfo {author} {\bibfnamefont {M.}~\bibnamefont {Hering}},
  \bibinfo {author} {\bibfnamefont {J.}~\bibnamefont {Reuther}}, \ and\
  \bibinfo {author} {\bibfnamefont {S.}~\bibnamefont {Trebst}},\ }\bibfield
  {title} {\enquote {\bibinfo {title} {{\it Quantum Spin Liquids in Frustrated
  Spin-1 Diamond Antiferromagnets}},}\ }\href {\doibase
  10.1103/PhysRevLett.120.057201} {\bibfield  {journal} {\bibinfo  {journal}
  {Phys. Rev. Lett.}\ }\textbf {\bibinfo {volume} {120}},\ \bibinfo {pages}
  {057201} (\bibinfo {year} {2018}{\natexlab{b}})}\BibitemShut {NoStop}%
\bibitem [{\citenamefont {Hering}\ and\ \citenamefont
  {Reuther}(2017)}]{Hering-2017}%
  \BibitemOpen
  \bibfield  {author} {\bibinfo {author} {\bibfnamefont {M.}~\bibnamefont
  {Hering}}\ and\ \bibinfo {author} {\bibfnamefont {J.}~\bibnamefont
  {Reuther}},\ }\bibfield  {title} {\enquote {\bibinfo {title} {{\it Functional
  Renormalization Group Analysis of Dzyaloshinsky-Moriya and Heisenberg Spin
  Interactions on the Kagome Lattice}},}\ }\href {\doibase
  10.1103/PhysRevB.95.054418} {\bibfield  {journal} {\bibinfo  {journal} {Phys.
  Rev. B}\ }\textbf {\bibinfo {volume} {95}},\ \bibinfo {pages} {054418}
  (\bibinfo {year} {2017})}\BibitemShut {NoStop}%
\bibitem [{\citenamefont {Iqbal}\ \emph {et~al.}(2017)\citenamefont {Iqbal},
  \citenamefont {M\"uller}, \citenamefont {Riedl}, \citenamefont {Reuther},
  \citenamefont {Rachel}, \citenamefont {Valent\'{\i}}, \citenamefont
  {Gingras}, \citenamefont {Thomale},\ and\ \citenamefont
  {Jeschke}}]{Iqbal-2017}%
  \BibitemOpen
  \bibfield  {author} {\bibinfo {author} {\bibfnamefont {Y.}~\bibnamefont
  {Iqbal}}, \bibinfo {author} {\bibfnamefont {T.}~\bibnamefont {M\"uller}},
  \bibinfo {author} {\bibfnamefont {K.}~\bibnamefont {Riedl}}, \bibinfo
  {author} {\bibfnamefont {J.}~\bibnamefont {Reuther}}, \bibinfo {author}
  {\bibfnamefont {S.}~\bibnamefont {Rachel}}, \bibinfo {author} {\bibfnamefont
  {R.}~\bibnamefont {Valent\'{\i}}}, \bibinfo {author} {\bibfnamefont
  {M.~J.~P.}\ \bibnamefont {Gingras}}, \bibinfo {author} {\bibfnamefont
  {R.}~\bibnamefont {Thomale}}, \ and\ \bibinfo {author} {\bibfnamefont
  {H.~O.}\ \bibnamefont {Jeschke}},\ }\bibfield  {title} {\enquote {\bibinfo
  {title} {{\it Signatures of a Gearwheel Quantum Spin Liquid in a
  Spin-$\frac{1}{2}$ Pyrochlore Molybdate Heisenberg Antiferromagnet}},}\
  }\href {\doibase 10.1103/PhysRevMaterials.1.071201} {\bibfield  {journal}
  {\bibinfo  {journal} {Phys. Rev. Mater.}\ }\textbf {\bibinfo {volume} {1}},\
  \bibinfo {pages} {071201} (\bibinfo {year} {2017})}\BibitemShut {NoStop}%
\bibitem [{\citenamefont {{Chillal}}\ \emph {et~al.}(2017)\citenamefont
  {{Chillal}}, \citenamefont {{Iqbal}}, \citenamefont {{Jeschke}},
  \citenamefont {{Rodriguez-Rivera}}, \citenamefont {{Bewley}}, \citenamefont
  {{Manuel}}, \citenamefont {{Khalyavin}}, \citenamefont {{Steffens}},
  \citenamefont {{Thomale}}, \citenamefont {{Islam}}, \citenamefont
  {{Reuther}},\ and\ \citenamefont {{Lake}}}]{Chillal-2017}%
  \BibitemOpen
  \bibfield  {author} {\bibinfo {author} {\bibfnamefont {S.}~\bibnamefont
  {{Chillal}}}, \bibinfo {author} {\bibfnamefont {Y.}~\bibnamefont {{Iqbal}}},
  \bibinfo {author} {\bibfnamefont {H.~O.}\ \bibnamefont {{Jeschke}}}, \bibinfo
  {author} {\bibfnamefont {J.~A.}\ \bibnamefont {{Rodriguez-Rivera}}}, \bibinfo
  {author} {\bibfnamefont {R.}~\bibnamefont {{Bewley}}}, \bibinfo {author}
  {\bibfnamefont {P.}~\bibnamefont {{Manuel}}}, \bibinfo {author}
  {\bibfnamefont {D.}~\bibnamefont {{Khalyavin}}}, \bibinfo {author}
  {\bibfnamefont {P.}~\bibnamefont {{Steffens}}}, \bibinfo {author}
  {\bibfnamefont {R.}~\bibnamefont {{Thomale}}}, \bibinfo {author}
  {\bibfnamefont {A.~T.~M.~N.}\ \bibnamefont {{Islam}}}, \bibinfo {author}
  {\bibfnamefont {J.}~\bibnamefont {{Reuther}}}, \ and\ \bibinfo {author}
  {\bibfnamefont {B.}~\bibnamefont {{Lake}}},\ }\bibfield  {title} {\enquote
  {\bibinfo {title} {{\it A Quantum Spin Liquid Based on a New
  Three-Dimensional Lattice}},}\ }\href@noop {} {\bibfield  {journal} {\bibinfo
   {journal} {ArXiv e-prints}\ } (\bibinfo {year} {2017})},\ \Eprint
  {http://arxiv.org/abs/1712.07942} {arXiv:1712.07942 [cond-mat.str-el]}
  \BibitemShut {NoStop}%
\bibitem [{\citenamefont {Kele\ifmmode~\mbox{\c{s}}\else \c{s}\fi{}}\ and\
  \citenamefont {Zhao}(2018)}]{Keles-2018}%
  \BibitemOpen
  \bibfield  {author} {\bibinfo {author} {\bibfnamefont {A.}~\bibnamefont
  {Kele\ifmmode~\mbox{\c{s}}\else \c{s}\fi{}}}\ and\ \bibinfo {author}
  {\bibfnamefont {E.}~\bibnamefont {Zhao}},\ }\bibfield  {title} {\enquote
  {\bibinfo {title} {{\it Absence of Long-Range Order in a Triangular Spin
  System with Dipolar Interactions}},}\ }\href {\doibase
  10.1103/PhysRevLett.120.187202} {\bibfield  {journal} {\bibinfo  {journal}
  {Phys. Rev. Lett.}\ }\textbf {\bibinfo {volume} {120}},\ \bibinfo {pages}
  {187202} (\bibinfo {year} {2018})}\BibitemShut {NoStop}%
\bibitem [{\citenamefont {Iqbal}\ \emph
  {et~al.}(2018{\natexlab{a}})\citenamefont {Iqbal}, \citenamefont {M\"uller},
  \citenamefont {Jeschke}, \citenamefont {Thomale},\ and\ \citenamefont
  {Reuther}}]{Iqbal-2018}%
  \BibitemOpen
  \bibfield  {author} {\bibinfo {author} {\bibfnamefont {Y.}~\bibnamefont
  {Iqbal}}, \bibinfo {author} {\bibfnamefont {T.}~\bibnamefont {M\"uller}},
  \bibinfo {author} {\bibfnamefont {H.~O.}\ \bibnamefont {Jeschke}}, \bibinfo
  {author} {\bibfnamefont {R.}~\bibnamefont {Thomale}}, \ and\ \bibinfo
  {author} {\bibfnamefont {J.}~\bibnamefont {Reuther}},\ }\bibfield  {title}
  {\enquote {\bibinfo {title} {{\it Stability of the Spiral Spin Liquid in
  ${\mathrm{MnSc}}_{2}{\mathrm{S}}_{4}$}},}\ }\href {\doibase
  10.1103/PhysRevB.98.064427} {\bibfield  {journal} {\bibinfo  {journal} {Phys.
  Rev. B}\ }\textbf {\bibinfo {volume} {98}},\ \bibinfo {pages} {064427}
  (\bibinfo {year} {2018}{\natexlab{a}})}\BibitemShut {NoStop}%
\bibitem [{\citenamefont {Khomskii}(2010)}]{Khomskii-2010}%
  \BibitemOpen
  \bibfield  {author} {\bibinfo {author} {\bibfnamefont {Daniel~I.}\
  \bibnamefont {Khomskii}},\ }\href
  {https://www.cambridge.org/core/books/basic-aspects-of-the-quantum-theory-of-solids/6FAA00743ABE9CD0774352627FA6CB96}
  {\emph {\bibinfo {title} {Basic Aspects of the Quantum Theory of Solids.
  Order and Elementary Excitations}}}\ (\bibinfo  {publisher} {Cambridge
  University Press},\ \bibinfo {address} {Cambridge, England},\ \bibinfo {year}
  {2010})\BibitemShut {NoStop}%
\bibitem [{\citenamefont {Shannon}\ \emph {et~al.}(2010)\citenamefont
  {Shannon}, \citenamefont {Penc},\ and\ \citenamefont
  {Motome}}]{Shannon-2010}%
  \BibitemOpen
  \bibfield  {author} {\bibinfo {author} {\bibfnamefont {N.}~\bibnamefont
  {Shannon}}, \bibinfo {author} {\bibfnamefont {K.}~\bibnamefont {Penc}}, \
  and\ \bibinfo {author} {\bibfnamefont {Y.}~\bibnamefont {Motome}},\
  }\bibfield  {title} {\enquote {\bibinfo {title} {{\it Nematic,
  Vector-Multipole, and Plateau-Liquid States in the Classical $\mathrm{O}(3)$
  Pyrochlore Antiferromagnet with Biquadratic Interactions in Applied Magnetic
  Field}},}\ }\href {\doibase 10.1103/PhysRevB.81.184409} {\bibfield  {journal}
  {\bibinfo  {journal} {Phys. Rev. B}\ }\textbf {\bibinfo {volume} {81}},\
  \bibinfo {pages} {184409} (\bibinfo {year} {2010})}\BibitemShut {NoStop}%
\bibitem [{\citenamefont {Andreev}\ and\ \citenamefont
  {Grishchuk}(1984)}]{Andreev-1984}%
  \BibitemOpen
  \bibfield  {author} {\bibinfo {author} {\bibfnamefont {A.~F.}\ \bibnamefont
  {Andreev}}\ and\ \bibinfo {author} {\bibfnamefont {I.~A.}\ \bibnamefont
  {Grishchuk}},\ }\bibfield  {title} {\enquote {\bibinfo {title} {{\it Spin
  Nematics}},}\ }\href
  {http://www.jetp.ac.ru/cgi-bin/e/index/e/60/2/p267?a=list} {\bibfield
  {journal} {\bibinfo  {journal} {JETP Lett.}\ }\textbf {\bibinfo {volume}
  {60}},\ \bibinfo {pages} {267} (\bibinfo {year} {1984})}\BibitemShut
  {NoStop}%
\bibitem [{\citenamefont {Taillefumier}\ \emph {et~al.}(2017)\citenamefont
  {Taillefumier}, \citenamefont {Benton}, \citenamefont {Yan}, \citenamefont
  {Jaubert},\ and\ \citenamefont {Shannon}}]{Taillefumier-2017}%
  \BibitemOpen
  \bibfield  {author} {\bibinfo {author} {\bibfnamefont {M.}~\bibnamefont
  {Taillefumier}}, \bibinfo {author} {\bibfnamefont {O.}~\bibnamefont
  {Benton}}, \bibinfo {author} {\bibfnamefont {H.}~\bibnamefont {Yan}},
  \bibinfo {author} {\bibfnamefont {L.~D.~C.}\ \bibnamefont {Jaubert}}, \ and\
  \bibinfo {author} {\bibfnamefont {N.}~\bibnamefont {Shannon}},\ }\bibfield
  {title} {\enquote {\bibinfo {title} {{\it Competing Spin Liquids and Hidden
  Spin-Nematic Order in Spin Ice with Frustrated Transverse Exchange}},}\
  }\href {\doibase 10.1103/PhysRevX.7.041057} {\bibfield  {journal} {\bibinfo
  {journal} {Phys. Rev. X}\ }\textbf {\bibinfo {volume} {7}},\ \bibinfo {pages}
  {041057} (\bibinfo {year} {2017})}\BibitemShut {NoStop}%
\bibitem [{\citenamefont {Millard}\ and\ \citenamefont
  {Leff}(1971)}]{Millard-1971}%
  \BibitemOpen
  \bibfield  {author} {\bibinfo {author} {\bibfnamefont {Kenneth}\ \bibnamefont
  {Millard}}\ and\ \bibinfo {author} {\bibfnamefont {Harvey~S.}\ \bibnamefont
  {Leff}},\ }\bibfield  {title} {\enquote {\bibinfo {title} {{\it Infinite-Spin
  Limit of the Quantum Heisenberg Model}},}\ }\href {\doibase
  10.1063/1.1665664} {\bibfield  {journal} {\bibinfo  {journal} {J. Math. Phys.
  (N.Y.)}\ }\textbf {\bibinfo {volume} {12}},\ \bibinfo {pages} {1000}
  (\bibinfo {year} {1971})}\BibitemShut {NoStop}%
\bibitem [{\citenamefont {Lieb}(1973)}]{Lieb-1973}%
  \BibitemOpen
  \bibfield  {author} {\bibinfo {author} {\bibfnamefont {Elliott~H.}\
  \bibnamefont {Lieb}},\ }\bibfield  {title} {\enquote {\bibinfo {title} {{\it
  The Classical Limit of Quantum Spin Systems}},}\ }\href {\doibase
  10.1007/BF01646493} {\bibfield  {journal} {\bibinfo  {journal} {Commun. Math.
  Phys.}\ }\textbf {\bibinfo {volume} {31}},\ \bibinfo {pages} {327} (\bibinfo
  {year} {1973})}\BibitemShut {NoStop}%
\bibitem [{\citenamefont {Luttinger}\ and\ \citenamefont
  {Tisza}(1946)}]{Luttinger-1946}%
  \BibitemOpen
  \bibfield  {author} {\bibinfo {author} {\bibfnamefont {J.~M.}\ \bibnamefont
  {Luttinger}}\ and\ \bibinfo {author} {\bibfnamefont {L.}~\bibnamefont
  {Tisza}},\ }\bibfield  {title} {\enquote {\bibinfo {title} {{\it Theory of
  Dipole Interaction in Crystals}},}\ }\href {\doibase 10.1103/PhysRev.70.954}
  {\bibfield  {journal} {\bibinfo  {journal} {Phys. Rev.}\ }\textbf {\bibinfo
  {volume} {70}},\ \bibinfo {pages} {954} (\bibinfo {year} {1946})}\BibitemShut
  {NoStop}%
\bibitem [{\citenamefont {Luttinger}(1951)}]{Luttinger-1951}%
  \BibitemOpen
  \bibfield  {author} {\bibinfo {author} {\bibfnamefont {J.~M.}\ \bibnamefont
  {Luttinger}},\ }\bibfield  {title} {\enquote {\bibinfo {title} {{\it A Note
  on the Ground State in Antiferromagnetics}},}\ }\href {\doibase
  10.1103/PhysRev.81.1015} {\bibfield  {journal} {\bibinfo  {journal} {Phys.
  Rev.}\ }\textbf {\bibinfo {volume} {81}},\ \bibinfo {pages} {1015} (\bibinfo
  {year} {1951})}\BibitemShut {NoStop}%
\bibitem [{\citenamefont {Kaplan}\ and\ \citenamefont
  {Menyuk}(2007)}]{Kaplan-2007}%
  \BibitemOpen
  \bibfield  {author} {\bibinfo {author} {\bibfnamefont {T.~A.}\ \bibnamefont
  {Kaplan}}\ and\ \bibinfo {author} {\bibfnamefont {N.}~\bibnamefont
  {Menyuk}},\ }\bibfield  {title} {\enquote {\bibinfo {title} {{\it Spin
  Ordering in Three-Dimensional Crystals with Strong Competing Exchange
  Interactions}},}\ }\href {\doibase 10.1080/14786430601080229} {\bibfield
  {journal} {\bibinfo  {journal} {Philos. Mag.}\ }\textbf {\bibinfo {volume}
  {87}},\ \bibinfo {pages} {3711} (\bibinfo {year} {2007})}\BibitemShut
  {NoStop}%
\bibitem [{\citenamefont {Kimchi}\ and\ \citenamefont
  {Vishwanath}(2014)}]{Kimchi-2014}%
  \BibitemOpen
  \bibfield  {author} {\bibinfo {author} {\bibfnamefont {I.}~\bibnamefont
  {Kimchi}}\ and\ \bibinfo {author} {\bibfnamefont {A.}~\bibnamefont
  {Vishwanath}},\ }\bibfield  {title} {\enquote {\bibinfo {title} {{\it
  \mbox{Kitaev}-\mbox{Heisenberg} Models for Iridates on the Triangular,
  Hyperkagome, Kagome, fcc, and Pyrochlore Lattices}},}\ }\href {\doibase
  10.1103/PhysRevB.89.014414} {\bibfield  {journal} {\bibinfo  {journal} {Phys.
  Rev. B}\ }\textbf {\bibinfo {volume} {89}},\ \bibinfo {pages} {014414}
  (\bibinfo {year} {2014})}\BibitemShut {NoStop}%
\bibitem [{\citenamefont {Bertaut}(1961)}]{Bertaut-1961}%
  \BibitemOpen
  \bibfield  {author} {\bibinfo {author} {\bibfnamefont {E.F.}\ \bibnamefont
  {Bertaut}},\ }\bibfield  {title} {\enquote {\bibinfo {title} {{\it
  Configurations Magnétiques. Méthode de Fourier}},}\ }\href {\doibase
  10.1016/0022-3697(61)90105-6} {\bibfield  {journal} {\bibinfo  {journal} {J.
  Phys. Chem. Solids}\ }\textbf {\bibinfo {volume} {21}},\ \bibinfo {pages}
  {256 -- 279} (\bibinfo {year} {1961})}\BibitemShut {NoStop}%
\bibitem [{\citenamefont {{Nussinov}}(2001)}]{Nussinov-2004}%
  \BibitemOpen
  \bibfield  {author} {\bibinfo {author} {\bibfnamefont {Z.}~\bibnamefont
  {{Nussinov}}},\ }\bibfield  {title} {\enquote {\bibinfo {title} {{\it
  Commensurate and Incommensurate $O(n)$ Spin Systems: Novel Even-Odd Effects,
  A Generalized Mermin-Wagner-Coleman Theorem, and Ground States}},}\
  }\href@noop {} {\bibfield  {journal} {\bibinfo  {journal} {ArXiv e-prints}\ }
  (\bibinfo {year} {2001})},\ \Eprint {http://arxiv.org/abs/cond-mat/0105253}
  {arXiv:cond-mat/0105253 [cond-mat.stat-mech]} \BibitemShut {NoStop}%
\bibitem [{\citenamefont {Yaresko}(2008)}]{Yaresko-2008}%
  \BibitemOpen
  \bibfield  {author} {\bibinfo {author} {\bibfnamefont {A.~N.}\ \bibnamefont
  {Yaresko}},\ }\bibfield  {title} {\enquote {\bibinfo {title} {{\it Electronic
  band structure and exchange coupling constants in $A{\mathrm{Cr}}_{2}{X}_{4}$
  spinels ($\mathrm{A}=\mathrm{Zn}$, $\mathrm{Cd}$, $\mathrm{Hg}$;
  $\mathrm{X}=\mathrm{O}$, $\mathrm{S}$, $\mathrm{Se}$)}},}\ }\href {\doibase
  10.1103/PhysRevB.77.115106} {\bibfield  {journal} {\bibinfo  {journal} {Phys.
  Rev. B}\ }\textbf {\bibinfo {volume} {77}},\ \bibinfo {pages} {115106}
  (\bibinfo {year} {2008})}\BibitemShut {NoStop}%
\bibitem [{\citenamefont {Unger}\ \emph {et~al.}(1975)\citenamefont {Unger},
  \citenamefont {G\"obel}, \citenamefont {Treitinger},\ and\ \citenamefont
  {Pink}}]{Unger-1975}%
  \BibitemOpen
  \bibfield  {author} {\bibinfo {author} {\bibfnamefont {W.K.}\ \bibnamefont
  {Unger}}, \bibinfo {author} {\bibfnamefont {H.}~\bibnamefont {G\"obel}},
  \bibinfo {author} {\bibfnamefont {L.}~\bibnamefont {Treitinger}}, \ and\
  \bibinfo {author} {\bibfnamefont {H.}~\bibnamefont {Pink}},\ }\bibfield
  {title} {\enquote {\bibinfo {title} {{\it Magnetic Susceptibility of
  Semiconducting
  $\mathrm{Cu}_{1-x}\mathrm{In}_{x}\mathrm{Cr}_{2}\mathrm{S}_{4}$ Spinels}},}\
  }\href {\doibase https://doi.org/10.1016/0378-4363(75)90052-2} {\bibfield
  {journal} {\bibinfo  {journal} {Physica (Amsterdam)B+C}\ }\textbf {\bibinfo
  {volume} {80}},\ \bibinfo {pages} {62} (\bibinfo {year} {1975})}\BibitemShut
  {NoStop}%
\bibitem [{\citenamefont {Okamoto}\ \emph {et~al.}(2013)\citenamefont
  {Okamoto}, \citenamefont {Nilsen}, \citenamefont {Attfield},\ and\
  \citenamefont {Hiroi}}]{Okamoto-2013}%
  \BibitemOpen
  \bibfield  {author} {\bibinfo {author} {\bibfnamefont {Y.}~\bibnamefont
  {Okamoto}}, \bibinfo {author} {\bibfnamefont {G.~J.}\ \bibnamefont {Nilsen}},
  \bibinfo {author} {\bibfnamefont {J.~P.}\ \bibnamefont {Attfield}}, \ and\
  \bibinfo {author} {\bibfnamefont {Z.}~\bibnamefont {Hiroi}},\ }\bibfield
  {title} {\enquote {\bibinfo {title} {{\it Breathing Pyrochlore Lattice
  Realized in $A$-Site Ordered Spinel Oxides
  ${\mathrm{LiGaCr}}_{4}{\mathrm{O}}_{8}$ and
  ${\mathrm{LiInCr}}_{4}{\mathrm{O}}_{8}$}},}\ }\href {\doibase
  10.1103/PhysRevLett.110.097203} {\bibfield  {journal} {\bibinfo  {journal}
  {Phys. Rev. Lett.}\ }\textbf {\bibinfo {volume} {110}},\ \bibinfo {pages}
  {097203} (\bibinfo {year} {2013})}\BibitemShut {NoStop}%
\bibitem [{\citenamefont {Tanaka}\ \emph {et~al.}(2014)\citenamefont {Tanaka},
  \citenamefont {Yoshida}, \citenamefont {Takigawa}, \citenamefont {Okamoto},\
  and\ \citenamefont {Hiroi}}]{Tanaka-2014}%
  \BibitemOpen
  \bibfield  {author} {\bibinfo {author} {\bibfnamefont {Y.}~\bibnamefont
  {Tanaka}}, \bibinfo {author} {\bibfnamefont {M.}~\bibnamefont {Yoshida}},
  \bibinfo {author} {\bibfnamefont {M.}~\bibnamefont {Takigawa}}, \bibinfo
  {author} {\bibfnamefont {Y.}~\bibnamefont {Okamoto}}, \ and\ \bibinfo
  {author} {\bibfnamefont {Z.}~\bibnamefont {Hiroi}},\ }\bibfield  {title}
  {\enquote {\bibinfo {title} {{\it Novel Phase Transitions in the Breathing
  Pyrochlore Lattice: $^{7}\mathrm{Li}\text{\ensuremath{-}}\mathrm{NMR}$ on
  ${\mathrm{LiInCr}}_{4}{\mathrm{O}}_{8}$ and
  ${\mathrm{LiGaCr}}_{4}{\mathrm{O}}_{8}$}},}\ }\href {\doibase
  10.1103/PhysRevLett.113.227204} {\bibfield  {journal} {\bibinfo  {journal}
  {Phys. Rev. Lett.}\ }\textbf {\bibinfo {volume} {113}},\ \bibinfo {pages}
  {227204} (\bibinfo {year} {2014})}\BibitemShut {NoStop}%
\bibitem [{\citenamefont {Okamoto}\ \emph {et~al.}(2015)\citenamefont
  {Okamoto}, \citenamefont {Nilsen}, \citenamefont {Nakazono},\ and\
  \citenamefont {Hiroi}}]{Okamoto-2015}%
  \BibitemOpen
  \bibfield  {author} {\bibinfo {author} {\bibfnamefont {Y.}~\bibnamefont
  {Okamoto}}, \bibinfo {author} {\bibfnamefont {G.~J.}\ \bibnamefont {Nilsen}},
  \bibinfo {author} {\bibfnamefont {T.}~\bibnamefont {Nakazono}}, \ and\
  \bibinfo {author} {\bibfnamefont {Z.}~\bibnamefont {Hiroi}},\ }\bibfield
  {title} {\enquote {\bibinfo {title} {{\it Magnetic Phase Diagram of the
  Breathing Pyrochlore Antiferromagnet
  $\mathrm{LiGa}_{1-x}\mathrm{In}_{x}\mathrm{Cr}_{4}\mathrm{O}_{8}$}},}\ }\href
  {\doibase 10.7566/JPSJ.84.043707} {\bibfield  {journal} {\bibinfo  {journal}
  {J. Phys. Soc. Jpn.}\ }\textbf {\bibinfo {volume} {84}},\ \bibinfo {pages}
  {043707} (\bibinfo {year} {2015})}\BibitemShut {NoStop}%
\bibitem [{\citenamefont {Nilsen}\ \emph {et~al.}(2015)\citenamefont {Nilsen},
  \citenamefont {Okamoto}, \citenamefont {Masuda}, \citenamefont
  {Rodriguez-Carvajal}, \citenamefont {Mutka}, \citenamefont {Hansen},\ and\
  \citenamefont {Hiroi}}]{Nilsen-2015}%
  \BibitemOpen
  \bibfield  {author} {\bibinfo {author} {\bibfnamefont {G.~J.}\ \bibnamefont
  {Nilsen}}, \bibinfo {author} {\bibfnamefont {Y.}~\bibnamefont {Okamoto}},
  \bibinfo {author} {\bibfnamefont {T.}~\bibnamefont {Masuda}}, \bibinfo
  {author} {\bibfnamefont {J.}~\bibnamefont {Rodriguez-Carvajal}}, \bibinfo
  {author} {\bibfnamefont {H.}~\bibnamefont {Mutka}}, \bibinfo {author}
  {\bibfnamefont {T.}~\bibnamefont {Hansen}}, \ and\ \bibinfo {author}
  {\bibfnamefont {Z.}~\bibnamefont {Hiroi}},\ }\bibfield  {title} {\enquote
  {\bibinfo {title} {{\it Complex Magnetostructural Order in the Frustrated
  Spinel ${\mathrm{LiInCr}}_{4}{\mathrm{O}}_{8}$}},}\ }\href {\doibase
  10.1103/PhysRevB.91.174435} {\bibfield  {journal} {\bibinfo  {journal} {Phys.
  Rev. B}\ }\textbf {\bibinfo {volume} {91}},\ \bibinfo {pages} {174435}
  (\bibinfo {year} {2015})}\BibitemShut {NoStop}%
\bibitem [{\citenamefont {Li}\ \emph {et~al.}(2016)\citenamefont {Li},
  \citenamefont {Li}, \citenamefont {Kim}, \citenamefont {Balents},
  \citenamefont {Yu},\ and\ \citenamefont {Chen}}]{Li-2016}%
  \BibitemOpen
  \bibfield  {author} {\bibinfo {author} {\bibfnamefont {F.-Y.}\ \bibnamefont
  {Li}}, \bibinfo {author} {\bibfnamefont {Y.-D.}\ \bibnamefont {Li}}, \bibinfo
  {author} {\bibfnamefont {Y.~B.}\ \bibnamefont {Kim}}, \bibinfo {author}
  {\bibfnamefont {L.}~\bibnamefont {Balents}}, \bibinfo {author} {\bibfnamefont
  {Y.}~\bibnamefont {Yu}}, \ and\ \bibinfo {author} {\bibfnamefont
  {G.}~\bibnamefont {Chen}},\ }\bibfield  {title} {\enquote {\bibinfo {title}
  {{\it Weyl Magnons in Breathing Pyrochlore Antiferromagnets}},}\ }\href
  {http://dx.doi.org/10.1038/ncomms12691} {\bibfield  {journal} {\bibinfo
  {journal} {Nat. Commun.}\ }\textbf {\bibinfo {volume} {7}},\ \bibinfo {pages}
  {12691} (\bibinfo {year} {2016})}\BibitemShut {NoStop}%
\bibitem [{\citenamefont {Lee}\ \emph {et~al.}(2016)\citenamefont {Lee},
  \citenamefont {Do}, \citenamefont {Lee}, \citenamefont {Choi}, \citenamefont
  {Lee}, \citenamefont {Choi}, \citenamefont {Reyes}, \citenamefont {Kuhns},
  \citenamefont {Ozarowski},\ and\ \citenamefont {Choi}}]{Lee-2016}%
  \BibitemOpen
  \bibfield  {author} {\bibinfo {author} {\bibfnamefont {S.}~\bibnamefont
  {Lee}}, \bibinfo {author} {\bibfnamefont {S.-H.}\ \bibnamefont {Do}},
  \bibinfo {author} {\bibfnamefont {W.-J.}\ \bibnamefont {Lee}}, \bibinfo
  {author} {\bibfnamefont {Y.~S.}\ \bibnamefont {Choi}}, \bibinfo {author}
  {\bibfnamefont {M.}~\bibnamefont {Lee}}, \bibinfo {author} {\bibfnamefont
  {E.~S.}\ \bibnamefont {Choi}}, \bibinfo {author} {\bibfnamefont {A.~P.}\
  \bibnamefont {Reyes}}, \bibinfo {author} {\bibfnamefont {P.~L.}\ \bibnamefont
  {Kuhns}}, \bibinfo {author} {\bibfnamefont {A.}~\bibnamefont {Ozarowski}}, \
  and\ \bibinfo {author} {\bibfnamefont {K.-Y.}\ \bibnamefont {Choi}},\
  }\bibfield  {title} {\enquote {\bibinfo {title} {{\it Multistage Symmetry
  Breaking in the Breathing Pyrochlore Lattice
  ${\mathrm{Li}(\mathrm{Ga},\mathrm{In})\mathrm{Cr}}_{4}{\mathrm{O}}_{8}$}},}\
  }\href {\doibase 10.1103/PhysRevB.93.174402} {\bibfield  {journal} {\bibinfo
  {journal} {Phys. Rev. B}\ }\textbf {\bibinfo {volume} {93}},\ \bibinfo
  {pages} {174402} (\bibinfo {year} {2016})}\BibitemShut {NoStop}%
\bibitem [{\citenamefont {Saha}\ \emph {et~al.}(2016)\citenamefont {Saha},
  \citenamefont {Fauth}, \citenamefont {Avdeev}, \citenamefont {Kayser},
  \citenamefont {Kennedy},\ and\ \citenamefont {Sundaresan}}]{Saha-2016}%
  \BibitemOpen
  \bibfield  {author} {\bibinfo {author} {\bibfnamefont {R.}~\bibnamefont
  {Saha}}, \bibinfo {author} {\bibfnamefont {F.}~\bibnamefont {Fauth}},
  \bibinfo {author} {\bibfnamefont {M.}~\bibnamefont {Avdeev}}, \bibinfo
  {author} {\bibfnamefont {P.}~\bibnamefont {Kayser}}, \bibinfo {author}
  {\bibfnamefont {B.~J.}\ \bibnamefont {Kennedy}}, \ and\ \bibinfo {author}
  {\bibfnamefont {A.}~\bibnamefont {Sundaresan}},\ }\bibfield  {title}
  {\enquote {\bibinfo {title} {{\it Magnetodielectric Effects in $A$-site
  Cation-Ordered Chromate Spinels
  $\mathrm{Li}M\mathrm{C}{\mathrm{r}}_{4}{\mathrm{O}}_{8}$ ($M=\mathrm{Ga}$
  $\mathrm{and}$ $\mathrm{In}$)}},}\ }\href {\doibase
  10.1103/PhysRevB.94.064420} {\bibfield  {journal} {\bibinfo  {journal} {Phys.
  Rev. B}\ }\textbf {\bibinfo {volume} {94}},\ \bibinfo {pages} {064420}
  (\bibinfo {year} {2016})}\BibitemShut {NoStop}%
\bibitem [{\citenamefont {Aoyama}\ and\ \citenamefont
  {Kawamura}(2016)}]{Aoyama-2016}%
  \BibitemOpen
  \bibfield  {author} {\bibinfo {author} {\bibfnamefont {K.}~\bibnamefont
  {Aoyama}}\ and\ \bibinfo {author} {\bibfnamefont {H.}~\bibnamefont
  {Kawamura}},\ }\bibfield  {title} {\enquote {\bibinfo {title} {{\it
  Spin-Lattice-Coupled Order in Heisenberg Antiferromagnets on the Pyrochlore
  Lattice}},}\ }\href {\doibase 10.1103/PhysRevLett.116.257201} {\bibfield
  {journal} {\bibinfo  {journal} {Phys. Rev. Lett.}\ }\textbf {\bibinfo
  {volume} {116}},\ \bibinfo {pages} {257201} (\bibinfo {year}
  {2016})}\BibitemShut {NoStop}%
\bibitem [{\citenamefont {Wawrzy\ifmmode~\acute{n}\else \'{n}\fi{}czak}\ \emph
  {et~al.}(2017)\citenamefont {Wawrzy\ifmmode~\acute{n}\else \'{n}\fi{}czak},
  \citenamefont {Tanaka}, \citenamefont {Yoshida}, \citenamefont {Okamoto},
  \citenamefont {Manuel}, \citenamefont {Casati}, \citenamefont {Hiroi},
  \citenamefont {Takigawa},\ and\ \citenamefont {Nilsen}}]{Wawrzy-2017}%
  \BibitemOpen
  \bibfield  {author} {\bibinfo {author} {\bibfnamefont {R.}~\bibnamefont
  {Wawrzy\ifmmode~\acute{n}\else \'{n}\fi{}czak}}, \bibinfo {author}
  {\bibfnamefont {Y.}~\bibnamefont {Tanaka}}, \bibinfo {author} {\bibfnamefont
  {M.}~\bibnamefont {Yoshida}}, \bibinfo {author} {\bibfnamefont
  {Y.}~\bibnamefont {Okamoto}}, \bibinfo {author} {\bibfnamefont
  {P.}~\bibnamefont {Manuel}}, \bibinfo {author} {\bibfnamefont
  {N.}~\bibnamefont {Casati}}, \bibinfo {author} {\bibfnamefont
  {Z.}~\bibnamefont {Hiroi}}, \bibinfo {author} {\bibfnamefont
  {M.}~\bibnamefont {Takigawa}}, \ and\ \bibinfo {author} {\bibfnamefont
  {G.~J.}\ \bibnamefont {Nilsen}},\ }\bibfield  {title} {\enquote {\bibinfo
  {title} {{\it Classical Spin Nematic Transition in
  ${\mathrm{LiGa}}_{0.95}{\mathrm{In}}_{0.05}{\mathrm{Cr}}_{4}{\mathrm{O}}_{8}$}},}\
  }\href {\doibase 10.1103/PhysRevLett.119.087201} {\bibfield  {journal}
  {\bibinfo  {journal} {Phys. Rev. Lett.}\ }\textbf {\bibinfo {volume} {119}},\
  \bibinfo {pages} {087201} (\bibinfo {year} {2017})}\BibitemShut {NoStop}%
\bibitem [{\citenamefont {Okamoto}\ \emph {et~al.}(2017)\citenamefont
  {Okamoto}, \citenamefont {Nakamura}, \citenamefont {Miyake}, \citenamefont
  {Takeyama}, \citenamefont {Tokunaga}, \citenamefont {Matsuo}, \citenamefont
  {Kindo},\ and\ \citenamefont {Hiroi}}]{Takeyama-2017}%
  \BibitemOpen
  \bibfield  {author} {\bibinfo {author} {\bibfnamefont {Y.}~\bibnamefont
  {Okamoto}}, \bibinfo {author} {\bibfnamefont {D.}~\bibnamefont {Nakamura}},
  \bibinfo {author} {\bibfnamefont {A.}~\bibnamefont {Miyake}}, \bibinfo
  {author} {\bibfnamefont {S.}~\bibnamefont {Takeyama}}, \bibinfo {author}
  {\bibfnamefont {M.}~\bibnamefont {Tokunaga}}, \bibinfo {author}
  {\bibfnamefont {A.}~\bibnamefont {Matsuo}}, \bibinfo {author} {\bibfnamefont
  {K.}~\bibnamefont {Kindo}}, \ and\ \bibinfo {author} {\bibfnamefont
  {Z.}~\bibnamefont {Hiroi}},\ }\bibfield  {title} {\enquote {\bibinfo {title}
  {{\it Magnetic Transitions under Ultrahigh Magnetic Fields of up to 130 T in
  the Breathing Pyrochlore Antiferromagnet
  ${\mathrm{LiInCr}}_{4}{\mathrm{O}}_{8}$}},}\ }\href {\doibase
  10.1103/PhysRevB.95.134438} {\bibfield  {journal} {\bibinfo  {journal} {Phys.
  Rev. B}\ }\textbf {\bibinfo {volume} {95}},\ \bibinfo {pages} {134438}
  (\bibinfo {year} {2017})}\BibitemShut {NoStop}%
\bibitem [{\citenamefont {Okamoto}\ \emph {et~al.}(2018)\citenamefont
  {Okamoto}, \citenamefont {Mori}, \citenamefont {Katayama}, \citenamefont
  {Miyake}, \citenamefont {Tokunaga}, \citenamefont {Matsuo}, \citenamefont
  {Kindo},\ and\ \citenamefont {Takenaka}}]{Okamoto-2018}%
  \BibitemOpen
  \bibfield  {author} {\bibinfo {author} {\bibfnamefont {Y.}~\bibnamefont
  {Okamoto}}, \bibinfo {author} {\bibfnamefont {M.}~\bibnamefont {Mori}},
  \bibinfo {author} {\bibfnamefont {N.}~\bibnamefont {Katayama}}, \bibinfo
  {author} {\bibfnamefont {A.}~\bibnamefont {Miyake}}, \bibinfo {author}
  {\bibfnamefont {M.}~\bibnamefont {Tokunaga}}, \bibinfo {author}
  {\bibfnamefont {A.}~\bibnamefont {Matsuo}}, \bibinfo {author} {\bibfnamefont
  {K.}~\bibnamefont {Kindo}}, \ and\ \bibinfo {author} {\bibfnamefont
  {K.}~\bibnamefont {Takenaka}},\ }\bibfield  {title} {\enquote {\bibinfo
  {title} {{\it Magnetic and Structural Properties of A-Site Ordered Chromium
  Spinel Sulfides: Alternating Antiferromagnetic and Ferromagnetic Interactions
  in the Breathing Pyrochlore Lattice}},}\ }\href {\doibase
  10.7566/JPSJ.87.034709} {\bibfield  {journal} {\bibinfo  {journal} {J. Phys.
  Soc. Jpn.}\ }\textbf {\bibinfo {volume} {87}},\ \bibinfo {pages} {034709}
  (\bibinfo {year} {2018})}\BibitemShut {NoStop}%
\bibitem [{\citenamefont {Pokharel}\ \emph {et~al.}(2018)\citenamefont
  {Pokharel}, \citenamefont {May}, \citenamefont {Parker}, \citenamefont
  {Calder}, \citenamefont {Ehlers}, \citenamefont {Huq}, \citenamefont
  {Kimber}, \citenamefont {Arachchige}, \citenamefont {Poudel}, \citenamefont
  {McGuire}, \citenamefont {Mandrus},\ and\ \citenamefont
  {Christianson}}]{Pokharel-2018}%
  \BibitemOpen
  \bibfield  {author} {\bibinfo {author} {\bibfnamefont {G.}~\bibnamefont
  {Pokharel}}, \bibinfo {author} {\bibfnamefont {A.~F.}\ \bibnamefont {May}},
  \bibinfo {author} {\bibfnamefont {D.~S.}\ \bibnamefont {Parker}}, \bibinfo
  {author} {\bibfnamefont {S.}~\bibnamefont {Calder}}, \bibinfo {author}
  {\bibfnamefont {G.}~\bibnamefont {Ehlers}}, \bibinfo {author} {\bibfnamefont
  {A.}~\bibnamefont {Huq}}, \bibinfo {author} {\bibfnamefont {S.~A.~J.}\
  \bibnamefont {Kimber}}, \bibinfo {author} {\bibfnamefont {H.~Suriya}\
  \bibnamefont {Arachchige}}, \bibinfo {author} {\bibfnamefont
  {L.}~\bibnamefont {Poudel}}, \bibinfo {author} {\bibfnamefont {M.~A.}\
  \bibnamefont {McGuire}}, \bibinfo {author} {\bibfnamefont {D.}~\bibnamefont
  {Mandrus}}, \ and\ \bibinfo {author} {\bibfnamefont {A.~D.}\ \bibnamefont
  {Christianson}},\ }\bibfield  {title} {\enquote {\bibinfo {title} {{\it
  Negative Thermal Expansion and Magnetoelastic Coupling in the Breathing
  Pyrochlore Lattice Material ${\mathrm{LiGaCr}}_{4}{\mathrm{S}}_{8}$}},}\
  }\href {\doibase 10.1103/PhysRevB.97.134117} {\bibfield  {journal} {\bibinfo
  {journal} {Phys. Rev. B}\ }\textbf {\bibinfo {volume} {97}},\ \bibinfo
  {pages} {134117} (\bibinfo {year} {2018})}\BibitemShut {NoStop}%
\bibitem [{\citenamefont {Ezawa}(2018)}]{Ezawa-2018}%
  \BibitemOpen
  \bibfield  {author} {\bibinfo {author} {\bibfnamefont {M.}~\bibnamefont
  {Ezawa}},\ }\bibfield  {title} {\enquote {\bibinfo {title} {{\it Higher-Order
  Topological Insulators and Semimetals on the Breathing Kagome and Pyrochlore
  Lattices}},}\ }\href {\doibase 10.1103/PhysRevLett.120.026801} {\bibfield
  {journal} {\bibinfo  {journal} {Phys. Rev. Lett.}\ }\textbf {\bibinfo
  {volume} {120}},\ \bibinfo {pages} {026801} (\bibinfo {year}
  {2018})}\BibitemShut {NoStop}%
\bibitem [{\citenamefont {Benton}\ and\ \citenamefont
  {Shannon}(2015)}]{Benton-2015}%
  \BibitemOpen
  \bibfield  {author} {\bibinfo {author} {\bibfnamefont {O.}~\bibnamefont
  {Benton}}\ and\ \bibinfo {author} {\bibfnamefont {N.}~\bibnamefont
  {Shannon}},\ }\bibfield  {title} {\enquote {\bibinfo {title} {{\it Ground
  State Selection and Spin-Liquid Behaviour in the Classical Heisenberg Model
  on the Breathing Pyrochlore Lattice}},}\ }\href {\doibase
  10.7566/JPSJ.84.104710} {\bibfield  {journal} {\bibinfo  {journal} {J. Phys.
  Soc. Jpn.}\ }\textbf {\bibinfo {volume} {84}},\ \bibinfo {pages} {104710}
  (\bibinfo {year} {2015})}\BibitemShut {NoStop}%
\bibitem [{\citenamefont {Kimura}\ \emph {et~al.}(2014)\citenamefont {Kimura},
  \citenamefont {Nakatsuji},\ and\ \citenamefont {Kimura}}]{Kimura-2014}%
  \BibitemOpen
  \bibfield  {author} {\bibinfo {author} {\bibfnamefont {K.}~\bibnamefont
  {Kimura}}, \bibinfo {author} {\bibfnamefont {S.}~\bibnamefont {Nakatsuji}}, \
  and\ \bibinfo {author} {\bibfnamefont {T.}~\bibnamefont {Kimura}},\
  }\bibfield  {title} {\enquote {\bibinfo {title} {{\it Experimental
  Realization of a Quantum Breathing Pyrochlore Antiferromagnet}},}\ }\href
  {\doibase 10.1103/PhysRevB.90.060414} {\bibfield  {journal} {\bibinfo
  {journal} {Phys. Rev. B}\ }\textbf {\bibinfo {volume} {90}},\ \bibinfo
  {pages} {060414} (\bibinfo {year} {2014})}\BibitemShut {NoStop}%
\bibitem [{\citenamefont {Rau}\ \emph {et~al.}(2016)\citenamefont {Rau},
  \citenamefont {Wu}, \citenamefont {May}, \citenamefont {Poudel},
  \citenamefont {Winn}, \citenamefont {Garlea}, \citenamefont {Huq},
  \citenamefont {Whitfield}, \citenamefont {Taylor}, \citenamefont {Lumsden},
  \citenamefont {Gingras},\ and\ \citenamefont {Christianson}}]{Rau-2016}%
  \BibitemOpen
  \bibfield  {author} {\bibinfo {author} {\bibfnamefont {J.~G.}\ \bibnamefont
  {Rau}}, \bibinfo {author} {\bibfnamefont {L.~S.}\ \bibnamefont {Wu}},
  \bibinfo {author} {\bibfnamefont {A.~F.}\ \bibnamefont {May}}, \bibinfo
  {author} {\bibfnamefont {L.}~\bibnamefont {Poudel}}, \bibinfo {author}
  {\bibfnamefont {B.}~\bibnamefont {Winn}}, \bibinfo {author} {\bibfnamefont
  {V.~O.}\ \bibnamefont {Garlea}}, \bibinfo {author} {\bibfnamefont
  {A.}~\bibnamefont {Huq}}, \bibinfo {author} {\bibfnamefont {P.}~\bibnamefont
  {Whitfield}}, \bibinfo {author} {\bibfnamefont {A.~E.}\ \bibnamefont
  {Taylor}}, \bibinfo {author} {\bibfnamefont {M.~D.}\ \bibnamefont {Lumsden}},
  \bibinfo {author} {\bibfnamefont {M.~J.~P.}\ \bibnamefont {Gingras}}, \ and\
  \bibinfo {author} {\bibfnamefont {A.~D.}\ \bibnamefont {Christianson}},\
  }\bibfield  {title} {\enquote {\bibinfo {title} {{\it Anisotropic Exchange
  within Decoupled Tetrahedra in the Quantum Breathing Pyrochlore
  ${\mathrm{Ba}}_{3}{\mathrm{Yb}}_{2}{\mathrm{Zn}}_{5}{\mathrm{O}}_{11}$}},}\
  }\href {\doibase 10.1103/PhysRevLett.116.257204} {\bibfield  {journal}
  {\bibinfo  {journal} {Phys. Rev. Lett.}\ }\textbf {\bibinfo {volume} {116}},\
  \bibinfo {pages} {257204} (\bibinfo {year} {2016})}\BibitemShut {NoStop}%
\bibitem [{\citenamefont {Savary}\ \emph {et~al.}(2016)\citenamefont {Savary},
  \citenamefont {Wang}, \citenamefont {Kee}, \citenamefont {Kim}, \citenamefont
  {Yu},\ and\ \citenamefont {Chen}}]{Savary-2016}%
  \BibitemOpen
  \bibfield  {author} {\bibinfo {author} {\bibfnamefont {L.}~\bibnamefont
  {Savary}}, \bibinfo {author} {\bibfnamefont {X.}~\bibnamefont {Wang}},
  \bibinfo {author} {\bibfnamefont {H.-Y.}\ \bibnamefont {Kee}}, \bibinfo
  {author} {\bibfnamefont {Y.~B.}\ \bibnamefont {Kim}}, \bibinfo {author}
  {\bibfnamefont {Y.}~\bibnamefont {Yu}}, \ and\ \bibinfo {author}
  {\bibfnamefont {G.}~\bibnamefont {Chen}},\ }\bibfield  {title} {\enquote
  {\bibinfo {title} {{\it Quantum Spin Ice on the Breathing Pyrochlore
  Lattice}},}\ }\href {\doibase 10.1103/PhysRevB.94.075146} {\bibfield
  {journal} {\bibinfo  {journal} {Phys. Rev. B}\ }\textbf {\bibinfo {volume}
  {94}},\ \bibinfo {pages} {075146} (\bibinfo {year} {2016})}\BibitemShut
  {NoStop}%
\bibitem [{\citenamefont {Anderson}(1956)}]{Anderson-1956}%
  \BibitemOpen
  \bibfield  {author} {\bibinfo {author} {\bibfnamefont {P.~W.}\ \bibnamefont
  {Anderson}},\ }\bibfield  {title} {\enquote {\bibinfo {title} {{\it Ordering
  and Antiferromagnetism in Ferrites}},}\ }\href {\doibase
  10.1103/PhysRev.102.1008} {\bibfield  {journal} {\bibinfo  {journal} {Phys.
  Rev.}\ }\textbf {\bibinfo {volume} {102}},\ \bibinfo {pages} {1008} (\bibinfo
  {year} {1956})}\BibitemShut {NoStop}%
\bibitem [{\citenamefont {Henley}(2010)}]{Henley-2010}%
  \BibitemOpen
  \bibfield  {author} {\bibinfo {author} {\bibfnamefont {Christopher~L.}\
  \bibnamefont {Henley}},\ }\bibfield  {title} {\enquote {\bibinfo {title}
  {{\it The Coulomb Phase in Frustrated Systems}},}\ }\href {\doibase
  10.1146/annurev-conmatphys-070909-104138} {\bibfield  {journal} {\bibinfo
  {journal} {Annu. Rev. Condens. Matter Phys.}\ }\textbf {\bibinfo {volume}
  {1}},\ \bibinfo {pages} {179} (\bibinfo {year} {2010})}\BibitemShut {NoStop}%
\bibitem [{\citenamefont {Stillinger}\ and\ \citenamefont
  {Cotter}(1973)}]{Stillinger-1973}%
  \BibitemOpen
  \bibfield  {author} {\bibinfo {author} {\bibfnamefont {F.~H.}\ \bibnamefont
  {Stillinger}}\ and\ \bibinfo {author} {\bibfnamefont {M.~A.}\ \bibnamefont
  {Cotter}},\ }\bibfield  {title} {\enquote {\bibinfo {title} {{\it Local
  Orientational Order in Ice}},}\ }\href {\doibase 10.1063/1.1679535}
  {\bibfield  {journal} {\bibinfo  {journal} {J. Chem. Phys.}\ }\textbf
  {\bibinfo {volume} {58}},\ \bibinfo {pages} {2532} (\bibinfo {year}
  {1973})}\BibitemShut {NoStop}%
\bibitem [{\citenamefont {Youngblood}\ and\ \citenamefont
  {Axe}(1981)}]{Youngblood-1981}%
  \BibitemOpen
  \bibfield  {author} {\bibinfo {author} {\bibfnamefont {R.~W.}\ \bibnamefont
  {Youngblood}}\ and\ \bibinfo {author} {\bibfnamefont {J.~D.}\ \bibnamefont
  {Axe}},\ }\bibfield  {title} {\enquote {\bibinfo {title} {{\it Polarization
  Fluctuations in Ferroelectric Models}},}\ }\href {\doibase
  10.1103/PhysRevB.23.232} {\bibfield  {journal} {\bibinfo  {journal} {Phys.
  Rev. B}\ }\textbf {\bibinfo {volume} {23}},\ \bibinfo {pages} {232} (\bibinfo
  {year} {1981})}\BibitemShut {NoStop}%
\bibitem [{\citenamefont {Henley}(1992, APS March Meeting)}]{Henley-1992}%
  \BibitemOpen
  \bibfield  {author} {\bibinfo {author} {\bibfnamefont {C.~L.}\ \bibnamefont
  {Henley}},\ }\bibfield  {title} {\enquote {\bibinfo {title} {{\it
  Polarization Fluctuations in Ferroelectric Models}},}\ }\href@noop {}
  {\bibfield  {journal} {\bibinfo  {journal} {Bull. Am. Phys. Soc.}\ }\textbf
  {\bibinfo {volume} {37}},\ \bibinfo {pages} {441} (\bibinfo {year} {1992, APS
  March Meeting})}\BibitemShut {NoStop}%
\bibitem [{\citenamefont {Huse}\ \emph {et~al.}(2003)\citenamefont {Huse},
  \citenamefont {Krauth}, \citenamefont {Moessner},\ and\ \citenamefont
  {Sondhi}}]{Huse-2003}%
  \BibitemOpen
  \bibfield  {author} {\bibinfo {author} {\bibfnamefont {D.~A.}\ \bibnamefont
  {Huse}}, \bibinfo {author} {\bibfnamefont {W.}~\bibnamefont {Krauth}},
  \bibinfo {author} {\bibfnamefont {R.}~\bibnamefont {Moessner}}, \ and\
  \bibinfo {author} {\bibfnamefont {S.~L.}\ \bibnamefont {Sondhi}},\ }\bibfield
   {title} {\enquote {\bibinfo {title} {{\it Coulomb and Liquid Dimer Models in
  Three Dimensions}},}\ }\href {\doibase 10.1103/PhysRevLett.91.167004}
  {\bibfield  {journal} {\bibinfo  {journal} {Phys. Rev. Lett.}\ }\textbf
  {\bibinfo {volume} {91}},\ \bibinfo {pages} {167004} (\bibinfo {year}
  {2003})}\BibitemShut {NoStop}%
\bibitem [{\citenamefont {Isakov}\ \emph {et~al.}(2004)\citenamefont {Isakov},
  \citenamefont {Gregor}, \citenamefont {Moessner},\ and\ \citenamefont
  {Sondhi}}]{Isakov-2004}%
  \BibitemOpen
  \bibfield  {author} {\bibinfo {author} {\bibfnamefont {S.~V.}\ \bibnamefont
  {Isakov}}, \bibinfo {author} {\bibfnamefont {K.}~\bibnamefont {Gregor}},
  \bibinfo {author} {\bibfnamefont {R.}~\bibnamefont {Moessner}}, \ and\
  \bibinfo {author} {\bibfnamefont {S.~L.}\ \bibnamefont {Sondhi}},\ }\bibfield
   {title} {\enquote {\bibinfo {title} {{\it Dipolar Spin Correlations in
  Classical Pyrochlore Magnets}},}\ }\href {\doibase
  10.1103/PhysRevLett.93.167204} {\bibfield  {journal} {\bibinfo  {journal}
  {Phys. Rev. Lett.}\ }\textbf {\bibinfo {volume} {93}},\ \bibinfo {pages}
  {167204} (\bibinfo {year} {2004})}\BibitemShut {NoStop}%
\bibitem [{\citenamefont {Hopkinson}\ \emph {et~al.}(2007)\citenamefont
  {Hopkinson}, \citenamefont {Isakov}, \citenamefont {Kee},\ and\ \citenamefont
  {Kim}}]{Hopkinson-2007}%
  \BibitemOpen
  \bibfield  {author} {\bibinfo {author} {\bibfnamefont {J.~M.}\ \bibnamefont
  {Hopkinson}}, \bibinfo {author} {\bibfnamefont {S.~V.}\ \bibnamefont
  {Isakov}}, \bibinfo {author} {\bibfnamefont {H.-Y.}\ \bibnamefont {Kee}}, \
  and\ \bibinfo {author} {\bibfnamefont {Y.~B.}\ \bibnamefont {Kim}},\
  }\bibfield  {title} {\enquote {\bibinfo {title} {{\it Classical
  Antiferromagnet on a Hyperkagome Lattice}},}\ }\href {\doibase
  10.1103/PhysRevLett.99.037201} {\bibfield  {journal} {\bibinfo  {journal}
  {Phys. Rev. Lett.}\ }\textbf {\bibinfo {volume} {99}},\ \bibinfo {pages}
  {037201} (\bibinfo {year} {2007})}\BibitemShut {NoStop}%
\bibitem [{\citenamefont {Fennell}\ \emph {et~al.}(2007)\citenamefont
  {Fennell}, \citenamefont {Bramwell}, \citenamefont {McMorrow}, \citenamefont
  {Manuel},\ and\ \citenamefont {Wildes}}]{Fennell-2007}%
  \BibitemOpen
  \bibfield  {author} {\bibinfo {author} {\bibfnamefont {T.}~\bibnamefont
  {Fennell}}, \bibinfo {author} {\bibfnamefont {S.~T.}\ \bibnamefont
  {Bramwell}}, \bibinfo {author} {\bibfnamefont {D.~F.}\ \bibnamefont
  {McMorrow}}, \bibinfo {author} {\bibfnamefont {P.}~\bibnamefont {Manuel}}, \
  and\ \bibinfo {author} {\bibfnamefont {A.~R.}\ \bibnamefont {Wildes}},\
  }\bibfield  {title} {\enquote {\bibinfo {title} {{\it Pinch Points and
  Kasteleyn Transitions in Kagome Ice}},}\ }\href
  {http://dx.doi.org/10.1038/nphys632} {\bibfield  {journal} {\bibinfo
  {journal} {Nat. Phys.}\ }\textbf {\bibinfo {volume} {3}},\ \bibinfo {pages}
  {566} (\bibinfo {year} {2007})}\BibitemShut {NoStop}%
\bibitem [{\citenamefont {Champion}(2001)}]{Champion-2001}%
  \BibitemOpen
  \bibfield  {author} {\bibinfo {author} {\bibfnamefont {J.~D.~M.}\
  \bibnamefont {Champion}},\ }\href@noop {} {\emph {\bibinfo {title} {\rm Ph.D.
  thesis}}}\ (\bibinfo  {publisher} {University of London},\ \bibinfo {year}
  {2001})\BibitemShut {NoStop}%
\bibitem [{\citenamefont {Champion}\ \emph {et~al.}(2003)\citenamefont
  {Champion}, \citenamefont {Harris}, \citenamefont {Holdsworth}, \citenamefont
  {Wills}, \citenamefont {Balakrishnan}, \citenamefont {Bramwell},
  \citenamefont {\ifmmode \check{C}\else \v{C}\fi{}i\ifmmode~\check{z}\else
  \v{z}\fi{}m\'ar}, \citenamefont {Fennell}, \citenamefont {Gardner},
  \citenamefont {Lago}, \citenamefont {McMorrow}, \citenamefont
  {Orend\'a\ifmmode~\check{c}\else \v{c}\fi{}}, \citenamefont
  {Orend\'a\ifmmode~\check{c}\else \v{c}\fi{}ov\'a}, \citenamefont {Paul},
  \citenamefont {Smith}, \citenamefont {Telling},\ and\ \citenamefont
  {Wildes}}]{Champion-2003}%
  \BibitemOpen
  \bibfield  {author} {\bibinfo {author} {\bibfnamefont {J.~D.~M.}\
  \bibnamefont {Champion}}, \bibinfo {author} {\bibfnamefont {M.~J.}\
  \bibnamefont {Harris}}, \bibinfo {author} {\bibfnamefont {P.~C.~W.}\
  \bibnamefont {Holdsworth}}, \bibinfo {author} {\bibfnamefont {A.~S.}\
  \bibnamefont {Wills}}, \bibinfo {author} {\bibfnamefont {G.}~\bibnamefont
  {Balakrishnan}}, \bibinfo {author} {\bibfnamefont {S.~T.}\ \bibnamefont
  {Bramwell}}, \bibinfo {author} {\bibfnamefont {E.}~\bibnamefont {\ifmmode
  \check{C}\else \v{C}\fi{}i\ifmmode~\check{z}\else \v{z}\fi{}m\'ar}}, \bibinfo
  {author} {\bibfnamefont {T.}~\bibnamefont {Fennell}}, \bibinfo {author}
  {\bibfnamefont {J.~S.}\ \bibnamefont {Gardner}}, \bibinfo {author}
  {\bibfnamefont {J.}~\bibnamefont {Lago}}, \bibinfo {author} {\bibfnamefont
  {D.~F.}\ \bibnamefont {McMorrow}}, \bibinfo {author} {\bibfnamefont
  {M.}~\bibnamefont {Orend\'a\ifmmode~\check{c}\else \v{c}\fi{}}}, \bibinfo
  {author} {\bibfnamefont {A.}~\bibnamefont {Orend\'a\ifmmode~\check{c}\else
  \v{c}\fi{}ov\'a}}, \bibinfo {author} {\bibfnamefont {D.~McK.}\ \bibnamefont
  {Paul}}, \bibinfo {author} {\bibfnamefont {R.~I.}\ \bibnamefont {Smith}},
  \bibinfo {author} {\bibfnamefont {M.~T.~F.}\ \bibnamefont {Telling}}, \ and\
  \bibinfo {author} {\bibfnamefont {A.}~\bibnamefont {Wildes}},\ }\bibfield
  {title} {\enquote {\bibinfo {title} {{\it
  ${\mathrm{Er}}_{2}{\mathrm{Ti}}_{2}{\mathrm{O}}_{7}:$ Evidence of quantum
  order by disorder in a frustrated antiferromagnet}},}\ }\href {\doibase
  10.1103/PhysRevB.68.020401} {\bibfield  {journal} {\bibinfo  {journal} {Phys.
  Rev. B}\ }\textbf {\bibinfo {volume} {68}},\ \bibinfo {pages} {020401}
  (\bibinfo {year} {2003})}\BibitemShut {NoStop}%
\bibitem [{\citenamefont {Champion}\ and\ \citenamefont
  {Holdsworth}(2004)}]{Champion-2004}%
  \BibitemOpen
  \bibfield  {author} {\bibinfo {author} {\bibfnamefont {J.~D.~M.}\
  \bibnamefont {Champion}}\ and\ \bibinfo {author} {\bibfnamefont {P.~C.~W.}\
  \bibnamefont {Holdsworth}},\ }\bibfield  {title} {\enquote {\bibinfo {title}
  {{\it Soft Modes in the Easy Plane Pyrochlore Antiferromagnet}},}\ }\href
  {http://stacks.iop.org/0953-8984/16/i=11/a=013} {\bibfield  {journal}
  {\bibinfo  {journal} {J. Phys. Condens. Matter}\ }\textbf {\bibinfo {volume}
  {16}},\ \bibinfo {pages} {S665} (\bibinfo {year} {2004})}\BibitemShut
  {NoStop}%
\bibitem [{\citenamefont {Zhitomirsky}\ \emph {et~al.}(2012)\citenamefont
  {Zhitomirsky}, \citenamefont {Gvozdikova}, \citenamefont {Holdsworth},\ and\
  \citenamefont {Moessner}}]{Zhitomirsky-2012}%
  \BibitemOpen
  \bibfield  {author} {\bibinfo {author} {\bibfnamefont {M.~E.}\ \bibnamefont
  {Zhitomirsky}}, \bibinfo {author} {\bibfnamefont {M.~V.}\ \bibnamefont
  {Gvozdikova}}, \bibinfo {author} {\bibfnamefont {P.~C.~W.}\ \bibnamefont
  {Holdsworth}}, \ and\ \bibinfo {author} {\bibfnamefont {R.}~\bibnamefont
  {Moessner}},\ }\bibfield  {title} {\enquote {\bibinfo {title} {{\it Quantum
  Order by Disorder and Accidental Soft Mode in
  ${\mathrm{Er}}_{2}{\mathrm{Ti}}_{2}{\mathrm{O}}_{7}$}},}\ }\href {\doibase
  10.1103/PhysRevLett.109.077204} {\bibfield  {journal} {\bibinfo  {journal}
  {Phys. Rev. Lett.}\ }\textbf {\bibinfo {volume} {109}},\ \bibinfo {pages}
  {077204} (\bibinfo {year} {2012})}\BibitemShut {NoStop}%
\bibitem [{\citenamefont {McClarty}\ \emph {et~al.}(2014)\citenamefont
  {McClarty}, \citenamefont {Stasiak},\ and\ \citenamefont
  {Gingras}}]{McClarty-2014}%
  \BibitemOpen
  \bibfield  {author} {\bibinfo {author} {\bibfnamefont {P.~A.}\ \bibnamefont
  {McClarty}}, \bibinfo {author} {\bibfnamefont {P.}~\bibnamefont {Stasiak}}, \
  and\ \bibinfo {author} {\bibfnamefont {M.~J.~P.}\ \bibnamefont {Gingras}},\
  }\bibfield  {title} {\enquote {\bibinfo {title} {{\it Order-by-Disorder in
  the $XY$ Pyrochlore Antiferromagnet}},}\ }\href {\doibase
  10.1103/PhysRevB.89.024425} {\bibfield  {journal} {\bibinfo  {journal} {Phys.
  Rev. B}\ }\textbf {\bibinfo {volume} {89}},\ \bibinfo {pages} {024425}
  (\bibinfo {year} {2014})}\BibitemShut {NoStop}%
\bibitem [{\citenamefont {Castelnovo}\ \emph {et~al.}(2012)\citenamefont
  {Castelnovo}, \citenamefont {Moessner},\ and\ \citenamefont
  {Sondhi}}]{Castelnovo-2012}%
  \BibitemOpen
  \bibfield  {author} {\bibinfo {author} {\bibfnamefont {C.}~\bibnamefont
  {Castelnovo}}, \bibinfo {author} {\bibfnamefont {R.}~\bibnamefont
  {Moessner}}, \ and\ \bibinfo {author} {\bibfnamefont {S.~L.}\ \bibnamefont
  {Sondhi}},\ }\bibfield  {title} {\enquote {\bibinfo {title} {{\it Spin Ice,
  Fractionalization, and Topological Order}},}\ }\href {\doibase
  10.1146/annurev-conmatphys-020911-125058} {\bibfield  {journal} {\bibinfo
  {journal} {Annu. Rev. Condens. Matter Phys.}\ }\textbf {\bibinfo {volume}
  {3}},\ \bibinfo {pages} {35} (\bibinfo {year} {2012})}\BibitemShut {NoStop}%
\bibitem [{\citenamefont {Yan}\ \emph {et~al.}(2017)\citenamefont {Yan},
  \citenamefont {Benton}, \citenamefont {Jaubert},\ and\ \citenamefont
  {Shannon}}]{Yan-2017}%
  \BibitemOpen
  \bibfield  {author} {\bibinfo {author} {\bibfnamefont {H.}~\bibnamefont
  {Yan}}, \bibinfo {author} {\bibfnamefont {O.}~\bibnamefont {Benton}},
  \bibinfo {author} {\bibfnamefont {L.}~\bibnamefont {Jaubert}}, \ and\
  \bibinfo {author} {\bibfnamefont {N.}~\bibnamefont {Shannon}},\ }\bibfield
  {title} {\enquote {\bibinfo {title} {{\it Theory of Multiple-Phase
  Competition in Pyrochlore Magnets with Anisotropic Exchange with Application
  to ${\mathrm{Yb}}_{2}{\mathrm{Ti}}_{2}{\mathrm{O}}_{7},
  {\mathrm{Er}}_{2}{\mathrm{Ti}}_{2}{\mathrm{O}}_{7}$, and
  ${\mathrm{Er}}_{2}{\mathrm{Sn}}_{2}{\mathrm{O}}_{7}$}},}\ }\href {\doibase
  10.1103/PhysRevB.95.094422} {\bibfield  {journal} {\bibinfo  {journal} {Phys.
  Rev. B}\ }\textbf {\bibinfo {volume} {95}},\ \bibinfo {pages} {094422}
  (\bibinfo {year} {2017})}\BibitemShut {NoStop}%
\bibitem [{\citenamefont {Benton}\ \emph {et~al.}(2012)\citenamefont {Benton},
  \citenamefont {Sikora},\ and\ \citenamefont {Shannon}}]{Benton-2012}%
  \BibitemOpen
  \bibfield  {author} {\bibinfo {author} {\bibfnamefont {O.}~\bibnamefont
  {Benton}}, \bibinfo {author} {\bibfnamefont {O.}~\bibnamefont {Sikora}}, \
  and\ \bibinfo {author} {\bibfnamefont {N.}~\bibnamefont {Shannon}},\
  }\bibfield  {title} {\enquote {\bibinfo {title} {{\it Seeing the Light:
  Experimental Signatures of Emergent Electromagnetism in a Quantum Spin
  Ice}},}\ }\href {\doibase 10.1103/PhysRevB.86.075154} {\bibfield  {journal}
  {\bibinfo  {journal} {Phys. Rev. B}\ }\textbf {\bibinfo {volume} {86}},\
  \bibinfo {pages} {075154} (\bibinfo {year} {2012})}\BibitemShut {NoStop}%
\bibitem [{\citenamefont {Benton}\ \emph {et~al.}(2016)\citenamefont {Benton},
  \citenamefont {Jaubert}, \citenamefont {Yan},\ and\ \citenamefont
  {Shannon}}]{Benton-2016}%
  \BibitemOpen
  \bibfield  {author} {\bibinfo {author} {\bibfnamefont {O.}~\bibnamefont
  {Benton}}, \bibinfo {author} {\bibfnamefont {L.~D.~C.}\ \bibnamefont
  {Jaubert}}, \bibinfo {author} {\bibfnamefont {H.}~\bibnamefont {Yan}}, \ and\
  \bibinfo {author} {\bibfnamefont {N.}~\bibnamefont {Shannon}},\ }\bibfield
  {title} {\enquote {\bibinfo {title} {{\it A Spin-Liquid with Pinch-Line
  Singularities on the Pyrochlore Lattice}},}\ }\href
  {http://dx.doi.org/10.1038/ncomms11572} {\bibfield  {journal} {\bibinfo
  {journal} {Nat. Commun.}\ }\textbf {\bibinfo {volume} {7}},\ \bibinfo {pages}
  {11572} (\bibinfo {year} {2016})}\BibitemShut {NoStop}%
\bibitem [{\citenamefont {Onoda}\ and\ \citenamefont
  {Tanaka}(2011)}]{Onoda-2011}%
  \BibitemOpen
  \bibfield  {author} {\bibinfo {author} {\bibfnamefont {S.}~\bibnamefont
  {Onoda}}\ and\ \bibinfo {author} {\bibfnamefont {Y.}~\bibnamefont {Tanaka}},\
  }\bibfield  {title} {\enquote {\bibinfo {title} {{\it Quantum Fluctuations in
  the Effective Pseudospin-$\frac{1}{2}$ Model for Magnetic Pyrochlore
  Oxides}},}\ }\href {\doibase 10.1103/PhysRevB.83.094411} {\bibfield
  {journal} {\bibinfo  {journal} {Phys. Rev. B}\ }\textbf {\bibinfo {volume}
  {83}},\ \bibinfo {pages} {094411} (\bibinfo {year} {2011})}\BibitemShut
  {NoStop}%
\bibitem [{\citenamefont {Lee}\ \emph {et~al.}(2012)\citenamefont {Lee},
  \citenamefont {Onoda},\ and\ \citenamefont {Balents}}]{Lee-2012}%
  \BibitemOpen
  \bibfield  {author} {\bibinfo {author} {\bibfnamefont {S.}~\bibnamefont
  {Lee}}, \bibinfo {author} {\bibfnamefont {S.}~\bibnamefont {Onoda}}, \ and\
  \bibinfo {author} {\bibfnamefont {L.}~\bibnamefont {Balents}},\ }\bibfield
  {title} {\enquote {\bibinfo {title} {{\it Generic Quantum Spin Ice}},}\
  }\href {\doibase 10.1103/PhysRevB.86.104412} {\bibfield  {journal} {\bibinfo
  {journal} {Phys. Rev. B}\ }\textbf {\bibinfo {volume} {86}},\ \bibinfo
  {pages} {104412} (\bibinfo {year} {2012})}\BibitemShut {NoStop}%
\bibitem [{\citenamefont {Benton}\ \emph {et~al.}(2018)\citenamefont {Benton},
  \citenamefont {Jaubert}, \citenamefont {Singh}, \citenamefont {Oitmaa},\ and\
  \citenamefont {Shannon}}]{Benton-2018}%
  \BibitemOpen
  \bibfield  {author} {\bibinfo {author} {\bibfnamefont {O.}~\bibnamefont
  {Benton}}, \bibinfo {author} {\bibfnamefont {L.~D.~C.}\ \bibnamefont
  {Jaubert}}, \bibinfo {author} {\bibfnamefont {R.~R.~P.}\ \bibnamefont
  {Singh}}, \bibinfo {author} {\bibfnamefont {J.}~\bibnamefont {Oitmaa}}, \
  and\ \bibinfo {author} {\bibfnamefont {N.}~\bibnamefont {Shannon}},\
  }\bibfield  {title} {\enquote {\bibinfo {title} {{\it Quantum Spin Ice with
  Frustrated Transverse Exchange: From a $\ensuremath{\pi}$-Flux Phase to a
  Nematic Quantum Spin Liquid}},}\ }\href {\doibase
  10.1103/PhysRevLett.121.067201} {\bibfield  {journal} {\bibinfo  {journal}
  {Phys. Rev. Lett.}\ }\textbf {\bibinfo {volume} {121}},\ \bibinfo {pages}
  {067201} (\bibinfo {year} {2018})}\BibitemShut {NoStop}%
\bibitem [{\citenamefont {Iqbal}\ \emph
  {et~al.}(2018{\natexlab{b}})\citenamefont {Iqbal}, \citenamefont {Poilblanc},
  \citenamefont {Thomale},\ and\ \citenamefont {Becca}}]{Iqbal-2018breathing}%
  \BibitemOpen
  \bibfield  {author} {\bibinfo {author} {\bibfnamefont {Y.}~\bibnamefont
  {Iqbal}}, \bibinfo {author} {\bibfnamefont {D.}~\bibnamefont {Poilblanc}},
  \bibinfo {author} {\bibfnamefont {R.}~\bibnamefont {Thomale}}, \ and\
  \bibinfo {author} {\bibfnamefont {F.}~\bibnamefont {Becca}},\ }\bibfield
  {title} {\enquote {\bibinfo {title} {{\it Persistence of the Gapless Spin
  Liquid in the Breathing Kagome Heisenberg Antiferromagnet}},}\ }\href
  {\doibase 10.1103/PhysRevB.97.115127} {\bibfield  {journal} {\bibinfo
  {journal} {Phys. Rev. B}\ }\textbf {\bibinfo {volume} {97}},\ \bibinfo
  {pages} {115127} (\bibinfo {year} {2018}{\natexlab{b}})}\BibitemShut
  {NoStop}%
\bibitem [{\citenamefont {Repellin}\ \emph {et~al.}(2017)\citenamefont
  {Repellin}, \citenamefont {He},\ and\ \citenamefont
  {Pollmann}}]{Repellin-2017}%
  \BibitemOpen
  \bibfield  {author} {\bibinfo {author} {\bibfnamefont {C.}~\bibnamefont
  {Repellin}}, \bibinfo {author} {\bibfnamefont {Y.-C.}\ \bibnamefont {He}}, \
  and\ \bibinfo {author} {\bibfnamefont {F.}~\bibnamefont {Pollmann}},\
  }\bibfield  {title} {\enquote {\bibinfo {title} {{\it Stability of the
  Spin-$\frac{1}{2}$ Kagome Ground State with Breathing Anisotropy}},}\ }\href
  {\doibase 10.1103/PhysRevB.96.205124} {\bibfield  {journal} {\bibinfo
  {journal} {Phys. Rev. B}\ }\textbf {\bibinfo {volume} {96}},\ \bibinfo
  {pages} {205124} (\bibinfo {year} {2017})}\BibitemShut {NoStop}%
\bibitem [{\citenamefont {Chen}(2017)}]{Chen-2017}%
  \BibitemOpen
  \bibfield  {author} {\bibinfo {author} {\bibfnamefont {G.}~\bibnamefont
  {Chen}},\ }\bibfield  {title} {\enquote {\bibinfo {title} {{\it Spectral
  Periodicity of the Spinon Continuum in Quantum Spin Ice}},}\ }\href {\doibase
  10.1103/PhysRevB.96.085136} {\bibfield  {journal} {\bibinfo  {journal} {Phys.
  Rev. B}\ }\textbf {\bibinfo {volume} {96}},\ \bibinfo {pages} {085136}
  (\bibinfo {year} {2017})}\BibitemShut {NoStop}%
\bibitem [{\citenamefont {Kugler}\ and\ \citenamefont {von
  Delft}(2018{\natexlab{a}})}]{Kugler-2018a}%
  \BibitemOpen
  \bibfield  {author} {\bibinfo {author} {\bibfnamefont {F.~B.}\ \bibnamefont
  {Kugler}}\ and\ \bibinfo {author} {\bibfnamefont {J.}~\bibnamefont {von
  Delft}},\ }\bibfield  {title} {\enquote {\bibinfo {title} {{\it Multiloop
  Functional Renormalization Group That Sums Up All Parquet Diagrams}},}\
  }\href {\doibase 10.1103/PhysRevLett.120.057403} {\bibfield  {journal}
  {\bibinfo  {journal} {Phys. Rev. Lett.}\ }\textbf {\bibinfo {volume} {120}},\
  \bibinfo {pages} {057403} (\bibinfo {year} {2018}{\natexlab{a}})}\BibitemShut
  {NoStop}%
\bibitem [{\citenamefont {Kugler}\ and\ \citenamefont {von
  Delft}(2018{\natexlab{b}})}]{Kugler-1998b}%
  \BibitemOpen
  \bibfield  {author} {\bibinfo {author} {\bibfnamefont {F.~B.}\ \bibnamefont
  {Kugler}}\ and\ \bibinfo {author} {\bibfnamefont {J.}~\bibnamefont {von
  Delft}},\ }\bibfield  {title} {\enquote {\bibinfo {title} {{\it Multiloop
  Functional Renormalization Group for General Models}},}\ }\href {\doibase
  10.1103/PhysRevB.97.035162} {\bibfield  {journal} {\bibinfo  {journal} {Phys.
  Rev. B}\ }\textbf {\bibinfo {volume} {97}},\ \bibinfo {pages} {035162}
  (\bibinfo {year} {2018}{\natexlab{b}})}\BibitemShut {NoStop}%
\bibitem [{\citenamefont {Kugler}\ and\ \citenamefont {von
  Delft}(2018{\natexlab{c}})}]{Kugler-2018c}%
  \BibitemOpen
  \bibfield  {author} {\bibinfo {author} {\bibfnamefont {F.~B.}\ \bibnamefont
  {Kugler}}\ and\ \bibinfo {author} {\bibfnamefont {J.}~\bibnamefont {von
  Delft}},\ }\bibfield  {title} {\enquote {\bibinfo {title} {{\it Derivation of
  exact flow equations from the self-consistent parquet relations}},}\ }\href
  {http://stacks.iop.org/1367-2630/20/i=12/a=123029} {\bibfield  {journal}
  {\bibinfo  {journal} {New J. Phys.}\ }\textbf {\bibinfo {volume} {20}},\
  \bibinfo {pages} {123029} (\bibinfo {year} {2018}{\natexlab{c}})}\BibitemShut
  {NoStop}%
\bibitem [{\citenamefont {Sklan}\ and\ \citenamefont
  {Henley}(2013)}]{Sklan-2013}%
  \BibitemOpen
  \bibfield  {author} {\bibinfo {author} {\bibfnamefont {S.~R.}\ \bibnamefont
  {Sklan}}\ and\ \bibinfo {author} {\bibfnamefont {C.~L.}\ \bibnamefont
  {Henley}},\ }\bibfield  {title} {\enquote {\bibinfo {title} {{\it Nonplanar
  Ground States of Frustrated Antiferromagnets on an Octahedral Lattice}},}\
  }\href {\doibase 10.1103/PhysRevB.88.024407} {\bibfield  {journal} {\bibinfo
  {journal} {Phys. Rev. B}\ }\textbf {\bibinfo {volume} {88}},\ \bibinfo
  {pages} {024407} (\bibinfo {year} {2013})}\BibitemShut {NoStop}%
\bibitem [{\citenamefont {Domenge}\ \emph {et~al.}(2005)\citenamefont
  {Domenge}, \citenamefont {Sindzingre}, \citenamefont {Lhuillier},\ and\
  \citenamefont {Pierre}}]{Domenge-2005}%
  \BibitemOpen
  \bibfield  {author} {\bibinfo {author} {\bibfnamefont {J.-C.}\ \bibnamefont
  {Domenge}}, \bibinfo {author} {\bibfnamefont {P.}~\bibnamefont {Sindzingre}},
  \bibinfo {author} {\bibfnamefont {C.}~\bibnamefont {Lhuillier}}, \ and\
  \bibinfo {author} {\bibfnamefont {L.}~\bibnamefont {Pierre}},\ }\bibfield
  {title} {\enquote {\bibinfo {title} {{\it Twelve Sublattice Ordered Phase in
  the ${J}_{1}$-${J}_{2}$ Model on the Kagom\'e Lattice}},}\ }\href {\doibase
  10.1103/PhysRevB.72.024433} {\bibfield  {journal} {\bibinfo  {journal} {Phys.
  Rev. B}\ }\textbf {\bibinfo {volume} {72}},\ \bibinfo {pages} {024433}
  (\bibinfo {year} {2005})}\BibitemShut {NoStop}%
\bibitem [{\citenamefont {Messio}\ \emph {et~al.}(2011)\citenamefont {Messio},
  \citenamefont {Lhuillier},\ and\ \citenamefont {Misguich}}]{Messio-2011}%
  \BibitemOpen
  \bibfield  {author} {\bibinfo {author} {\bibfnamefont {L.}~\bibnamefont
  {Messio}}, \bibinfo {author} {\bibfnamefont {C.}~\bibnamefont {Lhuillier}}, \
  and\ \bibinfo {author} {\bibfnamefont {G.}~\bibnamefont {Misguich}},\
  }\bibfield  {title} {\enquote {\bibinfo {title} {{\it Lattice Symmetries and
  Regular Magnetic Orders in Classical Frustrated Antiferromagnets}},}\ }\href
  {\doibase 10.1103/PhysRevB.83.184401} {\bibfield  {journal} {\bibinfo
  {journal} {Phys. Rev. B}\ }\textbf {\bibinfo {volume} {83}},\ \bibinfo
  {pages} {184401} (\bibinfo {year} {2011})}\BibitemShut {NoStop}%
\bibitem [{\citenamefont {Sindzingre}\ \emph {et~al.}(2009)\citenamefont
  {Sindzingre}, \citenamefont {Seabra}, \citenamefont {Shannon},\ and\
  \citenamefont {Momoi}}]{Sindzingre-2009}%
  \BibitemOpen
  \bibfield  {author} {\bibinfo {author} {\bibfnamefont {P.}~\bibnamefont
  {Sindzingre}}, \bibinfo {author} {\bibfnamefont {L.}~\bibnamefont {Seabra}},
  \bibinfo {author} {\bibfnamefont {N.}~\bibnamefont {Shannon}}, \ and\
  \bibinfo {author} {\bibfnamefont {T.}~\bibnamefont {Momoi}},\ }\bibfield
  {title} {\enquote {\bibinfo {title} {{\it Phase Diagram of the Spin-$1/2$ $J
  _1$-$J_2$-$J_3$ Heisenberg Model on the Square Lattice with Ferromagnetic
  $J_{1}$}},}\ }\href {http://stacks.iop.org/1742-6596/145/i=1/a=012048}
  {\bibfield  {journal} {\bibinfo  {journal} {J. Phys. Conf. Ser.}\ }\textbf
  {\bibinfo {volume} {145}},\ \bibinfo {pages} {012048} (\bibinfo {year}
  {2009})}\BibitemShut {NoStop}%
\bibitem [{\citenamefont {Chubukov}(1984)}]{Chubukov-1984}%
  \BibitemOpen
  \bibfield  {author} {\bibinfo {author} {\bibfnamefont {A.~V.}\ \bibnamefont
  {Chubukov}},\ }\bibfield  {title} {\enquote {\bibinfo {title} {{\it On the
  Quantum Effects in Helimagnets}},}\ }\href
  {http://stacks.iop.org/0022-3719/17/i=36/a=008} {\bibfield  {journal}
  {\bibinfo  {journal} {J. Phys. C}\ }\textbf {\bibinfo {volume} {17}},\
  \bibinfo {pages} {L991} (\bibinfo {year} {1984})}\BibitemShut {NoStop}%
\bibitem [{\citenamefont {Nagaev}(1984)}]{Nagaev-1984}%
  \BibitemOpen
  \bibfield  {author} {\bibinfo {author} {\bibfnamefont {\'E.L.}\ \bibnamefont
  {Nagaev}},\ }\bibfield  {title} {\enquote {\bibinfo {title} {{\it First-Order
  Magnetic Phase Transitions and Meta-Magnetism of Quantum Origin}},}\ }\href
  {http://www.jetpletters.ac.ru/ps/1301/article_19652.shtml} {\bibfield
  {journal} {\bibinfo  {journal} {JETP Lett.}\ }\textbf {\bibinfo {volume}
  {39}},\ \bibinfo {pages} {484} (\bibinfo {year} {1984})}\BibitemShut
  {NoStop}%
\bibitem [{\citenamefont {M\"uller}\ \emph
  {et~al.}(2017{\natexlab{a}})\citenamefont {M\"uller}, \citenamefont
  {Lohmann}, \citenamefont {Richter}, \citenamefont {Menchyshyn},\ and\
  \citenamefont {Derzhko}}]{Mueller-2017}%
  \BibitemOpen
  \bibfield  {author} {\bibinfo {author} {\bibfnamefont {P.}~\bibnamefont
  {M\"uller}}, \bibinfo {author} {\bibfnamefont {A.}~\bibnamefont {Lohmann}},
  \bibinfo {author} {\bibfnamefont {J.}~\bibnamefont {Richter}}, \bibinfo
  {author} {\bibfnamefont {O.}~\bibnamefont {Menchyshyn}}, \ and\ \bibinfo
  {author} {\bibfnamefont {O.}~\bibnamefont {Derzhko}},\ }\bibfield  {title}
  {\enquote {\bibinfo {title} {{\it Thermodynamics of the Pyrochlore Heisenberg
  Ferromagnet with Arbitrary Spin $S$}},}\ }\href {\doibase
  10.1103/PhysRevB.96.174419} {\bibfield  {journal} {\bibinfo  {journal} {Phys.
  Rev. B}\ }\textbf {\bibinfo {volume} {96}},\ \bibinfo {pages} {174419}
  (\bibinfo {year} {2017}{\natexlab{a}})}\BibitemShut {NoStop}%
\bibitem [{\citenamefont {Lohmann}\ \emph {et~al.}(2014)\citenamefont
  {Lohmann}, \citenamefont {Schmidt},\ and\ \citenamefont
  {Richter}}]{Lohmann-2014}%
  \BibitemOpen
  \bibfield  {author} {\bibinfo {author} {\bibfnamefont {A.}~\bibnamefont
  {Lohmann}}, \bibinfo {author} {\bibfnamefont {H.-J.}\ \bibnamefont
  {Schmidt}}, \ and\ \bibinfo {author} {\bibfnamefont {J.}~\bibnamefont
  {Richter}},\ }\bibfield  {title} {\enquote {\bibinfo {title} {{\it
  Tenth-Order High-Temperature Expansion for the Susceptibility and the
  Specific Heat of Spin-$s$ Heisenberg Models with Arbitrary Exchange Patterns:
  Application to Pyrochlore and Kagome Magnets}},}\ }\href {\doibase
  10.1103/PhysRevB.89.014415} {\bibfield  {journal} {\bibinfo  {journal} {Phys.
  Rev. B}\ }\textbf {\bibinfo {volume} {89}},\ \bibinfo {pages} {014415}
  (\bibinfo {year} {2014})}\BibitemShut {NoStop}%
\bibitem [{\citenamefont {Schmalfu\ss{}}\ \emph {et~al.}(2005)\citenamefont
  {Schmalfu\ss{}}, \citenamefont {Richter},\ and\ \citenamefont
  {Ihle}}]{Schmalfuss-2005}%
  \BibitemOpen
  \bibfield  {author} {\bibinfo {author} {\bibfnamefont {D.}~\bibnamefont
  {Schmalfu\ss{}}}, \bibinfo {author} {\bibfnamefont {J.}~\bibnamefont
  {Richter}}, \ and\ \bibinfo {author} {\bibfnamefont {D.}~\bibnamefont
  {Ihle}},\ }\bibfield  {title} {\enquote {\bibinfo {title} {{\it Green's
  Function Theory of Quasi-Two-Dimensional Spin-Half Heisenberg Ferromagnets:
  Stacked Square versus Stacked Kagom\'e Lattices}},}\ }\href {\doibase
  10.1103/PhysRevB.72.224405} {\bibfield  {journal} {\bibinfo  {journal} {Phys.
  Rev. B}\ }\textbf {\bibinfo {volume} {72}},\ \bibinfo {pages} {224405}
  (\bibinfo {year} {2005})}\BibitemShut {NoStop}%
\bibitem [{\citenamefont {M{\"u}ller}\ \emph {et~al.}(2015)\citenamefont
  {M{\"u}ller}, \citenamefont {Richter}, \citenamefont {Hauser},\ and\
  \citenamefont {Ihle}}]{Mueller-2015}%
  \BibitemOpen
  \bibfield  {author} {\bibinfo {author} {\bibfnamefont {P.}~\bibnamefont
  {M{\"u}ller}}, \bibinfo {author} {\bibfnamefont {J.}~\bibnamefont {Richter}},
  \bibinfo {author} {\bibfnamefont {A.}~\bibnamefont {Hauser}}, \ and\ \bibinfo
  {author} {\bibfnamefont {D.}~\bibnamefont {Ihle}},\ }\bibfield  {title}
  {\enquote {\bibinfo {title} {{\it Thermodynamics of the Frustrated
  $J_1$-$J_2$ Heisenberg Ferromagnet on the Body-Centered Cubic Lattice with
  Arbitrary Spin}},}\ }\href {\doibase 10.1140/epjb/e2015-60113-7} {\bibfield
  {journal} {\bibinfo  {journal} {Eur. Phys. J. B}\ }\textbf {\bibinfo {volume}
  {88}},\ \bibinfo {pages} {159} (\bibinfo {year} {2015})}\BibitemShut
  {NoStop}%
\bibitem [{\citenamefont {M\"uller}\ \emph
  {et~al.}(2017{\natexlab{b}})\citenamefont {M\"uller}, \citenamefont
  {Richter},\ and\ \citenamefont {Ihle}}]{Mueller-2017a}%
  \BibitemOpen
  \bibfield  {author} {\bibinfo {author} {\bibfnamefont {P.}~\bibnamefont
  {M\"uller}}, \bibinfo {author} {\bibfnamefont {J.}~\bibnamefont {Richter}}, \
  and\ \bibinfo {author} {\bibfnamefont {D.}~\bibnamefont {Ihle}},\ }\bibfield
  {title} {\enquote {\bibinfo {title} {{\it Thermodynamics of Frustrated
  Ferromagnetic Spin-$\frac{1}{2}$ Heisenberg Chains: Role of Interchain
  Coupling}},}\ }\href {\doibase 10.1103/PhysRevB.95.134407} {\bibfield
  {journal} {\bibinfo  {journal} {Phys. Rev. B}\ }\textbf {\bibinfo {volume}
  {95}},\ \bibinfo {pages} {134407} (\bibinfo {year}
  {2017}{\natexlab{b}})}\BibitemShut {NoStop}%
\bibitem [{\citenamefont {Troyer}\ \emph {et~al.}(2004)\citenamefont {Troyer},
  \citenamefont {Alet},\ and\ \citenamefont {Wessel}}]{Troyer-2004}%
  \BibitemOpen
  \bibfield  {author} {\bibinfo {author} {\bibfnamefont {Matthias}\
  \bibnamefont {Troyer}}, \bibinfo {author} {\bibfnamefont {Fabien}\
  \bibnamefont {Alet}}, \ and\ \bibinfo {author} {\bibfnamefont {Stefan}\
  \bibnamefont {Wessel}},\ }\bibfield  {title} {\enquote {\bibinfo {title}
  {{\it Histogram Methods for Quantum Systems: From Reweighting to Wang-Landau
  Sampling}},}\ }\href {\doibase 10.1590/S0103-97332004000300008} {\bibfield
  {journal} {\bibinfo  {journal} {Braz. J. Phys.}\ }\textbf {\bibinfo {volume}
  {34}},\ \bibinfo {pages} {377} (\bibinfo {year} {2004})}\BibitemShut
  {NoStop}%
\bibitem [{\citenamefont {Wessel}(2010)}]{Wessel-2010}%
  \BibitemOpen
  \bibfield  {author} {\bibinfo {author} {\bibfnamefont {S.}~\bibnamefont
  {Wessel}},\ }\bibfield  {title} {\enquote {\bibinfo {title} {{\it Critical
  Entropy of Quantum Heisenberg Magnets on Simple-Cubic Lattices}},}\ }\href
  {\doibase 10.1103/PhysRevB.81.052405} {\bibfield  {journal} {\bibinfo
  {journal} {Phys. Rev. B}\ }\textbf {\bibinfo {volume} {81}},\ \bibinfo
  {pages} {052405} (\bibinfo {year} {2010})}\BibitemShut {NoStop}%
\bibitem [{\citenamefont {Soldatov}\ \emph {et~al.}(2017)\citenamefont
  {Soldatov}, \citenamefont {Nefedev}, \citenamefont {Komura},\ and\
  \citenamefont {Okabe}}]{Soldatov-2017}%
  \BibitemOpen
  \bibfield  {author} {\bibinfo {author} {\bibfnamefont {Konstantin}\
  \bibnamefont {Soldatov}}, \bibinfo {author} {\bibfnamefont {Konstantin}\
  \bibnamefont {Nefedev}}, \bibinfo {author} {\bibfnamefont {Yukihiro}\
  \bibnamefont {Komura}}, \ and\ \bibinfo {author} {\bibfnamefont {Yutaka}\
  \bibnamefont {Okabe}},\ }\bibfield  {title} {\enquote {\bibinfo {title} {{\it
  Large-Scale Calculation of Ferromagnetic Spin Systems on the Pyrochlore
  Lattice}},}\ }\href {\doibase https://doi.org/10.1016/j.physleta.2016.12.039}
  {\bibfield  {journal} {\bibinfo  {journal} {Phys. Lett. A}\ }\textbf
  {\bibinfo {volume} {381}},\ \bibinfo {pages} {707} (\bibinfo {year}
  {2017})}\BibitemShut {NoStop}%
\bibitem [{\citenamefont {Peczak}\ \emph {et~al.}(1991)\citenamefont {Peczak},
  \citenamefont {Ferrenberg},\ and\ \citenamefont {Landau}}]{Peczak-1991}%
  \BibitemOpen
  \bibfield  {author} {\bibinfo {author} {\bibfnamefont {P.}~\bibnamefont
  {Peczak}}, \bibinfo {author} {\bibfnamefont {A.~M.}\ \bibnamefont
  {Ferrenberg}}, \ and\ \bibinfo {author} {\bibfnamefont {D.~P.}\ \bibnamefont
  {Landau}},\ }\bibfield  {title} {\enquote {\bibinfo {title} {{\it
  High-Accuracy Monte Carlo Study of the Three-Dimensional Classical Heisenberg
  Ferromagnet}},}\ }\href {\doibase 10.1103/PhysRevB.43.6087} {\bibfield
  {journal} {\bibinfo  {journal} {Phys. Rev. B}\ }\textbf {\bibinfo {volume}
  {43}},\ \bibinfo {pages} {6087} (\bibinfo {year} {1991})}\BibitemShut
  {NoStop}%
\bibitem [{\citenamefont {Chen}\ \emph {et~al.}(1993)\citenamefont {Chen},
  \citenamefont {Ferrenberg},\ and\ \citenamefont {Landau}}]{Chen-1993}%
  \BibitemOpen
  \bibfield  {author} {\bibinfo {author} {\bibfnamefont {K.}~\bibnamefont
  {Chen}}, \bibinfo {author} {\bibfnamefont {A.~M.}\ \bibnamefont
  {Ferrenberg}}, \ and\ \bibinfo {author} {\bibfnamefont {D.~P.}\ \bibnamefont
  {Landau}},\ }\bibfield  {title} {\enquote {\bibinfo {title} {{\it Static
  Critical Behavior of Three-Dimensional Classical Heisenberg Models: A
  High-Resolution Monte Carlo Study}},}\ }\href {\doibase
  10.1103/PhysRevB.48.3249} {\bibfield  {journal} {\bibinfo  {journal} {Phys.
  Rev. B}\ }\textbf {\bibinfo {volume} {48}},\ \bibinfo {pages} {3249}
  (\bibinfo {year} {1993})}\BibitemShut {NoStop}%
\bibitem [{\citenamefont {Hutak}\ \emph {et~al.}(2018)\citenamefont {Hutak},
  \citenamefont {M\"uller}, \citenamefont {Richter}, \citenamefont
  {Krokhmalskii},\ and\ \citenamefont {Derzhko}}]{Hutak-2018}%
  \BibitemOpen
  \bibfield  {author} {\bibinfo {author} {\bibfnamefont {T.}~\bibnamefont
  {Hutak}}, \bibinfo {author} {\bibfnamefont {P.}~\bibnamefont {M\"uller}},
  \bibinfo {author} {\bibfnamefont {J.}~\bibnamefont {Richter}}, \bibinfo
  {author} {\bibfnamefont {T.}~\bibnamefont {Krokhmalskii}}, \ and\ \bibinfo
  {author} {\bibfnamefont {O.}~\bibnamefont {Derzhko}},\ }\bibfield  {title}
  {\enquote {\bibinfo {title} {{\it The Spin-1/2 Heisenberg Ferromagnet on the
  Pyrochlore Lattice: A Green's Function Study}},}\ }\href {\doibase
  10.5488/CMP.21.33705} {\bibfield  {journal} {\bibinfo  {journal} {Condens.
  Matter Phys.}\ }\textbf {\bibinfo {volume} {21}},\ \bibinfo {pages} {33705}
  (\bibinfo {year} {2018})}\BibitemShut {NoStop}%
\bibitem [{\citenamefont {Tyablikov}(1967)}]{TyablikovBook}%
  \BibitemOpen
  \bibfield  {author} {\bibinfo {author} {\bibfnamefont {S.V.}\ \bibnamefont
  {Tyablikov}},\ }\href {https://www.springer.com/gp/book/9781489970916} {\emph
  {\bibinfo {title} {Methods in the Quantum Theory of Magnetism}}}\ (\bibinfo
  {publisher} {Plenum, New York},\ \bibinfo {year} {1967})\BibitemShut
  {NoStop}%
\bibitem [{\citenamefont {Curnoe}(2007)}]{Curnoe-2007}%
  \BibitemOpen
  \bibfield  {author} {\bibinfo {author} {\bibfnamefont {S.~H.}\ \bibnamefont
  {Curnoe}},\ }\bibfield  {title} {\enquote {\bibinfo {title} {{\it Quantum
  Spin Configurations in
  ${\mathrm{Tb}}_{2}{\mathrm{Ti}}_{2}{\mathrm{O}}_{7}$}},}\ }\href {\doibase
  10.1103/PhysRevB.75.212404} {\bibfield  {journal} {\bibinfo  {journal} {Phys.
  Rev. B}\ }\textbf {\bibinfo {volume} {75}},\ \bibinfo {pages} {212404}
  (\bibinfo {year} {2007})}\BibitemShut {NoStop}%
\bibitem [{\citenamefont {Ross}\ \emph {et~al.}(2011)\citenamefont {Ross},
  \citenamefont {Savary}, \citenamefont {Gaulin},\ and\ \citenamefont
  {Balents}}]{Ross-2011}%
  \BibitemOpen
  \bibfield  {author} {\bibinfo {author} {\bibfnamefont {K.~A.}\ \bibnamefont
  {Ross}}, \bibinfo {author} {\bibfnamefont {L.}~\bibnamefont {Savary}},
  \bibinfo {author} {\bibfnamefont {B.~D.}\ \bibnamefont {Gaulin}}, \ and\
  \bibinfo {author} {\bibfnamefont {L.}~\bibnamefont {Balents}},\ }\bibfield
  {title} {\enquote {\bibinfo {title} {{\it Quantum Excitations in Quantum Spin
  Ice}},}\ }\href {\doibase 10.1103/PhysRevX.1.021002} {\bibfield  {journal}
  {\bibinfo  {journal} {Phys. Rev. X}\ }\textbf {\bibinfo {volume} {1}},\
  \bibinfo {pages} {021002} (\bibinfo {year} {2011})}\BibitemShut {NoStop}%
\bibitem [{\citenamefont {Petit}\ \emph {et~al.}(2016)\citenamefont {Petit},
  \citenamefont {Lhotel}, \citenamefont {Guitteny}, \citenamefont {Florea},
  \citenamefont {Robert}, \citenamefont {Bonville}, \citenamefont {Mirebeau},
  \citenamefont {Ollivier}, \citenamefont {Mutka}, \citenamefont {Ressouche},
  \citenamefont {Decorse}, \citenamefont {Ciomaga~Hatnean},\ and\ \citenamefont
  {Balakrishnan}}]{Petit-2016}%
  \BibitemOpen
  \bibfield  {author} {\bibinfo {author} {\bibfnamefont {S.}~\bibnamefont
  {Petit}}, \bibinfo {author} {\bibfnamefont {E.}~\bibnamefont {Lhotel}},
  \bibinfo {author} {\bibfnamefont {S.}~\bibnamefont {Guitteny}}, \bibinfo
  {author} {\bibfnamefont {O.}~\bibnamefont {Florea}}, \bibinfo {author}
  {\bibfnamefont {J.}~\bibnamefont {Robert}}, \bibinfo {author} {\bibfnamefont
  {P.}~\bibnamefont {Bonville}}, \bibinfo {author} {\bibfnamefont
  {I.}~\bibnamefont {Mirebeau}}, \bibinfo {author} {\bibfnamefont
  {J.}~\bibnamefont {Ollivier}}, \bibinfo {author} {\bibfnamefont
  {H.}~\bibnamefont {Mutka}}, \bibinfo {author} {\bibfnamefont
  {E.}~\bibnamefont {Ressouche}}, \bibinfo {author} {\bibfnamefont
  {C.}~\bibnamefont {Decorse}}, \bibinfo {author} {\bibfnamefont
  {M.}~\bibnamefont {Ciomaga~Hatnean}}, \ and\ \bibinfo {author} {\bibfnamefont
  {G.}~\bibnamefont {Balakrishnan}},\ }\bibfield  {title} {\enquote {\bibinfo
  {title} {{\it Antiferroquadrupolar Correlations in the Quantum Spin Ice
  Candidate ${\mathrm{Pr}}_{2}{\mathrm{Zr}}_{2}{\mathrm{O}}_{7}$}},}\ }\href
  {\doibase 10.1103/PhysRevB.94.165153} {\bibfield  {journal} {\bibinfo
  {journal} {Phys. Rev. B}\ }\textbf {\bibinfo {volume} {94}},\ \bibinfo
  {pages} {165153} (\bibinfo {year} {2016})}\BibitemShut {NoStop}%
\bibitem [{\citenamefont {Onoda}\ and\ \citenamefont
  {Tanaka}(2010)}]{Onoda-2010}%
  \BibitemOpen
  \bibfield  {author} {\bibinfo {author} {\bibfnamefont {S.}~\bibnamefont
  {Onoda}}\ and\ \bibinfo {author} {\bibfnamefont {Y.}~\bibnamefont {Tanaka}},\
  }\bibfield  {title} {\enquote {\bibinfo {title} {{\it Quantum Melting of Spin
  Ice: Emergent Cooperative Quadrupole and Chirality}},}\ }\href {\doibase
  10.1103/PhysRevLett.105.047201} {\bibfield  {journal} {\bibinfo  {journal}
  {Phys. Rev. Lett.}\ }\textbf {\bibinfo {volume} {105}},\ \bibinfo {pages}
  {047201} (\bibinfo {year} {2010})}\BibitemShut {NoStop}%
\bibitem [{\citenamefont {Benton}(2014)}]{Benton-2014}%
  \BibitemOpen
  \bibfield  {author} {\bibinfo {author} {\bibfnamefont {O.}~\bibnamefont
  {Benton}},\ }\href@noop {} {\emph {\bibinfo {title} {\rm Ph.D. thesis}}}\
  (\bibinfo  {publisher} {University of Bristol},\ \bibinfo {year}
  {2014})\BibitemShut {NoStop}%
\bibitem [{\citenamefont {Essafi}\ \emph {et~al.}(2017)\citenamefont {Essafi},
  \citenamefont {Benton},\ and\ \citenamefont {Jaubert}}]{Essafi-2017}%
  \BibitemOpen
  \bibfield  {author} {\bibinfo {author} {\bibfnamefont {K.}~\bibnamefont
  {Essafi}}, \bibinfo {author} {\bibfnamefont {O.}~\bibnamefont {Benton}}, \
  and\ \bibinfo {author} {\bibfnamefont {L.~D.~C.}\ \bibnamefont {Jaubert}},\
  }\bibfield  {title} {\enquote {\bibinfo {title} {{\it Generic
  Nearest-Neighbor Kagome Model: XYZ and Dzyaloshinskii-Moriya Couplings with
  Comparison to the Pyrochlore-Lattice Case}},}\ }\href {\doibase
  10.1103/PhysRevB.96.205126} {\bibfield  {journal} {\bibinfo  {journal} {Phys.
  Rev. B}\ }\textbf {\bibinfo {volume} {96}},\ \bibinfo {pages} {205126}
  (\bibinfo {year} {2017})}\BibitemShut {NoStop}%
\bibitem [{\citenamefont {Hermele}\ \emph {et~al.}(2004)\citenamefont
  {Hermele}, \citenamefont {Fisher},\ and\ \citenamefont
  {Balents}}]{Hermele-2004}%
  \BibitemOpen
  \bibfield  {author} {\bibinfo {author} {\bibfnamefont {M.}~\bibnamefont
  {Hermele}}, \bibinfo {author} {\bibfnamefont {M.~P.~A.}\ \bibnamefont
  {Fisher}}, \ and\ \bibinfo {author} {\bibfnamefont {L.}~\bibnamefont
  {Balents}},\ }\bibfield  {title} {\enquote {\bibinfo {title} {{\it Pyrochlore
  Photons: The $U(1)$ spin Liquid in a $S=\frac{1}{2}$ Three-Dimensional
  Frustrated Magnet}},}\ }\href {\doibase 10.1103/PhysRevB.69.064404}
  {\bibfield  {journal} {\bibinfo  {journal} {Phys. Rev. B}\ }\textbf {\bibinfo
  {volume} {69}},\ \bibinfo {pages} {064404} (\bibinfo {year}
  {2004})}\BibitemShut {NoStop}%
\bibitem [{\citenamefont {Savary}\ and\ \citenamefont
  {Balents}(2012)}]{Savary-2012a}%
  \BibitemOpen
  \bibfield  {author} {\bibinfo {author} {\bibfnamefont {L.}~\bibnamefont
  {Savary}}\ and\ \bibinfo {author} {\bibfnamefont {L.}~\bibnamefont
  {Balents}},\ }\bibfield  {title} {\enquote {\bibinfo {title} {{\it Coulombic
  Quantum Liquids in Spin-$1/2$ Pyrochlores}},}\ }\href {\doibase
  10.1103/PhysRevLett.108.037202} {\bibfield  {journal} {\bibinfo  {journal}
  {Phys. Rev. Lett.}\ }\textbf {\bibinfo {volume} {108}},\ \bibinfo {pages}
  {037202} (\bibinfo {year} {2012})}\BibitemShut {NoStop}%
\bibitem [{\citenamefont {Savary}\ \emph {et~al.}(2012)\citenamefont {Savary},
  \citenamefont {Ross}, \citenamefont {Gaulin}, \citenamefont {Ruff},\ and\
  \citenamefont {Balents}}]{Savary-2012b}%
  \BibitemOpen
  \bibfield  {author} {\bibinfo {author} {\bibfnamefont {L.}~\bibnamefont
  {Savary}}, \bibinfo {author} {\bibfnamefont {K.~A.}\ \bibnamefont {Ross}},
  \bibinfo {author} {\bibfnamefont {B.~D.}\ \bibnamefont {Gaulin}}, \bibinfo
  {author} {\bibfnamefont {J.~P.~C.}\ \bibnamefont {Ruff}}, \ and\ \bibinfo
  {author} {\bibfnamefont {L.}~\bibnamefont {Balents}},\ }\bibfield  {title}
  {\enquote {\bibinfo {title} {{\it Order by Quantum Disorder in
  ${\mathrm{Er}}_{2}{\mathrm{Ti}}_{2}{\mathrm{O}}_{7}$}},}\ }\href {\doibase
  10.1103/PhysRevLett.109.167201} {\bibfield  {journal} {\bibinfo  {journal}
  {Phys. Rev. Lett.}\ }\textbf {\bibinfo {volume} {109}},\ \bibinfo {pages}
  {167201} (\bibinfo {year} {2012})}\BibitemShut {NoStop}%
\bibitem [{\citenamefont {Savary}\ and\ \citenamefont
  {Balents}(2013)}]{Savary-2013}%
  \BibitemOpen
  \bibfield  {author} {\bibinfo {author} {\bibfnamefont {L.}~\bibnamefont
  {Savary}}\ and\ \bibinfo {author} {\bibfnamefont {L.}~\bibnamefont
  {Balents}},\ }\bibfield  {title} {\enquote {\bibinfo {title} {{\it Spin
  Liquid Regimes at Nonzero Temperature in Quantum Spin Ice}},}\ }\href
  {\doibase 10.1103/PhysRevB.87.205130} {\bibfield  {journal} {\bibinfo
  {journal} {Phys. Rev. B}\ }\textbf {\bibinfo {volume} {87}},\ \bibinfo
  {pages} {205130} (\bibinfo {year} {2013})}\BibitemShut {NoStop}%
\bibitem [{\citenamefont {Hao}\ \emph {et~al.}(2014)\citenamefont {Hao},
  \citenamefont {Day},\ and\ \citenamefont {Gingras}}]{Hao-2014}%
  \BibitemOpen
  \bibfield  {author} {\bibinfo {author} {\bibfnamefont {Z.}~\bibnamefont
  {Hao}}, \bibinfo {author} {\bibfnamefont {A.~G.~R.}\ \bibnamefont {Day}}, \
  and\ \bibinfo {author} {\bibfnamefont {M.~J.~P.}\ \bibnamefont {Gingras}},\
  }\bibfield  {title} {\enquote {\bibinfo {title} {{\it Bosonic Many-Body
  Theory of Quantum Spin Ice}},}\ }\href {\doibase 10.1103/PhysRevB.90.214430}
  {\bibfield  {journal} {\bibinfo  {journal} {Phys. Rev. B}\ }\textbf {\bibinfo
  {volume} {90}},\ \bibinfo {pages} {214430} (\bibinfo {year}
  {2014})}\BibitemShut {NoStop}%
\bibitem [{\citenamefont {Fu}\ \emph {et~al.}(2017)\citenamefont {Fu},
  \citenamefont {Rau}, \citenamefont {Gingras},\ and\ \citenamefont
  {Perkins}}]{Fu-2017}%
  \BibitemOpen
  \bibfield  {author} {\bibinfo {author} {\bibfnamefont {J.}~\bibnamefont
  {Fu}}, \bibinfo {author} {\bibfnamefont {J.~G.}\ \bibnamefont {Rau}},
  \bibinfo {author} {\bibfnamefont {M.~J.~P.}\ \bibnamefont {Gingras}}, \ and\
  \bibinfo {author} {\bibfnamefont {N.~B.}\ \bibnamefont {Perkins}},\
  }\bibfield  {title} {\enquote {\bibinfo {title} {{\it Fingerprints of Quantum
  Spin Ice in Raman Scattering}},}\ }\href {\doibase
  10.1103/PhysRevB.96.035136} {\bibfield  {journal} {\bibinfo  {journal} {Phys.
  Rev. B}\ }\textbf {\bibinfo {volume} {96}},\ \bibinfo {pages} {035136}
  (\bibinfo {year} {2017})}\BibitemShut {NoStop}%
\bibitem [{\citenamefont {Wen}(2002)}]{Wen-2002}%
  \BibitemOpen
  \bibfield  {author} {\bibinfo {author} {\bibfnamefont {X.-G.}\ \bibnamefont
  {Wen}},\ }\bibfield  {title} {\enquote {\bibinfo {title} {{\it Quantum Orders
  and Symmetric Spin Liquids}},}\ }\href {\doibase 10.1103/PhysRevB.65.165113}
  {\bibfield  {journal} {\bibinfo  {journal} {Phys. Rev. B}\ }\textbf {\bibinfo
  {volume} {65}},\ \bibinfo {pages} {165113} (\bibinfo {year}
  {2002})}\BibitemShut {NoStop}%
\bibitem [{\citenamefont {Wen}(1991)}]{Wen-1991}%
  \BibitemOpen
  \bibfield  {author} {\bibinfo {author} {\bibfnamefont {X.~G.}\ \bibnamefont
  {Wen}},\ }\bibfield  {title} {\enquote {\bibinfo {title} {{\it Mean-Field
  Theory of Spin-Liquid States with Finite Energy Gap and Topological
  Orders}},}\ }\href {\doibase 10.1103/PhysRevB.44.2664} {\bibfield  {journal}
  {\bibinfo  {journal} {Phys. Rev. B}\ }\textbf {\bibinfo {volume} {44}},\
  \bibinfo {pages} {2664} (\bibinfo {year} {1991})}\BibitemShut {NoStop}%
\bibitem [{\citenamefont {Wen}(1990)}]{Wen-1990}%
  \BibitemOpen
  \bibfield  {author} {\bibinfo {author} {\bibfnamefont {X.~G.}\ \bibnamefont
  {Wen}},\ }\bibfield  {title} {\enquote {\bibinfo {title} {{\it Topological
  Orders in Rigid States}},}\ }\href {\doibase 10.1142/S0217979290000139}
  {\bibfield  {journal} {\bibinfo  {journal} {Int. J. Mod. Phys. B}\ }\textbf
  {\bibinfo {volume} {04}},\ \bibinfo {pages} {239} (\bibinfo {year}
  {1990})}\BibitemShut {NoStop}%
\bibitem [{\citenamefont {Bieri}\ \emph {et~al.}(2016)\citenamefont {Bieri},
  \citenamefont {Lhuillier},\ and\ \citenamefont {Messio}}]{Bieri-2016}%
  \BibitemOpen
  \bibfield  {author} {\bibinfo {author} {\bibfnamefont {S.}~\bibnamefont
  {Bieri}}, \bibinfo {author} {\bibfnamefont {C.}~\bibnamefont {Lhuillier}}, \
  and\ \bibinfo {author} {\bibfnamefont {L.}~\bibnamefont {Messio}},\
  }\bibfield  {title} {\enquote {\bibinfo {title} {{\it Projective Symmetry
  Group Classification of Chiral Spin Liquids}},}\ }\href {\doibase
  10.1103/PhysRevB.93.094437} {\bibfield  {journal} {\bibinfo  {journal} {Phys.
  Rev. B}\ }\textbf {\bibinfo {volume} {93}},\ \bibinfo {pages} {094437}
  (\bibinfo {year} {2016})}\BibitemShut {NoStop}%
\bibitem [{\citenamefont {Huang}\ \emph {et~al.}(2017)\citenamefont {Huang},
  \citenamefont {Kim},\ and\ \citenamefont {Lu}}]{Huang-2017}%
  \BibitemOpen
  \bibfield  {author} {\bibinfo {author} {\bibfnamefont {B.}~\bibnamefont
  {Huang}}, \bibinfo {author} {\bibfnamefont {Y.~B.}\ \bibnamefont {Kim}}, \
  and\ \bibinfo {author} {\bibfnamefont {Y.-M.}\ \bibnamefont {Lu}},\
  }\bibfield  {title} {\enquote {\bibinfo {title} {{\it Interplay of
  Nonsymmorphic Symmetry and Spin-Orbit Coupling in Hyperkagome Spin Liquids:
  Applications to ${\mathrm{Na}}_{4}{\mathrm{Ir}}_{3}{\mathrm{O}}_{8}$}},}\
  }\href {\doibase 10.1103/PhysRevB.95.054404} {\bibfield  {journal} {\bibinfo
  {journal} {Phys. Rev. B}\ }\textbf {\bibinfo {volume} {95}},\ \bibinfo
  {pages} {054404} (\bibinfo {year} {2017})}\BibitemShut {NoStop}%
\bibitem [{\citenamefont {Huang}\ \emph {et~al.}(2018)\citenamefont {Huang},
  \citenamefont {Choi}, \citenamefont {Kim},\ and\ \citenamefont
  {Lu}}]{Huang-2018}%
  \BibitemOpen
  \bibfield  {author} {\bibinfo {author} {\bibfnamefont {B.}~\bibnamefont
  {Huang}}, \bibinfo {author} {\bibfnamefont {W.}~\bibnamefont {Choi}},
  \bibinfo {author} {\bibfnamefont {Y.~B.}\ \bibnamefont {Kim}}, \ and\
  \bibinfo {author} {\bibfnamefont {Y.-M.}\ \bibnamefont {Lu}},\ }\bibfield
  {title} {\enquote {\bibinfo {title} {{\it Classification and Properties of
  Quantum Spin Liquids on the Hyperhoneycomb Lattice}},}\ }\href {\doibase
  10.1103/PhysRevB.97.195141} {\bibfield  {journal} {\bibinfo  {journal} {Phys.
  Rev. B}\ }\textbf {\bibinfo {volume} {97}},\ \bibinfo {pages} {195141}
  (\bibinfo {year} {2018})}\BibitemShut {NoStop}%
\bibitem [{\citenamefont {Lu}\ \emph {et~al.}(2011)\citenamefont {Lu},
  \citenamefont {Ran},\ and\ \citenamefont {Lee}}]{Lu-2011}%
  \BibitemOpen
  \bibfield  {author} {\bibinfo {author} {\bibfnamefont {Y.-M.}\ \bibnamefont
  {Lu}}, \bibinfo {author} {\bibfnamefont {Y.}~\bibnamefont {Ran}}, \ and\
  \bibinfo {author} {\bibfnamefont {P.~A.}\ \bibnamefont {Lee}},\ }\bibfield
  {title} {\enquote {\bibinfo {title} {{\it ${\mathbb{Z}}_{2}$ Spin Liquids in
  the $S=\frac{1}{2}$ Heisenberg Model on the Kagome Lattice: A Projective
  Symmetry-Group Study of Schwinger Fermion Mean-Field States}},}\ }\href
  {\doibase 10.1103/PhysRevB.83.224413} {\bibfield  {journal} {\bibinfo
  {journal} {Phys. Rev. B}\ }\textbf {\bibinfo {volume} {83}},\ \bibinfo
  {pages} {224413} (\bibinfo {year} {2011})}\BibitemShut {NoStop}%
\bibitem [{\citenamefont {Lu}(2016)}]{Lu-2016}%
  \BibitemOpen
  \bibfield  {author} {\bibinfo {author} {\bibfnamefont {Y.-M.}\ \bibnamefont
  {Lu}},\ }\bibfield  {title} {\enquote {\bibinfo {title} {{\it Symmetric
  ${Z}_{2}$ Spin Liquids and Their Neighboring Phases on Triangular
  Lattice}},}\ }\href {\doibase 10.1103/PhysRevB.93.165113} {\bibfield
  {journal} {\bibinfo  {journal} {Phys. Rev. B}\ }\textbf {\bibinfo {volume}
  {93}},\ \bibinfo {pages} {165113} (\bibinfo {year} {2016})}\BibitemShut
  {NoStop}%
\bibitem [{\citenamefont {Yunoki}\ and\ \citenamefont
  {Sorella}(2006)}]{Yunoki-2006}%
  \BibitemOpen
  \bibfield  {author} {\bibinfo {author} {\bibfnamefont {S.}~\bibnamefont
  {Yunoki}}\ and\ \bibinfo {author} {\bibfnamefont {S.}~\bibnamefont
  {Sorella}},\ }\bibfield  {title} {\enquote {\bibinfo {title} {{\it Two Spin
  Liquid Phases in the Spatially Anisotropic Triangular Heisenberg Model}},}\
  }\href {\doibase 10.1103/PhysRevB.74.014408} {\bibfield  {journal} {\bibinfo
  {journal} {Phys. Rev. B}\ }\textbf {\bibinfo {volume} {74}},\ \bibinfo
  {pages} {014408} (\bibinfo {year} {2006})}\BibitemShut {NoStop}%
\bibitem [{\citenamefont {Becca}\ and\ \citenamefont
  {Sorella}(2017)}]{BeccaBook}%
  \BibitemOpen
  \bibfield  {author} {\bibinfo {author} {\bibfnamefont {F.}~\bibnamefont
  {Becca}}\ and\ \bibinfo {author} {\bibfnamefont {S.}~\bibnamefont
  {Sorella}},\ }\href
  {https://www.cambridge.org/core/books/quantum-monte-carlo-approaches-for-correlated-systems/EB88C86BD9553A0738BDAE400D0B2900}
  {\emph {\bibinfo {title} {Quantum Monte Carlo Approaches for Correlated
  Systems}}}\ (\bibinfo  {publisher} {Cambridge University Press, Cambridge,
  England},\ \bibinfo {year} {2017})\BibitemShut {NoStop}%
\bibitem [{\citenamefont {Hu}\ \emph {et~al.}(2013)\citenamefont {Hu},
  \citenamefont {Becca}, \citenamefont {Parola},\ and\ \citenamefont
  {Sorella}}]{Hu-2013}%
  \BibitemOpen
  \bibfield  {author} {\bibinfo {author} {\bibfnamefont {W.-J.}\ \bibnamefont
  {Hu}}, \bibinfo {author} {\bibfnamefont {F.}~\bibnamefont {Becca}}, \bibinfo
  {author} {\bibfnamefont {A.}~\bibnamefont {Parola}}, \ and\ \bibinfo {author}
  {\bibfnamefont {S.}~\bibnamefont {Sorella}},\ }\bibfield  {title} {\enquote
  {\bibinfo {title} {{\it Direct Evidence for a Gapless ${Z}_{2}$ Spin Liquid
  by Frustrating N\'eel Antiferromagnetism}},}\ }\href {\doibase
  10.1103/PhysRevB.88.060402} {\bibfield  {journal} {\bibinfo  {journal} {Phys.
  Rev. B}\ }\textbf {\bibinfo {volume} {88}},\ \bibinfo {pages} {060402}
  (\bibinfo {year} {2013})}\BibitemShut {NoStop}%
\bibitem [{\citenamefont {Becca}\ \emph {et~al.}(2015)\citenamefont {Becca},
  \citenamefont {Hu}, \citenamefont {Iqbal}, \citenamefont {Parola},
  \citenamefont {Poilblanc},\ and\ \citenamefont {Sorella}}]{Iqbal-2015b}%
  \BibitemOpen
  \bibfield  {author} {\bibinfo {author} {\bibfnamefont {F.}~\bibnamefont
  {Becca}}, \bibinfo {author} {\bibfnamefont {W.-J.}\ \bibnamefont {Hu}},
  \bibinfo {author} {\bibfnamefont {Y.}~\bibnamefont {Iqbal}}, \bibinfo
  {author} {\bibfnamefont {A.}~\bibnamefont {Parola}}, \bibinfo {author}
  {\bibfnamefont {D.}~\bibnamefont {Poilblanc}}, \ and\ \bibinfo {author}
  {\bibfnamefont {S.}~\bibnamefont {Sorella}},\ }\bibfield  {title} {\enquote
  {\bibinfo {title} {{\it Lanczos Steps to Improve Variational Wave
  Functions}},}\ }\href {http://stacks.iop.org/1742-6596/640/i=1/a=012039}
  {\bibfield  {journal} {\bibinfo  {journal} {J. Phys. Conf. Ser.}\ }\textbf
  {\bibinfo {volume} {640}},\ \bibinfo {pages} {012039} (\bibinfo {year}
  {2015})}\BibitemShut {NoStop}%
\bibitem [{\citenamefont {{Hering}}\ \emph {et~al.}(2018)\citenamefont
  {{Hering}}, \citenamefont {{Sonnenschein}}, \citenamefont {{Iqbal}},\ and\
  \citenamefont {{Reuther}}}]{Hering-2018}%
  \BibitemOpen
  \bibfield  {author} {\bibinfo {author} {\bibfnamefont {M.}~\bibnamefont
  {{Hering}}}, \bibinfo {author} {\bibfnamefont {J.}~\bibnamefont
  {{Sonnenschein}}}, \bibinfo {author} {\bibfnamefont {Y.}~\bibnamefont
  {{Iqbal}}}, \ and\ \bibinfo {author} {\bibfnamefont {J.}~\bibnamefont
  {{Reuther}}},\ }\bibfield  {title} {\enquote {\bibinfo {title} {{\it
  Characterization of Quantum Spin Liquids and Their Spinon Band Structures via
  Functional Renormalization}},}\ }\href@noop {} {\bibfield  {journal}
  {\bibinfo  {journal} {ArXiv e-prints}\ } (\bibinfo {year} {2018})},\ \Eprint
  {http://arxiv.org/abs/1806.05021} {arXiv:1806.05021 [cond-mat.str-el]}
  \BibitemShut {NoStop}%
\bibitem [{\citenamefont {Krizan}\ and\ \citenamefont
  {Cava}(2015)}]{Krizan-2015}%
  \BibitemOpen
  \bibfield  {author} {\bibinfo {author} {\bibfnamefont {J.~W.}\ \bibnamefont
  {Krizan}}\ and\ \bibinfo {author} {\bibfnamefont {R.~J.}\ \bibnamefont
  {Cava}},\ }\bibfield  {title} {\enquote {\bibinfo {title} {{\it
  $\mathrm{NaCaN}{\mathrm{i}}_{2}{\mathrm{F}}_{7}$$\mathrm{:}$ A frustrated
  high-temperature pyrochlore antiferromagnet with $S~\mathrm{=1}$
  $\mathrm{Ni}^{2+}$}},}\ }\href {\doibase 10.1103/PhysRevB.92.014406}
  {\bibfield  {journal} {\bibinfo  {journal} {Phys. Rev. B}\ }\textbf {\bibinfo
  {volume} {92}},\ \bibinfo {pages} {014406} (\bibinfo {year}
  {2015})}\BibitemShut {NoStop}%
\bibitem [{\citenamefont {Plumb}\ \emph {et~al.}(2019)\citenamefont {Plumb},
  \citenamefont {Changlani}, \citenamefont {Scheie}, \citenamefont {Zhang},
  \citenamefont {Krizan}, \citenamefont {Rodriguez-Rivera}, \citenamefont
  {Qiu}, \citenamefont {Winn}, \citenamefont {Cava},\ and\ \citenamefont
  {Broholm}}]{Plumb-2017}%
  \BibitemOpen
  \bibfield  {author} {\bibinfo {author} {\bibfnamefont {K.~W.}\ \bibnamefont
  {Plumb}}, \bibinfo {author} {\bibfnamefont {Hitesh~J.}\ \bibnamefont
  {Changlani}}, \bibinfo {author} {\bibfnamefont {A.}~\bibnamefont {Scheie}},
  \bibinfo {author} {\bibfnamefont {S.}~\bibnamefont {Zhang}}, \bibinfo
  {author} {\bibfnamefont {J.~W.}\ \bibnamefont {Krizan}}, \bibinfo {author}
  {\bibfnamefont {J.~A.}\ \bibnamefont {Rodriguez-Rivera}}, \bibinfo {author}
  {\bibfnamefont {Y.}~\bibnamefont {Qiu}}, \bibinfo {author} {\bibfnamefont
  {B.}~\bibnamefont {Winn}}, \bibinfo {author} {\bibfnamefont {R.~J.}\
  \bibnamefont {Cava}}, \ and\ \bibinfo {author} {\bibfnamefont {C.~L.}\
  \bibnamefont {Broholm}},\ }\bibfield  {title} {\enquote {\bibinfo {title}
  {{\it Continuum of Quantum Fluctuations in a Three-Dimensional S
  $\mathrm{=1}$ Heisenberg Magnet}},}\ }\href {\doibase
  10.1038/s41567-018-0317-3} {\bibfield  {journal} {\bibinfo  {journal} {Nat.
  Phys.}\ }\textbf {\bibinfo {volume} {15}},\ \bibinfo {pages} {54} (\bibinfo
  {year} {2019})}\BibitemShut {NoStop}%
\bibitem [{\citenamefont {Krizan}\ and\ \citenamefont
  {Cava}(2014)}]{Krizan-2014}%
  \BibitemOpen
  \bibfield  {author} {\bibinfo {author} {\bibfnamefont {J.~W.}\ \bibnamefont
  {Krizan}}\ and\ \bibinfo {author} {\bibfnamefont {R.~J.}\ \bibnamefont
  {Cava}},\ }\bibfield  {title} {\enquote {\bibinfo {title} {{\it
  ${\mathrm{NaCaCo}}_{2}{\mathrm{F}}_{7}$$\mathrm{:}$ A Single-Crystal
  High-Temperature Pyrochlore Antiferromagnet}},}\ }\href {\doibase
  10.1103/PhysRevB.89.214401} {\bibfield  {journal} {\bibinfo  {journal} {Phys.
  Rev. B}\ }\textbf {\bibinfo {volume} {89}},\ \bibinfo {pages} {214401}
  (\bibinfo {year} {2014})}\BibitemShut {NoStop}%
\bibitem [{\citenamefont {Ross}\ \emph {et~al.}(2017)\citenamefont {Ross},
  \citenamefont {Brown}, \citenamefont {Cava}, \citenamefont {Krizan},
  \citenamefont {Nagler}, \citenamefont {Rodriguez-Rivera},\ and\ \citenamefont
  {Stone}}]{Ross-2017}%
  \BibitemOpen
  \bibfield  {author} {\bibinfo {author} {\bibfnamefont {K.~A.}\ \bibnamefont
  {Ross}}, \bibinfo {author} {\bibfnamefont {J.~M.}\ \bibnamefont {Brown}},
  \bibinfo {author} {\bibfnamefont {R.~J.}\ \bibnamefont {Cava}}, \bibinfo
  {author} {\bibfnamefont {J.~W.}\ \bibnamefont {Krizan}}, \bibinfo {author}
  {\bibfnamefont {S.~E.}\ \bibnamefont {Nagler}}, \bibinfo {author}
  {\bibfnamefont {J.~A.}\ \bibnamefont {Rodriguez-Rivera}}, \ and\ \bibinfo
  {author} {\bibfnamefont {M.~B.}\ \bibnamefont {Stone}},\ }\bibfield  {title}
  {\enquote {\bibinfo {title} {{\it Single-Ion Properties of the
  ${S}_{\mathrm{eff}}$ = $\frac{1}{2}$ XY Antiferromagnetic Pyrochlores
  $\mathrm{Na}{A}^{\ensuremath{'}}{\mathrm{Co}}_{2}{\mathrm{F}}_{7}$
  (${A}^{\ensuremath{'}}={\mathrm{Ca}}^{2+}, {\mathrm{Sr}}^{2+}$)}},}\ }\href
  {\doibase 10.1103/PhysRevB.95.144414} {\bibfield  {journal} {\bibinfo
  {journal} {Phys. Rev. B}\ }\textbf {\bibinfo {volume} {95}},\ \bibinfo
  {pages} {144414} (\bibinfo {year} {2017})}\BibitemShut {NoStop}%
\bibitem [{\citenamefont {Sarkar}\ \emph {et~al.}(2017)\citenamefont {Sarkar},
  \citenamefont {Krizan}, \citenamefont {Br\"uckner}, \citenamefont {Andrade},
  \citenamefont {Rachel}, \citenamefont {Vojta}, \citenamefont {Cava},\ and\
  \citenamefont {Klauss}}]{Sarkar-2017}%
  \BibitemOpen
  \bibfield  {author} {\bibinfo {author} {\bibfnamefont {R.}~\bibnamefont
  {Sarkar}}, \bibinfo {author} {\bibfnamefont {J.~W.}\ \bibnamefont {Krizan}},
  \bibinfo {author} {\bibfnamefont {F.}~\bibnamefont {Br\"uckner}}, \bibinfo
  {author} {\bibfnamefont {E.~C.}\ \bibnamefont {Andrade}}, \bibinfo {author}
  {\bibfnamefont {S.}~\bibnamefont {Rachel}}, \bibinfo {author} {\bibfnamefont
  {M.}~\bibnamefont {Vojta}}, \bibinfo {author} {\bibfnamefont {R.~J.}\
  \bibnamefont {Cava}}, \ and\ \bibinfo {author} {\bibfnamefont {H.-H.}\
  \bibnamefont {Klauss}},\ }\bibfield  {title} {\enquote {\bibinfo {title}
  {{\it Spin Freezing in the Disordered Pyrochlore Magnet
  ${\mathrm{NaCaCo}}_{2}{\mathrm{F}}_{7}$$\mathrm{:}$ NMR Studies and Monte
  Carlo Simulations}},}\ }\href {\doibase 10.1103/PhysRevB.96.235117}
  {\bibfield  {journal} {\bibinfo  {journal} {Phys. Rev. B}\ }\textbf {\bibinfo
  {volume} {96}},\ \bibinfo {pages} {235117} (\bibinfo {year}
  {2017})}\BibitemShut {NoStop}%
\bibitem [{\citenamefont {Sanders}\ \emph {et~al.}(2017)\citenamefont
  {Sanders}, \citenamefont {Krizan}, \citenamefont {Plumb}, \citenamefont
  {McQueen},\ and\ \citenamefont {Cava}}]{Sanders-2016}%
  \BibitemOpen
  \bibfield  {author} {\bibinfo {author} {\bibfnamefont {M.~B.}\ \bibnamefont
  {Sanders}}, \bibinfo {author} {\bibfnamefont {J.~W.}\ \bibnamefont {Krizan}},
  \bibinfo {author} {\bibfnamefont {K.~W.}\ \bibnamefont {Plumb}}, \bibinfo
  {author} {\bibfnamefont {T.~M.}\ \bibnamefont {McQueen}}, \ and\ \bibinfo
  {author} {\bibfnamefont {R.~J.}\ \bibnamefont {Cava}},\ }\bibfield  {title}
  {\enquote {\bibinfo {title} {{\it ${\mathrm{NaSrMn}}_{2}{\mathrm{F}}_{7}$,
  ${\mathrm{NaCaFe}}_{2}{\mathrm{F}}_{7}$, and
  ${\mathrm{NaSrFe}}_{2}{\mathrm{F}}_{7}$$\mathrm{:}$ Novel Single Crystal
  Pyrochlore Antiferromagnets}},}\ }\href
  {http://stacks.iop.org/0953-8984/29/i=4/a=045801} {\bibfield  {journal}
  {\bibinfo  {journal} {J. Phys. Condens. Matter}\ }\textbf {\bibinfo {volume}
  {29}},\ \bibinfo {pages} {045801} (\bibinfo {year} {2017})}\BibitemShut
  {NoStop}%
\bibitem [{\citenamefont {Andrade}\ \emph {et~al.}(2018)\citenamefont
  {Andrade}, \citenamefont {Hoyos}, \citenamefont {Rachel},\ and\ \citenamefont
  {Vojta}}]{Andrade-2017}%
  \BibitemOpen
  \bibfield  {author} {\bibinfo {author} {\bibfnamefont {E.~C.}\ \bibnamefont
  {Andrade}}, \bibinfo {author} {\bibfnamefont {J.~A.}\ \bibnamefont {Hoyos}},
  \bibinfo {author} {\bibfnamefont {S.}~\bibnamefont {Rachel}}, \ and\ \bibinfo
  {author} {\bibfnamefont {M.}~\bibnamefont {Vojta}},\ }\bibfield  {title}
  {\enquote {\bibinfo {title} {{\it Cluster-Glass Phase in Pyrochlore $XY$
  Antiferromagnets with Quenched Disorder}},}\ }\href {\doibase
  10.1103/PhysRevLett.120.097204} {\bibfield  {journal} {\bibinfo  {journal}
  {Phys. Rev. Lett.}\ }\textbf {\bibinfo {volume} {120}},\ \bibinfo {pages}
  {097204} (\bibinfo {year} {2018})}\BibitemShut {NoStop}%
\end{thebibliography}

%

\end{document}